\documentclass[12pt,twoside]{article}
\usepackage{amssymb}
\usepackage{amsmath}
\usepackage{mathabx}
\usepackage{latexsym}
\usepackage{longtable}
\usepackage{epsfig}
\usepackage{graphicx,bbm,psfrag}
\graphicspath{{images/}}
\usepackage{cite}
\usepackage{etoolbox}
\apptocmd{\thebibliography}{\raggedright}{}{}

\DeclareMathAlphabet{\mathsfit}{T1}{\sfdefault}{\mddefault}{\sldefault}
\SetMathAlphabet{\mathsfit}{bold}{T1}{\sfdefault}{\bfdefault}{\sldefault}

\begin{document}
\begin{titlepage}
\vspace*{5cm}
\begin{centering}
\Large{\textbf{Hydrodynamic Scales of Integrable Many-Particle Systems}}
\bigskip\bigskip\bigskip
\end{centering}
\begin{center} 
{\large{Herbert Spohn}}\bigskip\bigskip\\
Departments of Mathematics and Physics, Technical University Munich,\smallskip\\
Boltzmannstr. 3, 85747 Garching, Germany
\end{center}
\vspace{3cm}
\begin{flushright}
December 6, 2023
\end{flushright}
\vspace{6cm}
\begin{center}
\textit{To my grandson Lio Spohn for his continuing support}
\end{center}
\end{titlepage}

\tableofcontents
\newpage
\noindent
{\large\textbf{Preface}}\\\\
In August 1974 for the first time I attended  a summer school, which happened to be number three in the series on
``Fundamental Problems in Statistical Mechanics". The school took place at the  Agricultural University of Wageningen.
We were approximately 70 participants from 25 countries. The school lasted nearly three weeks with 11 lecture courses, each four hours long, and various more specialized seminars. As to be expected, present were the big shots, as Eddy Cohen who with moderate success tried to slow down the speaker through posing questions. Besides lectures, the truly exciting part of the school was to meet fellow youngsters who had similar interests and struggled more or less  with the  same difficulties. 
    
At the time, critical phenomena and the just invented RG methods were the overwhelming topic. Fortunately the Dutch physics community has a long tradition in Statistical Mechanics and therefore a wide range of topics were covered. I vividly recall the lectures by Nico van Kampen
on ``Stochastic differential equations''. Joe Ford lectured on ``The statistical mechanics of classical analytic dynamics", the early days of deterministic chaos. In fact, at the end of his lectures Joe mentioned the Toda lattice describing the just confirmed integrability.
Perhaps I should have had listened with more care. For me the strongest impact had the lectures presented by  Piet Kasteleyn
on ``Exactly solvable lattice models", explaining the fascinating link between equilibrium statistical mechanics and models from quantum many-body physics. 

My second encounter with integrable systems is related to the study of Dyson Brownian motion, which is an integrable stochastic particle system. At the time,  to me the model  was an intriguing example for the hydrodynamics of a many-particle system with long range forces. The third encounter was triggered by the KPZ revolution, which brought me in contact with other corners of integrable systems.
Around 2016 I first learned about the activities investigating the hydrodynamic scales for integrable quantum many-body systems. I could not resist.
Of course, major insights had been accomplished already. But, apparently, classical integrable  many-particle systems were in a state of dormancy. This is how my enterprise got started.

During the ongoing project, I had many insightful comments and good advice. Gratefully acknowledged are   Mark Adler, Amol Aggarwal, Vir Bulchandani, Xiangyu Cao,   Kedar Damle, Avijit Das, Percy Deift,  
Jacopo De Nardis, Atharv Deokule,  Abhishek Dhar, Maurizio Fagotti,  Pablo Ferrari, Patrik Ferrari, Chiara Franceschini,
Tamara Grava, Alice Guionnet,  David Huse, Thomas Kappeler, Karol Kozlowski, Thomas Kriecherbauer, Manas Kulkarni,  Anupam Kundu, Aritra Kundu, 
Gaultier Lambert, Joel Lebowitz, Guido Mazzuca, Ken McLaughlin, Christian Mendl, Pierre van Moerbecke, Joel Moore, Fumihiko Nakano, Neil O'Connell,
Stefano Olla, Lorenzo Piroli, Bal\'{a}zs Pozsgay, 
Michael Pr\"{a}hofer, Sylvain Prolhac, Tomaz Prosen, Keiji Saito, Makiko Sasada, Tomohiro Sasamoto, 
Naoto Shiraishi, J\"{o}rg Teschner,  Khan Duy Trinh,
Simone Warzel, and Takato Yoshimura.      

Special thanks are due to Benjamin Doyon. Our encounter at Pont-\`{a}-Mousson is well remembered. 

When working on the manuscript I had the opportunity to stay for extended time spans at the Mathematical Research Institute at Berkeley,
the Galileo Galilei Institute at Firenze, the Newton Institute at Cambridge, and the International Center for Theoretical Sciences at Bengaluru, in chronological order. I am most thankful for such generous invitations. 
\newpage
\section{Overview}
\label{sec1}
Hydrodynamics is based on the observation that the motion of a large assembly of strongly interacting particles is constrained by local conservation laws.
As a result, local equilibrium is established over an initial time span to be followed by a much longer time window when  local equilibrium parameters are governed by the hydrodynamic evolution equations. The initial time span could shrink to microscopic times when the 
system starts out already in local equilibrium. It is a matter of fact that a vast amount of interesting physics is covered by the hydrodynamic approach. 
Historically the best known example are simple fluids for which hydrodynamics is synonymous with fluid dynamics. 

Already for simple fluids the hydrodynamic approach carries the seed for further extensions, since the equilibrium phase diagram is richly structured. Most common is the occurrence of a liquid-gas
phase transition. This discrete order parameter has now to be added as a further parameter characterizing local equilibrium.  For example, gas and fluid phase may spatially coexist and
the respective interface is then an additional slow degree of freedom, to be included in the macroscopic dynamics.   

Close to critical points, conventional
hydrodynamics has to be augmented by more refined theories. At lower temperatures, generically a solid 
phase stabilizes. Due to slow relaxation of solids, for the dynamics of the solid-gas interface mostly non-hydrodynamic modelling is used.
Bosonic particles at low temperatures will form a condensate. One then employs a hydrodynamic two-fluid model, which governs the 
superfluid interacting with the normal fluid. Going beyond short-range interactions, magneto-hydrodynamics describes the motion of fluids made up of charged particles,
also including the Maxwell field as additional dynamical degrees of freedom. Relativistic hydrodynamics becomes relevant for extreme events such as
 the formation of superdense neutron stars, relativistic jets, and Gamma ray bursts. Each topic mentioned is part of a vast enterprise with ongoing research.
 
Approximately  seven years ago a novel item was added to our list under the name of \textit{generalized hydrodynamics} (GHD). Physically perhaps not as far reaching as other areas mentioned, generalized hydrodynamics relies on an amazing twist. The novel topic is concerned with integrable many-particle models
 for which the number of conserved fields is proportional to system size, in sharp contrast to the models listed before which have 
 only a few conserved fields, for example number, momentum, energy, plus broken symmetries in case of simple fluids. At first glance the mere idea of a hydrodynamic description of the time evolution of such an integrable system sounds like an intrinsic contradiction.  After all, establishing local equilibrium relies on chaotic dynamics which is just the opposite of integrability.
 But the huge number of degrees of freedom helps. Since integrable many-particle systems have  an extensive number of  local conservation laws,
 local equilibrium must now be characterized by a correspondingly large number of chemical potentials. It is this feature which is called ``generalized". In the limit of infinite system size, the hydrodynamic 
 fields are labelled by a parameter taking integer values, $n = 0,1,\ldots$, or possibly by more complicated labelling schemes. As a consequence, writing down the coupled set of hyperbolic conservation laws is already a major obstacle. 
 
 The notion of \textit{generalized Gibbs ensemble} (GGE) was introduced earlier and studied systematically in a related context, known as quantum quench. But the issue is generic. One starts from a spatially homogeneous random state and wants to identify the random state reached  after a long time. More physically, one prepares a homogeneous state of a particular hamiltonian dynamics and then abruptly changes the dynamics (the quench). If the quench dynamics is not integrable, generically one expects the system to thermalize with parameters
determined by the conserved fields when averaged over the initial state. But an integrable system has many conserved fields and the final state will  depend on an extensive set of parameters. Such asymptotic states are called GGE. 

For many-particle systems to be integrable requires fine-tuned interactions. Nevertheless the list is not so short.
 The first and still much studied model is the Lieb-Liniger $\delta$-Bose gas from 1963. The Toda lattice was discovered in 1967 and its integrability being firmly established seven years later through the construction of a Lax matrix. Further examples are the XXZ spin chain, the one-dimensional spin-$\tfrac{1}{2}$ Fermi-Hubbard model,
 the classical particle models of Calogero, and continuum wave equations as Korteweg-de Vries, nonlinear 
 Schr\"{o}dinger, and sinh-Gordon.
In fact, the central goal of our notes is to argue that
\begin{center} \textsf{On a hydrodynamic scale  all integrable many-particle systems are structurally alike}.
\end{center}
 Given the diversity of microscopic models such a claim is surprisingly bold. On the other hand, as to be discussed, the route to tackle the hydrodynamic scale  will depend on the specific model. The precise meaning of our claim will unfold. But to provide at least a very preliminary glimpse, in all models under study the two-body scattering shift will be a crucial piece of  the hydrodynamic description.
 
As familiar from fluids, Euler equations refer to the ballistic scale, which is characterized by space and time to be of same
order of magnitude. Formally, entropy is locally conserved. Transport properties arise at longer diffusive time scales and are included in the hydrodynamic equations through the 
Navier-Stokes correction. For integrable systems the same distinctions apply, at least in principle. This topic will be briefly touched upon in the very last chapter of our notes. Otherwise, hydrodynamics is understood as ballistic Euler type scaling.   
 
 From a broader perspective, in deriving the equations governing the motion on the hydrodynamic scale one faces several difficulties.\medskip\\
 (i) For a given system, the local conservation laws have to be listed in terms of which  GGEs can be constructed. 
 In a somewhat vague sense, this list has to be complete, since hydrodynamically all fields are expected to be coupled to each other.\smallskip\\
 (ii) The structure of the generalized free energy has to be understood, including its first order derivatives which are linked to the GGE averaged conserved fields.\smallskip\\
 (iii) To complete the hydrodynamic equations, one has to know the GGE averaged currents as a functional of the GGE averaged conserved fields.\medskip
     
My exposition is \textit{not} a review, even though much of the relevant literature will be cited. 
To establish a  guiding backbone, the classical Toda lattice is discussed in considerable detail. 
Particularly introduced are two distinct strategies (1) a closed system with a linearly varying pressure 
and (2) the canonical transformation to scattering coordinates. The first method will also be applied to the 
Ablowitz-Ladik discretized nonlinear Schr\"{o}dinger equation and the second one serves well for the Calogero fluid. 
The key quantum models accounted for will be the Lieb-Liniger $\delta$-Bose gas and the quantized
Toda lattice, for both models relying on the Bethe ansatz as strategy.

Let me refrain from further comments on the content of my notes and rather turn to some remarks on the history of the subject.
The  one-dimensional system of classical hard rods was studied around 1970. The system is integrable, since in a collision momenta are merely exchanged.
Conserved is any one-particle sum function depending only on  the momenta. Due to the hard core the hydrodynamic fields are nonlinearly coupled, in sharp contrast to an ideal gas. Jerry Percus first derived the hydrodynamic equations. In the 1980ies Roland Dobrushin and collaborators analyzed in much greater detail the time evolution of hard rods. He well understood the hydrodynamic 
 perspective, including the issue of Navier-Stokes corrections. But at the time no tools were available for handling more intricate models. In retrospect, the true simplification of hard rods is a two-particle scattering shift which is independent of the  incoming quasiparticle velocities.  
 
 Another early line of research concerns the Korteweg-de Vries equation which is an integrable nonlinear wave equation accessible through the inverse scattering transform. In the mid 1990ies Vladimir E. Zhakarov studied a low density gas of solitons and derived the respective kinetic equation for the spacetime dependence of the soliton counting function.
 The extension to a dense soliton gas, to say the respective hydrodynamic equations, has been obtained by Gennady El in 2003.
 
Generalized hydrodynamics as a systematic research activity relies on a breakthrough advance in 2016 independently by the two groups:
B. Bertini, M. Collura, J. De Nardis, M. Fagotti and O.A. Castro-Alvaredo, B. Doyon, T. Yoshimura.
They discovered a general scheme of how to write down the average currents, thereby covering classical field theories and  quantum many-body systems.  Only with such an input the equations of generalized hydrodynamics could be written with confidence, herewith opening the door to applications of physical interest. Such detailed studies strongly support our claim that on a hydrodynamic scale all integrable many-particle models look alike. 

In condensed matter physics the notion quantum many-body is widely used. Many-particle system 
is more natural for models from classical mechanics. Both notions are employed interchangeably and comprise also 
classical and quantum field theories.  \bigskip\\
\textit{How the text is structured}. Our material is arranged in 15 Chapters with sections varying in number. Longer subsections might be separated by boldface headers. No quotations are provided in the main text. Instead, at the end of each chapter, one finds "Notes and references", which roughly means a bibliography with extended comments. In addition, there are Inserts separated from the main text by $\blackdiamond\hspace{-1pt}\blackdiamond$~\textit{Header}. $\cdots$ $\blackdiamond\hspace{-1pt}\blackdiamond$. 
Typically an Insert deals with a closely related topic, which however can be touched upon only superficially. Also some more technical derivations have been shifted to Inserts. The idea is that at first reading an Insert can be skipped,
except for conventions on notation.
\bigskip
 \begin{center}
 \textbf{Notes and references}
 \end{center}
 \begin{center}
  \textbf{Overview}
\end{center} 
 The Proceedings of the Wageningen summer school have been edited by E.D.G. Cohen  \cite{C75}. My work on Dyson Brownian motion is published in Spohn \cite{S87}. The KPZ revolution is covered in many articles from which only the overviews Corwin \cite{C12}, Quastel and  Spohn \cite{QS15}, Spohn \cite{S17} and Takeuchi \cite{T17} are quoted.\bigskip
 \newpage
 \begin{center}
\textbf{Section 1}
\end{center} 
 A useful account on the developments prior to generalized hydrodynamics can be found in a special volume on  ``Quantum Integrability in Out-of-Equilibrium Systems'' edited by Calabrese et al. \cite{CEM16}. Its central theme are quantum quenches starting from a spatially homogeneous initial state.
 
 The notion ``scattering shift" is convenient but less widely used. It refers to the fact that, when two particles undergo a scattering motion,  each trajectory is asymptotically of the form $v_j t +\phi_j^\pm$, $j=1,2$, as $t \to \pm\infty$, i.e. free motion linear in time and on top a constant displacement as first order correction, which is the scattering shift. In quantum mechanical two-body scattering,  the wave function for the relative motion is asymptotically of the form $\exp\big[\mathrm{i}(-kx +\tfrac{1}{2}k^2t + \theta(k))\big]$. Then a narrow wave packet, centered at  
$k_0$ in momentum space, travels in physical space with velocity   $k_0$ and is displaced by $\theta'(k_0)$. $\theta$ is the phase shift, while its derivative,  $\theta'$, is the scattering shift.
 
 Percus \cite{P69} wrote down what he called a kinetic equation. In his prior work, in collaboration with Lebowitz and Sykes \cite{LPS68}, the exact spacetime two-point function of hard rods in thermal equilibrium was obtained.  Boldrighini, Dobrushin, and Suhov \cite{BDS83}
 prove, under fairly general assumptions on the initial probability measure, the validity of the Euler equation for a system of hard rods under ballistic scaling.
 They also established that the hydrodynamic equation has smooth solutions. Dobrushin  \cite{D89} is his vision on hydrodynamic limits. In the context of the Korteweg-de Vries equation, Zakharov \cite{Z71} studied a low density gas of solitons with Poisson distributed centers and statistically independent soliton velocities. He argued for a Boltzmann type kinetic equation. The hydrodynamic extension to a dense soliton fluid has been accomplished in El \cite{E03}, El and  Kamchatnov \cite{EK05}, see also
 El et al.  \cite{EKPZ11}, Carbone  et al. \cite{CDE16}. Soliton-based hydrodynamics will be explained in Chapter 10 with more references to be added.
 The upswing of generalized hydrodynamics can be traced back to the two independent seminal contributions Castro-Alvaredo et al. \cite{CDY16} and Bertini  et al.\cite{BCDF16}, in which a general scheme for the computation of GGE averaged currents is presented, GGE being the standard acronym for ``generalized Gibbs ensemble''.  Thus Euler type equations could be written down based on  convincing theoretical reasoning.
 \newpage
 \section{Dynamics of the classical Toda lattice}
\label{sec2}
\setcounter{equation}{0} 
As well known at the time when Morikazu Toda started his studies on the lattice with exponential interactions, shallow water waves in long channels have peculiar dynamical properties. One observes solitary waves and soliton collisions. In the latter, two incoming solitons  emerge with  their original shape after an intricate dynamical process. The Korteweg-de Vries (KdV) equation, a one-dimensional nonlinear wave equation, provides an accurate theoretical description of these phenomena. In 1967 Toda investigated whether discrete wave equations also might have solitary type dynamics. A point in case is the standard wave equation and the harmonic lattice as its discretization, both of them linear equations. With ingenious insight Toda discovered that a lattice with exponential interactions, later known as Toda lattice, exhibits the same dynamical features as the KdV equation. It took another seven years until it had been firmly established that the $N$-particle Toda lattice is indeed integrable with $N+1$ conservation laws.

In dimensionless form, the hamiltonian of the Toda lattice reads 
\begin{equation}\label{2.1}
H_\mathrm{to} = \sum_{j \in \mathbb{Z}}\big( \tfrac{1}{2}p_j^2 + \mathrm{e}^{-(q_{j+1} - q_j)}\big),
\end{equation}
where $j \in \mathbb{Z}$ is the particle label and $(q_j,p_j)$ are position and momentum of the $j$-th particle. 
One could introduce a particle mass, $m$, coupling strength, $g$, and decay parameter, $\gamma$,  as $(p_j^2/2m) + g \exp(-\gamma(q_{j+1}- q_j))$ with $m,g,\gamma >0$.  But through rescaling spacetime 
the standard form \eqref{2.1} is recovered. The Toda chain has no free parameters. Following Newton the equations of motion are
\begin{equation}\label{2.2}
\frac{\mathrm{d}}{\mathrm{d}t}q_j = p_j,\quad \frac{\mathrm{d}}{\mathrm{d}t}p_j =   \mathrm{e}^{-(q_{j} - q_{j-1})} -\mathrm{e}^{-(q_{j+1} - q_j)}, \quad j \in \mathbb{Z}.
\end{equation}
Physically this equation can be viewed in two different ways. (i) The displacements $q_j(t), j \in \mathbb{Z}$, are regarded as the lattice discretization
of a continuum wave field $q(x,t)$, $x \in \mathbb{R}$. We call this the lattice, or field theory, picture. (ii) The fluid picture is to literally
view $q_j(t), j = 1,\ldots,N$, as positions of particles moving on the real line. However they do not interact  pairwise as would be the case for a real fluid. The hamiltonian is not invariant under relabelling of particles. Toda was mostly thinking of a lattice discretization. But the fluid picture is easier to visualize. Of course, there is only a single set of equations of motion and 
one can switch back and forth between the two options.

At this stage, a review of the vast research on the Toda lattice can neither be supplied nor is it intended. We would have to refer to original research articles, monographs, reviews, and textbooks. 
However, it can be safely summarized that almost exclusively problems have been studied for which physically the chain is at zero temperature.  Examples are multi-soliton solutions and the spatial spreading of local perturbations of an initially  periodic particle configuration. In contrast, our focus are random initial 
data with an energy proportional to system size and away from a ground state energy. A paradigmatic set-up would be the thermal state at
some non-zero temperature. For such an enterprise novel techniques are required. In particular, the issue of large system size has to be properly understood.
 \begin{figure}[!t]
\centering
\includegraphics[width=1.00\columnwidth]{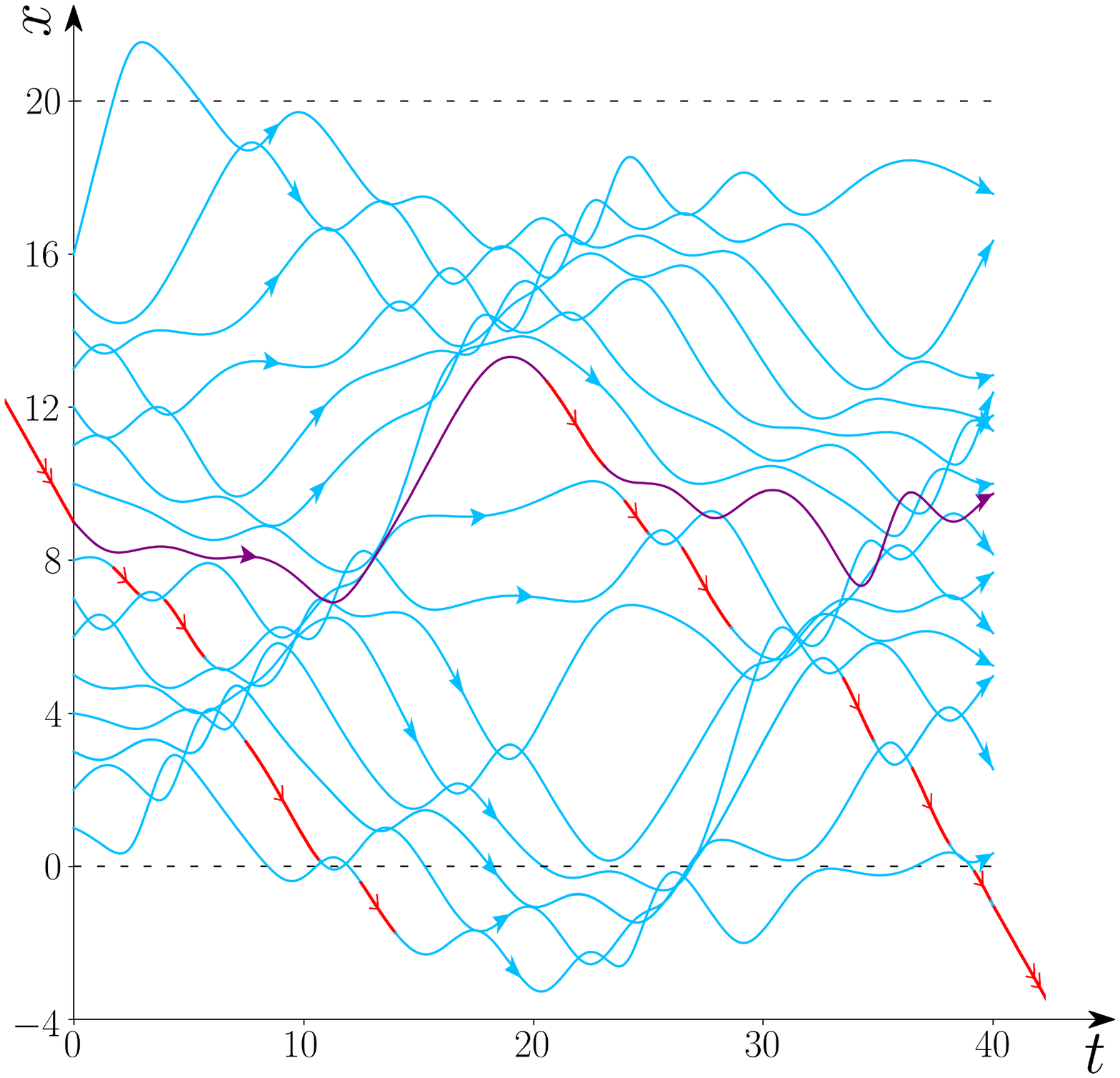}
\caption{Displayed are the trajectories, $q_j(t)$, of 16 Toda particles governed by $H_{\mathrm{cell},16}$ in a box of size $\ell = 20$. Very roughly the stretch $\nu = 1.25$ and inverse temperature $\beta = 1$, which corresponds to  medium pressure $P= 0.7$. Clearly visible are two-particle collisions, but also more complicated intertwined structures develop. The trajectory  of the 9-th particle is shown in magenta. Quasiparticles are introduced by the condition to 
approximately maintain their velocity, except for collisions. One choice for a quasiparticle trajectory with the same starting point as the 9-th particle is shown in red.   
From \cite{S20}.}
 \label{fig1}
 \end{figure}
 
 As illustration in Figure 1 we show a numerical simulation of the particle dynamics and note that 
 the widely used visualization as a sequence of two-body collision seems to be of limited value.

\subsection{Locally conserved fields and their currents}\label{sec2.1}
Our first task is to elucidate the integrable structure of the Toda lattice. For this purpose we introduce the \textit{stretch}
\begin{equation}\label{2.3}
r_j = q_{j+1} - q_j,
\end{equation}
also the free distance, or free volume, between particles $j$ and $j+1$, and the \textit{Flaschka variables}
\begin{equation}\label{2.4}
a_j = \mathrm{e}^{-r_j/2},\quad b_j =  p_j.
\end{equation}
The stretch can have either sign, while $a_j >0$. The $a$'s and $b$'s are conventional notation, but we will avoid the duplication of symbols  by using only the momentum $p_j$.  In terms of these variables the equations of motion read
\begin{equation}\label{2.5}
\frac{d}{dt} a_j = \tfrac{1}{2}a_j(p_j - p_{j+1}),\quad \frac{d}{dt} p_j = a_{j-1}^2 - a_{j}^2.
\end{equation}
Hence $a_j$ couples to the right neighbor and $p_j$ to the left one. In principle, we could have set $a_j = \tau  \mathrm{e}^{-r_j/2}$ and $b_j = \tau p_j $, which amounts to a mere time change. Flaschka picked $\tau = \tfrac{1}{2}$.
Below we will explain why $\tau = 1$ is singled out in our context.

For the purpose of thermodynamics, one first considers the lattice $[1,\ldots,N]$ with periodic boundary conditions. 
In Flaschka variables they amount to the obvious condition 
\begin{equation}\label{2.6}
a_{0} = a_N,\quad  p_{N+1} = p_1.
\end{equation}
This choice is also called the \textit{closed} or \textit{periodic} chain, for which the phase space is $\Gamma_{\!N}^\circ  = (\mathbb{R}_+ \times \mathbb{R})^N$. Note that
\begin{equation}\label{2.7} 
\frac{d}{dt}\sum_{j=1}^N \log a_j^2 = 0,
\end{equation}
since carrying out the time derivative yields a telescoping sum which vanishes by \eqref{2.6}. Hence
\begin{equation}\label{2.8} 
\sum_{j=1}^N r_j(t) = \ell, 
\end{equation}
with some constant $\ell$, which can have either sign. Under this constraint, an equivalent description is to consider the infinite Toda chain and to impose the initial conditions
\begin{equation}
\label{2.9} 
q_{j+N}= q_j +\ell, \quad p_{j+N} = p_j,
\end{equation}
 for some $\ell \in \mathbb{R}$ and all $j \in \mathbb{Z}$, which then holds at any time. Cutting the real line into cells, each of size $\ell$, in every cell there are $N$ particles and their dynamics is governed by the \textit{cell hamiltonian}
 \begin{equation}\label{2.10}
H_{\mathrm{cell},N} = \sum_{j=1}^N \tfrac{1}{2}p_j^2 + \sum_{j=1}^{N-1}\mathrm{e}^{-(q_{j+1} - q_j)}+ \mathrm{e}^{-(\ell - (q_N - q_1))}.
\end{equation}
 The positions are unconstrained. Total momentum is conserved and, under the dynamics generated by $H_{\mathrm{cell},N} $, 
 \begin{equation}\label{2.11}
\sum_{j=1}^N q_j(t) = \sum_{j=1}^N q_j(0) + t \sum_{j=1}^Np_j(0).
\end{equation}
 The center of mass moves with constant velocity.  The internal degrees of freedom are the stretches $r_1(t),\ldots,r_N(t)$ subject to the time-independent constraint \eqref{2.8}.
The stretches move in a potential which increases exponentially  in all directions.

In the literature, periodic boundary conditions are often stated as
$q_{N+1} = q_1$, which corresponds to the special case $\ell = 0$. Physically the parameter $\ell$ is of crucial importance because through it the stretch per particle, $\ell/N$, is controlled.  In the fluid picture
the unit cell has length $\ell$ and contains $N$ particles. The physical particle density is $N/|\ell|$. But often it is more natural to work with the signed particle density
$N/\ell$, which can be negative.

Out of the Flaschka variables one forms the tridiagonal \textit{Lax matrix}, $L_N$,
\begin{equation}\label{2.12} 
L_N = 
\begin{pmatrix}
p_1 & a_1&0  &\cdots&a_N\\
a_1 & p_2 & a_2 & \ddots&0\\
0&a_2& p_3&\ddots&\vdots\\
\vdots&\ddots &\ddots&\ddots&a_{N-1}\\
a_N&0&\cdots &a_{N-1}&p_N\\
\end{pmatrix}, 
\end{equation}
for $N\geq 3$, and its partner matrix 
\begin{equation}\label{2.12a} 
\qquad B_N = \frac{1}{2}
\begin{pmatrix}
0 & -a_1& 0 &\cdots& a_N\\
a_1 & 0 &-a_2&\ddots  &0\\
0&a_2 & 0&\ddots&\vdots\\
\vdots&\ddots&\ddots& \ddots&-a_{N-1}\\
- a_N&0&\cdots &a_{N-1}&0\\
\end{pmatrix}.
\end{equation}
$L_N,B_N$ is called a \textit{Lax pair}.  While $L$ is the common notation for a Lax matrix, the partner matrix is also denoted by $A$ and $M$.
Here we follow the convention of Toda Toda \cite{T89}. Clearly, $L_N$ is symmetric and $B_N$  skew symmetric, $(L_N)^\mathrm{T} = L_N$, $(B_N)^\mathrm{T} = - B_N$, with $^\mathrm{T}$ denoting the transpose of a matrix.  Later on for the adjoint of an operator also the more common $^*$ will be used.
From the equations of motion \eqref{2.5}, one verifies that
\begin{equation}\label{2.13} 
\frac{d}{dt} L_N = [B_N,L_N]
\end{equation}
with $[\cdot,\cdot]$ denoting the commutator, $[B_N,L_N] = B_NL_N -L_NB_N$. Since $B_N$ is skew symmetric, $L_N(t)$ is isospectral to $ L_N =  L_N(0)$. Thus the eigenvalues of $L_N$ are conserved. Actually any matrix $B_N$ would do, because
\begin{equation}\label{2.14} 
\frac{d}{dt} \mathrm{tr}\big[(L_N)^n\big] = \sum_{j=0}^{n-1}\mathrm{tr}\big[(L_N)^j[B_N,L_N](L_N)^{n-1-j}\big] =\mathrm{tr}\big[B_N,(L_N)^n\big] = 0
\end{equation}
for all $n$.
\bigskip\\
$\blackdiamond\hspace{-1pt}\blackdiamond$~\textit{Vector notation}.\hspace{1pt} In our text various $N$-vectors will appear. The standard notation $x\in \mathbb{R}^N$, $x = (x_1,\ldots,x_N)$, is adopted. The  $N$-dimensional volume element is denoted by $\mathrm{d}^N\hspace{-1pt}x$. Also $x\in \mathbb{R}$ will be used. The distinction should be obvious from the context. \hfill$\blackdiamond\hspace{-1pt}\blackdiamond$\bigskip\\
$\blackdiamond\hspace{-1pt}\blackdiamond$~\textit{Notation for time-dependence}. For time-dependent quantities,
as $X_j(t)$, we use the convention $X_j(0) = X_j$ and refer to $X_j$ as time-zero field. While this notation is convenient,
it might be ambiguous. For example in \eqref{2.2} we should have used $q_j(t)$ with initial condition $q_j$.  We anticipate that the exact meaning will be clear from the context.\hfill$\blackdiamond\hspace{-1pt}\blackdiamond$\bigskip\\
$\blackdiamond\hspace{-1pt}\blackdiamond$~\textit{Phase spaces}. It is recommendable to keep track of phase spaces. We will use $\Gamma$ as a generic symbol and $\Gamma_n$ to indicate a phase space of dimension $2n$. More specifically the notation is 
$\Gamma_{\!N} =  \mathbb{R}^N \times \mathbb{R}^N$,  $\Gamma_{\!N}^\circ =\mathbb{R}_+^N \times \mathbb{R}^N$, $\Gamma_{\!N}^\diamond = \mathbb{R}_+^{(N-1)} \times \mathbb{R}^N$,
    and $\Gamma_{\!N}^\triangleright =  \mathbb{W}_N\times \mathbb{R}^N$
with Weyl chamber $\mathbb{W}_N = \{q_1 < \ldots<q_N\}$.
 \hfill$\blackdiamond\hspace{-1pt}\blackdiamond$\\

Let us write the eigenvalue problem for $L_N$ as 
\begin{equation}\label{2.15} 
L_N \psi_\alpha = \lambda_\alpha \psi_\alpha
\end{equation}
with $\alpha = 1,\dots,N$. Then 
$\lambda_\alpha(a,p)$ is some function on phase space which  does not change under the Toda time evolution. 
However $\lambda_\alpha$ is a highly nonlocal function, in general. For example, considering  the dependence of $\lambda_\alpha$
only on $p_1$
and $p_{N/2}$, this will not split into a sum as $g(p_1) + \tilde{g}(p_{N/2})$ even approximately. Physically more relevant are \textit{local 
conservation laws}. For the Toda lattice they are easily obtained through forming the trace as
\begin{equation}\label{2.16} 
Q^{[n],N} = \mathrm{tr}\big[(L_N)^n\big] = \sum_{j=1}^N  (\lambda_j)^n = \sum_{j=1}^N ((L_N)^n)_{j,j}
= \sum_{j=1}^N Q^{[n],N}_j.
\end{equation}
The second identity confirms that $Q^{[n],N}$ is conserved and the fourth identity that the density $Q^{[n],N}_j$ is local. Indeed, taking $n < N/2 $
and expanding as
\begin{equation}\label{2.17}  
Q^{[n],N}_j = \sum_{j_1=1}^N\dots \sum_{j_{n-1}=1}^N (L_N)_{j,j_1}(L_N)_{j_1,j_2}\dots (L_N)_{j_{n-1},j},
\end{equation}
the density $Q^{[n],N}_j$ depends only on the variables $\{a_{j-i},p_{j-i},\ldots,a_{j+i},p_{j+i},i = 0,\ldots,n-1\}$, modulo $N$. 

The two-sided infinite volume limit of $L_N$ is the tridiagonal Lax matrix $L$, and correspondingly the partner matrix $B$, which are now operators acting on the Hilbert space $\ell_2(\mathbb{Z})$ of square-integrable two-sided sequences over the lattice $\mathbb{Z}$. The Lax pair still satisfies
\begin{equation}\label{2.18} 
\frac{d}{dt} L = [B,L].
\end{equation}
Of course $\mathrm{tr}[L^n]$ makes no sense, literally. But the infinite volume density  
\begin{equation}\label{2.19} 
Q^{[n]}_j = L_{j,j} 
\end{equation}
is well defined.  $Q^{[n]}_j$ is a finite polynomial in the variables $\{a_i,p_i, |i - j| \leq n-1\}$, which can be grasped more easily  by using the random walk expansion derived from the $N=\infty$ version of \eqref{2.17}. The walk is on $\mathbb{Z}$ with $n$ steps of step size $0,\pm1$, starting and ending at $j$.  A step from $i $ to $i$ carries the variable $p_i$ and the step  either from $i$ to $i+1$ or from $i+1$ to $i$ carries the variable $a_i$. For a given walk one forms the product along the path, which is a monomial of degree $n$, compare with Figure 2. $Q^{[n]}_j$ is then obtained by summing over all admissible walks.
By translation invariance of the model, $Q^{[n]}_j$ and $Q^{[n]}_{j+i}$ are identical polynomials, except that the particle label is shifted by $i$.  Just as for the hamiltonian \eqref{2.1}, formally we still write
\begin{equation}\label{2.20} 
Q^{[n]} =\sum_{j\in\mathbb{Z}}Q^{[n]}_j,  
\end{equation}
 where the superscript $[n]$ ranges over positive integers. 
\begin{figure}[!t]
\centering
\includegraphics[width=8cm]{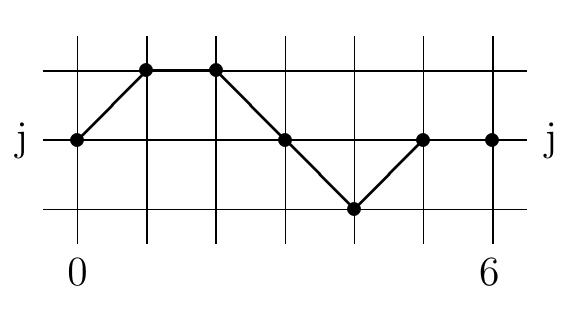}
\caption{A random walk path contributing to $Q^{[6]}_j$ and, according to the rules, carrying the weight  $a_jp_{j+1}a_ja_{j-1}a_{j-1} p_j = a_{j-1}^2a_{j}^2 p_jp_{j+1}$.}
\label{fig2}
\end{figure}
 
 As remarked already, 
in addition the stretch is locally conserved, which carries the label $0$, 
 \begin{equation}\label{2.21} 
Q^{[0]}_j = r_j, \qquad Q^{[0],N} = \sum_{j=1}^N r_j.
\end{equation}
The three lowest order fields have an immediate physical interpretation as stretch, momentum, and energy density,
 \begin{equation}\label{2.22} 
Q^{[0]}_j = r_j, \quad Q^{[1]}_j = p_j, \quad \tfrac{1}{2} Q^{[2]}_j = \tfrac{1}{2}\big(p_j^2 + a_j^2 + a_{j-1}^2\big).
\end{equation}
 Obviously, there cannot be physical names for higher orders. In the community of quantum integrable systems, $Q^{[n]}$ is called the 
 $n$-th \textit{conserved charge} or merely the $n$-th \textit{charge}, which serves as a concise notion but carries no specific physical meaning. We will use local conserved field and local charge interchangeably.
 
 Since the Toda hamiltonian has a local energy density, any locally conserved field must satisfy a continuity equation,
 in other words the lattice version of a local conservation law. We consider the infinite lattice and compute
 \begin{equation}\label{2.23}
\frac{d}{dt} Q_{j}^{[n]} = (BL^n - L^nB)_{j,j} =  a_{j-1}(L^n)_{j,j-1} - a_j (L^n)_{j+1,j} = J_{j}^{[n]} - J_{j+1}^{[n]}.
\end{equation} 
Hence $ J_{j+1}^{[n]}$ is the current of the $n$-th conserved field from $j$ to  $j+1$ and $ J_{j}^{[n]}$ the current from 
$j-1$ to  $j$. Defining the lower triangular matrix $L^{\scriptscriptstyle \downarrow}$ by $(L^{\scriptscriptstyle \downarrow})_{j+1,j}=a_j$ for all $j$ 
and $(L^{\scriptscriptstyle \downarrow})_{i,j} = 0$ otherwise, a more concise expression is
\begin{equation}\label{2.24}
J_{j}^{[n]}  = (L^nL^{\scriptscriptstyle \downarrow})_{j,j},
\end{equation} 
$n=1,2,\ldots\,$.
For the stretch
\begin{equation}\label{2.25}
\frac{d}{dt} Q_{j}^{[0]}  = -p_j +p_{j+1}, \quad J_{j}^{[0]}  = - p_j,
\end{equation} 
and hence
\begin{equation}\label{2.26}
J_{j}^{[0]}  = - Q_{j}^{[1]}.
\end{equation} 
This innocent looking equation will have surprising consequences.\bigskip\\
$\blackdiamond\hspace{-1pt}\blackdiamond$~\textit{Ambiguity of densities}.\hspace{1pt} In the way just presented, the densities, both for field and current, seem to be unique. This is not the case, however. We have made a particular choice which will be useful when investigating the hydrodynamic scale of the Toda lattice. 
A further common scheme is to require the density to depend on a minimal number of lattice sites.  In terms of the random walk representation described above, this amounts to a density  $\tilde{Q}_j^{[n],N}$ defined by  summing over all closed  admissible paths with minimum equal to $j$.
As an example,  for $n =4$ the minimal version of the density is given by
\begin{equation}\label{2.27}
\tilde{Q}_{j}^{[4]} = p_j^4 + 4 (p_j^2 +p_jp_{j+1} +p_{j+1}^2)a_{j}^2  +2 a_j^4 +4a_j^2a_{j+1}^2.
\end{equation} 

Rather than trying to dwell on generalities, we illustrate the issue by considering the energy density, $n=2$. From \eqref{2.17} we have
 \begin{equation}\label{2.28}
Q_{j}^{[2]} = p_j^2 +a_{j-1}^2+ a_j^2
\end{equation} 
with the current density
\begin{equation}\label{2.29}
J_{j}^{[2]}  = a_{j-1}^2(p_{j-1} +p_{j}).
\end{equation} 
The minimal version of the energy density would be
\begin{equation}\label{2.30}
\tilde{Q}_{j}^{[2]} = p_j^2 + 2 a_j^2,
\end{equation} 
having the current density
\begin{equation}\label{2.31}
\tilde{J}_{j}^{[2]}  = 2 a_{j-1}^2 p_{j}.
\end{equation} 
At infinite volume, we consider the spatial sums $\sum_{j=1}^N Q_{j}^{[2]} = Q^{[2],N\infty}$ and $\sum_{j=1}^N \tilde{Q}_{j}^{[2]} = 
\tilde{Q}^{[2],N\infty}$. They  differ only by a boundary term and  hence the spatial average
$N^{-1}\big( Q^{[2],N\infty}- \tilde{Q}^{[2],N\infty}\big) = \mathcal{O}(1/N)$. On the other hand, the corresponding total currents, $J^{[2],N\infty}$ and $\tilde{J}^{[2],N\infty}$, differ by $\mathcal{O}(N)$, but
\begin{equation}\label{2.32}
J_{j}^{[2]} - \tilde{J}_{j}^{[2]} = \frac{d}{dt}(a_{j-1})^2.
\end{equation} 
The difference is a total time derivative and thus vanishes when averaged over a time-stationary probability measure,
e.g. thermal average. As conclusion, while there is ambiguity on the microscopic scale, upon averaging over
large spacetime cells this amounts to only small surface type correction terms. In particular, the hydrodynamic equations for the Toda lattice  do not depend on the particular choice of microscopic densities.  \hfill$\blackdiamond\hspace{-1pt}\blackdiamond$
\subsection{Action-angle variables, notions of integrability}\label{sec2.2}
\subsubsection{Conventional integrability of the Toda lattice}
\label{sec2.2.1}
 A  cornerstone of hamiltonian dynamics is the abstract characterization of integrable systems. 
Just to recall, given is some hamiltonian, $H$, on a phase space $\Gamma_n$ of dimension $2n$. The dynamics generated by $H$ is called integrable, 
if there are $n$ differentiable functions, , the \textit{action variables}, on phase space which have the following properties: (i) They are conserved, which means that the Poisson brackets $\{I_j,H\} = 0$.
(ii) They are in involution, i.e. $\{I_i,I_j\} =0$ for all $i,j=1,\ldots,n$. (iii)  $I_1,\ldots,I_n$ span a $n$-dimensional hyper-surface in $\Gamma_n$, which is compact and connected. In particularly, no scattering orbits are permitted. The hypersurface is assumed to be invariant under the hamiltonian flow generated by $I_j$, for all $j$. So to speak, as regards to the dynamics generated by $I_j$, the hypersurface has is no boundary. Then the Arnold-Liouville theorem states that
there exists a canonical transformation to action variables $(I_1,\ldots,I_n)$ and the canonically conjugate \textit{angle variables} 
$(\vartheta_1,\ldots,\vartheta_n)\in \mathbb{T}^n$,  $\mathbb{T}^n$ the $n$-dimensional torus, such that the transformed hamiltonian $\tilde{H}$ depends only on $I$.
In these variables the dynamics trivializes as
\begin{equation}\label{2.33}
\frac{d}{dt}\vartheta_j = \omega_j, \qquad \omega_j = \partial_{I_j} \tilde{H}(I)
\end{equation} 
for $j = 1,\ldots,n$.
The angle $\vartheta_j$ moves on the unit torus with frequency $\omega_j$. 

This characterization applies also to the closed Toda chain. 
We consider a lattice of $N$ sites, the phase space $\Gamma_{\!N}^\circ = (\mathbb{R}_+ \times \mathbb{R})^N$, and  the evolution \eqref{2.5} in terms of the Flaschka variables $a= (a_1,\ldots,a_N)$ and momenta $p =(p_1,\ldots,p_N)$.
These are not canonical variables. However, instead of the usual Poisson bracket, one can introduce a nonstandard Poisson bracket 
by first defining the $N\times N$ matrix 
\begin{equation}\label{2.34}
\qquad A_N = \frac{1}{2}
\begin{pmatrix}
-a_1& 0&\cdots  &0&a_N\\
a_1 & -a_2 &0  & \ddots&0\\
0 &a_2& -a_3&\ddots& \vdots\\
\vdots&\ddots&\ddots &\ddots&0\\
0 &\cdots&0 &a_{N-1}&-a_N
\end{pmatrix}
\end{equation}
and adjusting  the Poisson bracket to
\begin{equation}\label{2.35}
\{f,g\} = \langle \nabla _p f, A_N \nabla_a g\rangle  - \langle \nabla _p g, A_N \nabla_a f\rangle
\end{equation} 
with $\langle\cdot,\cdot \rangle$ denoting the inner product in $\mathbb{R}^N$. For the usual Poisson bracket, $A_N$ would 
be the identity matrix. In the Flaschka variables
\begin{equation}\label{2.36}
H_{\mathrm{to},N} = \sum_{j =1}^N\big( \tfrac{1}{2}p_j^2 + a_j^2\big)
\end{equation}
and the equations of motion  \eqref{2.5} can be written in hamiltonian form as 
\begin{equation}\label{2.37}
\frac{d}{dt} p_j = \{p_j,H_{\mathrm{to},N} \}, \qquad \frac{d}{dt} a_j = \{a_j,H_{\mathrm{to},N} \}.
\end{equation}

The matrix $A_N$ is of rank $N-1$, the oblique projection for the eigenvalue $0$ being
\begin{equation}\label{2.38}
 |a_1^{-1},a_2^{-1},\ldots,a_N^{-1}\rangle \langle 1,1,\ldots,1|,
\end{equation}
i.e. the corresponding left eigenvector of $A_N$ equals  $-2\nabla_a Q^{[0],N}$ and the right one $\nabla_p Q^{[1],N}$. 
 The Poisson structure \eqref{2.35} is degenerate. But it can be turned  non-degenerate  simply by fixing
the two conservation laws as $Q^{[0],N} = c_0$, $Q^{[1],N} = c_1$ with an arbitrary choice of the real parameters $c_0,c_1$. The new phase space becomes $\Gamma_{\!{N-1}}^\circ$ and the dynamical evolution equations  involve only  the variables
$(a_1,\ldots,a_{N-1}, p_1,\ldots,p_{N-1})$, which are a hamiltonian system with a non-degenerate Poisson bracket structure. 
The phase space for the action-angle variables can be chosen as $\Gamma_{N-1}$ 
with coordinates $(x_1,\ldots,x_{N-1}, y_1,\ldots,y_{N-1})$ and corresponding action-angle variables $(\vartheta,I)$, with $I_j \in \mathbb{R}_+, \vartheta_j \in [0,2\pi]$, $j = 1,\ldots,N-1$, as
\begin{equation}\label{2.39}
 x_j = \sqrt{I_j} \cos \vartheta_j,\quad   y_j = \sqrt{I_j} \sin \vartheta_j, 
 \end{equation}
which are known as global Birkhoff coordinates.
 As proved in 2008 by A. Henrici and T. Kappeler, there is a canonical transformation $\Phi: \Gamma_{\!{N-1}}^\circ \to  \Gamma_{N-1}
 $ such that the transformed hamiltonian,  $H_\mathrm{aa}$, depends only on the action variables, $H_\mathrm{aa} = H_\mathrm{aa}(I_1,\ldots,I_{N-1})$. 
In fact, as for us crucial property, $H_\mathrm{aa}$ is a strictly convex, real-analytic function. This means that the phases $\omega_j = \partial_{I_{j}}H_\mathrm{aa}$ are incommensurate Lebesgue almost surely. In other words, $H_\mathrm{aa}$ has no linear pieces, as would be the case for a system of harmonic oscillators. The Toda lattice is 
 phase-mixing:  starting with some probability measure on $\Gamma_{N-1}^\circ$ with  a continuous density function,  in the long time limit the density
 will become uniform on almost every torus of dimension $N-1$ with an amplitude computed from the initial density. 
 The ``almost every'' is required because of tori with commensurate frequencies, which however form a set of Lebesgue measure zero.\\\\
  $\blackdiamond\hspace{-1pt}\blackdiamond$~\textit{Pitfalls of classical integrability}. We consider particles on the real line governed by the standard hamiltonian 
  \begin{equation}\label{2.40}
H_{\mathrm{mec},N} = \sum_{j = 1}^N\tfrac{1}{2}p_j^2 + \sum_{i < j = 1}^N V_\mathrm{mec}(q_i - q_j).
\end{equation}
The mechanical interaction potential is assumed to be even, $V_\mathrm{mec}(x) = V_\mathrm{mec}(-x)$, and repulsive, $V_\mathrm{mec}'(x) <0$ for
$x >0$.  Furthermore the potential  decays at infinity such that $|V_\mathrm{mec}(x)| < x^{-\gamma}$ for large $x$ with some  $\gamma > 1$  and diverges at the origin, thereby ensuring that particles 
do not cross. Hence the phase space equals  $\Gamma_{\!N}^\triangleright $. Under such assumptions it is proved that asymptotic momenta exist, 
 \begin{equation}\label{2.41}
\lim_{t \to \pm\infty} p_j(t) = p_j^\pm,
\end{equation}
and also the respective scattering shifts 
 \begin{equation}\label{2.42}
\lim_{t \to \pm\infty} q_j(t)- p_j^{\pm }t = \phi_j^{\pm}, \qquad \phi_j^{\pm} \in \mathbb{R},
\end{equation}
 as functions of the initial conditions. This defines the scattering map $\Phi^{-1}: (q,p)   \mapsto (p^{\pm}, \phi^{\pm})$. The asymptotic 
 momenta are ordered as $p_1^{+} < \ldots<  p_N^{+}$ and the scattering map is one-to-one on  $\Gamma_{\!N}^\triangleright $.
 The scattering map is canonical, which implies that Poisson brackets are conserved. In particular,
  \begin{equation}\label{2.43}
\{p_i^{+},p_j^{+}\} = 0
\end{equation}
and 
\begin{equation}\label{2.44}
H_{\mathrm{mec},N}\circ\Phi = \sum_{j=1}^N \tfrac{1}{2} (p_j^{+})^2.
\end{equation}
For the past asymptotic momenta are anti-ordered and the corresponding properties hold as well.

Clearly, the family $\{p_j^{+}, j =1,\ldots,N\}$ is in involution and conserved. As premature reaction
the mechanical system could be classified as integrable. However, the Arnold-Liouville theorem  does not apply. Instead of quasi-periodic motion on tori, according to \eqref{2.44}, in action-angle variables the dynamics trivializes as $\phi^+_j + p_j^+t$. 
For periodic boundary conditions integrability can be tested through molecular dynamics simulations. 
For a generic choice of $V_\mathrm{mec}$ the particle system will thermalize in the long time limit. 
Such dynamics is very different from the one of the Toda fluid on a ring. $N$-body integrability requires fine-tuning of the interaction potential.  
One option would be to ask for a Lax matrix. Then, under our conditions, it is known that  the only choices are $V_\mathrm{mec}(x) = \sinh^{-2}(x)$ and  $V_\mathrm{mec}(x) = x^{-2}$,  see Chapter \ref{sec11} for more details. But this option is ad hoc without link to standard definitions. 
In the following subsection we will argue that for hydrodynamic purposes the natural defining property is a quasilocal density for the conserved fields. 
\hfill$\blackdiamond\hspace{-1pt}\blackdiamond$
 \subsubsection{Hydrodynamic perspective on integrability} 
 \label{sec2.2.2}
 The Toda lattice is a very peculiar dynamical system in the sense that it is integrable for every system size $N$, which we call  \textit{integrable many-particle} or   \textit{integrable many-body}. Now the large $N$ limit is in focus and from a physics perspective the conventional definition of integrability might have to be reconsidered. This is even more urgent, since the naive extension of classical integrability to quantum systems  fails. A further desideratum would be a notion referring directly to the infinite lattice. For hydrodynamics the central building blocks are local conservation laws, to be more precise the hamiltonian and the conservation laws are constructed from a strictly local density.  
We thus propose to call an infinitely extended system \textit{nonintegrable}, if it admits only a few strictly local conservation laws. The system is called \textit{integrable}, if it possesses an infinite number of linearly independent local conservation laws. 
 
Starting from a strictly local density supported on an interval of $m$ sites, the property to be the density of a conservation law refers to a phase space of dimension $2(n+\ell)$, in case the hamiltonian density is supported   
 on $\ell$ sites. The condition of being in involution is no longer mentioned, but it seems to hold in concrete examples, possibly after first adjusting either the classical phase space or the Poisson bracket. 
Note that local conservation laws have a linear structure, in the sense that the sum of two local conservation laws is again a local conservation law. 

For the Toda lattice the Lax matrix is tridiagonal implying strictly local conservation laws. Another integrable model for 
a fluid is the Calogero system with a $1/\sinh^2$ pair potential. This potential decays exponentially and  the  Lax matrix is 
fully occupied. Hence the conserved fields of the Calogero fluid cannot be strictly local. Physically, the hydrodynamic scale has to  include \textit{quasilocal} fields with densities having exponential tails. However, depending on the model, the precise borderline could be a subtle issue.  \bigskip\\
$\blackdiamond\hspace{-1pt}\blackdiamond$~\textit{Toda local conservation laws}.\hspace{1pt} The Toda chain is integrable in 
the hydrodynamic sense with densities of the locally conserved fields stated in \eqref{2.19} and \eqref{2.21}. However, as a stronger property, one would like to establish that there are no further local conservation laws.  Currently this is a conjecture and more studies would be needed. Still, a precise formulation is worthwhile. As before, periodic boundary conditions are understood.  We assume some general density function $f$ 
of support  $\kappa$, in other words $f(a_1,p_1,\ldots,a_\kappa,p_\kappa)$.  Then the shifted densities are $f_j(a,p)= f(a_{j+1},p_{j+1},\ldots,a_{j +\kappa }, p_{j+\kappa})$, $f_0 = f$, and the 
conditions for being a local conservation law read
\begin{equation}\label{2.45}
Q^{[f],N} = \sum_{j=1}^N f_j, \qquad  \{Q^{[f],N}, H_{\mathrm{to},N}\} = \sum_{j=1}^N\{ f_j, H_{\mathrm{to},N}\} =0
\end{equation}
for $N > 2\kappa$. If so, each of the Poisson brackets is a local function. By translation invariance the sum has to be telescoping and thus necessarily there exists a current function, $J_j$, of support of size $\kappa +2$, such that
$\{ f_j, H_{\mathrm{to},N}\} = J_j - J_{j+1}$. 
According to the already proven conventional integrability there must be some function, $G$, such that 
\begin{equation}\label{2.46}
Q^{[f],N} = G\big( Q^{[0],N} ,\ldots, Q^{[N],N} \big).
\end{equation}
Our conjecture claims that $G$ is necessarily  \medskip  linear.\\
\textit{Conjecture}: For fixed $\kappa$ and $N$ sufficiently large, there exists coefficients $c_0,c_1,\ldots,c_\kappa$ such that 
\begin{equation}\label{2.47}
Q^{[f],N} =  \sum_{m=0}^\kappa c_m Q^{[m],N} . 
\end{equation}
One argument in favor of the conjecture comes from a simple observation. Consider some locally conserved field, $Q^{[n],N}$.
Then $(Q^{[n],N})^2$ is also conserved, but no longer local. The condition of locality should be strong enough to force a linear function in \eqref{2.46}. \bigskip\hfill$\blackdiamond\hspace{-1pt}\blackdiamond$

The reader might find the evidence for linking integrability and conservation laws not particularly convincing. Agreed, but it should be noted that our definition translates one-to-one to quantum spin chains, and also continuum quantum models. In the former case there are two concrete results strongly supporting the hydrodynamic notion of integrability. Considered is the XYZ spin chain with couplings
$J_x,J_y,J_z$ and external magnetic field, $h$, pointing in the $z$-direction. In terms of 
the Pauli spin-$\tfrac{1}{2}$ matrices, $\sigma^x,\sigma^y, \sigma^z$, the hamiltonian reads
\begin{equation}\label{2.48}
H_\mathrm{XYZ} = \sum_{j \in \mathbb{Z}}\big( J_x  \sigma_j^x \sigma_{j+1}^x +  J_y\sigma^y_j \sigma_{j+1}^y + J_z\sigma^z_j \sigma^z_{j+1}  - h \sigma_j^z\big).
\end{equation}
This model is integrable for $h=0$ and for $h\neq 0$ in case $J_x = J_y$, the XXZ model with external field. The model is expected to be non-integrable for any other choice of parameters.
Now, for the case of non-zero coupling constants and $h\neq 0$, $J_x \neq J_y$, it is proved that there is only a single local conservation law, 
namely the hamiltonian $H_\mathrm{XYZ}$ itself. This is, so to speak, the fully chaotic case. Only energy is transported and
the chain thermalizes, in the sense that expectations of local observables converge to the thermal average in the long time limit.

On the other hand for $h = 0$ and nonvanishing couplings the model is integrable. Most effectively the strictly local conserved charges  are computed  through the boost operator. The second result states that, if the  coupling constants are non-vanishing, then any local conservation law is a finite linear combination of the already known conservation laws. This is the precise analogue of our conjecture. However, around 2014, for the XXZ chain with magnetic field it was discovered that our notion of locality is indeed too restrictive. 
The XXZ chain possesses in addition quasilocal charges  which have to be included in a hydrodynamic description, see Notes for details.

In passing we note that the XXX chain at $h=0$ is integrable and all three spin components, $\sigma_j^x, \sigma_j^y,\sigma_j^z$,
are conserved. However they do not commute with each other.  There is still a tower
of local charges, $Q_2, Q_3,\ldots$ in the usual notation, which commute with each other and  with each spin-component. Strictly speaking the involution property is violated.

 Our discussion raises some difficult issues. From the available evidence, there is a dichotomy, either a few conservation laws or 
infinitely many. One does not know whether there is a deep reason behind or merely reflects the limited class of models studied.
Personally I believe in the first option. In this context, 
particularly intriguing are nearly integrable systems. For finite $N$, the KAM theorem provides information on the stability of the invariant tori, confirming the coexistence of chaotic and integrable regions in phase space. But in our definition the limit $N \to \infty$ is taken first
and  integrable regions might be rare.  
 \subsection{Scattering theory}\label{sec2.3}
In hydrodynamics the system is confined and  thus interactions  persist without interruption, which in the long time limit then leads to
some sort of statistical equilibrium.  A dynamically distinct  set-up is scattering: in the distant past particles are in the incoming configuration, for which they are far apart and do not interact. A time span of multiple collision processes follows. In the far future the outgoing particles move 
freely  again. If the interaction potential is repulsive, no bound states can be formed. To illustrate the special features of scattering for a integrable many-body  systems, we first discuss a fluid consisting of hard rods, which will serve as an instructive example also later on.   
\subsubsection{Hard rod fluid}
\label{sec2.3.1}
We consider $N$ hard rods, rod length $\mathsfit{a} >0$, moving on the real line. The hamiltonian reads
\begin{equation}\label{2.49}
H_{\mathrm{hr},N}^\diamond = \sum_{j = 1}^N  \tfrac{1}{2}p_j^2 + \sum_{j = 1}^{N-1} V_\mathrm{hr}(q_{j+1} - q_j),
\end{equation}
with the hard rod potential $V_\mathrm{hr}(x) = \infty$ for $|x| < \mathsfit{a}/2$ and $V_\mathrm{hr}(x) = 0$ for $|x| \geq \mathsfit{a}/2$.
Hard rods collide elastically with their two neighbors, except for the border particles with only one neighbor. Obviously, the system is integrable with one-particle sum functions, $\sum_{j=1}^N f(p_j)$, being conserved.

Since particles cannot cross, we order $q_1(t) <q_2(t) < \ldots < q_N(t)$. For sufficiently long times towards future and past,
\begin{equation}\label{2.50}
q_j(t) = p_j^+ t +  \phi_j^+, \quad t \to \infty,\qquad q_j(t) = p_j^- t + \phi_j^-, \quad t \to -\infty,
\end{equation}
where $\phi_j^+$ is the forward in time and $\phi_j^-$ the backward \textit{scattering shift}. When comparing with the point dynamics, $\mathsfit{a} \to 0$, one concludes
\begin{equation}\label{2.51}
p_j^+ = p_{N-j+1}^-, \qquad p_N^- < \ldots < p_1^-, \quad p_1^+ < \ldots < p_N^+.
\end{equation}
The \textit{relative scattering shift} of particle $j$, $\kappa_j$,  is defined by the deviation relative to the point particle dynamics. Hence
\begin{equation}\label{2.52}
\kappa_j = \phi_(N-j+1)^+ -  \phi_{j}^-.
\end{equation}
Just considering the intersection points arising from $N$ straight spacetime lines, one concludes
 \begin{equation}\label{2.53}
\kappa_j = \sum_{1 \leq i \leq N, i\neq j} \mathrm{sgn}(p_j^- - p_i^-)\phi_\mathrm{hr}(p_j^- - p_i^-), \qquad \phi_\mathrm{hr}(w) = -\mathsfit{a},
\end{equation} 
when written for incoming momenta. A similar expression holds for outgoing momenta. As a hallmark of  integrable many-body systems, the relative scattering shift is the properly signed sum of two-particle scattering shifts. 

More intuitive is the notion of a \textit{quasiparticle}, which maintains its velocity through a collision. Since for hard rods the collision time vanishes, a quasiparticle moves along a straight line interrupted by jumps of size  $\mathsfit{a}$ either to the right or left.
Quasiparticles are ordered increasingly  according to the ingoing particle configuration. Then the relative scattering shift  $\kappa_{j}$ is the accumulated spatial shift of the $j$-th quasiparticle. 

Our definitions involve sign conventions, which vary from author to author. As illustrated in Figure 3, by our rules, a negative scattering shift means that through a collision the two incoming particles get pushed apart relative to the free particle motion, while for a positive scattering shift 
they get pulled closer.
Hence the two-particle scattering shift of the hard rod fluid equals $-\mathsfit{a}$, which corresponds to the relative scattering shift of particle 1 in case incoming momenta are ordered as $p_2 < p_1$.
 \begin{figure}[!t]
\centering
\includegraphics[width=0.6\columnwidth]{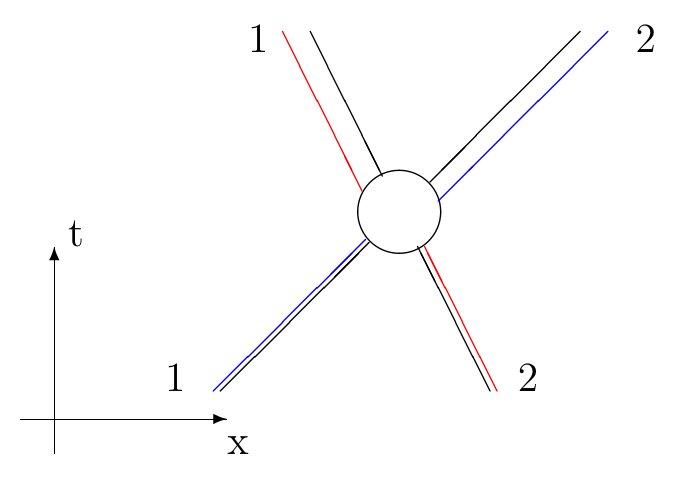}
\caption{Two-particle scattering. The numbers refer to particle labels. The red trajectory is quasiparticle 1 and blue quasiparticle 2. 
Quasiparticle 1 is shifted by $\phi_{12}$, negative in the example, and quasiparticle 2 by $-\phi_{12}$. }
\label{fig3}
\end{figure}

For physical rods the length is positive. But our rules make perfectly sense also for negative $\mathsfit{a}$. The two rods pass through each other 
 until they reach the now negative distance $\mathsfit{a}$. At that moment the velocities are exchanged. Since all rods have the same length, no ambiguities arise. Without much ado, the assumption $\mathsfit{a} >0$ will be dropped. However, in our verbal explanations
 we have to be more cautious.
\subsubsection{Two-particle Toda lattice}
\label{sec2.3.2}
Scattering is physically more intuitive in the fluid picture, i.e. particles move on the real line, also called the \textit{open} chain. 
For two particles, the equations of motion are
\begin{equation}\label{2.54}
\ddot{q}_1(t) = -\mathrm{e}^{q_1-q_2}, \quad \ddot{q}_2(t) = \mathrm{e}^{q_1-q_2}.
\end{equation} 
The relative motion, $q_2(t) - q_1(t)$, corresponds to a single particle subject to the potential $2\mathrm{e}^{-x}$.
We impose the asymptotic conditions $p_1(-\infty) = p_1^-$ and  $p_2(-\infty) = p^-_2$ with $p_2^- < p_1^-$. Adjusting the initial time such that ${q}_2(0) - {q}_1(0)$ is at the turning point, i.e. $\dot{q}_2(0) - \dot{q}_1(0) = 0$, the solution to \eqref{2.54} becomes
\begin{eqnarray}\label{2.55}
&&q_1(t) = \tfrac{1}{2}(p^-_1+p^-_2)t  - \log\big(\gamma^{-1} \cosh(\gamma t)\big), \nonumber\\[1ex]
&& q_2(t) = \tfrac{1}{2} (p^-_1+p^-_2)t  +\log \big(\gamma^{-1} \cosh(\gamma t)\big),
\end{eqnarray} 
with $\gamma = \tfrac{1}{2}(p^-_1 - p^-_2) >0$. The large time asymptotics is given by 
\begin{equation}\label{2.56}
q_1(t) = 
\begin{cases}p^-_1t  + \log|p^-_1 - p^-_2|,\\[1ex]
p^-_{2}t + \log|p^-_1 - p^-_2|,
\end{cases} 
q_2(t) = 
\begin{cases}p_2^-t - \log|p^-_1 - p^-_2|, &\quad t \to -\infty,\\[1ex]
p^-_{1}t - \ \log|p^-_1 - p^-_2|, &\quad t \to \infty.
\end{cases} 
\end{equation}
Since quasiparticle one is shifted by $ 2\log|p^-_1 - p^-_2|$ and quasiparticle two by $- 2 \log|p^-_1 - p^-_2|$, we conclude that the Toda two-particle relative scattering shift is given by
\begin{equation}\label{2.57}
2 \log |p^-_1 - p^-_2| = \phi_\mathrm{to}(p^-_1 - p^-_2) .
\end{equation} 
 The scattering shift has no definite sign. For $|p^-_1 - p^-_2| = 1$ the scattering shift vanishes. 
 For  $|p^-_1 - p^-_2| <1$ the scattering shift is negative, just as for hard rods. The trajectories of the two Toda particles look similar to the ones
 of hard rods, 
 but the hard rod zero collision time  is smeared to an exponential with rate $\gamma$. For $|p^-_1 - p^-_2| >1$ the 
 scattering shift is positive. The trajectories spatially cross each other, still approaching their asymptotic motion exponentially fast. 
 \subsubsection{$N$-particle Toda lattice}.
 \label{sec2.3.3}
 For $N$ particles the  hamiltonian of the open chain reads
\begin{equation}\label{2.58}
 H^\diamond_{\mathrm{to},N} = \sum_{j=1}^N \tfrac{1}{2}p_j^2 +  \sum_{j=1}^{N-1}\mathrm{e}^{-(q_{j+1} - q_j)},
\end{equation} 
 where the superscript $^\diamond$ is used to indicate the open chain. 
  In the Flaschka variables the equations of motion become 
 \begin{eqnarray}\label{2.59}
&&\hspace{0pt}\frac{d}{dt} a_j = \tfrac{1}{2}a_j(p_j - p_{j+1}),\quad j = 1,\ldots,N-1,\nonumber\\[1ex]
&&\hspace{0pt} \frac{d}{dt} p_j = a_{j-1}^2 - a_{j}^2, \quad j = 1,\ldots,N, 
\end{eqnarray}
with the boundary conditions $a_0 = 0, a_N = 0$. The Lax matrix $L^\diamond_N$ equals $L_N$ except for $a_N = 0$, correspondingly
for the partner matrix $B_N^\diamond$. The time evolution is encoded as
\begin{equation}\label{2.60} 
\frac{d}{dt} L_N^\diamond = [B_N^\diamond,L_N^\diamond].
\end{equation}
In general, modifying boundary conditions is likely to break integrability. But the open Toda chain is still integrable.

\begin{figure}[!b]
\centering
\includegraphics[width=0.9\columnwidth]{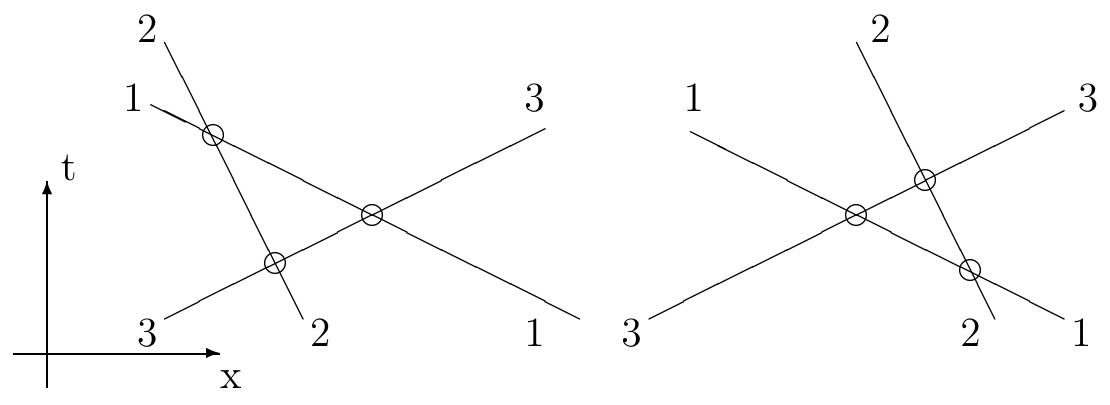}
\caption{Scattering shifts $\kappa_j$. The numbers refer to quasiparticle labels. In the left part the sequence of collisions is $\phi_{23},
\phi_{13},\phi_{12}$ and in the right part $\phi_{12},
\phi_{13},\phi_{23}$. Hence $\kappa_1 = \phi_{12} + \phi_{13}$,  $\kappa_2 = -\phi_{12} + \phi_{23}$, and $\kappa_3 = -\phi_{23} -\phi_{13}$, independently of the particular sequence. }
\label{fig4}
\end{figure}

The time zero phase point is denoted by $(q,p) \in \Gamma_{\!N}$. Since particles repel,
for sufficiently large $t$ one has $q_1(t) < \ldots <q_N(t)$ and $p_1(t) < \ldots <p_N(t)$. We order the eigenvalues of $L_N^\diamond$ as $\lambda_1 < \ldots<\lambda_N$, which defines the \textit{Weyl chamber} $\mathbb{W}_N$.
Then,  in the limit $t \to \infty$,
\begin{equation}\label{2.61} 
q_j(t) \simeq \lambda_{N -j +1} t + \phi_j^+
\end{equation}
with the forward scattering shift $\phi^+_j$. Correspondingly in the past,
\begin{equation}\label{2.62} 
q_j(t) \simeq \lambda_{j}t + \phi_j^-
\end{equation}
for $t \to -\infty$. The relative scattering shift is given by 
\begin{equation}\label{2.63}
\kappa_{j} = \sum_{1 \leq i \leq N, i\neq j} \mathrm{sgn}(\lambda_j - \lambda_i)\phi_\mathrm{to}(\lambda_j - \lambda_i),
\end{equation} 
compare with \eqref{2.53} for which the same conventions are used. In view of Figure 1 this result is very surprising. Despite the intricate pattern of multiple collisions, at very long times particles manage to have a scattering shift which is the weighted sum of two-particle scattering shifts. As one consequence of \eqref{2.63}, the scattering shift does not depend on the order of collisions. For quantum mechanical many-body systems this property is known as Yang-Baxter relation, see Chapter \ref{sec14}, and most commonly illustrated for three particles as in Figure 4. 

We return to the scattering map $\Phi^{-1}$ defined through \eqref{2.61} and for notational simplicity denote $\phi^+$ by $\phi$. As limit of canonical transformations the scattering map $\Phi^{-1}: (q,p) \mapsto (\lambda,\phi)$  is symplectic, which means that the asymptotic momenta and the scattering shifts are canonical coordinates. In other words,
 as functions on $\Gamma_{\!N}$
the Poisson brackets read
\begin{equation}\label{2.64} 
\{\lambda_i,\lambda_j\} = 0, \quad \{\phi_i,\phi_j\} = 0, \quad  \{\phi_i,\lambda_j\} = \delta_{ij}. 
\end{equation}
The variables $(\lambda,\phi)$ are the scattering analogue of  action-angle variables.
Only the angles vary over $\mathbb{R}$, rather than taking values on a torus. To make this distinction also verbally one should speak of  variables for asymptotic momenta and scattering shifts. But such practice becomes unwieldy and the common usage is  \textit{scattering coordinates} and more specifically 
\textit{action-angle variables}, keeping in mind that a scattering situation is discussed. 

Somewhat unexpectedly, for the Toda lattice the scattering map can be made explicit. The formulas simplify by considering the inverse transformation. We define the map 
$\Phi:  \Gamma_{\!N}^\triangleright \to \Gamma_{\!N}$, i.e. $\Phi:(\lambda, \phi) \mapsto (q,p)$ and  $ \Gamma_{\!N}^\triangleright =  \mathbb{W}_N\times \mathbb{R}^N$.
The map $\Phi$  is one-to-one,\hspace*{0pt} holomorphic, and given by
\begin{equation}\label{2.65}
q_j = \log\big(\sigma_{N+1-j}/\sigma_{N-j}\big), \qquad p_j = \big(\dot{\sigma}_{N+1-j}/\sigma_{N+1-j}\big)- 
 \big(\dot{\sigma}_{N-j}/\sigma_{N-j}\big).
\end{equation}
Here $\sigma_0 = 1$ and
\begin{equation}\label{2.66}
\sigma_k = \sum_{|I| =k}\exp\Big(\sum_{i \in I}\phi_i\Big) \prod_{i\in I, j \notin I}|\lambda_i - \lambda_j|^{-1}
\end{equation}
for $k= 1,\ldots,N$, where the first sum is over all subsets $I \subset \{1,\ldots,N\}$ of cardinality $|I|= k$.
The symbol $\dot{\sigma}_k$  refers to the Poisson bracket with the transformed hamiltonian,
\begin{equation}\label{2.67}
\dot{\sigma}_k = \big\{\sigma_k, \sum_{j=1}^N \tfrac{1}{2}\lambda_j^2\big\}.
\end{equation}
Later on, we will use these expressions to compute the generalized free energy, see Section \ref{sec9.2}.\bigskip
 \begin{center}
 \textbf{Notes and references}
 \end{center}
 \begin{center}
 \textbf{Section 2.0}
\end{center} 
The ground breaking discoveries are Toda  \cite{T67a,T67b}. The second edition of  the book Toda's ``Theory of Nonlinear Lattices'' \cite{T89} is still the most complete account up to 1989. Faddeev and Takhtajan  \cite{FT07} is a widely used standard monograph on classical integrable systems. A somewhat more elementary introduction is Arutyunov  \cite{A19}. Specifically the Toda lattice is reviewed by Kr\"{u}ger and  Teschl \cite{KT09}. Closer to hydrodynamics is the study of a 
  particular shock problem  by Venakides et al. \cite{VDO91}.  The fifty years anniversary volume edited by Bazhanov  et al. \cite{BDKT18} provides a glimpse on research in vastly diverse
  directions.
\newpage
\begin{center}
\textbf{Section 2.1}
\end{center}
Based on explicit soliton solutions, Toda conjectured integrability. For the case of three particles, Ford et al.  \cite{F73} obtained very supporting numerical Poincar\'{e} plots.
 The integrals of motion in full generality were obtained by H\'{e}non \cite{H74} by a tricky enumerative argument. H\'{e}non was worried about locality. 
 Flaschka  \cite{F74} had the advantage of working at the Courant Institute,
 at which Peter Lax \cite{L68} introduced his matrix in the context of the Korteweg-de Vries equation, see Section \ref{sec10.1}. 
  Once the Lax matrix had been discovered, locally conserved fields are easily constructed. Independently, the Lax matrix for the Toda lattice has been reported by Manakov \cite{M74}. Apparently, at the time and later on, currents were hardly in focus, one exception being Shastry and  Young \cite{SY10}, who study  the energy Drude weight in thermal equilibrium. 
 \bigskip
\begin{center}
\textbf{Section 2.2}
\end{center}
To find out the canonically conjugate angles is a much more technical enterprise, which was  accomplished in a series of papers by Henrici and Kappeler \cite{HK08a,HK08b, HK08c}, starting from an early proposal by  Flaschka  and McLaughlin \cite{FL76},
see Ferguson et al. \cite{FFL82} for complimentary aspects. The pitfalls are also discussed in Ruijsenaars \cite {R99}, Section  5. Lucid proofs of the stated properties are presented by Hubacher \cite{H89}. As a further result, 
the additivity \eqref{2.63} is deduced when assuming the property $p_j^+ = p_{N+1-j}^-$. In spirit this result ensures the existence of quasiparticles. However, the abstract proof does not
tell us for which interaction potentials the assumption holds.

For quantum systems, the notion of integrability is controversial, since the naive transcription of the classical notion would mean that every eigenprojection of the hamiltonian is conserved, in itself not such a helpful observation. We refer to Caux and Mossel   \cite{CM11} for an exhaustive discussion. The link between integrability and local conservation laws has been mostly pushed by the quantum community, see  Grabowski and Mathieu  \cite{GM94,GM95} for early work.  For classical systems, in general, this avenue still needs to be further developed. 
The mentioned results for the XYZ chain are prototypical for what one would like to achieve. The nonintegrable case is a result of
Shiraishi \cite{Sh19}. The integrable case,
$h =0$,  has been studied already by Grabowski and Mathieu \cite{GM94,GM95} with recent progress 
Nozawa and  Fukai \cite{NF20}.
Considering the XXZ chain with parameters $J_x =J_y = 1$, $J_z = \Delta$ and using only strictly local charges of the spin chain, one computes  the spin Drude weight at zero magnetization,
i.e. the persistent spin current, by using the Mazur formula. By spin inversion symmetry this weight turns out to be identically $0$. On the other hand for  $0 \leq \Delta < 1$, numerical evidence and exact steady state results for the boundary driven chain indicate that the Drude weight does not vanish.
The puzzle is resolved by the construction of quasilocal conserved charges in Mierzejewski  et al.  \cite{MPP15}, see also related work 
Ilievski et al.  \cite{IDWC15,IMPZ16}, for a broader perspective  Doyon  \cite{D17}, and the recent review
Ilievski  \cite{I22}. The Drude weight is nowhere continuous in its dependence on $\Delta$.

The relation between integrability and conservation laws has been investigated also in the context
of $1+1$-dimensional quantum field theories. Claimed is indeed a dichotomy, in the following sense:
If beyond the conservation of energy and momentum there is a single higher order locally conserved charge, then necessarily the theory is integrable, in the sense of possessing infinitely many conservation laws, see Coleman and Mandula \cite{CM67}, Iagolnitzer \cite{I78,I78a}, Parke \cite{P80}, and Doyon \cite{D08}.
\bigskip
\begin{center}
\textbf{Section 2.3}
\end{center}
In a beautiful piece of analysis Moser \cite{M75} proves the scattering shift for the $N$-particle Toda lattice. An account of his work can be found in  Toda \cite{T89}. 
From a different perspective, a more recent discussion are the notes of Deift et al.  \cite{De19} from his course  at the Courant Institute in Spring 2019. The action-angle map $\Phi$ was established by Ruijsenaars \cite{R90}, where also the stated properties are proved. Asymptotic momenta equal eigenvalues of the Lax matrix. In spirit, there is also  a corresponding algebraic identity
for the scattering shifts, which has been constructed in analogy to more accessible models.
But to check directly the validity of the Poisson bracket relations in \eqref{2.64}
seems to be completely out of reach. Thus as a major difficulty, one first has to establish agreement between algebraic
approach and scattering map. A more recent point of view is developed by Feh\'{e}r   \cite{F13}. 
 
 The distinction between integrable and nonintegrable scattering has been studied in detail for one-particle systems.
An example is the four hill potential $V_\mathrm{fh}(q_1,q_2) = q_1^2q_2^2\exp\big(- q_1^2 - q_2^2\big)$.
Chaotic scattering is reviewed by Seoane and Sanju\'{a}n \cite{SS13}.
\newpage
\section{Static properties}
\label{sec3}
\setcounter{equation}{0}
Considering the Toda lattice with periodic boundary conditions and a highly excited initial condition $q,p$, one would expect that in the long time limit a statistically stationary state is reached. For a simple fluid, this state would be thermal equilibrium. But the motion of  Toda particles is highly constrained through the conservation laws. Still, there are lots and lots of random like collisions. Following Boltzmann, a natural guess for the statistically stationary  state is a generalized microcanonical ensemble, namely the uniform measure on the $(N-1)$-dimensional torus $\mathbb{T}^{N-1}$ at fixed values of Lax eigenvalues $\lambda _1,\ldots,\lambda_N$ and of $Q^{[0],N}$, see the discussion in the beginning of Section \ref{2.2}.   The corresponding thermodynamics thus depends on $N+1$ extensive parameters. 
As for simple fluids, the first step towards hydrodynamic equations is a study of such generalized thermodynamics. 
\subsection{Generalized Gibbs ensembles}
\label{sec3.1}
For fixed number of lattice sites, the phase space is $(r, p) \in \Gamma_{\!N}$ with the a priori weight
\begin{equation}\label{3.1} 
\prod_{j=1}^N\mathrm{d}r_j \mathrm{d}p_j \delta\Big( \sum_{j=1}^N r_j- \ell\Big)
\end{equation} 
for some $\ell \in \mathbb{R}$, where we included already the microcanonical constraint resulting from the boundary conditions \eqref{2.9}. This measure is invariant under the flow generated by Eq. \eqref{2.5}.  Since particles are distinguishable, there is no factor of $1/N!$ in front.  The remaining conserved fields are taken into account  
through the grand canonical type Boltzmann weight
\begin{equation}\label{3.2} 
\exp\!\Big(  - \sum_{n=1}^N \mu_n Q^{[n],N} \Big).
\end{equation}
Here $\mu_1,\ldots,\mu_N$ are the intensive parameters. Only the low order ones have a physical interpretation, specifically
$\mu_2 = \tfrac{1}{2} \beta$ with $\beta$ the inverse temperature and $\mu_1$ as control parameter for the average total momentum.
As common in Statistical Mechanics we invoke the equivalence of ensembles to lift the delta constraint by the substitution
\begin{equation}\label{3.3} 
\delta\big( Q^{[0],N} - \ell\big)\quad \Rightarrow \quad \exp\!\big( - P Q^{[0],N}\big).
\end{equation}
 Such kind of equivalence has been extensively studied in rigorous statistical mechanics and presumably
some of the techniques can be used also for the Toda lattice. Along with other items, we have to leave this problem for future studies.
For a general anharmonic chain the physical pressure, $ \mathfrak{p}$, is defined as the average force between neighboring particles in thermal 
equilibrium. Using a simple integration by parts, one obtains the relation $ \mathfrak{p} = \beta^{-1}P$. We still refer to  $P$ as pressure, since it is the thermodynamic dual of the stretch. To have an integrable Boltzmann weight, $P> 0$ is required. In combination  the \textit{generalized Gibbs ensemble} (GGE) is defined through  
\begin{equation}\label{3.4} 
\prod_{j=1}^N\mathrm{d}r_j \mathrm{d}p_j \exp\!\Big( - P Q^{[0],N} - \sum_{n=1}^N \mu_n Q^{[n],N} \Big),
\end{equation}
which still has to be normalized. The GGE is invariant under the Toda dynamical flow.

To study properties of GGE the first natural step is to transform \eqref{3.4} to Flaschka variables.
For conciseness we introduce
\begin{equation}\label{3.5}
V(w) = \sum_{n=1}^\infty\mu_n w^n.
\end{equation}
The chemical potentials, $\mu_n$, are assumed to be independent of $N$. 
Then the transformed density reads
\begin{equation}\label{3.6} 
\exp\!\big(-\mathrm{tr}[V(L_N)]\big) \prod_{j=1}^{N}  \mathrm{d}p_j \prod_{j=1}^{N} \mathrm{d}a_j \frac{2}{a_j}
(a_j)^{2P},
\end{equation}
which is defined on the phase space of the Flaschka variables, i.e. on $\Gamma_{\!N}^\circ$. \bigskip\\
$\blackdiamond\hspace{-1pt}\blackdiamond$~\textit{Confining potential}.  A many-particle system is defined by 
a particular interaction potential, which throughout will be denoted by $V_\bullet $. For example $V_\mathrm{to}$ is the interaction potential 
of the Toda lattice and  $V_\mathrm{li}$ the $\delta$-potential of the Lieb-Liniger model. The potential \eqref{3.5} could be called a generalized chemical potential, a terminology  which however would entirely miss the central role of $V$. We wrote $V$ in terms of a power series. But there is no compelling reason to do so.
$V$ is simply a rather generic function on $\mathbb{R}$.  In the context of GGE, its purpose  is to properly confine the eigenvalues of $L_N$. 
We thus assume that $V$ is continuous and bounded linearly  from below as $V(w) \geq c_0 +c_1|w|$ with $c_1 >0$. Based on such  reasoning $V$ is called  \textit{confining potential}. It carries no relation to the interaction potential. Rather $V$ should be viewed as a thermodynamic variable, which
so to speak labels the GGEs.  The 
quadratic confining potential, $V(w) = \tfrac{1}{2}\beta w^2$, corresponds to thermal
equilibrium. 
\hfill$\blackdiamond\hspace{-1pt}\blackdiamond$\bigskip\\
$\blackdiamond\hspace{-1pt}\blackdiamond$~\textit{Infinite volume limit, exponential mixing, equivalence of ensembles}.\hspace{1pt} As for other Gibbs measures, one might want to know about the existence of the infinite volume limit for the normalized sequence of measures in \eqref{3.6}, the limit measure being independent of boundary conditions, and a bound on the decay of correlations. If the confining potential is a
finite polynomial with even strictly positive leading term, say $\kappa$, then the confining potential is bounded from below and such properties can be answered by using transfer matrix techniques.  In the language of statistical mechanics, $Q^{[n]}$ has a range of size $n$ and hence $\mathrm{tr}[V(L_N)]$ has range $\kappa$. 
One cuts $[1,\ldots,N]$ in blocks of size $\kappa$. The density in \eqref{3.6} can then be written as  an $(N/\kappa)$-fold power of the transfer matrix. This is just like the familiar case of the one-dimensional Ising model, in which case the transfer matrix is a $2\times 2$ matrix. For the Toda lattice the transfer matrix is given by an integral kernel with arguments in $\Gamma_{\kappa}^\circ$.
By the Perron-Frobenius theorem, the transfer matrix has a unique maximal eigenvalue, which is separated by a 
gap from the rest of the spectrum. With this input,
one concludes that there is a unique limit measure. In one dimension, phase transitions would  occur only if the interaction 
potential has a decay slower than (range)$^{-2}$, much slower than the case under consideration here. The spectral gap also ensures exponential decay of correlations. 
If $\kappa = \infty$ such methods fail completely. Other techniques will have to be developed, see Notes. 

A presumably more delicate issue is the \textit{equivalence of ensembles}. This refers to replacing the sharp constraint 
$\delta\big(H_N - \mathsfit{e}N\big)$ by $\exp\big(-\beta H_N\big)$. On the thermodynamic level, in the limit 
$N \to \infty$, this amounts to a Legendre transform between $\mathsfit{e}$ and $\beta$. As a stronger 
property the statistics of local observables is  the same provided $\mathsfit{e}$ and $\beta$ are related 
according to thermodynamics. For the Toda lattice we switched from $\nu N$ to $P$, a step which is expected to be accessible to current techniques. More delicate is the closed Toda chain as discussed in Section \ref{sec2.2}. In principle one should fix all action variables. The integral over the tori is trivial and one is left with $\sum_{j=1}^N V(\lambda_j)$. Abstractly this sum depends only on $I$, but  a more concrete  characterization does not seem to be available. Thus a fully microcanonical approach
will not be pursued any further. 
\hfill$\blackdiamond\hspace{-1pt}\blackdiamond$
\subsection{Lax matrix filter and local GGEs}
\label{sec3.2}
This section is somewhat premature, since so far the only model discussed is the Toda lattice.  Still, before entering in computational details, we should explain the physics underlying the notion of local GGE's as the central theme of hydrodynamic scales. Some of the material will be explained in greater detail  in the following. 

Let us start from the example of a classical ideal gas in one dimension, which consists of many particles moving along straight lines as $q_j(t) = q_j + p_j t$ for initial conditions $(q,p)$.
The momenta are conserved and hence independent of time. For general random initial data, one introduces the one-particle distribution function
\begin{equation}\label{3.7} 
\rho_1(x,t;w) = \big\langle \sum_{j}\delta(x- q_j(t)) \delta(w -p_j)\big \rangle_\mathrm{ini}
\end{equation}
average over the initial probability density function. For an ideal gas, $\rho_1(x,t;w) = \rho_1(x -wt,0;w)$.

For  the hydrodynamic scale of the ideal gas we will use the notation $\rho_\mathrm{Q}(x,t;w)$ which superficially looks rather similar to $\rho_1$.  Here $(x,t)$ refers  to a macroscopic spacetime point, which is the center of a microscopic cell of size $\ell$ containing $N(\ell)$ particles.
$N(\ell)/\ell$ is of order 1 and each cell contains a large number of particles. The index $\mathrm{Q}$ should be a reminder that $\rho_\mathrm{Q}(x,t;w)$
carries the information on the average conserved fields in the considered cell. $\rho_\mathrm{Q}(x,t;w)$ has a double
meaning. We define the empirical velocity distribution function 
through
\begin{equation}\label{3.8} 
\frac{1}{\ell} \sum_{j=1}^{N(\ell)} \delta(w -p_j) \simeq \rho_\mathrm{Q}(x,t;w),
\end{equation} 
where the sum is over all particles in the cell centered at $(x,t)$. The left hand side is random, but self-averaging in the sense that fluctuations vanish for large $N(\ell)$. In the limit, $\rho_\mathrm{Q}(x,t;w)$
is nonrandom and its integral with respect to $w$ is the macroscopic density $\bar{\rho}(x,t)$. In addition
$\rho_\mathrm{Q}(x,t;w)$ uniquely characterizes the local GGE governing the statistics of particles close to $(x,t)$. For an ideal gas velocities are independent with common
probability density function  $\rho_\mathrm{Q}(x,t;w)/\bar{\rho}(x,t)$. The positional distribution is independent of velocities, more precisely, a Poisson 
point process with density $\bar{\rho}(x,t)$, which means that inter-particle distances are independent and exponentially 
distributed. Note that in the limit of infinite scale separation, the local GGE lives on the entire line and is translation invariant. The construction \eqref{3.8} holds also for a hard rod fluid. Only now the local GGE has a positional distribution of particles satisfying the hard core constraint.

The at first sight truly surprising claim is that for classical integrable many-body systems the definition in \eqref{3.8} can still be used provided the
momenta $p_j$ are replaced by the eigenvalues $\lambda_j$ of the local Lax matrix $L(x,t)$. This is what we mean by a \textit{Lax matrix filter}. 
Local positions and momenta are noisy. But by inserting these data in the local Lax matrix and determining the local \textit{density of states} (DOS), magically, 
one filters the slowly varying degrees of freedom. Compared with the ideal gas,  
rather than sampling local velocities one has to sample the eigenvalues of the local Lax matrix. More precisely, the Lax matrix is random under the local GGE
and has an \textit{empirical} DOS according to 
\begin{equation}\label{3.9} 
\frac{1}{\ell} \sum_{j=1}^{N(\ell)} \delta(w -\lambda_j) .
\end{equation} 
The sum is over eigenvalues of the Lax matrix constructed from the particular fluid cell under consideration. As valid in great generality,  the DOS is self-averaging and thus has a deterministic limit, as before denoted by   $\rho_\mathrm{Q}(x,t;w)$, but now referring to the Lax DOS. The function $\rho_\mathrm{Q}(x,t;w)$ uniquely determines the underlying local GGE in the cell $(x,t)$. The sought for hydrodynamic equations are an evolution equation for   $\rho_\mathrm{Q}(x,t;w)$, which is of conservation type
\begin{equation}\label{3.10} 
\partial_t \rho_\mathrm{Q}(x,t;w)+\partial_x \rho_\mathrm{J}(x,t;w) = 0.
\end{equation} 
A major task will be to figure out the $\rho_\mathrm{J}(x,t;w)$ as a functional of  $\rho_\mathrm{Q}(x,t;w)$ at fixed fluid cell
$(x,t)$. Concrete examples will come. For some models the Lax matrix is unitary
and eigenvalues lie on the unit circle. Only through a more detailed analysis such properties can figured out.
\subsection{Generalized free energy}
\label{sec3.3}
The Toda partition function is defined by
\begin{equation}\label{3.11} 
Z_{\mathrm{to},N}(P,V)  = \int_{\Gamma_{\!N}^\circ} \prod_{j=1}^{N}  \mathrm{d}p_j  \prod_{j=1}^{N}\mathrm{d}a_j \frac{2}{a_j}
(a_j)^{2P} \exp\!\big(-\mathrm{tr}[V(L_N)]\big).
\end{equation}
Accordingly, normalizing  the expression in \eqref{3.6}, one arrives at the probability measure
\begin{equation}\label{3.12} 
\mu^\mathrm{GGE}_{P,V,N} (\mathrm{d}^N\hspace{-1pt}a\,\mathrm{d}^N\hspace{-1pt}p) = \frac{1}{Z_{\mathrm{to},N}(P,V)} \prod_{j=1}^{N}  \mathrm{d}p_j  \prod_{j=1}^{N}\mathrm{d}a_j \frac{2}{a_j}
(a_j)^{2P} \exp\!\big(-\mathrm{tr}[V(L_N)]\big).
\end{equation}
This measure is time-stationary under the dynamics \eqref{2.5}, since the a priori measure and the eigenvalues of $L_N$ do not change in time.
The  expectations of $\mu^\mathrm{GGE}_{P,V,N}$ will be denoted by $\langle \cdot \rangle_{P,V,N}$. As  central thermodynamic object,  the free energy per lattice site is defined by 
\begin{equation}\label{3.13} 
F_\mathrm{to}(P,V) = - \lim_{N \to \infty} \frac{1}{N}\log Z_{\mathrm{to},N}(P,V). 
\end{equation}
For hydrodynamics, a crucial input is the GGE average of the conserved fields. They can be computed as  derivative with respect to $P$ and  as variational  derivative with respect to $V$ of the
free energy. In terms of the eigenvalues of the Lax matrix, the averages can be written as
 \begin{equation}\label{3.14} 
\frac{1}{N} \langle Q^{[n],N} \rangle_{P,V,N} = \frac{1}{N} \big\langle \sum_{j=1}^N (\lambda_j)^n \big\rangle_{P,V,N}
= \int_\mathbb{R} \mathrm{d}w w^n  \langle\rho_{\mathrm{Q},N} (w)\rangle_{P,V,N}, 
\end{equation}
where
 \begin{equation}\label{3.15} 
\rho_{\mathrm{Q},N} (w) = \frac{1}{N}  \sum_{j=1}^N \delta(w - \lambda_j)
\end{equation}
is the empirical density of states of the Lax matrix $L_N$. Here empirical refers to the fact that the DOS is defined for every collection of eigenvalues $\{\lambda_1,\ldots,\lambda_N\}$. Thus $\rho_{\mathrm{Q},N} (w)$ is a random function under 
$\mu^\mathrm{GGE}_{P,V,N}$. More properly, $\rho_{\mathrm{Q},N}$ is a random probability measure supported on $N$ points each with weight $1/N$.

This observation suggests a novel perspective. The Lax matrix becomes a random matrix under $\mu^\mathrm{GGE}_{P,V,N}$.
Thermal equilibrium is particularly simple. Since $\mathrm{tr}[(L_N)^2]= \sum_{j=1}^N(p_j^2 + 2a_j^2)$, the diagonal and off-diagonal matrix elements of $L_N$ are families of independent 
identically distributed  (i.i.d.) random variables. For all other GGEs the matrix elements are correlated. The DOS encodes the complete statistical information on the conserved fields. In fact $\rho_{\mathrm{Q},N} (w)$ is self-averaging with fluctuations of order
$1/ \sqrt{N}$ and the limit
 \begin{equation}\label{3.16} 
 \lim_{N\to\infty} \rho_{\mathrm{Q},N} (w) = \rho_\mathrm{Q}(w)
\end{equation}
exists with some nonrandom limiting density $\rho_\mathrm{Q}$, which generically is a smooth function.\bigskip\\
$\blackdiamond\hspace{-1pt}\blackdiamond$~\textit{The Dumitriu-Edelman identity}.\hspace{1pt} In 2002 Dumitriu and Edelman studied the $\beta$-ensembles of random matrix theory.
To ease a comparison, I describe their result in the original notation,  in which some symbols will reappear with a different meaning. The proper translation
will be obvious, however. Their starting point is a $n \times n$, symmetric, tridiagonal matrix $T$ with real matrix elements $T_{j,j} = a_j$, $T_{j,j+1} = T_{j+1,j} =
b_j >0$, and zero otherwise, compare with $L^\diamond$ from  \eqref{2.60}, in particularly, $(a,b) \in \Gamma^\diamond_{\!n}
=  \mathbb{R}^{n} \times  \mathbb{R}_+^{n-1}$. The eigenvalues of $T$ are ordered as $\lambda_1 <\ldots.< \lambda_n$
and the first component of an eigenvector is denoted by $\psi_j(1) = q_j > 0$ with $|q| = 1$ imposed. We set $\mathrm{d}^na = \prod_{j=1}^n  \mathrm{d}a_j$, $\mathrm{d}^n\lambda = \prod_{j=1}^n  \mathrm{d}\lambda_j$ and 
$\mathrm{d}^nq$  the surface element of the unit $n$-sphere. As a general fact of Jacobi matrices with positive off-diagonal matrix elements, there is a one-to-one and onto map $\Psi: (a,b) \mapsto (\lambda, q)$. Dumitriu and Edelman managed to obtain some information on the Jacobian of $\Psi$.
Their identity has a free parameter $\beta > 0$ and reads
\begin{equation}\label{3.17} 
 \mathrm{e}^{-\mathrm{tr}[V(T)]}\,\mathrm{d}^n a\prod_{j=1}^{n-1}\frac{2}{b_j} (b_j)^{\beta j} \mathrm{d}b_j = \exp\Big(- \sum_{j=1}^n V(\lambda_j)\Big)\big(n!\tilde{\zeta}_n(\beta)
 \Delta (\lambda)^\beta\mathrm{d}^n\lambda\big) \Big(c_q^\beta \prod_{j=1}^n(q_j)^{\beta-1}\mathrm{d}^nq\Big).
\end{equation}
 Here 
  \begin{equation}\label{3.17a} 
 c_q^\beta =  2^{n -1} \Gamma(\tfrac{1}{2}\beta n) \Gamma( \tfrac{1}{2}\beta)^{- n}
 \end{equation}
 normalizes the third factor on the left to 1. In the second factor,
 $\Delta(\lambda)$ is the Vandermonde determinant
 \begin{equation}\label{3.18} 
\Delta(\lambda) = \prod_{1 \leq i<j\leq n}(\lambda_j - \lambda_i),
\end{equation}
while 
 \begin{equation}\label{3.19} 
\tilde{\zeta}_n(\beta) = \Gamma(\tfrac{1}{2}\beta n)^{-1} \Gamma(1+ \tfrac{1}{2} \beta)^n \prod_{j=1}^n \frac{\Gamma( \tfrac{1}{2} \beta j)}
{\Gamma(1 + \tfrac{1}{2} \beta j)} 
\end{equation}
is the proportionality constant. \hfill $\blackdiamond\hspace{-1pt}\blackdiamond$ \bigskip 

In the Dumitriu-Edelman identity \eqref{3.17} we substitute $n \leadsto N$, $a_j \leadsto p_j$, and $b_j \leadsto a_j$. Since the scale parameter 
mentioned below Eq. \eqref{2.4} has been set to $\tau =1$, one notes that $T = L_N^\diamond$. Therefore integrating both side of \eqref{3.17} over the entire phase space, first making the choice 
 \begin{equation}\label{3.20} 
\beta = \frac{2P}{N},
\end{equation}
yields 
\begin{eqnarray}\label{3.21} 
&&\hspace{-39pt}Z_{\mathrm{de},N}(P,V) = \int_{\Gamma^\diamond_{\!N}}\exp\!\big(-\mathrm{tr}[V(L^\diamond_N)]\big) \prod_{j=1}^{N}  \mathrm{d}p_j  
\prod_{j=1}^{N-1}\mathrm{d}a_j \frac{2}{a_j}
(a_j)^{2 (j/N)P} \nonumber\\
&&\hspace{18pt} = \zeta_N(P)\int_{\mathbb{R}^N} \mathrm{d}^N\hspace{-1pt} \lambda 
\exp\!\Big(- \sum_{j=1}^N V(\lambda_j) + \frac{P}{N}\sum_{i,j=1,i \neq j}^N \log|\lambda_i - \lambda_j|\Big)
\end{eqnarray}
with prefactor
\begin{equation}\label{3.22} 
\tilde{\zeta}_N(2P/N) = \zeta_N(P) = \Gamma(P)^{-1} \Gamma(1+ \tfrac{P}{N})^N \prod_{j=1}^N \frac{\Gamma(\tfrac{j}{N})}
{\Gamma(1 + \tfrac{j}{N})}.
\end{equation}
The integration over eigenvalues is not ordered, which takes care of the factor $N!$. Also, the integration over the $N$-sphere is normalized
to $1$. 

The term on the right side of \eqref{3.21} requires more explanations. For the normalization one obtains
\begin{equation}\label{3.23} 
 \lim_{N\to\infty}-\frac{1}{N}\log \zeta_N(P) = \log P -1 .
\end{equation}
Otherwise the partition function is the one of the \textit{repulsive one-dimensional log gas}. $V$ turns out to be  the confining potential of the log gas,
which is the real reason for our original choice of name. In the standard log gas the interaction strength is of order $1$,
which implies that the free energy is dominated by the energy term. But in our case the interaction strength is $1/N$,
which is the standard mean-field scaling. Such a problem can be handled through the study of a free energy functional. 
To distinguish from \eqref{3.15}, one introduces
\begin{equation}\label{3.24} 
\varrho_{N} (w) = \frac{1}{N}  \sum_{j=1}^N \delta(w - \lambda_j)
\end{equation}
with $\lambda$ now referring to  \eqref{3.21}.  Except for the diagonal contribution, the integrand of  \eqref{3.21} can be written as
\begin{equation}\label{3.25} 
\exp\Big[-N\Big(\int_\mathbb{R} \mathrm{d}w \varrho_{N} (w)V(w)  - P\int_\mathbb{R} \mathrm{d}w\int_\mathbb{R} \mathrm{d}w'
\varrho_{N} (w)\varrho_{N} (w')\log|w - w'|\Big)\Big].
\end{equation}
For given $\varrho_{N}$ the corresponding volume element of $\mathrm{d}^N\hspace{-1pt}\lambda$ is approximately
\begin{equation}\label{3.26} 
\exp \Big[- N\int_\mathbb{R} \mathrm{d}w  \varrho_{N} (w) \log \varrho_{N} (w) \Big].
\end{equation}
The large $N$ limit of the partition function is then determined by the mean-field free energy functional 
 \begin{equation}\label{3.27}
\mathcal{F}_\mathrm{de}^\circ(\varrho) =  \int _\mathbb{R}\mathrm{d}w \varrho(w) \Big(V(w)  + \log \varrho(w) +\log P   - P\int _\mathbb{R}\mathrm{d}w'   \varrho(w')\log|w - w'|  \Big).
\end{equation} 
The actual free energy is obtained by  minimizing over all $\varrho$ with $\varrho \geq 0$ and $\int_\mathbb{R} \mathrm{d}w \varrho(w) = 1$. As will be discussed below,
there is a unique minimizer, $\varrho^\star$, and thus
\begin{equation}\label{3.28} 
\lim_{N\to\infty}-\frac{1}{N}\log Z_{\mathrm{de},N}(P,V) = F_\mathrm{de}(P,V) = \mathcal{F}_\mathrm{de}^\circ(\varrho^\star) -1.
\end{equation}

To obtain the Toda free energy, we note that the Dumitriu-Edelman partition function has a pressure changing linearly with slope $2P/N$.
This is not exactly what is required, since the pressure is constant  for the Toda chain. But in a large segment $\ell$, still with size $\ell \ll N$, the pressure is constant to a very good approximation.
Since GGEs have good spatial mixing properties, local free energies merely add up and for the Dumitriu-Edelman free energy one concludes  
\begin{equation}\label{3.29} 
 F_{\mathrm{de}}(P,V) = \int_0^1\mathrm{d}u F_\mathrm{to}(uP,V)
\end{equation}
and hence 
\begin{equation}\label{3.30} 
F_\mathrm{to}(P,V) = \partial_P(PF_{\mathrm{de}}(P,V)).
\end{equation}

While the just presented derivation of $F_\mathrm{de}(P,V)$ is fairly standard, one may wonder about the missing steps.  A poor man's version will be explained in Section \ref{sec3.5}. Besides,
the topic  has been 
studied extensively with methods covering the case of interest. Less standard is the linear pressure ramp leading to the identity \eqref{3.29}.

In the free energy functional \eqref{3.27} the quadratic term has the kernel $\log|w - w'|$ that resulted from the 
Dumitriu-Edelman change of coordinates. 
The same expression came up already in Section \ref{sec2.3} in the context of scattering theory, which might be considered as purely accidental. 
In fact, we touched upon a generic feature of generalized free energies for integrable models. For the Toda lattice the connection to scattering theory will be elucidated in Section \ref{sec9.2}.
 
It turns out to be more convenient to absorb $P$ into $\varrho$ by setting $\rho = P\varrho$.  Then 
$P \mathcal{F}_\mathrm{de}^\circ(P^{-1}\rho)=  \mathcal{F}_\mathrm{de}(\rho)$ with 
the transformed free energy functional 
\begin{equation}\label{3.31}
\mathcal{F}_\mathrm{de}(\rho) =  \int _\mathbb{R}\mathrm{d}w \rho(w) \Big(V(w)  + \log \rho(w)    - \int _\mathbb{R}\mathrm{d}w'  \rho(w')  \log|w - w'| 
\Big).
\end{equation} 
$\mathcal{F}_\mathrm{de}$ has to be minimized under the constraint
\begin{equation}\label{3.32}
\rho(w) \geq 0,\quad  \int _\mathbb{R}\mathrm{d}w\rho(w) =P 
\end{equation}
with minimizer denoted by $\rho^\star$. Then
 \begin{equation}\label{3.33} 
 F_\mathrm{to}(P,V) =  \partial_P \mathcal{F}_\mathrm{de}(\rho^\star) -1.
 \end{equation}
 
The constraint \eqref{3.32} is removed by introducing the Lagrange multiplier $\mu$ as
  \begin{equation}\label{3.34} 
 \mathcal{F}_\mathrm{de}^\bullet(\rho) =  \mathcal{F}_\mathrm{de}(\rho) - \mu \int _\mathbb{R}\mathrm{d}w \rho(w).
 \end{equation}
A minimizer of  $\mathcal{F}_\mathrm{de}^\bullet(\rho)$ is denoted by  $\rho_{\mathsf{n}}$ and by  $\rho_{\mathsf{n},\mu}$ when keeping track of the $\mu$-dependence. It is determined as solution of the Euler-Lagrange equation
\begin{equation}\label{3.35} 
  V(w) +\log \rho_{\mathsf{n},\mu}(w)  - \mu -  2 \int_\mathbb{R} \mathrm{d}w'  \rho_{\mathsf{n},\mu}(w') \log|w-w'|  = 0.
 \end{equation}
The Lagrange parameter $\mu$ has to be adjusted such that
\begin{equation}\label{3.36} 
 P =  \int _\mathbb{R}\mathrm{d}w  \rho_{\mathsf{n},\mu}(w).
 \end{equation}
 In fact, it will be more convenient to work directly with $\rho_\mathsf{n}$. Note that the Lagrange parameter $\mu$ amounts to shifting the confining potential as $V - \mu$. Thus $\mu$ could be incorporated in the definition \eqref{3.5} 
 of $V$ as $\mu_0 = - \mu$. The minus sign is a standard convention for the chemical potential  dual to the particle number.
 
 To obtain the Toda free energy, we differentiate as
 \begin{equation}\label{3.37} 
\partial_P \mathcal{F}_\mathrm{de}(\rho^\star) =
  \int_\mathbb{R} \mathrm{d}w \partial_P\rho^\star(w)\Big( V(w)  + \log \rho^\star(w)    - 2 \int_\mathbb{R} \mathrm{d}w'  \rho^\star(w')\log|w-w'|\Big)
+1.
  \end{equation}
 Integrating \eqref{3.35} against  $\partial_P\rho^\star$ one arrives at
 \begin{equation}\label{3.38} 
 \partial_P \mathcal{F}_\mathrm{de}(\rho_{\mathsf{n},\mu}) = \mu +1
\end{equation}
 and thus 
 \begin{equation}\label{3.39} 
  F_\mathrm{to}(P,V) =  \mu(P,V).
 \end{equation}
Sharing with other integrable models, the Toda lattice has the property that its free energy 
is determined by a variational problem for densities over $\mathbb{R}$, in our case  normalized to $P$.\bigskip\\
$\blackdiamond\hspace{-1pt}\blackdiamond$~\textit{Densities}.\hspace{1pt}  Our generic symbol for a density is $\rho$, resp. $\varrho$. We use $\varrho$ for normalized densities, $\int \mathrm{d}w\varrho(w) = 1$, while $\rho$ has a context dependent normalization.
For Toda and other integrable systems the densities are defined on $\mathbb{R}$.  Later on we will also encounter models where the densities live on the unit circle. Density does not refer to position space. The appropriate picture is a density in distorted momentum space.
Since there are several densities, one has to distinguish them
by a label which in our notation appears through a lower index, as in $\rho_\mathrm{Q}$ and $\rho_\mathrm{J}$. We introduced already the density $\rho_{\mathsf{n}}$, also called \textit{number density}. But it will turn out to be convenient to introduce the further densities  $\rho_{\mathsf{p}}$, $\rho_{\mathsf{h}}$, $\rho_{\mathsf{s}}$, called \textit{particle, hole, space density}. The label is now in serif to avoid confusion with arguments. All densities are functionals of $V$, a dependence which is mostly suppressed in our notation. 
For example $\rho_{\mathsf{n}}$ functionally depends on $V-\mu$, hence it  is a function of $\mu$  and written as
$\rho_{\mathsf{n},\mu}$. $\rho_{\mathsf{n}}$
can be regarded  also as a function of $P$, since $\mu = \mu(P)$ for fixed $V$.
  \bigskip\hfill $\blackdiamond\hspace{-1pt}\blackdiamond$
\subsection{Lax density of states, TBA equation}
\label{sec3.4}
To obtain the hydrodynamic equations,  GGE averages of the conserved fields are required. As common practice in statistical mechanics, they are defined  in the infinite volume limit. For this purpose we adopt the volume $[-N,\ldots,N]$ with periodic boundary conditions and adjust our notation by using the label $2N+1$ instead of
$N$. Then, using translation invariance, in the infinite volume limit
 \begin{equation}\label{3.40} 
\lim_{N \to \infty} \frac{1}{2N+1}\langle Q^{[n],2N+1}\rangle_{P,V,2N+1} = 
\lim_{N \to \infty} \langle Q^{[n],2N+1}_0\rangle_{P,V,2N+1} = 
\langle Q^{[n]}_0\rangle_{P,V}.
\end{equation}  
On the right hand side $Q^{[n]}_0$ is a local function and $\langle \cdot\rangle_{P,V}$ refers to the average with respect to the infinite volume GGE at parameters $P,V$. Since in our context boundary terms should be negligible, we assume this 
measure to be well-defined and independent of boundary conditions. More pragmatically, only for particular observables, as $Q^{[n]}_0$ and $Q^{[n]}_0Q^{[m]}_j$, the infinite volume average has to exist.

To determine $\langle Q^{[n]}_0\rangle_{P,V}$, one can start from the microscopic definition above and use that  $Q^{[n],N}$ depends only on the eigenvalues of the Lax matrix. The other method, employed here, is to simply differentiate the infinite volume free energy. We start with $n=0$ and note that the average stretch
\begin{equation}\label{3.41} 
\nu= \langle Q^{[0]}_{0}\rangle_{P,V} = \partial_P F_\mathrm{to}(P,V) = \partial_P\mu(P,V) = \Big(\int _\mathbb{R}\mathrm{d}w  \partial_\mu\rho_\mathsf{n}(w) \Big)^{-1},
\end{equation}
where the last equality results from differentiating Eq. \eqref{3.36} as 
\begin{equation}\label{3.41a} 
1 = (\int \mathrm{d}w\partial_\mu\rho_{\mathsf{n},\mu}(w)) \partial_P\mu(P). 
\end{equation}
For $n\geq 1$ we perturb $V$ as $V_\kappa(w) = V(w) + \kappa w^n$ and differentiate the free energy at $\kappa = 0$.  Then 
\begin{equation}\label{3.42} 
 \langle Q^{[n]}_{0}\rangle_{P,V} = \partial_\kappa F_\mathrm{to}(P,V_\kappa)\big|_{\kappa = 0} = \partial_P \partial_\kappa \mathcal{F}_\mathrm{de}(\rho^\star(P,V_\kappa))\big|_{\kappa = 0} 
 \end{equation}
 and, first introducing the linearization of $\rho^\star$ as
\begin{equation}\label{3.43} 
\rho{^\star}' = \partial_\kappa \rho^\star(P,V_\kappa )\big|_{\kappa = 0},
\end{equation}
one obtains
\begin{eqnarray}\label{3.44} 
 && \hspace{-40pt}  \partial_\kappa \mathcal{F}_\mathrm{de}(\rho^\star(P,V_\kappa))\big|_{\kappa = 0} =
\int_\mathbb{R} \mathrm{d}w \rho^\star(w,P,V)w^n + \int_\mathbb{R} \mathrm{d}w  \rho{^\star}'(w)\Big(V(w) + \log \rho^\star(w,P,V)\nonumber\\
 &&\hspace{80pt} 
  - 2 \int_\mathbb{R} \mathrm{d}w' \rho^*(w',P,V) \log|w-w'|\Big) 
  \end{eqnarray}
using that $\int \mathrm{d}w \rho{^\star}'(w)= 0$.
 Integrating the Euler-Lagrange equation  \eqref{3.35} at $\mu = \mu(P)$ against $\rho{^\star}'$, the terms on the right side of \eqref{3.43} vanish and
 \begin{equation}\label{3.45} 
 \langle Q^{[n]}_{0}\rangle_{P,V} 
 =  \int_\mathbb{R}\mathrm{d}w \partial_P\rho^\star(w,P,V) w^n. 
\end{equation}
Thus the Lax DOS is given by
\begin{equation}\label{3.46} 
 \rho_\mathrm{Q}(w) = \partial_P\rho^\star(w).
\end{equation}
Naively one might have guessed that the Lax DOS equals $\varrho^\star$. But the slow linear variation of the pressure in the Dumitriu-Edelman identity amounts to a slightly deviating  result. 

In the literature the Euler-Lagrange equation \eqref{3.35} is written differently by formally defining a Boltzmann weight as
 \begin{equation}\label{3.47} 
 \rho_\mathsf{n}(w) = \mathrm{e}^{-\varepsilon(w)}
\end{equation}
with quasi-energy $\varepsilon(w)$. Then
\begin{equation}\label{3.48} 
 \varepsilon(w) = V(w) - \mu  -  2 \int_\mathbb{R} \mathrm{d}w'  \log|w-w'| \mathrm{e}^{-\varepsilon(w')},
 \end{equation}
In addition one also introduces the particle density through
 \begin{equation}\label{3.49} 
 \rho_{\mathsf{p},\mu}(w) = \partial_\mu\rho_{\mathsf{n},\mu}(w),
\end{equation}
 at the moment just a convenient terminology.
  
 The structure uncovered is familiar from the Yang-Yang thermodynamics of the Lieb-Liniger $\delta$-Bose gas,
 which is an integrable quantum many-body  system and solved by Bethe ansatz. 
 For quantum integrable systems the analogue of \eqref{3.48} is called TBA 
 (thermodynamic Bethe ansatz) equation. We will call \eqref{3.48} \textit{classical TBA equation} or simply TBA, despite the fact that no Bethe ansatz had been used in its derivation. Some patience is required to fully appreciate this analogy. 
Further evidence will be accumulated from an alternative route based on scattering coordinates, see Section \ref{sec9.2} for the Toda lattice
 and Section \ref{sec11.3} for the Calogero fluid. The quantum side of the analogy will be covered 
  in Chapter \ref{sec13} for the $\delta$-Bose gas and in Chapter \ref{sec14} for the quantum Toda lattice.

 Later on we will use some identities based on TBA. We collect them here, together with introducing standard notations.
 The Hilbert space of square integrable functions on the real line is denoted by $L^2(\mathbb{R}, \mathrm{d}w)$ with scalar product
 \begin{equation}\label{3.50} 
\langle f,g\rangle = \int_\mathbb{R} \mathrm{d}w f(w)^* g(w).
 \end{equation}
 We will work mostly with real functions and then the complex conjugation in \eqref{3.50} can be omitted. There will be many integrals 
 over $\mathbb{R}$ and a convenient  shorthand is simply
  \begin{equation}\label{3.51} 
\langle f \rangle = 
 \langle 1,f\rangle  = \int_\mathbb{R} \mathrm{d}w f(w).
 \end{equation}
 To distinguish, an average over some probability measure is denoted by  $ \langle \cdot\rangle_{P,V,N}$, carrying suitable subscripts.
 Starting  from the $Q^{[n]}$'s, so far a discrete basis has been used. Obviously any linear combination of conserved fields is still conserved
 and, as in other linear problems, the choice of basis is an important consideration. From the viewpoint of Lax DOS, the label 
 $n$ corresponds to the monomial $w^n$, which will continued to be used and is denoted by 
  \begin{equation}\label{3.52} 
 \varsigma_{n}(w) = w^n, 
  \end{equation}
including  $ \varsigma_{0}(w) =1$. More generally, the set of basis functions will depend on the particular  integrable model under consideration.

Let us define the integral operator 
\begin{equation}\label{3.53}
Tf(w) =  \int_\mathbb{R} \mathrm{d}w' \phi_\mathrm{to}( w-w') f(w') = 2 \int_\mathbb{R} \mathrm{d}w' \log |w-w'| f(w')
\end{equation}
with $w \in \mathbb{R}$. Then the  TBA equation can be rewritten as 
\begin{equation}\label{3.54} 
\varepsilon (w)  = V(w) -\mu   - (T \mathrm{e}^{-\varepsilon})(w).
 \end{equation}
 In addition one introduces the  dressing of a real-valued function $f$  through
\begin{equation}\label{3.55} 
f^\mathrm{dr} = f + T \rho_\mathsf{n} f^\mathrm{dr},\quad f^\mathrm{dr} = \big(1 - T\rho_\mathsf{n}\big)^{-1} f,
\end{equation}
where $\rho_\mathsf{n}$ is regarded as multiplication operator, i.e. $(\rho_\mathsf{n}f)(w) = \rho_\mathsf{n}(w)f(w) $.
With our improved notation, the Lax DOS \eqref{3.46} can be written as 
\begin{equation}\label{3.56} 
\rho_{\mathrm{Q}} =  \partial_P\rho_\mathsf{n} = ( \partial_P\mu)\partial_\mu \rho_\mathsf{n} = \nu \rho_\mathsf{p},
 \end{equation}
where \eqref{3.49} has been used. Since $\rho_{\mathrm{Q}}$ is normalized, 
\begin{equation}\label{3.57} 
 \nu\langle\rho_\mathsf{p}\rangle = 1.
 \end{equation}
Physically the central objects of the theory are $\nu$ and $\nu\rho_\mathsf{p}$, since they encode the GGE average of the conserved fields. Differentiating TBA with respect to $\mu$ we conclude
\begin{equation}\label{3.58} 
\rho_\mathsf{p}= (1 - \rho_\mathsf{n} T)^{-1} \rho_\mathsf{n} = \rho_\mathsf{n}(1 - T\rho_\mathsf{n})^{-1}\varsigma_0 = \rho_\mathsf{n} \varsigma_0^\mathrm{dr}, 
 \end{equation}
which expresses $\rho_\mathsf{p}$ as a functional of $\rho_\mathsf{n}$. This relation can be inverted to yield 
\begin{equation}\label{3.59} 
\rho_\mathsf{n}(w)= \frac{\rho_\mathsf{p}(w)}{1 +  T\rho_\mathsf{p}(w)}. 
 \end{equation}
For later purposes, we also state the definition
 \begin{equation}\label{3.60} 
q_n =   \langle Q^{[n]}_{0}\rangle_{P,V}  = \nu\langle\rho_\mathsf{p}\varsigma_n\rangle,
\end{equation}
$n \geq 1$, not to be confused with a particle position. 

Identities as \eqref{3.54} to \eqref{3.59} will reappear in other, either classical or quantum, integrable systems. Due to their wide use, 
these identities are referred to as \textit{TBA formalism} which in a specific way reflects the underlying free energy functional.
  \bigskip\\
$\blackdiamond\hspace{-1pt}\blackdiamond$~\textit{Uniqueness of solutions of the TBA equation}.\hspace{1pt}  For the Toda lattice at given $\mu$ the TBA equation has two solutions.
 At first glance this looks surprising. In fact, in a standard numerical solution scheme one follows a particular branch, say starting from small $P$, and encounters  
 an end-point at which instabilities arise. So, some explanations are in demand.
 
Firstly $\mu(P)$ is concave and its derivative, $\nu(P)$, is strictly decreasing. For example, in the case of thermal equilibrium 
 $\mu(P) = \log \sqrt{\beta/2\pi} +P\log \beta -\log\Gamma(P)$,  
which has a single maximum at $P=P_\mathrm{c}$.  The physics is rather obvious. At very small $P$ the average stretch is huge and diverges
as $P \to 0$. By increasing pressure the stretch is  decreased. Since there is no hard core, increasing $P$ even further the stretch becomes negative. In physical space, for small $P$, up to small random errors, the labelling of particles is increasing. But at large $P$ the labelling is reversed.
At $P_\mathrm{c}$ the stretch vanishes
and the typical distance between particles with adjacent index is of order $1/\sqrt{N}$.
The function  inverse to $\mu(P)$ has two branches, meaning that for given $\mu$ there are two values of $P$.

Now considering the densities, by construction $\rho_{\mathsf{n}} \geq 0$,  $\nu \rho_\mathsf{p} \geq 0$, $\nu \langle\rho_\mathsf{p}\rangle =1$.
$\rho_{\mathsf{n},\mu(P)} $ is pointwise increasing in $P$ and  varies smoothly through $P_\mathrm{c}$, so does $\nu \rho_{\mathsf{p},\mu(P)}$. On the other hand, $\rho_{\mathsf{p},\mu(P)}(w)$ diverges to $+\infty$
as $P$ approaches $P_\mathrm{c}$ from the left, globally flips to  $-\infty$ at $P_\mathrm{c}$, and then flattens out as $P \to \infty$.
\hfill$\blackdiamond\hspace{-1pt}\blackdiamond$
\subsubsection{Thermal equilibrium}.\hspace{1pt}
Thermal equilibrium corresponds to the quadratic confining potential $V(w) = \tfrac{1}{2} \beta w^2$
with $\beta$ the inverse temperature. 
Only for this particular case the diagonal entries of the Lax matrix, 
 $\{p_j, j \in \mathbb{Z}\}$,  are independent  with $p_j$ a Gaussian random variable of mean zero and variance $\beta^{-1}$.
Hence $\langle (p_j)^n \rangle_{P,\beta}  = 0$ for odd $n$ and  $\langle (p_j)^n \rangle_{P,\beta}  = (n-1)!!(\beta)^{-n/2}$ for even $n$. 
The off-diagonal entries, $\{a_j, j \in \mathbb{Z}\}$,  are also independent with $a_j$ a 
$\chi$ distributed random variable with parameter $2P$. In particular for the even moments $\langle (a_j)^{2n} \rangle_{P,\beta}  = 
P(P+1)\ldots(P+n -1)$, $n = 1,2,\ldots\,$. Due to independence, the free energy of the chain is easily computed with the result
\begin{equation}\label{3.61}
F_\mathrm{eq}(P,\beta) =  \log \sqrt{\beta/2\pi} +P\log \beta -\log\Gamma(P).
\end{equation}
However to figure out the entire DOS requires the TBA machinery. 

We start from the Euler-Lagrange equation for the free energy \eqref{3.27}, set $V(w) = \tfrac{1}{2} \beta w^2$, differentiate
with respect to $w$, and multiply the resulting expression by $\varrho^\star$. Then
\begin{equation}\label{3.62} 
( \beta w + \partial_w) \varrho^\star(w) -  2 P \int_\mathbb{R} \mathrm{d}w' \frac{1}{w - w'} \varrho^\star(w)\varrho^\star(w') = 0.
 \end{equation}
 Note that $\beta$  scales by setting
 \begin{equation}\label{3.63}
 \varrho^\star_{P,\beta}(w) = \sqrt{\beta} \varrho^\star_{P,1}(\sqrt{\beta}w).
  \end{equation}
 Hence, for simplicity, we set $\beta = 1$, omit the explicit dependence  on $P$, and denote  by $\varrho^\star$ the solution to \eqref{3.62}.
 
\begin{figure}[!t]
	\centering
	\includegraphics[width=.45\linewidth]{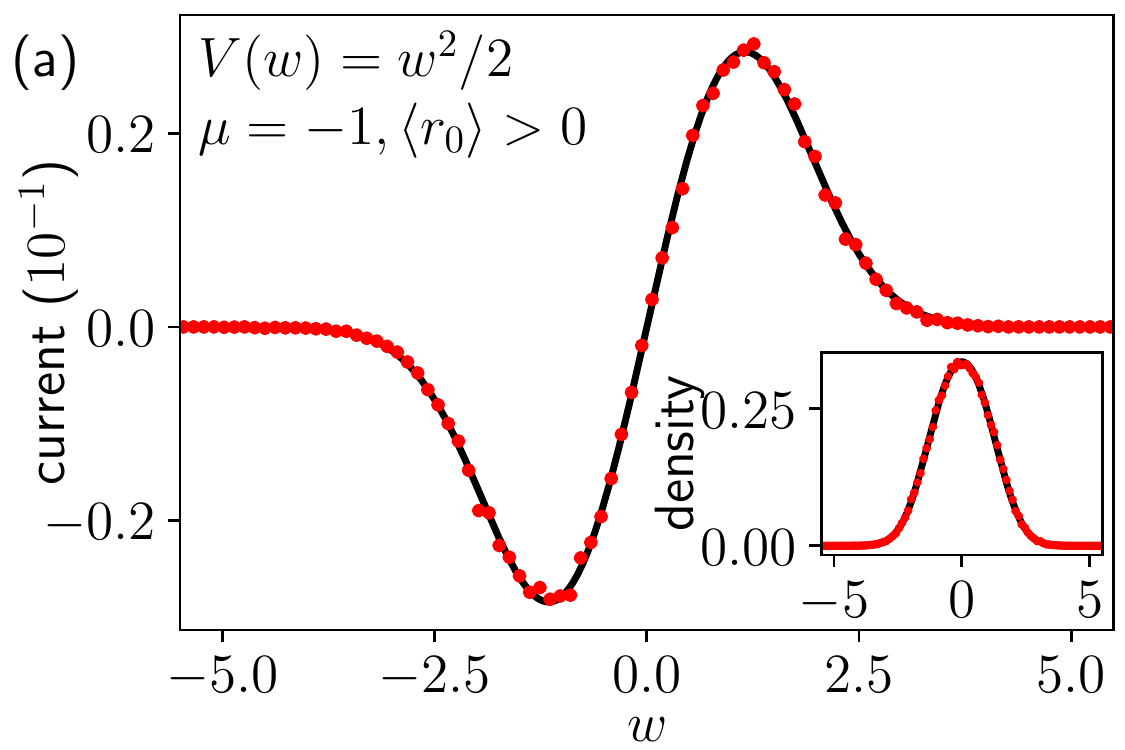}
	\includegraphics[width=.45\linewidth]{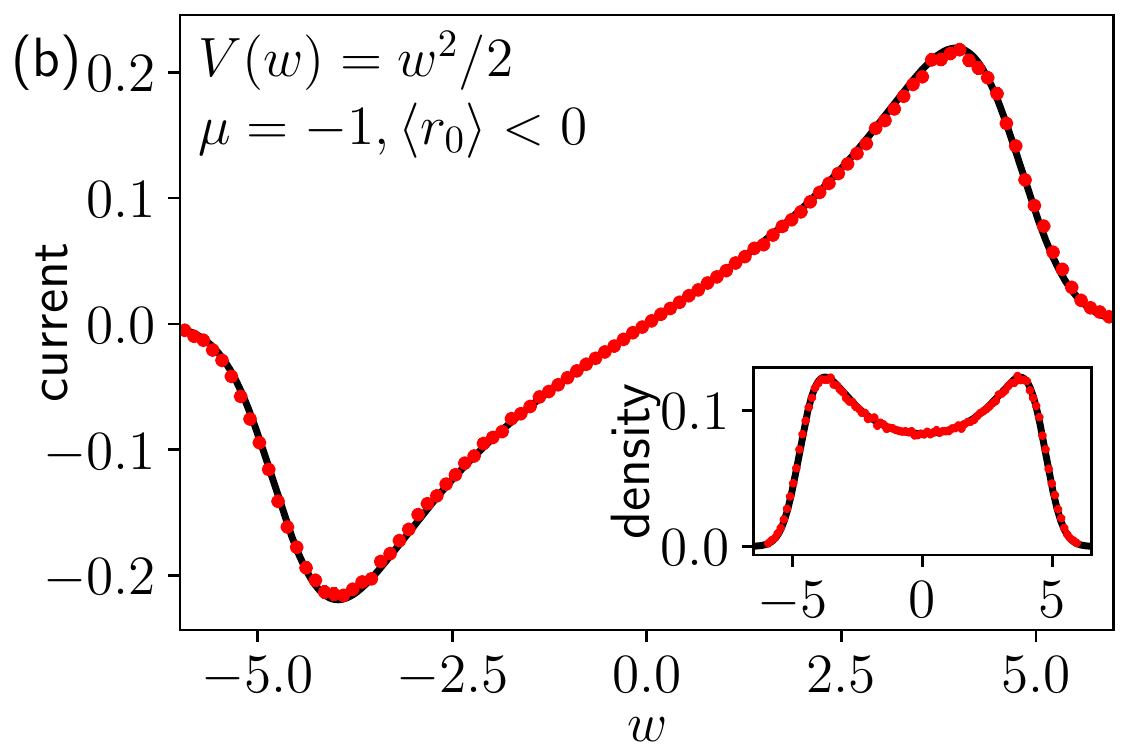} \\
	\includegraphics[width=.45\linewidth]{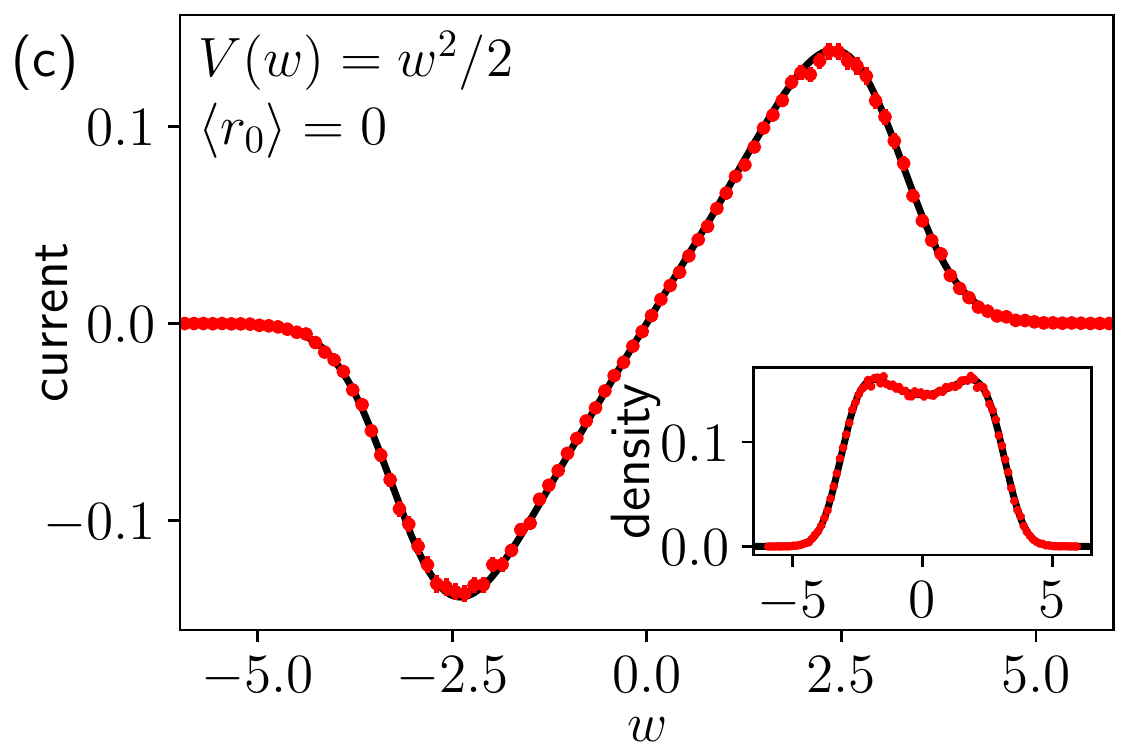}
	\includegraphics[width=.45\linewidth]{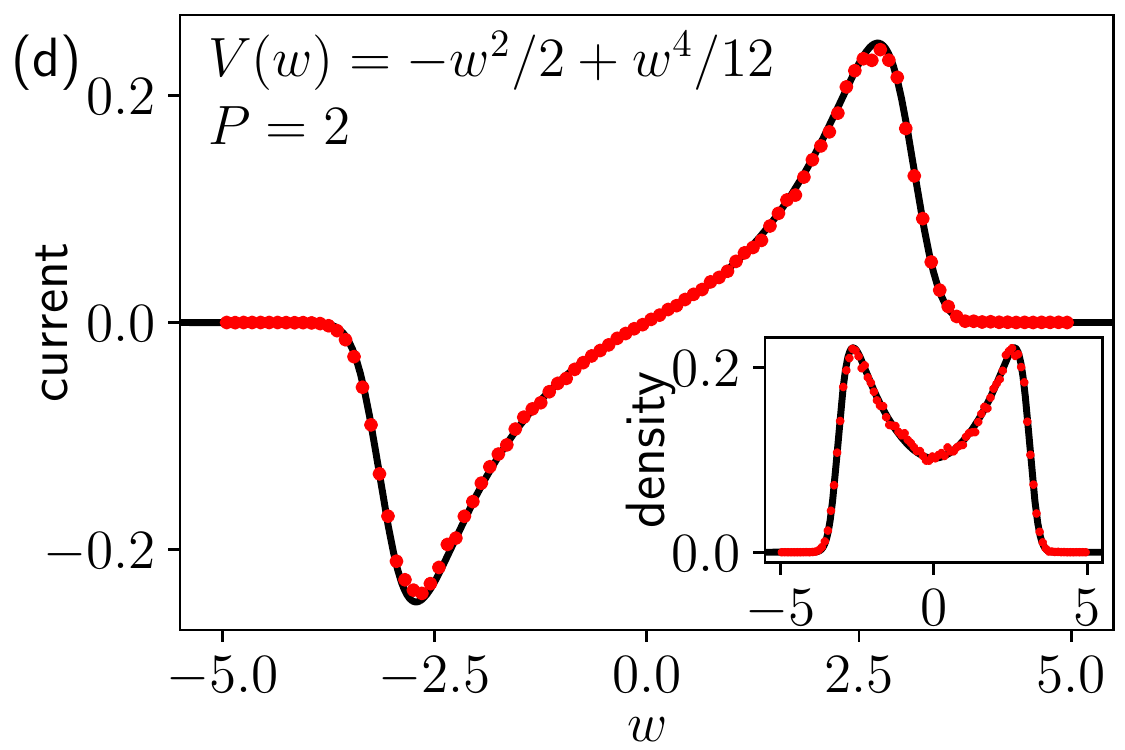}
\caption{For the Toda lattice shown is the current DOS $\rho_\mathrm{J}$ and as inset the corresponding DOS $\rho_\mathrm{Q}$. For panels (a), (b), (c) the Toda lattice is in thermal equilibrium with  $\beta = 1$ and $(\nu,P) =  (7.64,0.138), \,( -1.28, 4.10),\, (0, 1.461)$, respectively. Panel (c) corresponds to the critical pressure $P_\mathrm{c}$,  while (a) and (b) have the same value of $\mu$, thereby confirming that the TBA equation has two distinct solutions.  Panel (d) displays the same simulation for a quartic confining potential at $P = 2$, $\beta = 1$.  For the DOS, sampled are the eigenvalues of the Lax matrix at $N= 1024$ (red dots) and compared with the numerical solution of the TBA equation as based on the nonlinear Fokker-Planck equation \eqref{4.11}. For the current DOS, the expression \eqref{2.24} is sampled from the random Lax matrix and compared with $\rho_\mathrm{J}$ computed through the TBA formalism, see Eq. \eqref{6.17}. From Cao et al.\cite{CBS19}.}
\label{fig5}
\end{figure}
 
 Taking the Stieltjes transform, 
 \begin{equation}\label{3.64}
 g(z) =   \int_\mathbb{R} \mathrm{d}w\varrho{^\star}(w) \frac{1}{w -z},
  \end{equation}
yields the equation
\begin{equation}\label{3.65}
 zg(z) + \frac{d}{dz}g(z) + P g(z)^2 = -1 .
  \end{equation}
Setting $g(z) = u'(z)/u(z)$, Eq. \eqref{3.65} transforms to the linear second order differential equation
\begin{equation}\label{3.66}
 u''(z) + zu'(z)+ Pu(z) = 0.
  \end{equation}
Since $\varrho{^*}$ is a probability density, the $|z| \to \infty$ asymptotics,
 \begin{equation}\label{3.67}
 u(z) \simeq \frac{c_1}{z^P},
  \end{equation}
follows.
  Finally changing to the function $u(z) = \mathrm{e}^{z^2/4} y(z) $ one arrives at the Schr\"{o}dinger type equation
\begin{equation}\label{3.68}
 y''(z) + \big( P - \tfrac{1}{2} - \tfrac{1}{4} z^2\big)y(z) = 0,
  \end{equation}
  which can be solved in terms of parabolic cylinder functions. The appropriate linear combination is determined by the asymptotic condition \eqref{3.67}. Somewhat unusually, the Stieltjes transform can be still inverted and yields the fairly explicit expression
  \begin{equation}\label{3.69}
\varrho{^\star}(w) = \frac{\Gamma(P)\mathrm{e}^{-w^2/2}}{\sqrt{2 \pi}P|\hat{D}_P(w)|^2}, \qquad
\hat{D}_P(w) = 
\int_0^\infty \mathrm{d}t  \, t^{P - 1} \mathrm{e}^{\mathrm{i}wt} \mathrm{e}^{-\frac{1}{2}t^2}. 
\end{equation}
Particular examples are shown in Figure 5, which however are obtained from numerically solving a nonlinear Fokker-Planck equation rather than using \eqref{3.69}, see Chapter \ref{sec4}.

We reintroduce the dependence on $P,\beta$. For small $P$, the Lax off-diagonal matrix elements $a_j \to 0$ and hence in leading order $\varrho_{P,\beta}^\star(w) = (2\pi)^{-1/2} \exp[-\tfrac{1}{2}\beta w^2]$. 
On the other hand, one can integrate \eqref{3.62} against $\varsigma_n(w)$ to obtain a recursion relation for the even moments of $\varrho_{P,\beta}^\star$, $c_n = \langle \varrho_{P,\beta}^\star(w) w^{2n}\rangle$, 
\begin{equation}\label{3.70}
\beta c_n = (2n-1)c_{n-1}  +  \mathfrak{p}\beta \sum_{j=0}^{n-1}c_{n-1-j} c_j,
\end{equation} 
with  $n =1,2,\ldots\,$ and $c_0 = 1$, where we switched to the physical pressure $\mathfrak{p}= P/\beta$. For low temperatures, $\beta \to \infty$, the  term $(2n-1)c_{n-1}(\beta)$ can be neglected implying the asymptotic result $c_n(\beta ) \to c_n(\infty )$ with
\begin{equation}\label{3.71}
c_n(\infty) =  \mathfrak{p}^n   \frac{1}{n+1}
 \begin{pmatrix}
2n\\
n
\end{pmatrix}. 
\end{equation}  
On the right side one notes the Catalan numbers. Hence $\varrho_{\mathfrak{p}\infty,\infty}^\star$ is the normalized Wigner semi-circle probability density function
\begin{equation}\label{3.72}
\varrho_{\mathfrak{p}\infty,\infty}^\star(w) = \frac{1}{2\pi \mathfrak{p}} \sqrt{4 \mathfrak{p} - w^2} 
\end{equation}
for $|w| \leq 2\sqrt{\mathfrak{p}}$. To obtain the Lax DOS one still has to act with the operator $\partial_P P$. The low pressure Gaussian does not change. For low temperatures. upon applying the operator $\partial_\mathfrak{p} \mathfrak{p}$ one concludes the convergence 
 \begin{equation}\label{3.73}
\lim_{\beta\to\infty, P = \mathfrak{p}\beta}\rho_\mathrm{Q}(w) = \frac{1}{\pi \sqrt{4\mathfrak{p} - w^2}}
  \end{equation}
for $|w| \leq 2\sqrt{\mathfrak{p}}$
in the sense of convergence of moments. The Lax eigenvalues concentrate close to the borders $\pm 2\sqrt{\mathfrak{p}}$. In probability theory the density \eqref{3.73} is known as centered arcsine law. 

This convergence can also be obtained from the random walk expansion of $\langle Q^{[ n]}_0 \rangle_{\mathfrak{p}\beta,\beta}$. Then every path with at least one horizontal step carries a weight with at least one factor of  $\beta^{-1}$. Thus in the limit
one has to count only the number of simple random walks with $n$ steps starting at $j=0$ and ending at $j=0$, each step carrying the weight $\mathfrak{p}^{1/2}$. This number is nonzero only for even $n$, in which case
 $\langle Q^{[2 n]}_0 \rangle_{\mathfrak{p}\infty,\infty} = (n+1)c_{n(\infty)}$. 
\subsection{Mean-field techniques}
\label{sec3.5}
Not to interrupt the line of arguments, we only claimed that the free energy for the Dumitriu-Edelman partition function can be obtained by minimizing the free energy  functional $\mathcal{F}_\mathrm{de}^\circ$. In  fact, in all integrable models studied so far, both classical and quantum, there is such a variational characterization for the generalized free energy. Thus it is worthwhile  to elucidate the main characteristics independently from the specific applications we are interested in. From a probabilistic perspective the problem belongs to the theory of large deviations, as pioneered by  Donsker and Varadhan. For our purposes a more simple-minded statistical mechanics approach will do.  But beware: we only explain the main steps. More space would be needed for a complete proof in particular cases. 

We will outline what mathematicians call  a proof by compactness. A simple example would be a sequence of real numbers $y_1, y_2,\ldots$ with $|y_j| \leq a$, the condition of compactness. 
One wants to determine the large $j$ limit $y_j \to y_\infty$. Due to the uniform bound, one can choose a subsequence $\tilde{y}_j$ such that  $\tilde{y}_j \to \tilde{y}_\infty$ for $j \to \infty$. The crucial input is some information on $\tilde{y}_\infty$. In our context a
natural example  would  be to have some function $g:[-a,a] \to \mathbb{R}$ with the property that $g(\tilde{y}_\infty) =0$. Assuming in addition that
$g$ has a unique zero, $\bar{y}$, one concludes that $\lim_{j \to \infty} \tilde{y}_j  = \bar{y}$. Since this argument applies to any convergent subsequence, one arrives at $\lim_{j \to \infty} y_j  = \bar{y}$.  To obtain some information on the rate of convergence would require additional considerations.

We turn to the mean-field type problem by considering $N$ ``spins" $x_1,\ldots,x_N$ with $x_j \in \mathbb{R}$. The ``energy" of a spin configuration is taken to be of the particular form
\begin{equation}\label{3.74}
H_N(x) = \sum_{j=1}^N V(x_j) + \frac{1}{N}\sum_{i,j=1}^NW(x_i,x_j).
\end{equation}
Thereby one defines partition function  and Gibbs probability density relative to $\mathrm{d}^N\hspace{-1pt} x$,
\begin{equation}\label{3.75}
Z_N = \int_{\mathbb{R}^N}\mathrm{d}^N\hspace{-1pt}x \exp\big(- H_N(x)\big),\qquad \mu^N(x) = Z_N^{-1}\exp\big(- H_N(x)\big).
\end{equation}
The superscript $N$ refers to a probability density on $\mathbb{R}^N$ with standard definitions for free energy, energy, and entropy, 
\begin{equation}\label{3.76}
F(\mu^N) = - \log Z_N, \qquad E(\mu^N) = \langle H_N\rangle_{\mu^N}, \qquad S(\mu^N)= - \int_{\mathbb{R}^N}\mathrm{d}^N\hspace{-1pt}x 
\mu^N \log \mu^N.
\end{equation}

As before, the main goal is the large $N$ limit of the empirical density
\begin{equation}\label{3.77}
\varrho_N(w) =  \frac{1}{N}\sum_{j=1}^N \delta (w - x_j).
\end{equation}
The energy of our model is very special in the sense that it is merely a functional of the empirical density,
\begin{equation}\label{3.78}
H_N = N\Big( \int_\mathbb{R}\mathrm{d}w\varrho_N(w) \big( V(w) +  \int_\mathbb{R}\mathrm{d}w'\varrho_N(w')W(w,w')\big)\Big).
\end{equation}
The quadratic functional serves here only as illustration. More complicated functionals will appear in Chapters \ref{sec9} and \ref{sec11}.
Of course the potential $V$ has to be sufficiently confining.  In a first run the kernel $W(w,w') = W(w',w)$ could taken to be bounded.
As central observation, $\mu^N$ is invariant under permutation of indices.  As a striking consequence, for example, the two-point function $\langle f_1(x_i)f_2(x_j)\rangle_{\mu^N}$ does not depend on the choice of  $(i,j)$. Clearly, such  a large symmetry group will force special properties. 

The mentioned  compactness can be ensured by bounds as $Z_N \leq \mathrm{e}^{c_0 N}$. Thus there exists a subsequence,
denoted again as $\mu^N \to \mu_\infty$, where the limit  is a probability measure on $\mathbb{R}^\mathbb{N}$. Now by a theorem of Hewitt and Savage, in a simplified context earlier studied by De Finetti, the limit measure must be a convex combination of product probability measures. A pure product measure is written as
\begin{equation}\label{3.79}
\mu_\varrho = \varrho\times\varrho\times\ldots
\end{equation}
with $\varrho \geq 0$ and $\langle \varrho\rangle = 1$. Each factor has finite entropy and hence is of the form $\varrho(w) \mathrm{d} w$ with finite 
\begin{equation}\label{3.80}
\mathsfit{S}(\varrho) = - \int_\mathbb{R}\mathrm{d}w\varrho(w)\log \varrho(w).
\end{equation}
The entropy and energy per site equal 
\begin{equation}\label{3.81}
s(\mu_\varrho) = \mathsfit{S}(\varrho), \qquad e(\mu_\varrho) =  \langle V,\varrho\rangle +  \langle \varrho,W\varrho\rangle = \mathsfit{E}(\varrho).
\end{equation}
A convex combination of product measures is written as
\begin{equation}\label{3.82}
\mu = \int \nu(\mathrm{d}\varrho|\mu) \mu_\varrho.
\end{equation}
The set of such convex combinations is denoted by $\mathcal{S}$. Entropy and energy per site now turns to 
\begin{equation}\label{3.83}
s(\mu) = \int \nu(\mathrm{d}\varrho|\mu) \mathsfit{S}(\varrho),\qquad e(\mu) =  \int \nu(\mathrm{d}\varrho|\mu) \mathsfit{E}(\varrho).
\end{equation}
While the second identity is obvious, the first one looks wrong and is so at finite volume. But consider $\alpha \varrho_1\times\ldots \times \varrho_1 + (1-\alpha)
\varrho_2\times\ldots \times \varrho_2$, each product having $N$ factors. For large $N$ these densities are supported on essentially disjoint
sets in phase space and the integral defining the total entropy is in leading approximation $\alpha N\mathsfit{S}(\varrho_1) +
(1 -\alpha) N\mathsfit{S}(\varrho_2)$.

We now choose a (sub)sequence such that $\mu^N \to \mu_\infty$. Then the limit
\begin{equation}\label{3.84}
\lim_{N\to \infty} \frac{1}{N} E(\mu^N) = e(\mu_\infty)
\end{equation}
exists.
\smallskip\\
\textit{(i) lower bound}. For $\mu^N$ the marginal onto $[1,\ldots,n]$ is denoted by $\mu^N_n$ and correspondingly by $\mu_n$ for some infinite volume measure by $\mu$. We consider large $N$ and divide the volume into disjoint segments each of length $n$. Then, by subadditivity of entropy,
\begin{equation}\label{3.85}
S(\mu^N) \leq \Big\lfloor\frac{N}{n}\Big\rfloor S(\mu^N_n) + \mathcal{O}(1)
\end{equation}
with $\lfloor\cdot\rfloor$ indicating integer part. Therefore
\begin{equation}\label{3.86}
\lim_{N\to \infty} \frac{1}{N} S(\mu^N) \leq \lim_{n \to \infty}\frac{1}{n}s(\mu_{\infty,n}) = s(\mu_\infty)
\end{equation}
and, since $f = e -s$, 
\begin{equation}\label{3.87}
\lim_{N\to \infty} \frac{1}{N} F(\mu^N) \geq e(\mu_\infty)- s(\mu_\infty)= f(\mu_\infty).\smallskip
\end{equation}
\textit{(ii) upper bound}. By the standard finite volume variational principle for the free energy
\begin{equation}\label{3.88}
\lim_{N\to \infty}\frac{1}{N} F(\mu^N) \leq\lim_{N\to \infty}\min_{\mu' \in \mathcal{S}} \frac{1}{N}F(\mu'_N) = 
\min_{\mu' \in \mathcal{S}}f(\mu') \leq f(\mu_\infty).\smallskip
\end{equation}
Combining both bounds
\begin{equation}\label{3.89}
\lim_{N\to \infty}\frac{1}{N} F(\mu^N) = f(\mu_\infty) = 
\min_{\mu' \in \mathcal{S}} \int \nu(\mathrm{d}\varrho'|\mu') \mathcal{F}(\varrho'),
\end{equation}
where the one-particle free energy functional reads 
\begin{equation}\label{3.90}
\mathcal{F}(\varrho) = \big\langle\varrho, V +\log \varrho  + W\varrho\big\rangle.  
\end{equation}

In mean-field models from equilibrium statistical mechanics  generically there are several minimizers,  indicating a phase transition in the microscopic system.
For integrable many-body systems uniqueness seems to be the rule. 
Assuming such property to hold and denoting the unique minimizer by $\varrho^* $, our formulas 
simplify drastically, 
\begin{equation}\label{3.91}
\lim_{N\to \infty}\frac{1}{N} F(\mu^N) = \cal{F}(\varrho^*).
\end{equation}
Modulo technical points, this is the limit claimed in \eqref{3.28}. Furthermore, subsequence can be replaced by sequence
and 
\begin{equation}\label{3.92} 
\mu^N \to \mu_{\varrho^*}.
\end{equation}

The density of states, integrated against some test function $f$, equals $\langle f\varrho_N \rangle$ which is still random. From \eqref{3.92}
one deduces the limits
\begin{equation}\label{3.93}
\lim_{N\to \infty} \big\langle\langle f\varrho_N\rangle\big\rangle_{\mu^N} = \langle f\varrho^*\rangle,\qquad
\lim_{N\to \infty} \big\langle\big(\langle f\varrho_N\rangle\big)^2\big\rangle_{\mu^N} = \langle f\varrho^*\rangle^2,
\end{equation}
which establishes the law of large numbers for $\langle f\varrho_N\rangle$, in other words the almost sure convergence  of $\langle f\varrho_N \rangle$ to the deterministic limit $\langle f\varrho^*\rangle$ as $N \to \infty$.

In statistical mechanics mean-field usually refers to an approximation where a fluctuating field is replaced by its average.
In contrast,
for integrable many-body systems the respective mean-field problem is exact and arrives unexpectedly. For the Toda lattice the transformation of Dumitriu-Edelman
accomplishes the deal. As to be discussed, for the Calogero fluid the proof relies on  scattering coordinates and for the Ablowitz-Ladik system
on the transformation linked to CMV matrices. For quantum many-body systems a mean-field type problem naturally arrives through the Bethe ansatz, more generally through the asymptotic Bethe ansatz.

\newpage
\begin{center}
 \textbf{Notes and references}
 \end{center}
 \begin{center}
 \textbf{Section 3.0}
\end{center} 
At age twenty-four Ludwig Boltzmann \cite{B68a} wrote his fundamental contribution on  the microcanonical
  ensemble. He argued that this ensemble provides the natural description of the long time behavior of a mechanical system. In addition,
he uncovered the connection to thermodynamics. Recommended is his very readable letter-style account Boltzmann \cite{B68b}. The way how we teach equilibrium statistical mechanics today goes back to the must-read  book by J. Williard Gibbs \cite{G02}.\bigskip
\begin{center}
\textbf{Section 3.1}
\end{center} 
The generalized Gibbs ensemble came naturally into focus when studying the quench of integrable systems 
 starting from a translation invariant state, see the reviews Polkovnikov et al. \cite{PSS11}, Vidmar and  Rigol \cite{VR16}, the special volume edited by Calabrese et al. \cite{CEM16}, and the more recent short review with many references by
 Alba and Calabrese \cite {AC17}. In this context the conserved charges averaged over the initial state determine the parameters of the GGE obtained in the long time limit.
In some circles the notion ``generalized" was initially not so welcome, since there is a long tradition in the study of Gibbs measures 
for a general class of potentials, for a minute subset see  Ruelle \cite{R69}, Lanford \cite{L73}, Friedli and Velenik  \cite{FV17}. In our context ``generalized" means that for the given mechanical system one has a high dimensional  set of time-invariant measures all of them having the form anticipated by Statistical Mechanics. 

Equivalence of ensembles in its original meaning refers to the property that the resulting thermodynamic potentials are related to each other through the respective Legendre transform. This is a widely studied subject, with close relations to the theory of large deviations, Lanford \cite{L73}, Simon  \cite{Si16}, Friedli and Velenik \cite{FV17}. In our context we use a stronger version:  In the limit  of infinite volume, two distinct ensembles constructed from the same bulk hamiltonian yield  identical distributions of strictly local observables. To put it differently,  provided the respective parameters are transformed according to the rules of thermodynamics, correlations of local functions are identical. Such a stricter version generically fails  at phase transitions, for a detailed discussion see Georgii  \cite{G94,G95}.
\bigskip  
 \begin{center}
 \textbf{Section 3.2}
\end{center} 
 A general problem of nonequilibrium statistical mechanics is to determine the slowly varying  fields. The obvious candidates are the locally conserved fields. In higher dimensions the equilibrium phase diagram has more structure. For example, gas and fluid phase may coexist. The respective phase boundary is then also a slowly varying field. More generally, the order parameters of the thermodynamic phases have to be included in the list, see  Forster  \cite{F75}.
  Generalized hydrodynamics (GHD) follows the same approach. For the Toda lattice the Lax filter is a natural construction, which is applicable to other integrable systems whenever there is a Lax matrix available. 
\newpage
 \begin{center}
 \textbf{Section 3.3}
\end{center} 
 The generalized free energy of the Toda lattice has been obtained in Spohn \cite{S19}, see also Doyon \cite{D19a,D19b} for complimentary discussions. 
  Originally, Dumitriu and Edelman \cite{DE02} investigated how the fully occupied GUE random matrix transforms isospectrally to a 
  tridiagonal matrix. For this purpose they iteratively applied the Householder transform, which is a standard numerical scheme. Once Dumitriu  and Edelman had understood 
 how this scheme works for GUE random matrices, they realized that their algebra holds for arbitrary values of the parameter $\beta$ and not only for $\beta = 2$. In their context the quadratic confining potential appears naturally. Since  the transformation of volume elements is established, the extension to general confining potentials is straightforward.  The Dumitriu-Edelman identity has been used to obtain a fairly detailed information on the edge behavior of the general $\beta$-ensemble, see   Ramirez et al. \cite{RRV11} and Bourgade   et al. \cite{BEY14} for further developments. A standard reference on log gases is the monumental volume by  Forrester \cite{F10}. Variational problems with a logarithmic kernel are investigated by
Saff and Totik \cite{ST97}. 
 
 In a recent contribution Guionnet and Memin \cite{GM21} prove the generalized free energy to be  given by \eqref{3.30} and the 
 almost sure limit of the DOS as stated in \eqref{3.16} with limit $\nu \rho_\mathsf{p}$. Their theorem covers confining potentials such that $\lim_{|w|\to \infty} w^{-2n} V(w) =a_{\infty,n} >0$ for some positive integer $n$. In particularly, $V$ is not necessarily given by a convergent power series. From the hydrodynamic perspective, required would be also the inverse operation. One chooses $\rho_\mathsf{n}$ from a specified class of functions and concludes that $V$ is determined by Eq. \eqref{3.48}. 
 A further line of research, not yet covered at all, concerns the spatial structure of GGEs.
 This includes the decay of correlation functions and mixing properties. For such an analysis of use could  
 be Dobrushin's statistical mechanics theory, Dobrushin  \cite{D74}, developed specifically for one-dimensional systems. 
 \bigskip
 \begin{center}
 \textbf{Section 3.4}
\end{center} 
 The particular form of the TBA formalism can be better grasped upon reading Section \ref{sec13}. The double-valuedness of solutions to the TBA equation was pointed out to me by Bulchandani  et al. \cite{BCM19} with numerical confirmation in  Cao  et al. \cite{CBS19}. Numerical plots of the densities as they vary through $P_\mathrm{c}$ can be found in Mendl and Spohn \cite{MS21}.
The TBA equation \eqref{3.48} with quadratic confining potential was first obtained by Opper \cite{O85}, who investigated 
the classical limit of the quantum Toda chain. He already obtained the solution \eqref{3.69}. Our discussion is based on 
 Allez et al. \cite{ABG12}, a study of Dyson Brownian motion with weak interaction, see Chapter \ref{sec4} for details.   A related study has been carried out for $\beta$-Wishart ensembles by  Allez et al. \cite{ABM12}. The recursion relation \eqref{3.70} appears already in Duy and  Shirai  \cite{DS15} and Duy \cite{D18}. \bigskip
  \begin{center}
  \textbf{Section 3.5}
\end{center} 
 Out of the many texts and monographs on the theory of large deviations recommended is  the book by Dembo and  Zeitouni \cite{DZ10}.
 The complete proof based on permutation symmetry can be found in Messer and  Spohn \cite{MS82}. A more recent study are the lecture notes by Rougerie \cite{R15} who also presents an anologous discussion  of  quantum mean-field models.  De Finetti's theorems on exchangeable measures are discussed in Aldous \cite{A85}. The generality required in our context is covered by Hewitt and Savage \cite{HS55}.
\newpage
\section{Dyson Brownian motion}
\label{sec4}
\setcounter{equation}{0}
In the mid-sixties Freeman Dyson pioneered the study of random matrices, as initiated by Eugene Wigner and others 
as a phenomenological model for the complex energy spectra of highly excited nuclei. More specifically, Dyson studied the statistics of eigenvalues
for the Gaussian orthogonal, unitary, and symplectic ensembles. For this purpose he introduced a stochastic dynamics of eigenvalues with the property that its stationary distribution coincides with the distribution of  eigenvalues of the respective random matrix ensemble. Such a diffusion process is now called
Dyson Brownian motion. In this approach, one studies a physically more intuitive stochastic particle system,
which in the long time limit approaches the desired probability density of eigenvalues. Particles move in one dimension, their positions being denoted by
$\{x_j(t), j = 1,\ldots,N\}$, and are governed by the coupled stochastic differential equations,      
\begin{equation}\label{4.1}
dx_j(t) = -V'(x_j(t))dt + 2\beta\sum_{i = 1,i\neq j}^N \frac{1}{x_j(t) - x_i(t)} dt + \sqrt{2} db_j(t), 
\end{equation}
for  $j = 1,\ldots,N$ and with $\beta \geq 0$. The particles repel each other with a $1/x$ force and are confined by the  external potential $V$, which in fact will coincide with the generalized chemical potential  \eqref{3.5} of the Toda lattice -- in the first place the reason for using the same name and
symbol. $\{ b_j(t), j = 1,\ldots,N\}$ is a collection of independent standard Brownian motions.
For later purposes the generator, $\mathcal{L}_N$, of the coupled diffusion processes is defined as the linear operator
\begin{equation}\label{4.2}
\mathcal{L}_N \mathsfit{f}(x) = \sum_{j=1}^N\Big(\partial_{x_j}- V'(x_j) + 2\beta\sum_{i = 1,i\neq j}^N \frac{1}{x_j - x_i}\Big) \partial_{x_j}\mathsfit{f}(x), 
\end{equation}
acting on functions on configuration space, $\mathsfit{f}: \mathbb{R}^N \to \mathbb{R}$,
$x = (x_1,\ldots,x_N)$. For $t \geq 0$,
the kernel of the semigroup
\begin{equation}\label{4.3}
\mathrm{e}^{\mathcal{L}_Nt}(x,x') \mathrm{d}^N\hspace{-1pt} x' 
\end{equation}
is the transition probability, in other words, the probability density at time $t$ given the initial configuration $x$. 

The reader might suspect a luxury detour. But this is not the case. Dyson Brownian motion is a powerful method to numerically 
solve the TBA  equations of the Toda lattice. Even more importantly, as will be discussed, the GGE averaged currents are linked to
eigenvalue fluctuations. To study such properties  Dyson Brownian motion is a convenient tool.

Dyson investigated the particular values $\beta = 1,2,4$, which possess a large symmetry group. But recently there has 
been much progress also for general $\beta$. In our context, we will have to study the case of very small $\beta$,
specifically $\beta = \alpha/N$. In fact, $\alpha =P$ in the application to the Toda lattice. In \eqref{4.1} the drift is the gradient 
of a potential. Hence the diffusion process is reversible and its unique stationary measure is given by
\begin{equation}\label{4.4}
(Z_{\alpha,V,N})^{-1} \exp\!\Big(- \sum_{j=1}^N V(x_j) 
+\frac{\alpha}{N}\sum_{i,j =1, i \neq j}^N  \log| x_i - x_j| \Big),
\end{equation} 
 which agrees with the normalized version of the probability density function appearing  in \eqref{3.21}. Our strategy will be to first study the large $N$ limit of the dynamics and then deduce information on the stationary measure, which is our real interest. 
\subsection{Macroscopic equation, law of large numbers}
\label{sec4.1}
We choose some smooth test function $f$ and introduce the empirical
 density, $\varUpsilon_N(x,t)$, through
 \begin{equation}\label{4.5}
\varUpsilon_N(f,t) = \frac{1}{N} \sum_{j = 1}^N f(x_j(t))  = \int_\mathbb{R}\mathrm{d}x  \varUpsilon_N(x,t) f(x). 
\end{equation}
Then
\begin{eqnarray}\label{4.6} 
&&\hspace{-45pt}d\varUpsilon_N(f,t)= \varUpsilon_N( -V'\partial_x f + \partial_x^2f,t) dt\\
&&\hspace{10pt}+ \alpha \int_\mathbb{R}\mathrm{d}x \int_\mathbb{R} \mathrm{d}y
\frac{f'(x) - f'(y)}{x-y} \varUpsilon_N(x,t)\varUpsilon_N(y,t) dt + \frac{1}{N}  \sum_{j = 1}^N f'(x_j(t))\sqrt{2} db_j(t).\nonumber
 \end{eqnarray}
 When integrated in time, the process averaged square of the noise term becomes
\begin{equation}\label{4.7}
 \frac{2}{N^2} \sum_{j = 1}^N \int _0^t \mathrm{d} s \mathbb{E}\big(f(x_j(s))^2\big).
\end{equation}
Thus the noise is of order $1/\sqrt{N}$ and vanishes in the limit $N \to \infty$.

We assume a starting measure such that with probability one the initial empirical density has the limit
$\varrho_0$,
 \begin{equation}\label{4.8}
\lim_{N \to \infty} \varUpsilon_N(f,0) = \int_\mathbb{R} \mathrm{d}x\varrho_0(x) f(x). 
\end{equation} 
In the statistical physics literature this property is often referred to as ``self-averaging", meaning that the limit is non-random and no averaging is required.
Since in \eqref{4.6} only the drift term survives, one concludes  the limit
\begin{equation}\label{4.9}
\lim_{N \to \infty}  \varUpsilon_N(f,t) = \varUpsilon(f,t) = \int_\mathbb{R} \mathrm{d}x \varrho(x,t) f(x)
\end{equation} 
for all $t>0$, again with probability one. The limit density $ \varrho(x,t) $ then satisfies
\begin{eqnarray}\label{4.10} 
&&\hspace{-30pt}\frac{d}{dt} \int_\mathbb{R}\mathrm{d}x \varrho(x,t) f(x) \\
&& =  \int_\mathbb{R}\mathrm{d}x\varrho(x,t)(-V'\partial_x f + \partial_x^2f)(x)
+ \alpha \int_\mathbb{R}\mathrm{d}x \int_\mathbb{R} \mathrm{d}y
\frac{f'(x) - f'(y)}{x-y} \varrho(x,t) \varrho(y,t) \nonumber
\end{eqnarray}
together with the initial data $\varrho_0(x)$. Written pointwise, $\varrho(x,t)$ is governed by the nonlinear Fokker-Planck equation
\begin{equation}\label{4.11}
\partial_t \varrho(x,t) = \partial_x\big( V_\mathrm{eff}'(x,t)\varrho(x,t) + \partial_x  \varrho(x,t)\big).
\end{equation} 
The bare confining potential $V$ is modified to  an effective potential given by 
\begin{equation}\label{4.12}
V_\mathrm{eff}(x,t) = V(x) - \alpha (T \varrho)(x,t),
\end{equation} 
 the  integral operator $T$ being defined in \eqref{3.53}.

Our particular interest is the stationary Fokker-Planck equation 
\begin{equation}\label{4.13}
\partial_x\Big( V'(x)\varrho_\mathfrak{s}(x) - 2\alpha\int _\mathbb{R}\mathrm{d}y\frac{1}{x-y}\varrho_\mathfrak{s}(y) \varrho_\mathfrak{s}(x)  + \partial_x  \varrho_\mathfrak{s}(x)\Big) = 0,\quad
  \int _\mathbb{R}\mathrm{d}x\varrho_\mathfrak{s}(x) = 1,
  \end{equation} 
which under the constraint
\begin{equation}\label{4.13a}
\int_\mathbb{R}\mathrm{d}x\varrho_\mathfrak{s}(x) = 1,
 \end{equation} 
has a unique strictly positive solution $\varrho_\mathfrak{s}$.
Since Dyson Brownian motion is time-reversible, the large round bracket itself has to vanish. We compare to the mean-field free energy
functional 
$\mathcal{F}^\circ_\mathrm{de}(\varrho)$ in Eq. \eqref{3.27}, setting $P = \alpha$. Its minimizer $\varrho^\star$ satisfies the Euler-Lagrange equation
\begin{equation}\label{4.14} 
  V(x)  - \mu -  2\alpha \int_\mathbb{R} \mathrm{d}x'  \log|x-x'| \varrho^\star(x') +\log \varrho^\star(x) + 1 = 0,
  \end{equation}
where $w$ has been substituted by $x$. Differentiating with respect to $x$ and then multiplying by $\varrho^\star$ yield
\begin{equation}\label{4.15} 
  V'(x)\varrho^\star(x) -  2 \alpha \int_\mathbb{R} \mathrm{d}y \frac{1}{x-y} \varrho^\star(x)\varrho^\star(y) + \partial_x\varrho^\star(x) = 0.
 \end{equation}
Thus we conclude $\varrho_\mathfrak{s} = \varrho^\star$.

The nonlinear Fokker-Planck equation depends smoothly on $\alpha$ and so does $\varrho_\mathfrak{s}$.
The issue of two-valuedness appears only in the TBA equation. 
\subsection{Fluctuation theory}
\label{sec4.2}
The nonlinear Fokker-Planck equation is the result of a law of large numbers. Thus one would expect to have
a central limit type theorem which captures the order $1/\sqrt{N}$ fluctuations. Quite generically the fluctuations are governed
by a linear Langevin equation for the conserved field, in our case the density field. The drift term is determined by the linearized macroscopic equation. But in addition 
the dynamics generates an effective noise term, whose precise structure depends on the model.

Our interest is the stationary dynamics, hence the initial state of the $N$-particle system is the one in Eq. \eqref{4.4}. To study the fluctuations of the density it is convenient to  introduce the fluctuation field as 
\begin{equation}\label{4.16}
\phi_N(f,t) = \frac{1}{\sqrt{N}} \sum_{j=1}^N \big(f(x_j(t)) - \langle\varrho_\mathfrak{s} f \rangle\big) = \int_\mathbb{R} \mathrm{d}x f(x) \phi_N(x,t).
\end{equation}
As explained in more detail below,  there is a Gaussian random field $\phi(f,t)$, jointly in $f$ and $t$, such that in distribution
 \begin{equation}\label{4.17}
\lim_{N\to \infty} \phi_N(f,t) = \phi(f,t). 
\end{equation} 
Note that the limit is still random. The limit field   $\phi$ is governed by the linear Langevin equation
\begin{equation}\label{4.18}
\partial_t \phi(x,t) = \partial_x \big(D\phi(x,t) + \sqrt{2\varrho_\mathfrak{s}(x)}\xi(x,t)\big).
\end{equation}  
Here  $\xi(x,t)$ is normalized spacetime Gaussian white noise, $\mathbb{E}\big(\xi(x,t)\xi(x',t')\big) =
\delta(x - x')\delta(t-t')$ and $D$ the linear operator 
\begin{equation}\label{4.19}
D =  V'_\mathrm{eff} + \partial_x - \alpha \varrho_\mathfrak{s} T' .
\end{equation} 
Thus $\partial_x D$ is the Fokker-Planck evolution operator linearized  at $\varrho_\mathfrak{s}$. The effective potential 
$V_\mathrm{eff}$ is still defined as in \eqref{4.12} upon  substituting $\varrho(x,t)$  by $\varrho_\mathfrak{s}(x)$. 

The Gaussian process $\phi(x,t)$ is stationary in time, has  mean zero and is uniquely characterized by its covariance
$\mathbb{E}\big(\phi(x,t)\phi(x',t')\big)$. For the Toda lattice of interest is the spatial covariance 
\begin{equation}\label{4.19a}
\mathbb{E}\big(\phi(x,0)\phi(x',0)\big)
= C^\sharp(x,x'). 
\end{equation}
As a general property of linear Langevin equations with time-independent coefficients,
$C^\sharp$ is determined by 
\begin{equation}\label{4.20}
\langle D^*\partial_x f,C^\sharp g\rangle + \langle f,C^\sharp D^*\partial_x g\rangle = 2\langle \partial_x f, \varrho_\mathfrak{s} \partial_x g\rangle
\end{equation} 
with $D^*$ the adjoint operator to $D$. We claim that, as an operator, the solution is 
\begin{equation}\label{4.21}
C^\sharp = (1 - \alpha \varrho_\mathfrak{s} T)^{-1}\varrho_\mathfrak{s} - \big\langle(1 - \alpha\varrho_\mathfrak{s}T)^{-1}\varrho_\mathfrak{s}
\big\rangle^{-1}  \big|(1 - \alpha \varrho_\mathfrak{s} T)^{-1}\varrho_\mathfrak{s}\big\rangle \big\langle (1 - \alpha\varrho_\mathfrak{s}T)^{-1}\varrho_\mathfrak{s}\big|,
\end{equation} 
where for simplicity we use the Dirac notation $|\cdot\rangle\langle \cdot|$ for the one-dimensional projector. The subtracted  term ensures that the number of particles does not fluctuate, i.e. $C^\sharp \varsigma_0 = 0$.

To confirm the claim, we consider only the left most term of \eqref{4.20}, the other one following by symmetry, and have to show that  
\begin{equation}\label{4.22}
\langle \partial_x f,D \varrho_\mathfrak{s} (1 -\alpha T \varrho_\mathfrak{s})^{-1}g\rangle = \langle \partial_x f, \varrho_\mathfrak{s} \partial_x g\rangle.
\end{equation} 
Upon  replacing $g$ by $(1 -\alpha T \varrho_\mathfrak{s})g$, one arrives at
\begin{equation}\label{4.23}
\langle \partial_x f,D \varrho_\mathfrak{s} g\rangle = \langle \partial_x f, \varrho_\mathfrak{s} \partial_x (1 -\alpha T \varrho_\mathfrak{s})g\rangle.
\end{equation} 
$\varrho_\mathfrak{s}$ satisfies the stationary nonlinear Fokker-Planck equation \eqref{4.15},  
\begin{equation}\label{4.24}
\big(V_\mathrm{eff}'  +  \partial_x\big)\varrho_\mathfrak{s} = 0.
\end{equation} 
When inserted in \eqref{4.23} this leads to the condition 
\begin{equation}\label{4.25}
\varrho_\mathfrak{s}  \partial_x g -\alpha \varrho_\mathfrak{s} T' ( \varrho_\mathfrak{s}g) = \varrho_\mathfrak{s} \partial_x (1 -\alpha T \varrho_\mathfrak{s})g,
\end{equation} 
which is easily checked.  \bigskip\\
$\blackdiamond\hspace{-1pt}\blackdiamond$~\textit{Martingales and central limit theorem}.\hspace{-1pt}  In probability theory tremendous efforts
have been invested to
develop tools for proving the central limit theorem in case of dependent random variables. Such techniques
can be applied to our infinite-dimensional setting, which is required since the test functions form an infinite-dimensional linear space. 
Only the basic computational steps are explained here. A full proof would be too technical. 
For example, we would have to discuss the existence of solutions to \eqref{4.1}. In fact, if $\beta > 1$, trajectories never touch and the solution theory is standard. But for $\beta < 1$ the crossing probability is 
strictly positive and the existence of solutions becomes more intricate. The proof of the limit \eqref{4.17} is based on compactness. In an appropriate function space one ensures that along some subsequence the stochastic process of fluctuations has a limit. 
The limit process satisfies an equation for which uniqueness is established, which then implies convergence.
We consider only the stationary process, but the time-dependent case can be handled in a similar fashion. 

Using \eqref{4.2}, one arrives at
\begin{equation}\label{4.26}
\mathcal{L}_N\phi_N(f) = \phi_N( -V'\partial_x f + \partial_x^2f) 
+ \alpha \int_\mathbb{R}\mathrm{d}x \int_\mathbb{R} \mathrm{d}y
\frac{f'(x) - f'(y)}{x-y} \phi_N(x)\varUpsilon_N(y) = \varkappa_{[1],N}(f).
\end{equation}
For the quadratic variation we obtain
\begin{equation}\label{4.27}
\mathcal{L}_N\phi_N(f)^2  - 2\phi_N(f)\mathcal{L}_N\phi_N(f)= 2N^{-1}\sum_{j=1}^N f'(x_j)^2 = \varkappa_{[2],N}(f).
\end{equation}
As before,  $\varkappa_{[1],N}(f,t)$ is the function $\varkappa_{[1],N}(f)$ evaluated at the random configuration\\
$(x_1(t),\ldots,x_N(t))$, and the same for $\varkappa_{[2],N}(f,t)$. By the standard theory of Markov processes,
\begin{equation}\label{4.28}
M_{[1],N}(f,t) = \phi_N(f,t) - \phi_N(f,0) - \int_0^t \mathrm{d}s \varkappa_{[1],N}(f,s)
\end{equation}
and
\begin{equation}\label{4.29}
M_{[2],N}(f,t) =  M_{[1],N}(f,t)^2 -  \int_0^t \mathrm{d}s \,\varkappa_{[2],N}(f,s)
\end{equation}
are martingales. 
  
By compactness one first ensures that the limit in  \eqref{4.17} exists. Since the martingale property is preserved in the limit $N \to \infty$,  using the law of large numbers to handle $\varkappa_{[1],N}(f)$ and $\varkappa_{[2],N}(f)$, one concludes that  
\begin{eqnarray}\label{4.30}
&&\hspace{-30pt} M_{[1]}(f,t) = \phi(f,t) - \phi(f,0) -  \int_0^t \mathrm{d}s \,\phi( -V'\partial_x f + \partial_x^2f,s) \nonumber\\
&&\hspace{20pt}-  \int_0^t \mathrm{d}s \, \alpha \int_\mathbb{R}\mathrm{d}x \int_\mathbb{R} \mathrm{d}y
\frac{f'(x) - f'(y)}{x-y} \phi(x,t)\varrho_\mathfrak{s}(y)
\end{eqnarray}
 and   
\begin{equation}\label{4.31}
M_{[2]}(f,t) =  M_{[1]}(f,t)^2 - 2 t \int_\mathbb{R}\mathrm{d}x\varrho_\mathfrak{s}(x) f'(x)^2
\end{equation}
are still martingales under the limit process $\phi(f,t)$.  The unique solution to the latter martingale problem is the linear Langevin equation of  \eqref{4.18}, thereby
confirming the claim.
\hfill$\blackdiamond\hspace{-1pt}\blackdiamond$
\newpage
\begin{center}
 \textbf{Notes and references}
 \end{center}
 \begin{center}
 \textbf{Section 4.0}
\end{center} 
Freeman Dyson introduced his model in Dyson   \cite{D62}. An interesting account of the early history is his preface to The Oxford Handbook on Random Matrix Theory  edited by Akemann et al.   \cite{ABF11}. The Handbook serves as a useful source of information and allows one to capture the vastness of the subject. 
An introductory monograph on random matrices is  Anderson et al. \cite{AGZ10}.
 Dyson Brownian motion also serves as a technical tool in the study of universality for Wigner random matrices, see Erd\H{o}s et al. \cite{EYY12},
 and the edge behavior of the density of states of the beta random matrix ensembles, see Bourgade   et al.  \cite{BEY14}. \bigskip
  \begin{center}
 \textbf{Section 4.1}
\end{center} 
For  general $\beta > 0$, Dyson Brownian motion is properly constructed in C\'{e}pa and  L\'{e}pingle   \cite{CL97}, where also the law of large numbers is proved.\bigskip
\begin{center}
 \textbf{Section 4.2}
\end{center}
Fluctuation theory at strong coupling is established in  Israelsson \cite{I01}. The methods used can be extended to the case discussed here.
 Fluctuation theory for a wider class of stochastic particle systems is discussed in Spohn  \cite{S91} with more detailed arguments and references on the  martingale method. For thermal equilibrium, eigenvalue fluctuations are proved by Nakano and  Trinh \cite{NT18} by a moment method.
 Hardy and  Lambert  \cite{HL21} use arguments from optimal transport to more directly estimate the distance between the true density function and  the approximating Gaussian.
\newpage
\section{Hydrodynamics for hard rods}
\label{sec5}
\setcounter{equation}{0} 
So far our focus has been the generalized free energy. But hydrodynamics relies on further building blocks.
The hard rod system will serve to illustrate the method as such, thereby building a bridge towards the Toda lattice. 
\begin{figure}[!b]
\centering
\includegraphics[width=0.7\columnwidth]{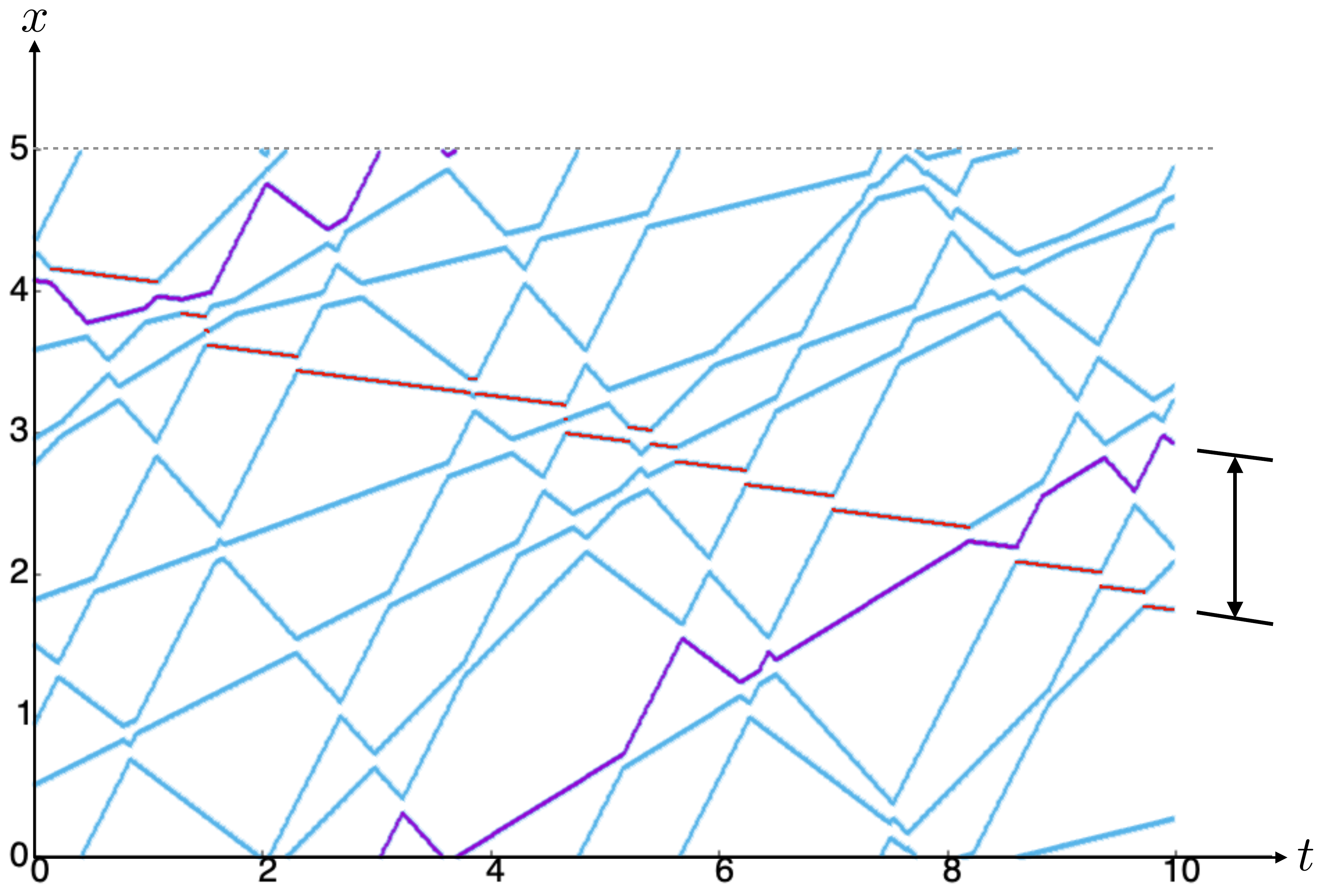}
\caption{Displayed are the trajectories, $q_j(t)$, of $9$ hard rod particles with length $0.3$ in a volume of size $5$. Very roughly the density $\bar{\rho} = 1.8$ and inverse temperature $\beta = 1$, which corresponds to a pressure $P=3.9 $. The trajectory  of the 8-th particle is shown in magenta and its quasiparticle trajectory in red. The scattering shift is indicated on the right. Courtesy of 
A. Kundu.}
\label{fig6}
\end{figure}
 \subsection{Hard rod fluid} 
 \label{sec5.1}
 The fluid consists of hard rods of length $\mathsfit{a}$, $\mathsfit{a} \in \mathbb{R}$, which interact through elastic collisions.
 The lattice version will be discussed in section \ref{sec5.2}. Positions are constrained by $r_j \geq  \mathsfit{a}$ and quasiparticles maintain their velocity through collisions. At a collision, i.e. $r_j = \mathsfit{a}$,
 $q_j$ jumps instantaneously  to $q_j  +  \mathsfit{a}$ and $q_{j+1}$ to $q_{j+1} -  \mathsfit{a}$. Besides the number of particles,
 any sum function of the velocities is conserved. This field is labelled by some general function, $\phi$, 
 and has a density given by
 \begin{equation}\label{5.1}
 Q^{[\phi]}(x) = \sum_{j \in \mathbb{Z}} \delta(q_j - x) \phi(p_j).
 \end{equation}
The particle number is included and corresponds to the constant function, $\phi(w) = 1$. The rod length can be negative. In fact, if $\mathsfit{a} <0$, then for sufficiently high pressure the free volume turns negative, a property  resembling the Toda lattice. 

For the current one has to differentiate the density with respect to $t$. Then there is the flow term resulting from the $\delta$-function and a collision term resulting  from  $\phi$. 
 For the latter
 we consider a small time $\mathrm{d}t>0$ and the change
$\phi(p_j(\mathrm{d}t)) - \phi(p_j)$. If $r_j -\mathsfit{a} < \delta$ with sufficiently small $\delta$ and if $p_{j+1}< p_j$, then $p_j(t) = p_{j+1}$.
Similarly, if $r_{j-1} -\mathsfit{a} < \delta$ and if $p_{j-1} >  p_j$, then $p_j(t) = p_{j-1}$, which implies
 \begin{eqnarray}\label{5.2}
&&\hspace{-34pt}\phi(p_j(\mathrm{d}t)) - \phi(p_j) 
= \delta(r_j - \mathsfit{a})(\phi(p_{j+1}) - \phi(p_j))\chi(\{p_{j+1}< p_j\})(p_j - p_{j+1}) \mathrm{d}t  \nonumber\\
&&\hspace{66pt}
+\delta(r_{j-1}- \mathsfit{a}) (\phi(p_{j-1}) - \phi(p_j))\chi(\{p_{j} < p_{j-1}\})(p_{j-1} - p_{j}) \mathrm{d}t,
\end{eqnarray}
where $\chi(\{\cdot\})$ is the indicator function of the set $\{\cdot\}$. Thus  
 \begin{eqnarray}\label{5.3}
 &&\hspace{-25pt}J^{[\phi]}(x) = \sum_{j \in \mathbb{Z}} \Big(\delta(q_j - x) p_j \phi(p_j) \\
 &&\hspace{0pt} +  \big(\theta(q_{j} -x) - \theta(q_{j-1} -x)\big) 
\delta(r_{j-1}- \mathsfit{a}) \big(\phi(p_{j-1}) - \phi(p_{j})\big)\chi(\{p_{_j} < p_{j-1}\})(p_{j-1} - p_{j})\Big)\nonumber
 \end{eqnarray}
 with $\theta$ the step function, $\theta(x) =0$ for $x\leq 0$ and $\theta(x) =1$ for $x > 0$. Under the hard rod dynamics the local conservation law has the standard form
 \begin{equation}\label{5.4}
\partial_t Q^{[\phi]}(x,t) +\partial_x J^{[\phi]}(x,t)=0.
\end{equation}
 
 A GGE is characterized by the one-particle distribution function $f(v) = \bar{\rho} h(v)$. Here $h \geq 0$ and $\langle h\rangle =1$. The velocities $\{p_j, j \in \mathbb{Z}\}$
 are a family of i.i.d. random variables with common probability density $h$.   
 The positions $\{q_j, j \in \mathbb{Z}\}$ are statistically invariant under spatial shifts and 
 $\{r_j = q_{j+1} - q_j \}$ are i.i.d., such that $r_j$ is exponentially distributed with density $P\exp\big(-P(x -\mathsfit{a})\big)\chi(\{x\geq \mathsfit{a}\})$, $P > 0$. Then its first moment $\langle x \rangle = \bar{\rho} = P/(1 + \mathsfit{a}P)$ and $\bar{\rho} \mathsfit{a} < 1$. Close packing is reached in the limit $P \to \infty$ with $\bar{\rho} =  1/\mathsfit{a}$.
In the sense of thermodynamics, $\bar{\rho}, h$ are extensive parameters, while $P,V$  are intensive parameters upon setting $h = \mathrm{e}^{-V}/\langle  \mathrm{e}^{-V}\rangle$.
 Slightly deviating from our standard  convention the GGE average is denoted by 
$\langle \cdot \rangle_{\bar{\rho},h}$. The required GGE averages are easily computed. For the fields one obtains 
\begin{equation}\label{5.5}
 \langle Q^{[\phi]}(0) \rangle_{\bar{\rho},h} = \bar{\rho} \langle  h \phi\rangle.
 \end{equation}
For the currents one uses that
\begin{equation}\label{5.6}
\int_\mathbb{R}\mathrm{d}v \int_\mathbb{R}\mathrm{d}w h(v) h(w)
\big(\phi(w) - \phi(v)\big) \chi(\{v< w\})(w-v) = \int_\mathbb{R}\mathrm{d}v \phi(v)(v-u)h(v)
\end{equation}
with the mean velocity $u = \langle vh(v)\rangle$. Then
\begin{equation}\label{5.7}
 \langle J^{[\phi]}(0) \rangle_{\bar{\rho},h} = \bar{\rho}\,\langle h  v^\mathrm{eff} \phi\rangle 
 \end{equation}
with the effective velocity
\begin{equation}\label{5.8}
  v^\mathrm{eff}(v)   = \frac{1}{1 - \mathsfit{a} \bar{\rho}} ( v -  \mathsfit{a} \bar{\rho} u).
\end{equation}

The hydrodynamic evolution equations are built from the average conserved fields and currents, i.e. the input \eqref{5.5} and \eqref{5.7}. For the initial state one assumes to have  locally the statistics of  some GGE. In our case, one example would be to maintain independence but let $f = \bar{\rho} h $ change very slowly  on the scale set  by the maximal correlation length. On a more formal level one introduces the dimensionless  parameter $\epsilon$,  $\epsilon \ll 1$, of a typical scale for the ratio ``microscopic length\,/\,macroscopic length''. Microscopically the physically relevant length is the  maximal correlation length of the initial GGE. Translated to the initial $f$, the scale separation is ensured through the functional form $f(\epsilon x; v)$. 
Given such random initial conditions, the hard rod fluid is evolving through free motion and collisions. Independence of momenta is lost immediately. Yet on the macroscopic scale nothing is moving,
 at least not for short times. Due to the conservation laws, changes can be observed only on microscopic times of order $\epsilon^{-1}t$.
Hydrodynamic scaling is ballistic, since  changes in space and in time are both $\mathcal{O}(\epsilon ^{-1})$ in microscopic units. GHD then postulates that    
even for such long times local GGE remains  approximately valid. If propagation of local GGE holds, one can write down the equations
governing the evolution of the parameters characterizing  local GGEs. 

The starting point is the exact microscopic conservation law \eqref{5.4}.
Since this law holds for any dynamical trajectory, one can average over an arbitrary ensemble, in particular the initial slowly varying state. 
Instead of averaging the time $t$ fields over initial conditions, we average the physical fields over the statistical state at time $t$, average denoted by $\langle \cdot \rangle_{t,\epsilon}$ with the subscript $\epsilon$ indicating the parameter of initial slow variation,
\begin{equation}\label{5.9}
\partial_t\langle Q^{[\phi]}(x)\rangle_{t,\epsilon} +\partial_x  \langle J^{[\phi]}(x)\rangle_{t,\epsilon} =0,
\end{equation}
which is still exact. Now, switching to macroscopic spacetime $(\epsilon^{-1}x, \epsilon^{-1}t)$ propagation of local GGE amounts to 
the approximation
\begin{equation}\label{5.10}
 \langle Q^{[\phi]}(\epsilon^{-1}x) \rangle_{\epsilon^{-1}t,\epsilon} \simeq \langle Q^{[\phi]}(0)\rangle_{f(x,t)},\qquad
 \langle J^{[\phi]}(\epsilon^{-1}x) \rangle_{\epsilon^{-1}t,\epsilon} \simeq \langle J^{[\phi]}(0) \rangle_{f(x,t)},
\end{equation}
where on the right hand side of each equation  $Q^{[\phi]}(x)$ is replaced by $Q^{[\phi]}(0)$, resp.  $J^{[\phi]}(x)$ by $J^{[\phi]}(0)$, because of the spatial translation invariance of the GGE average
$\langle \cdot \rangle_{f(x,t)}$. Using the effective velocity \eqref{5.7} evaluated at $(x,t)$, 
 we arrive at a closed equation for $f(t) = \bar{\rho}(t)h(t)$,
\begin{equation}\label{5.11}
\partial_t f(x,t;v) +\partial_x \big((1 - \mathsfit{a} \bar{\rho}(x,t))^{-1} ( v -  \mathsfit{a} \bar{\rho}(x,t) u(x,t))f(x,t;v)\big) = 0.
\end{equation}
Here the argument of $f$ indicates that $(x,t)$ is the spacetime patch under consideration and $v$ is the label of the conserved
field in that patch. Eq. \eqref{5.11} is the hydrodynamic equation of the hard rod fluid.  Through nonlinearity the hydrodynamic fields are coupled and decouple only in the ideal gas limit
$\mathsfit{a} \to 0$. For $\mathsfit{a} \geq 0$ the function $f(t)$ counts the number of particles in the macroscopic volume element $\mathrm{d}x \mathrm{d}v$ at time $t$. But for
$\mathsfit{a} < 0$, the density $\bar{\rho}(t)$ can have negative values, and so does $f(t)$.

On purpose we used the symbol $f$ to emphasize that this quantity is very close to the Boltzmann
$f$-function in the sense of  counting the number of particles in one-particle phase space. However the mathematical structure of the dynamical evolution is distinct from kinetic theory. On the Euler scale the Boltzmann collision term is not yet visible and 
Eq. \eqref{5.11} has some similarity with the Vlasov equation. In that case, the current at $(x,t)$ results from the force when averaged over particles, which is linear in $f$ and nonlocal in space. On the other hand for the hydrodynamic equation \eqref{5.11}, the current has a more complicated nonlinearity and is strictly local in $(x,t)$.

As an apparently general feature of generalized hydrodynamics, the hydrodynamic equation \eqref{5.11} can be written in a quasilinear form. For hard rods the transformation reads
\begin{equation}\label{5.12}
\rho_\mathsf{n}(v) = \frac{1}{\nu - \mathsfit{a}}f(v), \quad \nu \bar{\rho} = 1,
\end{equation}
and
\begin{equation}\label{5.13}
 	\partial_t \rho_\mathsf{n}(x,t;v) +v^\mathrm{eff}(x,t;v)\partial_x \rho_\mathsf{n}(x,t;v) = 0.
\end{equation}
The fields $\rho_\mathsf{n}(v)$ are the normal modes of the hard rod fluid equations and the convective equation \eqref{5.13} identifies their local propagation velocities as $v^\mathrm{eff}(v)$, which  depend nonlinearly on  $\rho_\mathsf{n}$ through \eqref{5.8}.

Hydrodynamic equations are expected to become exact in the limit $\epsilon \to 0$. But in practice, hydrodynamics becomes applicable much before the actual limit. The precise value of $\epsilon$ depends on the particular physical system and its initial conditions. The range of validity for hydrodynamics is mostly a qualitative rule of thumb, rather than a sharp error  \bigskip estimate.\\
$\blackdiamond\hspace{-1pt}\blackdiamond$~\textit{Hydrodynamics of simple fluids, the issue of mathematical rigor}.\hspace{1pt} In 1757 Leonhard Euler published his monumental
treatise on fluids, which in particular included the compressible Euler equations.  In this case the locally conserved fields are 
number, momentum, and total energy. Euler argued for the appropriate form of the currents on a phenomenological basis. It was a triumph of early Statistical Mechanics to understand that these currents can also be obtained from averaging the microscopic currents
over a Gibbs ensemble. The required technical tool is an identity known as virial theorem. Since the resulting equations 
agreed with what had been known already for a long time, the discussion was mostly confined to the textbook level.   
Still, the derivation was a further compelling evidence for the universal validity of Statistical Mechanics. For  integrable many-body
systems one cannot rely on phenomenological evidence. The only available method is to figure out the 
GGE averaged  currents as based on the microscopic model. 

A much harder question is the propagation of local GGEs. For the hard rod fluid Roland Dobrushin and collaborators 
proved such a result in the mid 80ies. But for more complicated integrable systems, in particular the Toda lattice, no
results are available yet. Due to their physical importance, for simple fluids there have been many attempts to establish the compressible Euler equations in a mathematically convincing way. From that perspective the best result is still the work of Olla, 
Varadhan, and Yau. They investigated the kind of dynamical mixing which would be needed for propagation of local equilibrium. In fact they modified the mechanical model by assuming that the deterministic collisions between particles are substituted by random ones, still respecting the mechanical conservation laws. Thus the macroscopic equations are the compressible version of Euler equations. 
For such stochastic dynamics, Olla, 
Varadhan, and Yau prove the required mixing. But for a mechanical system mixing would have to come from deterministic chaos. The resulting gap is so huge that progress is unlikely, even allowing for a long term perspective. 
\hfill$\blackdiamond\hspace{-1pt}\blackdiamond$
\subsection{Hard rod lattice}
\label{sec5.2}
Physically the fluid picture can be more easily visualized. But as in the case of the Toda chain, one can switch to the lattice version.
Then the dynamics is formulated in terms of the stretches $r_j$, where  $r_j(t) \geq \mathsfit{a}$ because of the hard core.
If $p_j(t) -p_{j+1}(t) < 0$, then $r_j(t)$ is decreasing and reaches collision at
$r_j(t_\mathrm{c}) = \mathsfit{a}$, $t_\mathrm{c}$ the collision time.   Instantaneously the momenta $(p_j,p_{j+1})$ are exchanged and $r_j(t)$ is increasing for
$t> t_\mathrm{c}$. Such local collisions happen for every spatial patch throughout time. Obviously, the densities of the conserved fields are
 \begin{equation}\label{5.14}
 Q^{[0]}_j = r_j, \qquad Q^{[\phi]}_j = \phi(p_j).
 \end{equation}
 Note that $\phi =1$ is just the constant function, hence not admissible. Thus, in contrast to the fluid, the index $0$ will have to be treated  differently 
 from $n\geq 1$. The currents are
\begin{equation}\label{5.15}
 J^{[0]}_j = -p_j
 \end{equation}
 and
 \begin{equation}\label{5.16}
 J^{[\phi]}_j = \delta(r_{j-1}- \mathsfit{a}) (\phi(p_{j-1}) - \phi(p_{j}))\chi(\{p_{j} < p_{j-1}\})(p_{j-1} - p_{j}).
 \end{equation} 
 Compared to \eqref{5.3}, there is no flow term and the factor $(\theta(q_{j} -x) - \theta(q_{j-1} -x))$ is missing.
 
 In a hard rod lattice GGE, the velocities are i.i.d. with probability density $h(v)$, $h \geq 0$, $\langle h \rangle = 1$ and 
 the $r_j$'s  are also i.i.d.  with  probability density $P\exp\big(-P(x -\mathsfit{a})\big)\chi(\{x\geq \mathsfit{a}\})$, $P>0$.  The average stretch is then
$\nu = P^{-1} +\mathsfit{a}$. In particular, $\nu - \mathsfit{a}>0$. Now $\nu, h$, are the extensive parameters, while $P,V$ are still intensive parameters. The GGE average is denoted by  $\langle \cdot \rangle_{\nu,h}$. The averaged conserved fields are given by 
\begin{equation}\label{5.17}
 \langle Q^{[0]}_0  \rangle_{\nu,h} = \nu, \qquad \langle Q^{[\phi]}_0  \rangle_{\nu,h} = \langle h \phi \rangle,
 \end{equation}
and the averaged currents by
\begin{eqnarray}\label{5.18}
&& \langle J^{[0]}_0 \rangle_{\nu,h} = - \langle h v\rangle = -u, \nonumber\\[1ex] 
&&\langle J^{[\phi]}_0 \rangle_{\nu,h} = (\nu - \mathsfit{a})^{-1}\big( \langle hv\phi\rangle - u\langle h \phi\rangle\big)
= \nu^{-1}\langle h(v^\mathrm{eff} - u) \phi\rangle,
 \end{eqnarray}
 where we used the same computation as in \eqref{5.6} and in Eq. \eqref{5.8} rewritten as  
$v^\mathrm{eff}(v)   = (\nu - \mathsfit{a})^{-1} ( \nu v -  \mathsfit{a}  u)$.
As a result the hydrodynamic equations of the hard rod lattice take the form
\begin{equation}\label{5.19}
\partial_t \nu(x,t) - \partial_x u(x,t) = 0, \quad
 \partial_t h(x,t;v) +\partial_x \big(\nu(x,t)^{-1}(v^\mathrm{eff}(x,t;v) -u(x,t))h(x,t;v)\big) = 0.
\end{equation}
Compared to \eqref{5.11}, there are now two equations rather than one.
But switching to normal modes through
\begin{equation}\label{5.20}
\rho_\mathsf{n}(v) = \frac{1}{\nu- \mathsfit{a}}h,
\end{equation}
one arrives again at the quasilinear equation \eqref {5.13}.
\subsection{TBA, collision rate ansatz}
 A central model-dependent feature of TBA is the two-particle scattering shift
as appearing in the definition of the $T$ operator. Now for hard rods the two-particle scattering shift equals $-\mathsfit{a}$.
To have the same structure as the Toda lattice would mean to replace $T$ in \eqref{3.48} by  
the projection operator 
 \begin{equation}\label{5.21}
 T = -\mathsfit{a} |1\rangle\langle 1|,
 \end{equation}
in other words $Tf(w) = - \mathsfit{a}\, \langle f\rangle$.
With this input the TBA equation for the hard rod lattice reads
\begin{equation}\label{5.22}
 V- \mu + \mathsfit{a} \,\langle \rho_\mathsf{n} \rangle +\log \rho_\mathsf{n} = 0, \quad \langle \rho_\mathsf{n} \rangle =P.
 \end{equation}
 Since $T$ is such a simple operator, the solution is still explicit with
 \begin{equation}\label{5.23}
\mu = \log P + \mathsfit{a} P - \log   \langle \mathrm{e}^{-V} \rangle
 \end{equation}
 and 
 \begin{equation}\label{5.24}
h =  \mathrm{e}^{-V} /\langle\mathrm{e}^{-V} \rangle, \quad \rho_\mathsf{n}  = (\nu - \mathsfit{a})^{-1}h,\quad  \rho_\mathsf{p}
 =  \nu^{-1}h.
 \end{equation}
The average of the conserved fields is determined by 
\begin{equation}\label{5.25}
\langle Q_0^{[\phi]}\rangle_{\nu,h}  =  \langle h\phi\rangle = \nu \langle \rho_\mathsf{p}\phi\rangle,
 \end{equation} 
 in agreement with \eqref{5.17}. Indeed, with adjusted $T$, the TBA formalism holds also for hard rods.

Qualitatively, the Toda lattice is best approximated by choosing $\mathsfit{a} <0$. The identities in \eqref{5.25} then provide a simple illustration for the uniqueness Insert below Eq. \eqref{3.60}. The stretch
$\nu(P) = (1/P) + \mathsfit{a}$ is monotone decreasing with $P_\mathrm{c} = -1/\mathsfit{a}$. Since $\nu - \mathsfit{a} >0$, in addition $\rho_\mathsf{n} \geq 0$.
However $\rho_\mathsf{p}$ globally flips sign at $P_\mathrm{c}$, while $\nu\rho_\mathsf{p} \geq 0$.

We now invert the logic and try to guess a formula for the GGE averaged currents of hard rods, through using a
conventional GHD argument. Employed is the notion of quasiparticles,  
most easily visualized in the fluid picture. A quasiparticle retains
 its velocity when undergoing a collision. While a physical particle rattles back and forth between its two neighbors, the quasiparticle moves with constant velocity interrupted by jumps of size $\pm\mathsfit{a}$ at collisions. We now prepare a GGE characterized
 by $\bar{\rho},h$ and initially start a tracer quasiparticle  at the origin with velocity $v$. Viewed on a somewhat coarser scale, when 
 time-averaging over  
 many collisions, the quasiparticle travels with a yet to be determined effective velocity $v^\mathrm{eff}(v)$, which turns out to be increasing in $v$.
 The tracer quasiparticle collides with a fluid quasiparticle of velocity $w$. Such fluid particles have density $\rho_\mathsf{p}(w)$.  If the collision is from the left, $v< w$, then the tracer jumps by
 $-\mathsfit{a}$.  Hence the collision rate is  $\rho_\mathsf{p}(w)(v^\mathrm{eff}(w) - v^\mathrm{eff}(v))$. On the other hand,   if the collision is from the right, $v>w$, then the tracer jumps by
 $\mathsfit{a}$ and the collision rate is  $\rho_\mathsf{p}(w)(v^\mathrm{eff}(v) - v^\mathrm{eff}(w))$. When integrating over all fluid quasiparticles, the bare tracer quasiparticle velocity is modified through collisions according to
 \begin{equation}\label{5.26}
 v^\mathrm{eff}(v) 
 = v - \mathsfit{a}\int_v^\infty \mathrm{d}w  \rho_\mathsf{p}(w)\big(v^\mathrm{eff}(w) - v^\mathrm{eff}(v)\big)
 + \mathsfit{a}\int_{-\infty} ^v \mathrm{d}w  \rho_\mathsf{p}(w)\big(v^\mathrm{eff}(v) - v^\mathrm{eff}(w)\big)
 \end{equation}
 and thus
 \begin{equation}\label{5.27}
 v^\mathrm{eff}(v) = v - \mathsfit{a}\int_\mathbb{R} \mathrm{d}w  \rho_\mathsf{p}(w)\big(v^\mathrm{eff}(w) - v^\mathrm{eff}(v)\big).
 \end{equation}
Equation \eqref{5.27} is the \textit{collision rate ansatz}.  More concisely in operator form,
 \begin{equation}\label{5.28}
 v^\mathrm{eff}(v)  = v + T(\rho_\mathsf{p}v^\mathrm{eff})(v) - (T\rho_\mathsf{p})(v) v^\mathrm{eff}(v).
 \end{equation}
For hard rods
the rate equation is easily solved with the result
\begin{equation}\label{5.29}
v^\mathrm{eff}(v)  =  v + \frac{\mathsfit{a}\bar{\rho}}{1 -  \mathsfit{a}\bar{\rho}} \int_\mathbb{R} \mathrm{d}w h(w) (v-w)   =      (1 - \mathsfit{a}\bar{\rho})^{-1}
( v -  \mathsfit{a} \bar{\rho} u).
\end{equation}
 Indeed, the average current equals $\bar{\rho} h v^\mathrm{eff} $, in agreement with the microscopic computation \eqref{5.8}.

With hindsight, the expression \eqref{5.27} for the effective velocity  carries already a recipe of how to extend to other integrable models. 
The two-particle scattering shift for the Toda fluid equals $2 \log|v - v'|$. Thus one might hope to obtain the correct average currents for the Toda fluid by considering the motion of a tracer quasiparticle. Analytically this amounts to 
substituting the operator $T$ in \eqref{5.28} by the operator $T$ from \eqref{3.53}. \bigskip
\begin{center}
 \textbf{Notes and references}
 \end{center}
 \begin{center}
 \textbf{Section 5.1}
\end{center} 
The hydrodynamic limit for hard rods, i.e. the asymptotic validity of \eqref{5.11}, is proved by  Boldrighini  et al. \cite{BDS83} under the assumption of a
  sufficiently random initial state. The limit holds with probability one. A short account is provided by Spohn  \cite{S91}. 
  A different proof is constructed by Ferrari et al. \cite{FF22}. From the perspective 
  of generalized hydrodynamics, hard rods are discussed by Doyon and  Spohn \cite{DS17a} and by Doyon  \cite{D19a}.  In experiments on integrable many-body systems,
  particles are moving in a one-dimensional tube and are confined by an external potential along the tube. Generically this breaks integrability. To elucidate more precisely the underlying mechanism, hard rods serve as an instructive model  in the contribution by Cao et al. \cite{CBM18}.
  
For a many-particle hamiltonian systems with short range interactions, C.B. Morrey launched an interesting early attempt,
Morrey \cite{M55}. The modification of adding randomness in collisions is studied by Olla et al. \cite{OVY93}.  The hydrodynamic limit on the Euler scale can be viewed as a stability result, 
  in the sense that on the ballistic scale there is no deviation yet from local equilibrium, resp. local GGE. 
  
  Starting in the early 1980ies it was realized that the hydrodynamic limit is also a challenging problem for stochastic lattice gases and interacting diffusions. The dynamics is overdamped in the sense that particles have only a position and evolve
  through random update rules. An early account is  Spohn  \cite{S91}.
\newpage
\section{Equations of generalized hydrodynamics }
\label{sec6}
\setcounter{equation}{0} 
\subsection{Average currents}
\label{sec6.1}
The microscopic version of the currents has been obtained already, see \eqref{2.24} and \eqref{2.26}. Now the goal is to compute their 
GGE average in the limit $N \to \infty$. As before the lattice size is $N$ and we use the pressure ensemble \eqref{3.6}. Then in terms of the spectral resolution of $L_N$ the $n$-th microscopic current reads
\begin{equation}\label{6.1}
J^{[n],N}  = \mathrm{tr}\big[(L_N)^nL_N^{\scriptscriptstyle \downarrow}\big] = N \langle  \rho_{\mathrm{J},N} \varsigma_n\rangle,
\end{equation}
 $n \geq 1$, with 
 \begin{equation}\label{6.2} 
\rho_{\mathrm{J},N} (w) = \frac{1}{N}  \sum_{j=1}^N \delta(w - \lambda_j)\Big(\sum_{i=1}^N a_i\psi_j(i) \psi_j(i+1)\Big).
\end{equation}
The $\delta$-peaks of the Lax density of states are  weighted by coefficients with arbitrary sign. In analogy, not to duplicate names, we call
$\rho_{\mathrm{J},N}$ the \textit{empirical current DOS}. For given GGE the current DOS is self-averaging and has the deterministic limit 
$\rho_\mathrm{J}$. In particular 
\begin{equation}\label{6.3}
\lim_{N\to \infty}\frac{1}{N} \langle J^{[n],N} \rangle_{P,V,N} = \langle J^{[n]}_0 \rangle_{P,V} = \langle\rho_{\mathrm{J}} \varsigma_n\rangle.
\end{equation}
Setting $n=0$ in \eqref{6.1} right hand side, one concludes $\langle \rho_\mathrm{J}\rangle = 0$. Hence, as physically expected, $\rho_\mathrm{J}$ cannot have a definite sign.
From the few numerical simulations available, very qualitatively   $\rho_\mathrm{J}(w)$ has the shape of $-\partial_w\rho_\mathrm{Q}(w)$.

Since the Dumitriu-Edelman identity worked so well for the conserved fields, one is tempted to use the same strategy 
for the currents. But now not only the distribution of eigenvalues is in demand. While we have an explicit form for the joint distribution of $\{\lambda_j,\psi_j(1), j = 1,\ldots,N\}$, to apply the map $\Phi$ of Dumitriu-Edelman to the weights $\sum_{i=1}^N a_i\psi_j(i)\psi_j(i+1)$ seems to be a complicated enterprise. A new 
approach is needed.
Unexpectedly, the key idea will come from the susceptibility matrices.

We start with the fields and define the infinite volume field-field correlator 
\begin{equation}\label{6.4} 
C_{m,n}(j-i) = \langle Q_j^{[m]}Q_i^{[n]} \rangle_{P,V}^\mathrm{c}
\end{equation}
for $m,n \geq 0$, where the superscript $^\mathrm{c}$ denotes truncation, respectively connected correlation, $\langle gf \rangle^\mathrm{c}_{P,V} = \langle gf \rangle_{P,V} -\langle g\rangle_{P,V}\langle f \rangle_{P,V}$. Truncated correlations decay rapidly to zero and the field-field susceptibility matrix is given by
\begin{equation}\label{6.5} 
C_{m,n} = \sum_{j\in \mathbb{Z}}C_{m,n}(j) = \langle Q^{[m]};Q^{[n]}\rangle_{P,V}, 
\end{equation}
where the second equality is merely a convenient notation. $C_{m,n}$ is the matrix of second derivatives of the generalized free energy. 
Correspondingly we introduce the field-current correlator and the field-current susceptibility matrix
\begin{equation}\label{6.6} 
B_{m,n}(j-i) = \langle J_j^{[m]} Q_i^{[n]} \rangle_{P,V}^\mathrm{c}, \quad B_{m,n} =  \sum_{j\in \mathbb{Z}} B_{m,n}(j).
\end{equation}
Despite its apparently asymmetric definition, $B$ satisfies
\begin{equation}\label{6.7} 
B_{m,n}(j) = B_{n,m}(-j).
\end{equation}
To prove, we employ the conservation law and spacetime stationarity to arrive at
\begin{eqnarray}\label{6.8} 
&&\hspace{-70pt}\partial_j \langle J_j^{[m]}(t) Q_0^{[n]}(0) \rangle_{P,V}^\mathrm{c} 
= -\partial_t  \langle Q_j^{[m]}(t) Q_0^{[n]}(0) \rangle_{P,V}^\mathrm{c}\nonumber\\[0.5ex]
&&\hspace{37pt} = -\partial_t \langle Q_0^{[m]}(0) Q_{-j}^{[n]}(-t) \rangle_{P,V}^\mathrm{c} 
= \partial_j \langle Q_0^{[m]}(0) J_{-j}^{[n]}(-t) \rangle_{P,V}^\mathrm{c},
\end{eqnarray}
denoting the difference operator by $\partial_jf(j) = f(j+1) - f(j)$.
Setting $t=0$, the difference $\langle J_j^{[m]} Q_0^{[n]} \rangle_{P,V}^\mathrm{c} -  \langle J_{-j}^{[n]} Q_0^{[m]} \rangle_{P,V}^\mathrm{c}$ is constant in $j$. Since truncated correlations decay to zero,
this constant has to vanish, which yields \eqref{6.7}. In particular, the field-current susceptibility matrix is symmetric,
\begin{equation}\label{6.9} 
B_{m,n} = B_{n,m}.
\end{equation}

Using  this symmetry and restricting to $n\geq 1$, we consider the $P$-derivative of the average current 
 \begin{equation}\label{6.10}
   \partial_P \langle( L^nL^{\scriptscriptstyle \downarrow})_{0,0} \rangle_{P,V} = - B_{n,0} = - B_{0,n}
   =  \langle Q^{[1]};Q^{[n]}\rangle_{P,V},
\end{equation}
since $J^{[0]} = - Q^{[1]}$ by \eqref{2.26}. We easily arrived at a very surprising identity. The $P$-derivative of the average current equals a particular
matrix element of the field-field susceptibility. Such susceptibility depends only on the eigenvalues of the Lax matrix.
In fact,  this quantity has been studied already in Section \ref{sec4.1} and one only has to adjust the results from there, setting $\alpha = P$.

As in the case of the free energy, since the pressure is varying as $1/N$, the fluctuation covariance is adding up, resulting in
\begin{equation}\label{6.11}
\langle \varsigma_1, C^\sharp \varsigma_n\rangle = \int_0^1 \mathrm{d}u \langle Q^{[1]};Q^{[n]} \rangle_{uP,V}
\end{equation}
with the operator $C^\sharp$ defined in \eqref{4.21}. Inserting from \eqref{6.10}, one concludes that
\begin{equation}\label{6.12}
\partial_P\big( \langle J^{[n]}_0 \rangle_{P,V} - P\langle \varsigma_1, C^\sharp \varsigma_n\rangle \big) = 0.
\end{equation}
The term in the round brackets has to be independent of $P$, in particular equal to its value at $P=0$. Since the covariance is bounded,
the second summand vanishes at $P= 0$. For  $\langle J^{[n]}_0 \rangle_{P,V}$, we use the random walk expansion \eqref{2.17}. Each walk contains at least one factor of $a_j$ and thus vanishes because under the a priori measure $a_j \to 0$ in the limit $P\to 0$. Therefore
\begin{equation}\label{6.13}
    \langle J^{[n]}_0 \rangle_{P,V} = P\langle \varsigma_1, C^\sharp \varsigma_n\rangle.
\end{equation}
Inserting from \eqref{4.21} and substituting as $\alpha \rho_\mathfrak{s} = \rho_\mathsf{n}$, 
\begin{equation}\label{6.14}
    \langle J^{[n]}_0 \rangle_{P,V} = \langle \varsigma_1, (1 -  \rho_\mathsf{n} T)^{-1}\rho_\mathsf{n} \varsigma_n\rangle
    -    \nu \langle \varsigma_1 (1 -  \rho_\mathsf{n} T)^{-1} \rho_\mathsf{n}\rangle  \langle \varsigma_n (1 -  \rho_\mathsf{n} T)^{-1} \rho_\mathsf{n}\rangle.
\end{equation}
Since $(1- \rho_\mathsf{n} T)^{-1} \rho_\mathsf{n}$ is a symmetric operator, one arrives at 
 \begin{equation}\label{6.15}
 \langle J^{[n]}_0 \rangle_{P,V} =   \langle \rho_\mathsf{n} \varsigma_1^\mathrm{dr}   \varsigma_n\rangle -  q_1\langle \rho_\mathsf{p}\varsigma_n\rangle,\qquad q_1 = \nu \langle \rho_\mathsf{p}\varsigma_1\rangle.
\end{equation}
Using \eqref{6.3},  one finally concludes
 \begin{equation}\label{6.16} 
\langle\rho_\mathrm{J} \varsigma_n \rangle =   \langle(\rho_\mathsf{n} \varsigma_1^\mathrm{dr}  -q_1 \rho_\mathsf{p})\varsigma_n\rangle
\end{equation}
valid for $n \geq 1$ and, in addition, 
\begin{equation}\label{6.16a} 
\langle\rho_\mathrm{J} \varsigma_0\rangle = - q_1.
\end{equation}
In the previous chapter we entertained the possibility that, by analogy, the Toda current can be written as
 \begin{equation}\label{6.17} 
\rho_\mathrm{J} = \rho_\mathsf{p}(v^\mathrm{eff} -q_1),
\end{equation}
compare with \eqref{5.18} and \eqref{5.24}.  Using the scattering shift of the Toda lattice, this proposal amounts to
\begin{equation}\label{6.18}
v^\mathrm{eff}(v) 
 = v +2 \int_\mathbb{R} \mathrm{d}w \log|v-w| \rho_\mathsf{p}(w)\big(v^\mathrm{eff}(w) - v^\mathrm{eff}(v)\big). 
\end{equation} 
If such an analogy holds, then $\rho_\mathsf{p} v^\mathrm{eff} = \rho_\mathsf{n} \varsigma_1^\mathrm{dr}$ and hence  
 \begin{equation}\label{6.19} 
 v^\mathrm{eff} = \frac{\varsigma_1^\mathrm{dr}}{\varsigma_0^\mathrm{dr}}\,.
\end{equation}
To confirm, we start from
\begin{eqnarray}\label{6.20}
&&\hspace{-40pt} T(\rho_\mathsf{p}v^\mathrm{eff}) - (T\rho_\mathsf{p})v^\mathrm{eff} = T(\rho_\mathsf{n}\varsigma_1^\mathrm{dr} ) 
-(T\rho_\mathsf{p}) v^\mathrm{eff}
 = (T\rho_\mathsf{n} -1 +1)(1- T\rho_\mathsf{n})^{-1}\varsigma_1  - (T\rho_\mathsf{p}) v^\mathrm{eff}\nonumber\\[0.7ex]
&&\hspace{68pt}= -\varsigma_1+ \varsigma_1^\mathrm{dr}  - \big(\rho_\mathsf{n}^{-1}\rho_\mathsf{p}-1\big)v^\mathrm{eff} 
 = v^\mathrm{eff}   -\varsigma_1,
\end{eqnarray}
as claimed.

At low pressure the Toda particles are far apart and interact mostly through isolated two-particle collisions. Quasiparticles are then defined as for hard rods, namely through  maintaining their velocity upon excluding the time span for collisions. The tracer quasiparticle jumps by a distance regulated by the Toda two-particle scattering shift.
 The validity of the collision rate ansatz \eqref{6.18} can be argued  as we did already for a hard rod fluid. However, for a dense Toda fluid with $P$ not too close to  $0$ 
 the notion of a quasiparticle is fuzzy, even more so the meaning of a tracer quasiparticle.  The validity of \eqref{6.18} 
 for arbitrary GGE parameters will be concluded only indirectly.
 
 Conceptually the collision rate ansatz can be read in a somewhat different way, as inspired by the time-of-flight method employed in recent experiments on the rapidity distribution for the $\delta$-Bose gas.
 One prepares the $N$-particle Toda system in a GGE and runs the dynamics over the time span $[0,t]$ with periodic 
 boundary conditions, $t$ of the same order as $N$. To identify the tracer quasiparticle, a past and future diagnostic step is added, to say in the time interval $[-\tau,0]$ backwards  and $[t,t +\tau]$ forwards the dynamics is executed with open boundary conditions, $\tau
 >0$ and of order 1. Then at times $-\tau,t+\tau$ the particles are well separated and have definite velocities. At time $-\tau$, we pick one particle with velocity $w$ and call it tracer
 quasiparticle. Then at time $t +\tau$, there is exactly one particle with the same velocity $w$. For $\tau \ll t$ the displacement of the tracer quasiparticle should be
 $ v^\mathrm{eff}(w) t$ in good approximation.
\subsection{Hydrodynamic equations}
\label{sec6.2}
On the hydrodynamic scale the local GGE is characterized by the stretch $\nu$ and the Lax DOS $\nu\rho_\mathsf{p}$, both of which now become spacetime dependent. Merely inserting the average currents, one arrives at the Euler type hydrodynamic evolution equations,
\begin{eqnarray}\label{6.21}
&&\partial_t \nu(x,t) -\partial_x q_1(x,t) = 0,\nonumber\\[0.5ex] 
&&\partial_t\langle\nu(x,t) \rho_\mathsf{p}(x,t)\varsigma_n \rangle + \partial_x\langle(v^\mathrm{eff}(x,t) - q_1(x,t))\rho_\mathsf{p}(x,t)\varsigma_n\rangle = 0
\end{eqnarray}
for $n\geq 1$. The latter equation extends to $n=0$, since the bracket under $\partial_t$ equals $1$ and, as can be concluded from \eqref{6.20}, under $\partial_x$ equals $0$. Therefore one can switch to the pointwise version of the hydrodynamic equations,
\begin{eqnarray}\label{6.22} 
&&\partial_t \nu(x,t) -\partial_x q_1(x,t) = 0,\nonumber\\[0.5ex]
&& \partial_t\big(\nu(x,t) \rho_\mathsf{p}(x,t;v)\big) + \partial_x\big((v^\mathrm{eff}(x,t;v) - q_1(x,t))\rho_\mathsf{p}(x,t;v)\big) = 0.
\end{eqnarray}

As a most remarkable feature of generalized hydrodynamics,  these equations can be  transformed explicitly to a quasilinear 
form. 
For this purpose we recall the identity  \eqref{3.59}, which expresses  $\rho_\mathsf{n}$ in terms of $\rho_\mathsf{p}$,  
\begin{equation}\label{6.23} 
\rho_\mathsf{n}=  \rho_\mathsf{p}(1+ (T\rho_\mathsf{p}))^{-1}.
\end{equation}
Then Eq. \eqref{6.22} acquires the normal form
\begin{equation}\label{6.24} 
 \partial_t \rho_\mathsf{n} + \nu^{-1}(v^\mathrm{eff} - q_1)\partial_x \rho_\mathsf{n} = 0.
\end{equation}
Thus the linearization operator is in fact merely a multiplication by $\nu^{-1}(v^\mathrm{eff} - q_1)$, in other words the operator is diagonal. 
For nonintegrable chains  the corresponding operator is a $3\times 3$ matrix, which in general is not diagonal. Its diagonalization would yield the normal modes.  The solution of the respective hyperbolic  conservation laws can develop shock discontinuities, since locally the solution may have to jump to another eigenvalue. In our case $\nu^{-1}(v^\mathrm{eff} - q_1)$
is expected to be a smooth function. Thus the spectrum of the linearization operator is continuous and has no eigenvalues. Hence for smooth initial data
solutions of \eqref{6.24} are expected to stay smooth. No mathematical analysis has been attempted to verify this conjecture.  
In fact, the situation could be more subtle since  $\nu$ can take either sign and $\nu^{-1}$ might be singular.

To verify \eqref{6.24} we  start from 
\begin{equation}\label{6.25} 
 \rho_\mathsf{p}\partial_t\nu  + \nu\partial_t\rho_\mathsf{p} + \partial_x\big((v^\mathrm{eff} - q_1)\rho_\mathsf{p}\big) = 0,
\end{equation}
which together with the continuity equation yields
\begin{equation}\label{6.26}
\nu\partial_t \rho_\mathsf{p} + \partial_x(v^\mathrm{eff} \rho_\mathsf{p})  - q_1 \partial_x \rho_\mathsf{p}= 0.
\end{equation}
Using this identity and \eqref{6.23}, one obtains
\begin{eqnarray}\label{6.27}
&&\hspace{-20pt} \nu\partial_t \rho_\mathsf{n} = \frac{\rho_\mathsf{n}}{\rho_\mathsf{p}}\big(-\partial_x(v^\mathrm{eff}\rho_\mathsf{p}) +q_1\partial_x \rho_\mathsf{p}\big)
-\frac{\rho_\mathsf{n}^2}{\rho_\mathsf{p}}\big(T(-\partial_x(v^\mathrm{eff}\rho_\mathsf{p}) + q_1\partial_x \rho_\mathsf{p})\big)\nonumber\\
&&\hspace{12pt} =  \frac{\rho_\mathsf{n}}{\rho_\mathsf{p}}\big(- \partial_x(v^\mathrm{eff}\rho_\mathsf{p}) + \rho_\mathsf{n} \partial_x T(v^\mathrm{eff}\rho_\mathsf{p})\big) + q_1 \partial_x\rho_\mathsf{n}.
\end{eqnarray}
The effective velocity solves the integral equation
\begin{equation}\label{6.28}
v^\mathrm{eff} = v + T(v^\mathrm{eff}\rho_\mathsf{p}) - (T\rho_\mathsf{p})v^\mathrm{eff}
\end{equation}
and, by inserting \eqref{6.23},
\begin{equation}\label{6.29}
\partial_x\big( \rho_\mathsf{n}^{-1} v^\mathrm{eff}\rho_\mathsf{p} \big) = \partial_x  T(v^\mathrm{eff}\rho_\mathsf{p}).
\end{equation}
Hence the first summand in \eqref{6.27} reads
\begin{equation}\label{6.30}
\frac{\rho_\mathsf{n}}{\rho_\mathsf{p}}\big(- \partial_x(v^\mathrm{eff}\rho_\mathsf{p}) + \rho_\mathsf{n} \partial_x( \rho_\mathsf{n}^{-1} v^\mathrm{eff}\rho_\mathsf{p} )\big)
= - v^\mathrm{eff}\partial_x \rho_\mathsf{n},
\end{equation}
thereby confirming \eqref{6.24}. 

The normal form \eqref{6.24} of the hydrodynamic equations is a crucial insight.  A priori, the coupled conservation laws might be so complicated that it becomes an impossible task to extract information of interest. Here the normal form is of considerable help. It allows to obtain some partially analytic solutions, as the domain wall problem and the linearized version to be discussed in the next Chapters. The normal form suggests also 
how to devise a numerical scheme for solving the hydrodynamic equations. 
In its most basic version, at the current time, $t$,
one keeps 
$\nu^{-1}(v^\mathrm{eff} - q_1)$ fixed and solves the resulting linear equation for one further time step $\mathrm{d}t$. With the so 
obtained $\rho_\mathsf{n}(t +\mathrm{d}t)$ one updates $\nu^{-1}(v^\mathrm{eff} - q_1)$ according to \eqref{6.19} for $v^\mathrm{eff}$, $\nu^{-1} = \langle \rho_\mathsf{p}\rangle$, $\nu^{-1}q_1 = \langle \rho_\mathsf{p}\varsigma_1\rangle$, and $\rho_\mathsf{p} = \rho_\mathsf{n}  \varsigma_0^\mathrm{dr}$.  
No explicit use of TBA is involved in this step. The iteration is then repeated many times until the desired final time is reached.

The collision rate equation suggests a distinct approach, called flea gas algorithm. The basic idea is to replace the true motion by an instantaneous exchange of velocities $v_1,v_2$ whenever  the two particles first reach the distance $2 \log|v_1-v_2|$. Thereby the simulation of the particle dynamics is massively speeded up and it is ensured that on large scales 
 the Toda generalized hydrodynamics is realized.  In fact, the true dynamics is more complicated, since a third particle or even more particles might interfere with the collision between particles 1 and 2. To properly  set up the particle model requires a separation into isolated particle clusters combined with an instantaneous many-body collision. For the Toda fluid, and also integrable quantum many-body  systems,
 one can use such a scheme as an alternative to direct numerical simulations of the hydrodynamic equations.  
 
 Physically, the initial conditions have to satisfy the constraints resulting from GGE. This amounts to $\nu \in \mathbb{R}$, $\nu \rho_\mathsf{p} \geq 0$, and $\rho_\mathsf{n}\geq 0$. Such conditions should be propagated by the Euler equations. For $\rho_\mathsf{n}$,
 this property follows from \eqref{6.24}. The prefactor of $\partial_x$ looks singular at $\nu = 0$ and the propagation of $\rho_\mathsf{p}$ properties is less obvious.
 A specific example is discussed in Chapter \ref{sec8}.  
 
 Comparing Eqs. \eqref{5.19} and \eqref{6.22} and identifying $h = \nu \rho_\mathsf{p}$, $u = q_1$, Toda and hard rods have structurally identical hydrodynamic evolution equations. On the hard rod side the hydrodynamic limit has been proved, thereby supporting the validity of the corresponding property for the Toda lattice.\bigskip 
\begin{center}
 \textbf{Notes and references}
 \end{center}
 \begin{center}
 \textbf{Section 6.1}
\end{center} 
The computation leading to \eqref{6.8} is taken from De Nardis et al. \cite{DBD19}, see also Karevski and Sch\"{u}tz \cite{KS19}. The symmetry of the 
  $B$-matrix is discussed in Grisi and Sch\"{u}tz
 \cite{GS11} and  Spohn \cite{S14}. Rather likely there is earlier work. The connection between the symmetry of the $B$-matrix 
 and the average current was first noted by Spohn   \cite{S20} and extended to quantum systems by  Yoshimura and Spohn \cite{YS20}. For integrable quantum systems, the appropriately 
 adjusted rate equation \eqref{6.18} has been discovered by Castro-Alvaredo et al.  \cite{CDY16}. Very quickly it was understood that 
 the scheme  originally developed in Bertini  et al. \cite{BCDF16} and Castro-Alvaredo et al. \cite{CDY16}  for  specific models has a much wider applicability, see Doyon et al.  \cite{DYC18}.
There are further attempts to microscopically justify the form of the average currents, to mention the ones  by Vu and Yoshimura \cite{VY19},
Borsi et al. \cite{BPP21},  Pristy\'{a}k and Pozsgay \cite{PP22},  and the review Cubero et al. \cite{CSY21}.  
 \bigskip
 \begin{center}
 \textbf{Section 6.2}
\end{center}  
 The transformation to normal form is established by Castro-Alvaredo et al.  \cite{CDY16}. The flea gas algorithm is discussed by 
   Doyon et al.  \cite{DYC18} and Mesty\'{a}n and Alba \cite{MA20} as an alternative approach to more standard PDE discretization schemes. A numerically oriented contribution is M{\o}ller et al. \cite{MS20a}. In  Bulchandani \cite{B17} it is argued that GHD equations are continuum integrable systems, which is related to the geometric viewpoint developed by Doyon et al. \cite{DSY18}.
\newpage
\section{Linearized hydrodynamics and GGE dynamical correlations}
\label{sec7} 
\setcounter{equation}{0}
In equilibrium statistical mechanics a central theme is to understand the structure of static correlations, with particular focus on critical points 
in thermodynamic parameter space  in the neighborhood  of which the correlation length is large on microscopic scales. A natural extension are time-dependent correlations, which can be viewed as the propagation of initially small perturbations in the equilibrium state.  Now conservation laws
and broken symmetries will play a crucial role. 
The most elementary approach is the Landau-Lifshitz theory which uses the link to  macroscopic equations
linearized at thermal equilibrium, as will be discussed below. Near critical points the more sophisticated techniques of critical dynamics  
would come into play.
\subsection{Equilibrium spacetime correlations for nonintegrable chains}
\label{sec7.1}
The general structure 
behind the Landau-Lifshitz theory can be explained already in the context of nonintegrable systems with a few conservation laws.
For concreteness we consider nonintegrable anharmonic chains.  In essence, the extension to integrable chains consists in
substituting the respective $3\times 3$ matrices by operators on Hilbert spaces with infinite dimension. To distinguish from the infinite dimensional case, we use $\vec{u}$ for $3$\,-vectors and $\mathsfit{A}$ for $3\times3$ matrices.

The infinitely extended chain is governed by the hamiltonian
\begin{equation}\label{7.1}
  H_\mathrm{ch}=\sum_{j \in \mathbb{Z}}\big( \tfrac{1}{2}p^2_j +V_\mathrm{ch}(r_j)\big)
\end{equation}
with equations of motion
\begin{equation}\label{7.2}
\frac{d}{dt}r_j=p_{j+1}-p_j\,,\qquad
\frac{d}{dt}p_j=V_\mathrm{ch}'(r_j)-V_\mathrm{ch}'(r_{j-1}).
\end{equation}
The chain potential, $V_\mathrm{ch}$, is bounded from below and increases  at least one-sided linearly 
at infinity so to have a finite partition function.
Stretch, momentum, and energy are defined by the local fields
\begin{equation}\label{7.3}
\vec{Q}_j= \big(r_j, p_j,2e_j\big), \qquad e_j=\tfrac{1}{2}\big(p^2_j + V_\mathrm{ch}(r_{j-1}) +V_\mathrm{ch}(r_j)\big),
\end{equation}
and the respective current densities are
\begin{equation}\label{7.4}
\vec{J}_j = \big( -p_j,-V_\mathrm{ch}'(r_{j-1}), - (p_{j-1} + p_j)V_\mathrm{ch}'(r_{j-1})\big).
\end{equation}
It is assumed that there are no further local conservation laws, which is the essence of non-integrability.

The canonical  equilibrium state factorizes with one factor given by
\begin{equation}\label{7.5}
Z_0(P,\bar{u},\beta)^{-1} \exp\!\big(-\beta \big(\tfrac{1}{2}(p_0- \bar{u})^2 +V_\mathrm{ch}(r_0)\big)  - Pr_0\big),
\end{equation}
where we introduced the intensive dual parameters $P, \bar{u} ,\beta$. $ \mathfrak{p} = P/\beta$ is the physical pressure, $\bar{u}$ the mean velocity,  and
$\beta$ the inverse temperature. The chain free energy per site  equals
\begin{equation}\label{7.6}
F_\mathrm{ch} = - \log Z_0(P,\bar{u},\beta). 
\end{equation}
Of course, more explicit expressions could be provided, but this is not needed at the moment. In accordance with conventions for the Toda chain, 
the indices run over $m,n = 0,1,2$, throughout. The static field-field correlator in infinite volume is defined by 
\begin{equation}\label{7.7}
\mathsfit{C}_{m,n}(j) = \langle Q_j^{[m]}Q_0^{[n]}\rangle_{P,\bar{u},\beta}^\mathrm{c} = \delta_{0j}
 \langle Q_0^{[m]}Q_0^{[n]}\rangle_{P,\bar{u},\beta}^\mathrm{c}
\end{equation}
and the static field-field susceptibility matrix by
\begin{equation}\label{7.8}
\mathsfit{C}_{m,n} = \sum_{j \in \mathbb{Z}}\mathsfit{C}_{m,n}(j) = \langle Q_0^{[m]}Q_0^{[n]}\rangle_{P,\bar{u},\beta}^\mathrm{c}.
\end{equation}
$\mathsfit{C}$ is the matrix of second derivatives of the chain free energy $F_\mathrm{ch}$. In the same fashion the field-current correlator is given by
\begin{equation}\label{7.9}
\mathsfit{B}_{m,n}(j) = \langle J_j^{[m]} Q_0^{[n]}\rangle_{P,\bar{u},\beta}^\mathrm{c}
\end{equation}
and the static field-current susceptibility matrix by
\begin{equation}\label{7.10}
\mathsfit{B}_{m,n} = \sum_{j \in \mathbb{Z}}\mathsfit{B}_{m,n}(j).
\end{equation}
The field-field spacetime correlator is the matrix
\begin{equation}\label{7.11}
\mathsfit{S}_{m,n}(j,t)  = \langle Q_j^{[m]}(t)Q_0^{[n]}(0)\rangle_{P,\bar{u},\beta}^\mathrm{c},
\end{equation}
which for $t \neq 0$ is no longer $\delta$-correlated in $j$. It is  convenient to also introduce its Fourier transform
\begin{equation}\label{7.12}
\hat{\mathsfit{S}}_{m,n}(k,t) = \sum_{j \in \mathbb{Z}}\mathrm{e}^{\mathrm{i}kj}\mathsfit{S}_{m,n}(j,t)
\end{equation}
with  $k \in [-\pi,\pi]$.

Switching to the macroscopic continuum scale, the average fields and currents are denoted by $\vec{\mathsfit{u}}= \langle \vec{Q}_0\rangle_{P,\bar{u},\beta}$,
$\vec{\mathsfit{j}}= \langle \vec{J}_0\rangle_{P,\bar{u},\beta}$. According to the rules of thermodynamics there is a one-to-one map between $\vec{\mathsfit{u}}$ and $(P,\bar{u},\beta)$. Therefore the hydrodynamic equations for the chain can be written as
\begin{equation}\label{7.13}
\partial_t \vec{\mathsfit{u}}(x,t) +  \partial_x \vec{\mathsfit{j}}(\vec{\mathsfit{u}}(x,t)) = 0
\end{equation}
with quasilinear form
\begin{equation}\label{7.14}
\partial_t \vec{\mathsfit{u}}(x,t) + \mathsfit{A}(\vec{\mathsfit{u}}(x,t))\partial_x \vec{\mathsfit{u}}(x,t) = 0.
\end{equation}
 The matrix $\mathsfit{A}$ is known as flux Jacobian and given through  
\begin{equation}\label{7.15}
\mathsfit{A}_{m,n}(\vec{\mathsfit{u}}) = \partial_{\mathsfit{u}_n}\langle J^{[m]}_0\rangle_{\vec{\mathsfit{u}}},
\end{equation}
which by the chain rule yields 
\begin{equation}\label{7.16}
\mathsfit{A} = \mathsfit{B}\mathsfit{C}^{-1}, 
\end{equation}
viewed as an identity in $(P,\bar{u},\beta)$.
The matrix $\mathsfit{A}$ is not symmetric in general. As established in \eqref{6.8}, essentially using only spacetime stationarity,  the matrix $\mathsfit{B}$ turns out to be symmetric. By construction $\mathsfit{C}$ is symmetric and also $\mathsfit{C}>0$. Thus
\begin{equation}\label{7.17}
 \mathsfit{A} = \mathsfit{C}^{\frac{1}{2}} \mathsfit{C}^{-\frac{1}{2}}\mathsfit{B}^{-1}\mathsfit{C}^{-\frac{1}{2}}\mathsfit{C}^{-\frac{1}{2}}, 
 \end{equation}
 which implies that $\mathsfit{A}$ has real eigenvalues, $c_\alpha$, $\alpha = 0,1,2$,  and a complete system of right, $|\psi_\alpha\rangle$, and left eigenvectors, $\langle \tilde{\psi}_\alpha  | $.

Following Landau and Lifshitz the average \eqref{7.11} is viewed as a small initial perturbation of equilibrium. Then to leading order the correlator  may be 
approximated by Eq. \eqref{7.14} linearized around a constant background with parameters  $P,\bar{u},\beta$. 
Denoting this perturbation by $\vec{u}$ the nonlinear hydrodynamic equation turns to its linearized version
\begin{equation}\label{7.18}
\partial_t \vec{u}(x,t) +  \mathsfit{A}\partial_x \vec{u}(x,t) = 0,
\end{equation}
where the flux Jacobian $\mathsfit{A}$ is now evaluated at the background, thus constant in spacetime. However the initial conditions, 
$\vec{u}(x)$, are random, their average being denoted by $\mathbb{E}$.  According to thermal equilibrium, because of independence, on the macroscopic scale the random field $\vec{u}(x)$ is Gaussian white noise with mean zero and covariance 
\begin{equation}\label{7.19}
\mathbb{E}\big( u_m(x)u_n(0)\big)= \mathsfit{C}_{m,n} \delta(x) = \mathsfit{S}^\sharp_{m,n}(x,0),
\end{equation}
which defines the continuum static correlator $\mathsfit{S}^\sharp(x,0)$. To obtain its spacetime version one has to solve \eqref{7.18}
with initial conditions \eqref{7.19}, which yields
\begin{equation}\label{7.20}
\mathbb{E}\big( u_m(x,t)u_n(0,0)\big) = \mathsfit{S}^\sharp_{m,n}(x,t), 
\end{equation}
 thereby defining the right hand side. In Fourier space
 \begin{equation}\label{7.21}
\hat{\mathsfit{S}}^\sharp_{m,n}(k,t) = (\mathrm{e}^{-\mathrm{i}k\mathsfit{A}t}\mathsfit{C})_{m,n}.
\end{equation}
Using the eigenvectors of  $\mathsfit{A}$, in position space the correlator \eqref{7.20} has the explicit form
\begin{equation}\label{7.22}
\mathsfit{S}^\sharp_{m,n}(x,t) = \sum_{\alpha = 0}^2 \delta(x - c_\alpha t) (|\psi_\alpha\rangle \langle \tilde{\psi}_\alpha | \mathsfit{C})_{m,n}.
\end{equation}

To establish the relation to the microscopic definition \eqref{7.11}, one introduces the dimensionless scale parameter $\epsilon$. Then under ballistic scaling it is expected 
that  for small $\epsilon$,
\begin{equation}\label{7.23}
\epsilon^{-1} \mathsfit{S}_{m,n}(\lfloor\epsilon^{-1}x\rfloor, \epsilon^{-1}t) \simeq \mathsfit{S}^\sharp_{m,n}(x,t).
\end{equation}
Here $\lfloor\cdot\rfloor$ denotes the integer part and the prefactor is chosen such that the sum rule 
\begin{equation}\label{7.24}
\sum_{j\in \mathbb{Z}} \mathsfit{S}(j,t) = \sum_{j\in \mathbb{Z}} \mathsfit{S}(j,0)
\end{equation}
holds. In Fourier space, more compactly, 
\begin{equation}\label{7.25}
\lim_{\epsilon \to 0}\hat{\mathsfit{S}}_{m,n}(\epsilon k,\epsilon^{-1}t) =  \hat{\mathsfit{S}}^\sharp_{m,n}(k,t).
\end{equation}

The eigenvectors of $\mathsfit{A}$ are called normal modes. The $\alpha$-th mode travels with velocity $c_\alpha$ and the 
initial condition determines the particular linear combination of normal modes as encoded by the susceptibility matrix $\mathsfit{C}$.  The spacetime correlator has three 
delta peaks moving ballistically. If the background has zero average momentum, then the eigenvalues are $\vec{c} =
(-c,0,c)$ with $c$ the isentropic speed of sound. There are two sound peaks with equal speed moving in opposite directions and a heat peak standing
still. For an integrable system, there are so to speak infinitely many peaks, each moving with its own velocity. We thus expect the corresponding time correlator to have a broad spectrum which expands ballistically.

In Figure 7 shown is molecular dynamics of a fluid with shoulder interaction potential. Clearly visible is the 
anomalous broadening of the delta-peaks with power laws $t^{3/5}$ and $t^{2/3}$.
\begin{figure}[!t]
	\centering
	\includegraphics[width=.45\linewidth]{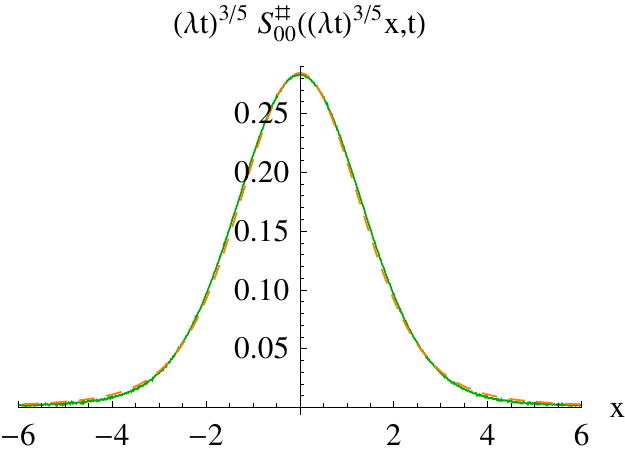}
	\includegraphics[width=.45\linewidth]
	{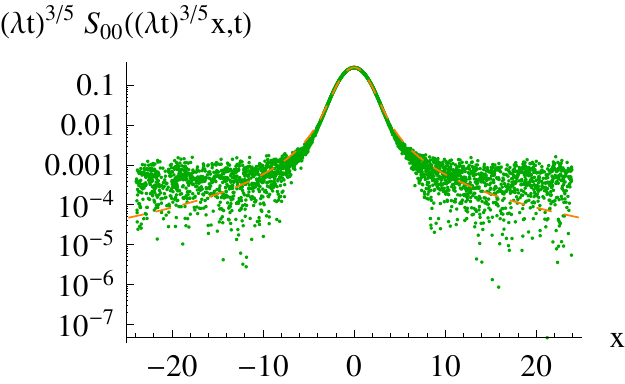} 
\\[3ex]
	 \includegraphics[width=.45\linewidth]{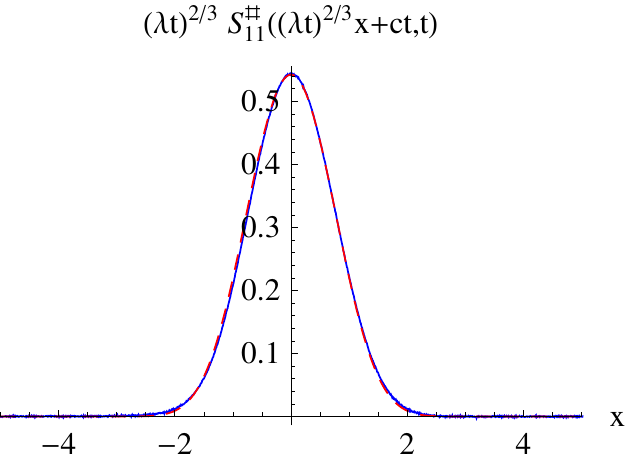}	 \includegraphics[width=.45\linewidth]
{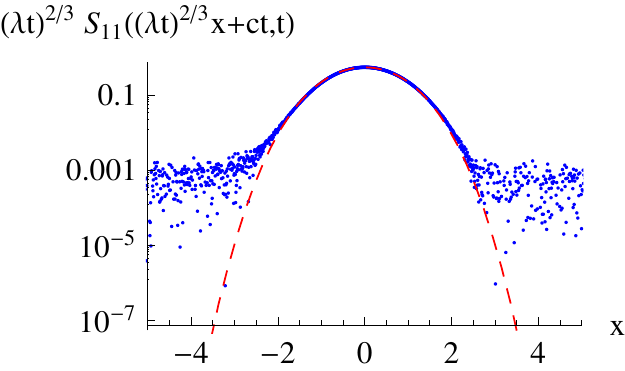}
	  \caption{Heat and sound peak of the equilibrium spacetime correlations for a fluid interacting through a shoulder potential. 
Shoulder means $V_\mathrm{sh}(x) = \infty$ for $|x| < \tfrac{1}{2}$, $V_\mathrm{sh}(x) = 1$ for $\tfrac{1}{2} <|x| < 1$, and $V_\mathrm{sh}(x) = 0$ for $1 <|x| $. Heat peak is on  top and sound peak, displaced by $ct$, below. System size is $N= 4096$, time $t = 1024$, $\beta = 2$, and $P = 1.2$ for heat respectively $P = 1$ for sound. The sound speed is $c = 1.74$. There is a single scale-parameter in comparison with theoretical predictions.
	 The right panels display the corresponding  logarithmic plot. Numerically, the shoulder potential has the  simplification that particle trajectories are piecewise linear. From \cite{MS14}.}
\label{fig7}
\end{figure}
\subsection{GGE spacetime correlations for the Toda lattice}
\label{sec7.2}
For the Toda chain we follow step by step the road map provided by the finite mode case, where we recall that  $\varsigma_n(w) = w^n$, including $n = 0$. The field-field and field-current correlator have been introduced before in \eqref{6.5}, resp. \eqref {6.6}.
Since we have computed already the GGE average of fields in \eqref{3.60} and of currents in \eqref{6.16}, the matrix elements of $C,B$
can be determined 
by differentiating these averages with respect to $P$ and $\mu_n$. The computation is somewhat lengthy and stated is merely the final result.
For the field-field correlator one obtains
\begin{eqnarray}\label{7.26}
&&\hspace{-20pt} C_{0,0} =\nu^3 \langle \rho_\mathsf{p} \varsigma_0^\mathrm{dr} \varsigma_0^\mathrm{dr} \rangle,\nonumber\\[0.5ex]
&&\hspace{-20pt}C_{0,n}= C_{n,0}= -\nu^2  \langle \rho_\mathsf{p} \varsigma_0^\mathrm{dr} (\varsigma_n- q_n\varsigma_0)^\mathrm{dr} \rangle,\nonumber\\[0.5ex]
&&\hspace{-20pt} C_{m,n}= \nu  \langle \rho_\mathsf{p} (\varsigma_m- q_m\varsigma_0)^\mathrm{dr} (\varsigma_n- q_n\varsigma_0)^\mathrm{dr} \rangle,
\end{eqnarray}
 $m,n \geq 1$, and for the field-current correlator
\begin{eqnarray}\label{7.27}
&&\hspace{-20pt} B_{0,0} = \nu^2 \langle \rho_\mathsf{p}(v^\mathrm{eff} -q_1) \varsigma_0^\mathrm{dr}   \varsigma_0^\mathrm{dr}\rangle,\nonumber\\[0.5ex]
&&\hspace{-20pt}B_{0,n}= B_{n,0}=  -\nu \langle \rho_\mathsf{p}(v^\mathrm{eff} -q_1)  \varsigma_0^\mathrm{dr} ( \varsigma_n- q_n \varsigma_0)^\mathrm{dr}\rangle,\nonumber\\[0.5ex]
&&\hspace{-20pt} B_{m,n} =  \langle \rho_\mathsf{p}(v^\mathrm{eff} -q_1)( \varsigma_m- q_m \varsigma_0)^\mathrm{dr} ( \varsigma_n- q_n \varsigma_0)^\mathrm{dr}\rangle.
\end{eqnarray}

Both matrices have a two-block structure. Rather than using the discrete index, $n = 1,2,\ldots$, we introduce some real valued function,
$\phi$.
Our linear space consists now of two-vectors $(r,\phi)$ with $r$ the coefficient for the index $0$ and $\phi$ for all other modes.
More formally, the appropriate linear space is $\mathbb{C} \oplus \big(L^2(\mathbb{R}, \mathrm{d}w)\ominus\{\varsigma_0\}\big)$,
the second summand consisting of all square-integrable functions orthogonal to $\varsigma_0$.
We also introduce  the linear operator, $F$, in Dirac notation,
\begin{equation}\label{7.28}
F = 
(1 - T\rho_\mathsf{n})^{-1}  - \nu | \varsigma_0^\mathrm{dr}\rangle\langle \rho_\mathsf{p}|.
\end{equation}
Note that $F\varsigma_0 = 0$. Then
\begin{equation}\label{7.29}
C = 
\begin{pmatrix}
\nu^3 \langle  \rho_\mathsf{p}  \varsigma_0^\mathrm{dr} \varsigma_0^\mathrm{dr}\rangle &-\nu^2 \big\langle  \rho_\mathsf{p} \varsigma_0^\mathrm{dr}\big|F\\[0.5ex]
-\nu^2 F^*\big| \rho_\mathsf{p}\varsigma_0^\mathrm{dr}\big\rangle& \nu F^* \rho_\mathsf{p} F
\end{pmatrix}.
\end{equation}
For some general operator $A$, as introduced before, $A^*$ stands for the adjoint operator of $A$
with respect to the standard $L^2$ inner product $\langle \cdot,\cdot\rangle$. In particular,
\begin{equation}\label{7.30}
F^* = (1 - \rho_\mathsf{n} T)^{-1} - \nu | \rho_\mathsf{p}\rangle\langle  \varsigma_0^\mathrm{dr}|
\end{equation} 
 With the same notation, the matrix $B$ is given by
\begin{equation}\label{7.31}
B = 
\begin{pmatrix}
\nu^2 \langle  \rho_\mathsf{p} (v^\mathrm{eff} -q_1)\varsigma_0^\mathrm{dr}\varsigma_0^\mathrm{dr}\rangle &-\nu \big\langle  \rho_\mathsf{p} (v^\mathrm{eff}-q_1)\varsigma_0^\mathrm{dr}\big|F\\[0.5ex]
-\nu F^*\big|  \rho_\mathsf{p}(v^\mathrm{eff} - q_1)\varsigma_0^\mathrm{dr}\big\rangle& F^* \rho_\mathsf{p}(v^\mathrm{eff} -q_1) F
\end{pmatrix}.
\end{equation}

Following the road map we are supposed to compute $\mathrm{e}^{At}C$ with $A = B C^{-1}$, which does not seem to be completely
straightforward. But instead one might guess the entire solution $S(x,t)$ by noting that $S(x,0) = \delta(x)C$ and $\partial_t S(x,t)\big|_{t=0}
= -\delta'(x)B$. Indeed, the two conditions can be satisfied by setting
\begin{equation}\label{7.32}
S(x,t) = 
   \begin{pmatrix}
\nu^3 \langle  \rho_\mathsf{p} \,\delta(x - t\nu^{-1}(v^\mathrm{eff} -q_1)) \varsigma_0^\mathrm{dr}\varsigma_0^\mathrm{dr}\rangle &-\nu^2 \big\langle  \rho_\mathsf{p}\,\delta(x - t\nu^{-1}(v^\mathrm{eff} -q_1))\varsigma_0^\mathrm{dr}\big|F\\[0.5ex]
-\nu^2 F^*\big| \rho_\mathsf{p}\,\delta(x - t\nu^{-1}(v^\mathrm{eff}  -q_1))\varsigma_0^\mathrm{dr}\big\rangle& \nu F^* \rho_\mathsf{p}\delta(x - t\nu^{-1}
(v^\mathrm{eff} -q_1)) F
\end{pmatrix}.
\end{equation}
As a control check, the correlator scales indeed ballistically,  
\begin{equation}\label{7.33}
S(x,t) = t^{-1}S(x/t,1).
\end{equation}
Our computation misses that matching simply the value and first derivative at $t=0$ does not determine the full solution. 
Leaving  for the moment a more convincing argument aside, let us reflect on the resulting predictions. For example, within the stated approximations, the spacetime stretch-stretch correlation function is predicted as
\begin{equation}\label{7.34}
S_{0,0}(x,t) = \nu^2 \int_\mathbb{R}\mathrm{d}w   \delta(x- t\nu^{-1}(v^\mathrm{eff}(w) - q_1)) \nu\rho_\mathsf{p}(w)\varsigma_0^\mathrm{dr}(w)^2,
\end{equation}
and similarly for the momentum-momentum correlation, 
\begin{equation}\label{7.35}
S_{1,1}(x,t) =  \int_\mathbb{R}\mathrm{d}w  \delta(x- t\nu^{-1}(v^\mathrm{eff}(w)-q_1)) \nu\rho_\mathsf{p}(w)
(\varsigma_1- q_1\varsigma_0)^\mathrm{dr}(w)^2.
\end{equation}
Both identities can be deduced most easily by going back to \eqref{7.26}  upon inserting the $\delta$-function.
Our result is in perfect analogy to the finite mode case. The modes are labelled by the spectral parameter $w$ and propagate with velocity
$\nu^{-1}(v^\mathrm{eff}(w)-q_1)$. The last factor provides the weights and depends on the particular choice for the 
matrix elements of $S(x,t)$.

To confirm the guess, one has to figure out the spectral representation of the operator $A = BC^{-1}$. 
As key observation,  in \eqref{6.23} we already introduced a nonlinear map which transforms the system of conservation laws into its quasilinear version in such a way that the  operator corresponding to $A$ is manifestly diagonal. This property is expected to persist 
when linearizing the map \eqref{6.23}. In the $\nu, \rho_\mathsf{p}$ variables the map is given by
\begin{equation}\label{7.36}
\rho_\mathsf{n} = \frac{ \rho_\mathsf{p}}{1 + T \rho_\mathsf{p}}.
\end{equation} 
Both sides are linearized as $\rho_\mathsf{n} +\epsilon g$, $\nu \rho_\mathsf{p}+\epsilon r$, $\nu +\epsilon \phi$, $\langle \phi \rangle = 0$. To first order in $\epsilon$
this yields the linear map   
$R: g \mapsto (r,\phi)$  given by  
\begin{equation}\label{7.37}
Rg = \nu
\begin{pmatrix} 
- \nu \langle g (\varsigma_0^\mathrm{dr})^2\rangle\\[0.5ex]
F^*(g \varsigma_0^\mathrm{dr})
\end{pmatrix}
\end{equation} 
with $F^*$ as in \eqref{7.30}. 
 Note that indeed $\langle F^* \phi\rangle = 0$.  Eq. \eqref{7.36} can be inverted as  
\begin{equation}\label{7.38}
 \rho_\mathsf{p} = (1- \rho_\mathsf{n} T)^{-1}\rho_\mathsf{n},
\end{equation} 
thereby deriving, by a similar argument as before,
\begin{equation}\label{7.39}
R^{-1} 
\begin{pmatrix} 
r\\
\phi
\end{pmatrix}
= (\nu \varsigma_0^\mathrm{dr})^{-1}(- \rho_\mathsf{n} r + (1-\rho_\mathsf{n} T )\phi). 
\end{equation} 
Indeed one checks that
\begin{equation}\label{7.40}
RR^{-1} = 1,\quad R^{-1}R = 1,
\end{equation}
the first $``1"$ standing for the identity operator as a $2\times 2$ block matrix and the second $``1"$ for the identity operator in the space of scalar functions.

The operators $C,B$ can be written in the new basis with the result
\begin{equation}\label{7.41}
R^{-1}CR =   \nu \big| \rho_\mathsf{p} \big\rangle \big\langle ( \varsigma_0^\mathrm{dr})^2\big| + \nu( \varsigma_0^\mathrm{dr})^{-1} \rho_\mathsf{p} FF^*\varsigma_0^\mathrm{dr}
\end{equation}
and 
\begin{eqnarray}\label{7.42}
&&\hspace{0pt}R^{-1}BR
 = \big| (v^\mathrm{eff}-q_1)\rho_\mathsf{p}\big\rangle\big\langle  (\varsigma_0^\mathrm{dr})^2\big|
   +   (v^\mathrm{eff} -q_1)(\varsigma_0^\mathrm{dr})^{-1}\rho_\mathsf{p}FF^* \varsigma_0^\mathrm{dr} \big)\nonumber\\[0.5ex]
&&\hspace{44pt} = \nu^{-1} (v^\mathrm{eff} - q_1) R^{-1}CR.  
\end{eqnarray}
As anticipated, in the new basis  $R^{-1}AR$ is simply multiplication by $\nu^{-1} (v^\mathrm{eff} - q_1)$.
We conclude that 
\begin{equation}\label{7.43}
\mathrm{e}^{At}C =  R\exp\!\big(\nu^{-1}(v^\mathrm{eff} - q_1)t\big)R^{-1}C.
\end{equation}
Working out the algebra, one arrives at 
\begin{equation}\label{7.44}
\mathrm{e}^{At}C =  \begin{pmatrix}
\nu^3 \langle  \rho_\mathsf{p} \exp\!\big(\nu^{-1}(v^\mathrm{eff} - q_1)t\big)   (\varsigma_0^\mathrm{dr})^2\rangle &- \big\langle  \rho_\mathsf{p}
\exp\!\big(\nu^{-1}(v^\mathrm{eff} - q_1)t\big)  \varsigma_0^\mathrm{dr}\big|F\\[0.5ex]
-\nu^2 F^*\big|  \rho_\mathsf{p} \exp\!\big(\nu^{-1}(v^\mathrm{eff} - q_1)t\big) \varsigma_0^\mathrm{dr}\big\rangle& \nu F^*
 \rho_\mathsf{p} \exp\!\big(\nu^{-1}(v^\mathrm{eff} - q_1)t\big) F
\end{pmatrix}.
\end{equation}
Adding the spatial dependence, first in Fourier space, the propagator $\mathrm{e}^{At}C$ is modified to 
$\mathrm{e}^{\mathrm{i}kAt}C$, which in position space yields \eqref{7.32}. 
For completeness the matrix $A$ is recorded as 
\begin{equation}\label{7.45}
A =  \nu^{-1}R(v^\mathrm{eff} - q_1)R^{-1} = 
\begin{pmatrix}
\langle \rho_\mathsf{n} (v^\mathrm{eff} -q_1) \varsigma_0^{\mathrm{dr}}\rangle &- \big\langle  (v^\mathrm{eff}- q_1) \varsigma_0^\mathrm{dr}
(1 -\rho_\mathsf{n} T)\big|\\[0.5ex]
-\nu^{-1} F^*\big| \rho_\mathsf{n} (v^\mathrm{eff} - q_1)\big\rangle& \nu^{-1}F^*(v^\mathrm{eff} -q_1)(1 -\rho_\mathsf{n} T)
\end{pmatrix}.
\end{equation}

In Figure 8 we report on molecular dynamics simulations of the Toda lattice in thermal equilibrium for $3$, out of $9$ published,  distinct parameter values. While there
is a large body of  work in providing microscopic confirmation of GHD in the context of a variety of models, the results 
below presumably constitute the most extensive check, so far. There are several points in favor. The initial statistical state
is sharply defined. Modulo numerically computing the quantities appearing in  the TBA formalism, there is the  concrete prediction \eqref{7.32} based on linearized GHD. The comparison uses \textit{no} adjustable parameter. Errors are smaller than $3.5\%$, which is of the order of statistical noise. For the shorter time $t=150$ (data not displayed)  
one still observes ballistic scaling for our range of parameters. However, as systematic deviation, the peaks are slightly lower than predicted by the theory. 

At the highest pressure the peaks are shifted towards the boundaries, which is even more pronounced for $\beta = 2.0, 
P= 3.53$ (data not displayed). This behavior reflects that for $\beta, P \to \infty $, at fixed ratio, the Toda DOS converges to the arcsine law which exhibits an inverse square root singularity at either boundary, compare with \eqref{3.73}.
To be emphasized is the middle frame at which $\nu = - 0.03$, hence $|\rho_\mathsf{f}| = 33$. Closer to $\nu = 0$ the TBA simulation becomes difficult. Of course for molecular dynamics one would
merely have to set $q_1 = q_N$ at which point the typical distance of particles is order $1/\sqrt{N}$,  which equals $0.02$ in our case. At such high densities the collision rate ansatz becomes dubious. But the simulation confirms that GHD remains valid even under such extreme conditions.
\begin{figure}[!t]
\centering
\includegraphics[width=0.69\textwidth]{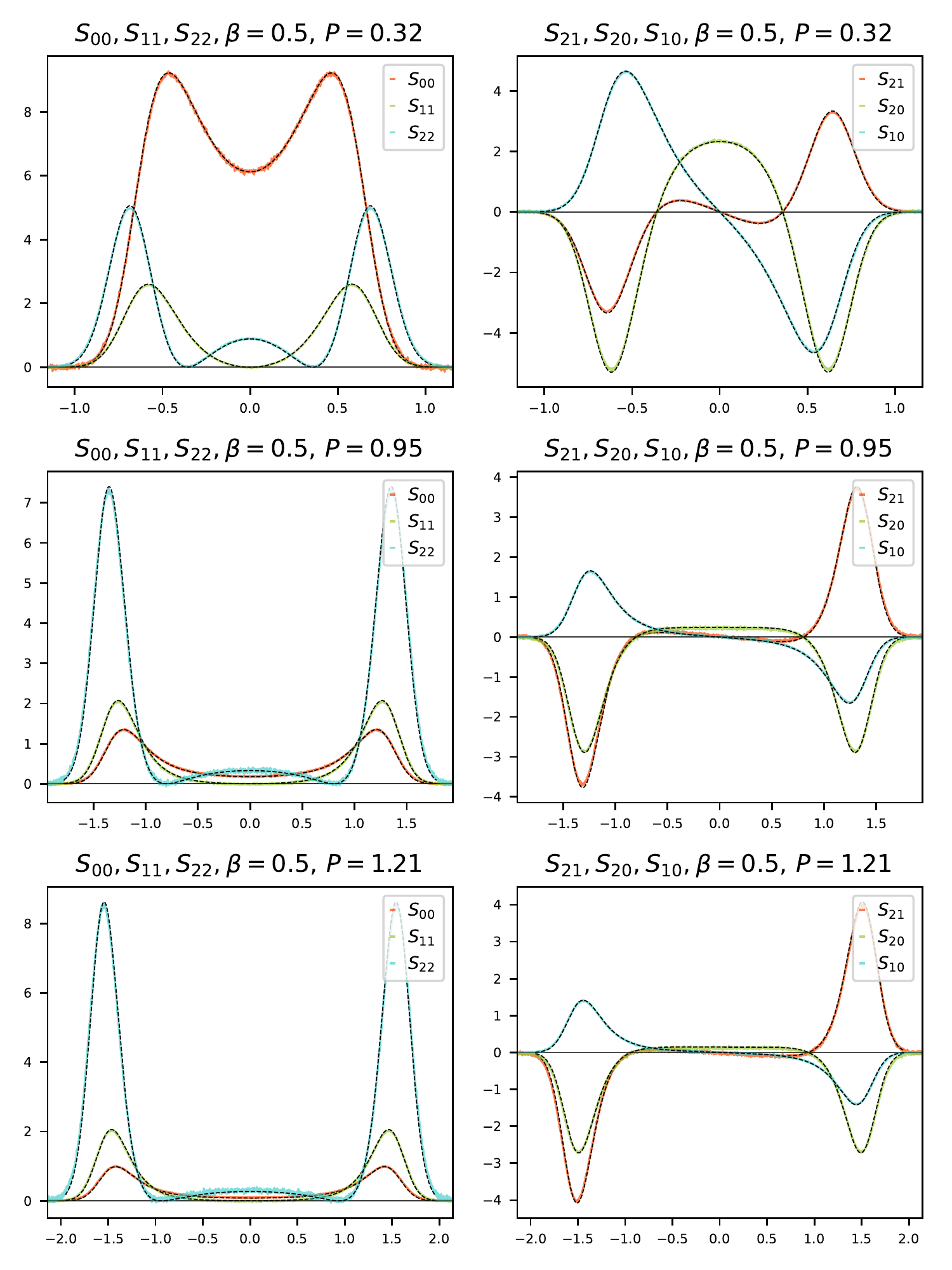}
\caption{Toda spacetime correlation functions in thermal equilibrium.  The dashed line is the GHD prediction \eqref{7.32} for $m,n= 0,1,2$ and the colored lines are molecular dynamics simulations. The system size is $N= 3000$ and time $t=600$. The parameters $\beta,P$ are indicated and $\nu = 2.58, -0.03, -0.42$. Thermal average is based on $3\times10^6$
 samples. From \cite{GKMMMS22}.
Toda spacetime correlation functions in thermal equilibrium.  The dashed line is the GHD prediction \eqref{7.32} for $m,n= 0,1,2$ and the colored lines are molecular dynamics simulations. The system size is $N= 3000$ and time $t=600$. The parameters $\beta,P$ are indicated and $\nu = 2.58, -0.03, -0.42$. The thermal average is over samples is $3\times10^6$. From \cite{GKMMMS22}.}
\label{fig8}
\end{figure}

Actually, the peak structure on the Euler scale is only the starting step of the Landau-Lifshitz theory. Its main focus is the broadening of the peaks due to dissipation and molecular noise. For a simple fluid in three dimensions the broadening is of order $\sqrt{t}$ with a Gaussian shape function. In one dimension the standard Landau-Lifshitz theory would make the same prediction. But the linear 
fluctuation theory has to be corrected by expanding the nonlinear Euler term to second order, which then results in a well-confirmed superdiffusive spreading as $t^{2/3}$. More details can be found in Section \ref{sec15.2}.
But even before there is a generic obstacle. For a well-isolated peak the broadening is easily observed. But for the Toda 
lattice the Euler term yields already  a smeared out background. So one has to specifically tune to physical situations for which the anticipated dissipative corrections can be detected.\\\\
$\blackdiamond\hspace{-1pt}\blackdiamond$~\textit{Equilibrium spacetime correlations for hard rods}.
For hard rods the $T$ operator can be inverted and identities as \eqref{7.32}
become more explicit. We consider thermal equilibrium, hence 
$h(w) = h_\beta(w) = \sqrt{\beta/2\pi}\exp(-\beta w^2/2)$, $q_1 = 0$, and
\begin{equation}\label{7.46}
v^\mathrm{eff}(v) = \frac{\nu}{\nu - \mathsfit{a}} v.
\end{equation}    
The dressing operator reads
\begin{equation}\label{7.47}
(1 - T \rho_\mathsf{n})^{-1} = 1 + \frac{\mathsfit{a}}{\nu} |\varsigma_0\rangle\langle h_\beta|
\end{equation}
and hence
\begin{equation}\label{7.48}
 \varsigma_0^\mathrm{dr} = 1 +   \frac{\mathsfit{a}}{\nu} ,\quad  \varsigma_1^\mathrm{dr}  =v ,\quad  \varsigma_2^\mathrm{dr} = v^2 + \frac{\mathsfit{a}}{\beta\nu}.
\end{equation}
Rescaling the spatial coordinate to $\mathsfit{x} = (\nu - \mathsfit{a})x/t$ one concludes
\begin{equation}\label{7.49}
\{S_{00}^\mathrm{hr}(x,t), S_{11}^\mathrm{hr}(x,t),S_{22}^\mathrm{hr}(x,t)\} = (\nu - \mathsfit{a})t^{-1}h_\beta(\mathsfit{x})\{1 +(\mathsfit{a}/\nu))^2, \mathsfit{x}^2, (\mathsfit{x}^2 + (\mathsfit{a}/\beta \nu))^2\}.
\end{equation}
The interesting term is the energy-energy spacetime correlation $S_{22}^\mathrm{hr}(x,t)$. For the ideal gas,  $\mathsfit{a} =0$, the scaling function 
$\mathsfit{x}^4h_\beta(\mathsfit{x})$ vanishes at $\mathsfit{x} =0$ and has a double peak. Turning on the interaction  the minimum at $\mathsfit{x}=0$ is shifted to $h_\beta(0)(\mathsfit{a}/\beta \nu)^4$, resulting in three peaks.
 \hfill$\blackdiamond\hspace{-1pt}\blackdiamond$\bigskip
\begin{center}
 \textbf{Notes and references}
 \end{center}
 \begin{center}
 \textbf{Section 7.0}
\end{center} 
The book by Forster \cite{F75} is still a very readable exposition of the fluctuation theory developed by Landau and Lifshitz \cite{LL75}.
 Critical dynamics is covered in the classic review of Hohenberg and  Halperin \cite{HH77}.  
  \newpage
  \begin{center}
  \textbf{Section 7.1}
\end{center}
The existence of the infinite volume dynamics is proved by Lanford et al. \cite{LLL77}. Fluctuation theory for simple fluids is discussed in  Spohn  \cite{S91}. 
 As a crucial distinction, the hydrodynamic equations for a few conservation laws may develop shocks. An instructive account is the most readable lecture course by Bressan \cite{B13}. In Das et al.   \cite{DDH20} a spin chain is considered, for which at low temperatures phase differences emerge as an almost conserved field, on top of the two strictly conserved fields.
 \bigskip
 \begin{center}
 \textbf{Section 7.2}
\end{center} 
 A more detailed account of GGE spacetime correlations can be found in Spohn \cite{S19b}.
  Early molecular dynamics simulations of the Toda lattice and related theoretical investigations have been carried out by Schneider and  Stoll \cite{SS80},  Schneider  \cite{S83}, 
  Diederich  \cite{D81}, and Cuccoli et al. \cite{CSTV93}.
 Molecular dynamics simulations of thermal spacetime correlations are reported by  Kundu and Dhar \cite{KD16}. Amongst other features, investigated are the low density regime and the low temperature harmonic approximation. A wider range of parameters is covered 
 by Mazzuca et al. \cite{GKMMMS22} including self- and cross-correlations of stretch, momentum, and energy. The numerical results are compared with the GHD prediction in Eq. \eqref{7.32}.   
  A more general perspective on fluctuations at the Euler scale and beyond is developed by Doyon and Myers  \cite{DM20}.
\newpage
\section{Domain wall initial states}
\label{sec8} 
\setcounter{equation}{0}
Spacetime correlations encode small deviations from a GGE. But the hydrodynamic scale covers also  initial states very far from stationarity, thereby exhibiting a rich portfolio of time-dependent behavior. 
From a theoretical perspective, a very natural and much studied nonequilibrium initial state is enforced by joining  two semi-infinite systems, the left half line in one GGE and the right half line in another GGE. The fields are spatially constant except for a  single jump at the origin. On the level of GHD, the solution scales exactly as 
$x/t$, a simplification which promises to yield exact solutions. In the mathematical theory of finitely many hyperbolic conservation laws, 
such set-up is known as Riemann problem. Its solution consists of  flat pieces, smooth rarefaction waves, and jump discontinuities, called shocks.
Their precise spatial sequence can be complicated and is, in principle, encoded  by the eigenvectors and eigenvalues of the matrix $\mathsfit{A}$ in their dependence on the thermodynamic parameters, compare with \eqref{7.15}. 
As discussed in Section \ref{sec6.2} in normal form the flux Jacobian $A$ is multiplication by $\nu^{-1} (v^\mathrm{eff} - q_1)$. This
form admits only rarefaction waves. No spontaneous generation of jump discontinuities is expected. 

Besides its intrinsic interest, we pick this example to also illustrate the predictive power of GHD. For the domain wall
problem, the primary goal is to figure out  macroscopic behavior,  i.e. how stretch, momentum, and energy vary spatially given their particular initial jump discontinuity. Apparently GHD is the only theoretical scheme through which such a macroscopic behavior can be computed.

We slightly rewrite Eq. \eqref{6.24} as
\begin{equation}\label{8.1} 
 \partial_t \rho_\mathsf{n}(x,t;v) + \tilde{v}^\mathrm{eff}(x,t;v) \partial_x  \rho_\mathsf{n}(x,t;v) = 0,\quad 
\end{equation}
with
\begin{equation}\label{8.1a}
\tilde{v}^\mathrm{eff}(v) = \nu^{-1}(v^\mathrm{eff}(v) - q_1). 
\end{equation}
Domain wall initial conditions read 
\begin{equation}\label{8.2}
 \rho_\mathrm{dw}(x,0;v) = \chi(\{x <0\}) \rho_{\mathsf{n}-}(v)+ \chi(\{x \geq 0\})\rho_{\mathsf{n}+}(v).
\end{equation}
Instead of $\rho_{\mathsf{n}\pm}$, physically it might be more natural to prescribe the DOS of the Lax matrix and the average stretch. But mathematically the normal form \eqref{8.1} together with \eqref{8.2} is more accessible. 

Since the solution to  \eqref{8.1}, \eqref{8.2} scales ballistically, we set $\rho_\mathrm{dw}(x,t;v) = \mathsf{g}(t^{-1}x;v)$ and $\tilde{v}^\mathrm{eff}(x,t;v) = \tilde{\mathsf{v}}^\mathrm{eff}(t^{-1}x;v)$. Without loss of generality one adopts $t=1$ and arrives at  
\begin{equation}\label{8.3}
( x - \tilde{\mathsf{v}}^\mathrm{eff}(x;v))\partial_x \mathsf{g}(x;v) =0, \qquad \lim_{x \to \pm\infty} \mathsf{g}(x;v) = \rho_{\mathsf{n}\pm}(v).
 \end{equation}
Therefore  $x \mapsto \mathsf{g}(x;v)$  for fixed $v$ has to be constant except for jumps at the zeros of $ x - \tilde{\mathsf{v}}^\mathrm{eff}(x;v)$ as function of $x$. 
At this stage, it is not clear which level of generality to adopt. Physically, one would expect to have a unique solution. Thus 
in the $(x,v)$-plane there should be a \textit{contact line} which divides the plane into two domains, one characterized to contain the set $\{-\infty\}\times \mathbb{R}$ and the other the set
$\{\infty\}\times \mathbb{R}$. The solution  $\mathsf{g}(x;v)$ is constant in either domain and jumps from $\rho_{\mathsf{n}-}(v)$ to $\rho_{\mathsf{n}+}(v)$ across the contact line. Uniqueness means a unique contact point for every $v$.
Then the contact line is represented as the graph of a  function defined by $v \mapsto \tilde{\phi}(v)$. As discussed below, for hard rods, $\tilde{\phi}$ is invertible  with  inverse denoted by $\phi$. Hence  the contact line can be written also as the graph of $x \mapsto\phi(x)$.  We present our argument for the latter case, since hard rods are reasonably close to the Toda lattice. More general contact lines could be handled in a similar fashion. 

With our assumptions, for every $x$ there is a unique contact point $v = \phi(x)$, which satisfies
\begin{equation}\label{8.4}
x - \tilde{\mathsf{v}}^\mathrm{eff}(x;v),
\end{equation}
and the solution ansatz reads
\begin{equation}\label{8.5}
g^\phi(v) = \chi(\{v > \phi\}) \rho_{\mathsf{n}-}(v)+ \chi(\{v \leq \phi\})\rho_{\mathsf{n}+}(v).
\end{equation}
The superscript $\phi$ will be used to generically  indicate that in the TBA formalism $\rho_\mathsf{n}$ is replaced by $g^\phi$.
  For example,  
\begin{equation}\label{8.6} 
f^{\mathrm{dr},\phi} = \big(1 - Tg^\phi\big)^{-1} f,\qquad \nu^\phi \langle \rho_\mathsf{p}^\phi\rangle =1,
\end{equation}
compare with \eqref{3.55} and \eqref{3.58}. Note that $g^{\pm\infty}(v) = \rho_{\mathsf{n}\pm}(v)$. In particularly, we define
\begin{equation}\label{8.7}
\tilde{\mathsf{v}}^{\mathrm{eff}, \phi}(v)= (\nu^\phi)^{-1} \Big(\frac{\varsigma_1^\phi(v)}{\varsigma_0^{\phi}(v)} - q_1^\phi\Big).
\end{equation}
From the solution ansatz it follows that
\begin{equation}\label{8.8}
\mathsf{g}(x;v) = g^{\phi(x)}(v), \qquad \tilde{\mathsf{v}}^\mathrm{eff}(x;v) = \tilde{\mathsf{v}}^{\mathrm{eff}, \phi(x)}(v).
\end{equation}
The condition \eqref{8.4} turns into
\begin{equation}\label{8.9}
x = \tilde{\mathsf{v}}^{\mathrm{eff},\phi(x)} (\phi(x)).
\end{equation}
Defining
\begin{equation}\label{8.10}
G(\phi) = \tilde{\mathsf{v}}^{\mathrm{eff},\phi} (\phi),
\end{equation}
we conclude that
\begin{equation}\label{8.11}
x = G(\phi(x)),\qquad \phi(x) = G^{-1}(x).
\end{equation}
The contact line is the inverse of the function $G$ in \eqref{8.10}. 
To determine $G$ numerically, one has to compute the dressing depending on $\phi$. This then yields
$\mathsf{g}(x;v)$ from which the observables of interest, as average stretch, momentum, and energy, at location $x$ are deduced.

Qualitatively $G$ is linear for large $\phi$. But this is a special feature of fluid-like models with a quadratic kinetic energy.
For the discrete linear wave equation with a general dispersion relation, because of the bounded momentum space, the concept of contact line as such remains untouched, but the line is no longer equal to the graph of a function. 

The predicted sharp step
appears to be oversimplified. Indeed on longer time scales the step is expected to be smeared roughly as an error function, compare with Section \ref{sec15.3}. 
\begin{figure}[!t]
	\centering
	\includegraphics[width=.45\linewidth]{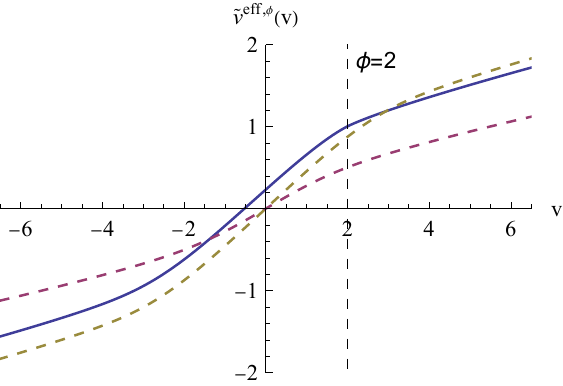}
	\includegraphics[width=.45\linewidth]{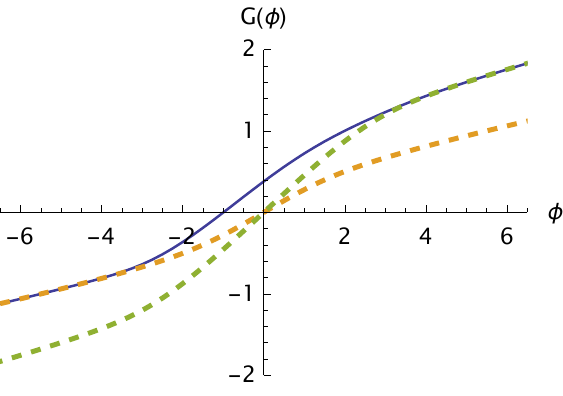}
\caption{Domain wall initial state for the Toda lattice at uniform inverse temperature $\beta = 1$, $P_- = \frac{1}{2}$, and $P_+ = 2$. Displayed in blue are $\tilde{\mathsf{v}}^{\mathrm{eff},\phi}(v)$ at $\phi = 2$ and the contact line $G(\phi)$.  On the left, the dotted  magenta line is $\tilde{\mathsf{v}}^\mathrm{eff}(v)$ computed by using  $\rho_{\mathsf{n}+}$ and the olive  line using $\rho_{\mathsf{n}-}$. Correspondingly on the right displayed is $\tilde{\mathsf{v}}^\mathrm{eff}(\phi)$ using either $\rho_{\mathsf{n}+}$ or  $\rho_{\mathsf{n}-}$. From \cite{MS21}.}
\label{fig9}
\end{figure}	
$\blackdiamond\hspace{-1pt}\blackdiamond$~\textit{Hard rod domain wall}.\hspace{1pt} For the hard rod lattice the function $G$ can still be computed
analytically. We recall  that $\nu$ is the average stretch, $\nu>\mathsfit{a}$, and $h(v)$ denotes the normalized velocity distribution,
 \begin{equation}\label{8.12} 
\rho_\mathsf{p}(v) = \nu^{-1} h(v),\quad \rho_\mathsf{n}(v) = (\nu- \mathsfit{a})^{-1} h(v), \quad q_1 = \int_\mathbb{R} \mathrm{d}w h(w)w. 
 \end{equation} 
In normal form the hard rod GHD reads
\begin{equation}\label{8.13} 
\partial_t \rho_\mathsf{n}(x,t;v) + (\rho(x,t) -\mathsfit{a})^{-1}(v - q_1(x,t))\partial_x \rho_\mathsf{n}(x,t;v) = 0, 
 \end{equation}
compare with the structurally identical Eq. \eqref{8.1}. The initial condition is the same as in \eqref{8.2} and the $G$-function is computed by the method explained above with the result
\begin{equation}\label{8.14} 
G_\mathrm{hr}(\phi) 
= \Big( \int_{\phi}^\infty \hspace{-4pt}\mathrm{d}v  \rho_{\mathsf{n}-}(v) + \int^{\phi}_{-\infty} \hspace{-4pt}\mathrm{d}v \rho_{\mathsf{n}+}(v)\Big) \phi
- \int_{\phi}^\infty \hspace{-4pt}\mathrm{d}v \,v \rho_{\mathsf{n}-}(v) - \int^{\phi}_{-\infty} \hspace{-4pt}\mathrm{d}v \,v \rho_{\mathsf{n}+}(v).
 \end{equation}
Clearly, $dG_\mathrm{hr}(\phi)/d\phi >0$. Thus  $G_\mathrm{hr}$ is invertible. For $\phi \to \infty$ one obtains 
\begin{equation}\label{8.15} 
G_\mathrm{hr}(\phi) \simeq \frac{1}{\nu_+ -\mathsfit{a}}(\phi - q_{1,+})
 \end{equation}
and correspondingly for $\phi \to - \infty$ with $\nu_+, q_{1+}$ replaced by $\nu_-, q_{1-}$. If the boundary velocity distributions have an
exponential decay, then the error in \eqref{8.13} is also exponentially small. The contact line is asymptotically linear with a well localized monotone interpolation.

If $\mathsfit{a} <0$, one can tune the boundary values such that $\nu_- < 0< \nu_+$. There is then some intermediate point, $x_0$ with $\nu(x_0) = 0$, a case studied in more detail at the end of Section \ref{sec9.1}. 
 \hfill$\blackdiamond\hspace{-1pt}\blackdiamond$\bigskip
\begin{center}
 \textbf{Notes and references}
 \end{center}
 \begin{center}
 \textbf{Section 8}
\end{center} 
This chapter is based on the study of Mendl and Spohn \cite{MS21}, where in particular numerical simulations of GHD with domain wall initial conditions are reported. Domain wall for hard rods are studied in Doyon and  Spohn  \cite{DS17}.  On the Euler scale the step at the contact line is sharp.  Earlier work of Bulchandani et al. \cite{BCM19} considers the case of low pressure, $ P_- < P_+ < P_\mathrm{c}$. Molecular dynamics is compared with predictions of GHD. 
Also quantum mechanically corrections are obtained to lowest order. 
 For the Toda lattice, diffusive corrections are discussed in Section \ref{sec15.3}. The broadening is convincingly observed in the XXZ model, see De Nardis et al. \cite{DBD18}. 

Domain wall initial conditions have been studied numerically for the discrete sinh-Gordon model by Bastianello et al. \cite{BDWY18}.
The most detailed investigations are available for the XXZ model, see  Piroli et al. \cite{PNCBF17} and Misguich et al. \cite{MMK17}. In this case the spectral parameter space contains in addition the type of string states and the contact line refers to a larger space. Gamayun et al.  \cite{GMI19} study domain wall initial conditions for a classical spin chain and its quantized version. 

Due to momentum conservation, the contact line of the Toda lattice is one-to-one and covers the full real line. On the other hand, for XXZ, and other discrete models, 
the contact line takes values only in a bounded $v$-interval, which implies  left and right edges in $x$-space up to which  boundary values remain
constant. The behavior near the edge often shows intricate oscillatory decay, which is beyond the hydrodynamic scale and has been elucidated in considerable detail for a variety of models, see 
Collura   et al. \cite{CLV18}, Bulchandani and Karrasch \cite{BK19}, and Grava et al. \cite{GKMM21}. 
\newpage
 \section{Toda fluid}
\label{sec9}
\setcounter{equation}{0}
Toda particles might as well be viewed to move on the real line, which is the fluid picture, see Fig. 1.
The name ``fluid" is somewhat misleading, since particles are distinguishable. Still the dynamical behavior is fluid-like.
  In a way, we  have to start from the beginning and redo the computation for average fields and their currents.   Since 
we rely on chain results, the discussion can be more compressed. To avoid duplicating symbols,
the fluid is distinguished from the chain through the index ``${}_\mathsf{f}$''. $x \in \mathbb{R}$ stands for the coordinate of the one-dimensional physical space.
\subsection{Euler equations}
\label{sec9.1}
For the Toda fluid conserved fields have a density given by  
\begin{equation}\label{9.1} 
Q^{[0]}_\mathsf{f}(x) = \sum_{j \in \mathbb{Z}} \delta(q_j - x),\qquad Q^{[n]}_\mathsf{f}(x) = \sum_{j \in \mathbb{Z}} 
\delta (q_j- x) Q^{[n]}_j
 \end{equation}
 for $\;n\geq1$. Taking their time derivative,
 \begin{eqnarray}\label{9.2} 
&&\frac{d}{dt}Q^{[0]}_\mathsf{f}(x) = \sum_{j \in \mathbb{Z}} \delta'(q_j - x)p_j,\nonumber\\
&&\frac{d}{dt}Q^{[n]}_\mathsf{f}(x) = \sum_{j \in \mathbb{Z}} \big(\delta'(q_j - x)p_jQ^{[n]}_j + \delta(q_j - x)(J_j^{[n]}
- J_{j+1}^{[n]})\big).
 \end{eqnarray}
Hence fluid current densities read
 \begin{eqnarray}\label{9.3} 
&&\hspace{0pt}J^{[0]}_\mathsf{f}(x) = \sum_{j \in \mathbb{Z}}\delta(q_j - x)p_j ,\\
&&\hspace{0pt} J^{[n]}_\mathsf{f}(x) = \sum_{j \in \mathbb{Z}}\big(\delta(q_j - x)p_jQ^{[n]}_j + \big(\theta(q_{j} -x) - \theta(q_{j-1} -x)\big)J_j^{[n]}\big)\nonumber
\end{eqnarray}
with $\theta$ the step function, $\theta(x) =0$ for $x\leq 0$ and $\theta(x) =1$ for $x > 0$. The index $0$ has been treated separately.
But in fact, $Q^{[n]}_j$  naturally extends to  $n=0$ by setting $Q^{[0]}_j = 1$. This property foreshadows a single hydrodynamic
equation.

\begin{figure}[!b]
\centering
\includegraphics[width=0.9\textwidth]{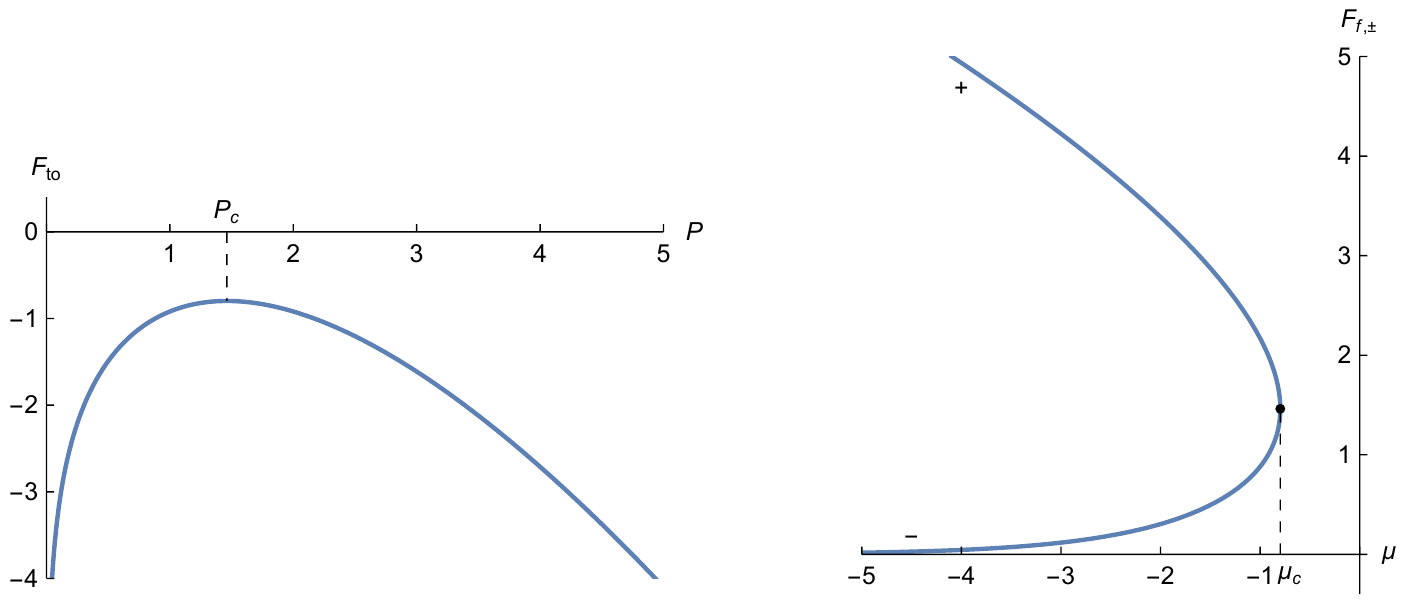}
\caption{ 
For thermal equilibrium at $\beta = 1$, to the left is the Toda lattice free energy
$F_\mathrm{to} = \mu$ as a function of the pressure $P$. To the right are the two branches of the fluid free energy 
$F_{\mathsf{f},\pm} = -P$ as a function of $\mu$, see \eqref{9.10}.}
\label{fig10}
\end{figure}

The next item is GGE, where we start from the canonical ensemble with partition function
\begin{equation}\label{9.4} 
Z_\mathrm{can}(N,\ell)= \int_{\Gamma_N}\prod_{j=1}^N\mathrm{d}r_j \mathrm{d}p_j \delta\big( Q^{[0],N} -\ell\big)\exp\!\big(-\mathrm{tr}[V(L_N)]\big). 
\end{equation} 
The ensemble is canonical, since volume, $q_{N+1} - q_1$, and number of particles are fixed. In our discussion the confining potential will play a passive role and is hence dropped from the notation, except when appearing as GGE index. For the chain we considered $\ell = \nu N$ and, by the equivalence of ensembles, switched to $ \exp\!\big( - P Q^{[0],N}\big)$
with the pressure $P$ dual to the stretch $\nu$. In the  limit $N\to \infty$  the free energy per site,
$F_\mathrm{to}(P)$, has been obtained already, compare with  \eqref{3.39}. Turning to the fluid we want to keep the volume fixed and allow for fluctuations in $N$.
The corresponding chemical potential is denoted by $\mu$. Then
\begin{equation}\label{9.5} 
Z_\mathsf{f}(\mu,\ell) = \sum_{N=1}^\infty \mathrm{e}^{\mu N} \int_{\Gamma_N} \prod_{j=1}^N\mathrm{d}r_j \mathrm{d}p_j \delta\big( Q^{[0],N} -\ell\big)\exp\!\big(-\mathrm{tr}[V(L_N)]\big),
\end{equation}
compare with \eqref{3.1}. One has to distinguish the cases $\ell >0$ and $\ell <0$, corresponding to  either increasing or decreasing particle labelling,
on average.
Thus the particle number $Q_\mathsf{f}^{[0],N} >0$, while the fluid density $\rho_\mathsf{f} = N/\ell$ can take either sign. Since $\nu = \ell/N$,
the relation  $\nu \rho_\mathsf{f}  = 1$ holds always. Close to $P_\mathrm{c}$, defined by $\nu(P_\mathrm{c}) = 0$, the fluid density $\rho_\mathsf{f}$ jumps from $\infty$ to 
$- \infty$ when increasing $P$. This looks singular but merely reflects a particular choice of coordinates.  $\nu(P)$ is a smooth function. 
Of interest is the limit
\begin{equation}\label{9.6} 
\lim_{\ell \to \pm \infty} -\frac{1}{\ell}\log Z_\mathsf{f}(\mu,\ell) = F_{\mathsf{f},\pm}(\mu),
\end{equation}
where $\pm$ refers to the sign of $\ell$. The fluid free energy has two branches labeled  by $\mathrm{sgn}(\ell)$. Correspondingly there are two distinct infinite volume GGE averages denoted by
$\langle \cdot \rangle_{\mu,\pm,V}$. For a convergent partition function, one has to require $\mu < \mu_\mathrm{max}$
for either branch with $\mu_\mathrm{max}= F_\mathrm{to}(P_\mathrm{c})$, as displayed in Figure
\ref{fig10}. 

Thermodynamically  the dual to $P$ is the stretch $\nu$ and to $\mu$ the fluid density $\rho_\mathsf{f}$. Let us denote by $\mathcal{L}F$ the Legendre
transform of some function $F$. By the equivalence of ensembles on the level of  free energies one concludes that 
\begin{equation}\label{9.7}
\lim_{N \to \infty} - N^{-1} \log Z_\mathsf{can}(N,\nu N) = \mathcal{L}F_\mathrm{to}(\nu)
\end{equation}
and 
 \begin{equation}\label{9.8}
\lim_{\ell \to \pm\infty} - \ell^{-1} \log Z_\mathrm{can}(\rho_\mathsf{f}\ell,\ell) = \mathcal{L}F_{\mathsf{f},\pm}(\rho_\mathsf{f}).
\end{equation}
Thus the two free energies are related by
\begin{equation}\label{9.9} 
\mathcal{L}F_\mathrm{to}(\nu) = \rho_\mathsf{f}^{-1}\mathcal{L}F_{\mathsf{f},\pm}(\rho_\mathsf{f}), \quad \nu \rho_\mathsf{f} =1,
\end{equation}
which implies
\begin{equation}\label{9.10} 
F_\mathrm{to}(P) = \mu,\quad - F_{\mathsf{f},\pm}(\mu) = P.
\end{equation}
Stated differently, $F_\mathrm{to}$ and $- F_{\mathsf{f},\pm}$ are inverse functions of each other. The first identity was already noted in
\eqref{3.39} and the argument of $F_{\mathsf{f},\pm}$ is in fact the parameter appearing in the TBA equation. $F_\mathrm{to}(P)$ is concave
with maximum at $P_\mathrm{c}$. On the other hand, $F_{\mathsf{f},+}$  is concave and   $F_{\mathsf{f},-}$ convex  with both graphs 
being smoothly joined at $\mu_\mathrm{max}$. 

The free energy of the Toda fluid can be written in such a way as to suggest the extension to a larger class of integrable models. Starting from \eqref{9.10} and using \eqref{3.40}, the free energy equals
\begin{equation}\label{9.11} 
F_\mathsf{f}(\mu,V) =  -\int_\mathbb{R} \mathrm{d}w \mathrm{e}^{-\varepsilon(w)},
\end{equation}
compare with \eqref{9.10} and \eqref{3.36}. Differentiating the TBA equation with respect to $\mu$ one arrives at 
\begin{equation}\label{9.12} 
-\partial_\mu F_\mathsf{f}(\mu,V) =  \int_\mathbb{R} \mathrm{d}w \rho_\mathsf{n}(w)(1- T\rho_\mathsfit{n})^{-1}\varsigma_0(w) = \nu^{-1} =
\rho_\mathsf{f}.
\end{equation}
Similarly by perturbing $V$ as $V +\kappa \varsigma_n$, one concludes  
\begin{equation}\label{9.13} 
\partial_\kappa F_\mathsf{f}(\mu,V+ \kappa\varsigma_n)|_{\kappa = 0} =  \int_\mathbb{R} \mathrm{d}w \rho_\mathsf{n}(w)(1- T\rho_\mathsf{n})^{-1}\varsigma_n(w) = \langle \rho_\mathsf{p}\varsigma_n\rangle = \langle Q_{\mathsf{f}}^{[n]} (0)\rangle_{\mu,V},
\end{equation}
which confirms the defining derivatives for the fluid free energy $F_\mathsf{f}$.

For the current averages we follow the strategy as developed for the lattice case by introducing the field-current susceptibility matrix through
\begin{equation}\label{9.14} 
B_{\mathsf{f},m,n} = \int_\mathbb{R} \mathrm{d}x \langle J_{\mathsf{f}}^{[m]} (x)Q_{\mathsf{f}}^{[n]} (0)\rangle^\mathrm{c}_{\mu,\pm,V}.
\end{equation}
By the same argument as leading to \eqref{6.9}, $B_\mathsf{f}$ is a symmetric matrix, $B_{\mathsf{f},m,n} = B_{\mathsf{f},n,m}$. 
Since $J_{\mathsf{f}}^{[0]} =   Q_{\mathsf{f}}^{[1]}$, one starts  from
\begin{equation}\label{9.15} 
 \partial_\mu \langle J_{\mathsf{f}}^{[n]} (0)\rangle_{\mu,\pm,V}  = B_{\mathsf{f},n,0} =  B_{\mathsf{f},0,n} =
 \int_\mathbb{R} \mathrm{d}x \langle Q_{\mathsf{f}}^{[1]} (x)Q_{\mathsf{f}}^{[n]} (0)\rangle^\mathrm{c}_{\mu,\pm,V}.
 \end{equation}
The last expression can be written as derivative,
\begin{equation}\label{9.16} 
B_{\mathsf{f},0,n} =  -\partial_\kappa  \langle Q_{\mathsf{f}}^{[n]} (0)\rangle_{\mu,\pm,V+ \kappa w}\big|_{\kappa = 0}
= -\partial_\kappa\langle \rho_{\mathsf{p},\kappa}  \varsigma_n \rangle\big|_{\kappa = 0},  
\end{equation}
where $\rho_{\mathsf{p},\kappa}$ is determined by the solution of the TBA equation
\begin{equation}\label{9.17} 
V +\kappa  \varsigma_1 -\mu - T \rho_{\mathsf{n},\kappa} + \log \rho_{\mathsf{n},\kappa}  =0.
\end{equation}
In \eqref{3.58} we differentiated TBA with respect to $\varsigma_0$. This time it is with respect to $\varsigma_1$.
Denoting by $'$ the derivative at $\kappa = 0$, one obtains
\begin{equation}\label{9.18} 
\rho_\mathsf{n}' = - \rho_\mathsf{n} (1 - T \rho_\mathsf{n})^{-1}  \varsigma_1
\end{equation}
and
\begin{equation}\label{9.19} 
\rho_\mathsf{p}' = \partial_\mu \rho_\mathsf{n}' = - \partial_\mu (\rho_\mathsf{n} \varsigma_1^\mathrm{dr})
= -\partial_\mu \big(\rho_\mathsf{p} v^\mathrm{eff}\big),
\end{equation}
where, as before, $v^\mathrm{eff}$ is defined in \eqref{6.19}. Therefore
\begin{equation}\label{9.20} 
\partial_\mu\big(\langle J_{\mathsf{f}}^{[n]} (0)\rangle_{\mu,\pm,V} - \langle \rho_\mathsf{p}v^\mathrm{eff} \varsigma_n \rangle\big) = 0,
\end{equation}
implying that the difference  does not depend on $\mu$. To fix the constant, 
the current vanishes in the limit of vanishing density of particles, in other words for $\mu \to -\infty$, and so does
$\rho_\mathsf{p}$ since $\nu \langle \rho_\mathsf{p}\rangle = 1$.  We conclude that
\begin{equation}\label{9.21}
\langle J_{\mathsf{f}}^{[n]} (0)\rangle_{\mu,\pm,V} = \langle \rho_\mathsf{p}v^\mathrm{eff} \varsigma_n \rangle,
\end{equation}
which holds for all $n$, including $n=0$. Compared to the lattice currents \eqref{6.17}, only the shift $\rho_\mathsf{p}q_1$ has been removed. 
This can be understood by starting from the fluid side. Then switching to the lattice dynamical equations,  the lateral motion is subtracted and such a term has to appear in the average lattice currents.

Since $B_\mathsf{f}$ is a symmetric matrix, there has to be a current potential, denoted by $G_\mathsf{f}$, such that
\begin{equation}\label{9.22} 
\partial_\kappa G_\mathsf{f}(\mu,V +\kappa \varsigma_n)|_{\kappa = 0} = \langle J_{\mathsf{f}}^{[n]} (0)\rangle_{\mu,V}.
\end{equation}
Indeed, one only has to slightly modify the free energy as 
\begin{equation}\label{9.23} 
G_\mathsf{f}(\mu,V) =  \int_\mathbb{R} \mathrm{d}w w \mathrm{e}^{-\varepsilon(w)}.
\end{equation}
Then, as before  by differentiating TBA, 
\begin{equation}\label{9.24} 
\partial_\kappa G_\mathsf{f}(\mu,V+ \kappa\varsigma_n)|_{\kappa = 0} =  \int_\mathbb{R} \mathrm{d}w w\rho_\mathsf{n}(w)(1- T\rho_\mathsf{n})^{-1}\varsigma_n(w) = \langle \varsigma_1^\mathrm{dr}\rho_\mathsf{n}\varsigma_n\rangle.
\end{equation}

Since in \eqref{9.21} the term $n=0$ no longer plays a special role, there is only a single hydrodynamic equation which reads
\begin{equation}\label{9.25} 
\partial_t\rho_\mathsf{p}(x,t;v) + \partial_x\big(v^\mathrm{eff}(x,t;v)\rho_\mathsf{p}(x,t;v)\big) = 0.
\end{equation}
Using \eqref{6.23} to transform to quasilinear form yields
\begin{equation}\label{9.26} 
\partial_t\rho_\mathsf{n}(x,t;v) + v^\mathrm{eff}(x,t;v)\partial_x\rho_\mathsf{n}(x,t;v) = 0,
\end{equation}
which is identical to the lattice version, except for the term accounting for the lateral motion.
By construction $\rho_\mathsf{n}(x,t;v) \geq 0$, a property which is correctly propagated by \eqref{9.26}.
On the other hand the overall sign of $\rho_\mathsf{p}(x,t;v)$ is governed by $\nu(x,t)$, to say, if  $\nu(x,t) > 0$, then
$\rho_\mathsf{p}(x,t;v)\geq 0$, and if $\nu(x,t) < 0$, then $\rho_\mathsf{p}(x,t;v)\leq 0$. An example for $\nu(x,t)$ changing sign
is discussed in the Insert just below. \bigskip\\
$\blackdiamond\hspace{-1pt}\blackdiamond$~\textit{High-low pressure domain wall}.\hspace{1pt}  As discussed in Chapter \ref{sec8},
a domain wall initial state consists of two half lines with homogeneous states $\rho_{\mathsf{n} -}$, resp. $\rho_{\mathsf{n}+}$. If $\nu_{-} < 0 <\nu_+$, then there is a sign change of the particle density $\rho_{\mathsf{p}}$. While in that Chapter we used the Toda lattice, it is instructive to view the 
high-low pressure domain wall from the perspective of a fluid. To be concrete, in the initial state
the particle with label $0$ is placed at the origin, $q_0 = 0$, the particles with negative indices correspond to a thermal state with parameters $(\nu_-, \beta)$ and those with positive indices to $(\nu_+, \beta)$. Thus the momenta are i.i.d.
Gaussian with mean zero and variance $\beta^{-1}$, while $\{q_{j+1} - q_j, j\leq -1\}$ are i.i.d. $ \chi_{2P_-}$ distributed random variables,
  and   $\{q_{j+1} - q_j, j\geq 0\}$ are i.i.d. $ \chi_{2P_+}$, compare with Chapter \ref{sec8}. The case of interest is $P_- > P_\mathrm{c} >P_+$, thus $\nu_-  < 0 < \nu_+$. Then, by assumption, $\langle q_{j+1} - q_j \rangle_{P_-} = \nu_- <0$ for $j\leq -1$, while  $\langle q_{j+1} - q_j \rangle_{P_+} = \nu_+   > 0$ for $j\geq 0$.    

A typical ordering is of the form $\dots > q_{-1} > q_0 =0 < q_1< \dots$\,. Thus initially the average particle density equals $|\nu_-|^{-1} + \nu_+^{-1}$ on $[0,\infty)$ and decays exponentially on $(-\infty,0]$. Under the dynamics, the point at which the ordered domain touches the anti-ordered domain is moving in time, its label being denoted by $\kappa(t)$. Close to  $q_{\kappa(t)}$ particles pile up. The domain boundary acts as bottleneck for particles. The position of the bottleneck is $ x_\mathrm{bn}(t) = q_{\kappa(t)}$. Numerical simulations suggest that $\kappa(t)$ and $x_\mathrm{bn}(t)$ change linearly in time, at least approximately. Also to the left of the bottleneck the particle density vanishes rapidly. As observed numerically, the scaling function $G$, see \eqref{8.11}, has a nonzero slope at $x_\mathrm{c}$. Thus, near the bottleneck the particles should be distributed according to equilibrium with a linearly varying pressure. From this property one infers that in physical space the particle density at the bottleneck diverges as an inverse square root. It would be of interest to better understand the precise particle statistics close to the bottleneck.
\hfill$\blackdiamond\hspace{-1pt}\blackdiamond$
\subsection{Generalized free energy - again}
\label{sec9.2}
The fluid picture suggests an alternative approach to obtain the generalized free energy, thereby circumventing entirely the Dumitriu-Edelman change of volume elements. 
The same method will be applied to Calogero-Moser models
as to be discussed in Chapter \ref{sec11}. As further spin-off, a closer structural similarity to the Lieb-Liniger 
$\delta$-Bose gas will be accomplished.  

One starts from the Toda fluid, $(q_j,p_j) \in\mathbb{R}^2$, $j = 1,\ldots,N$. The open chain is governed by the hamiltonian  of Eq. 
\eqref{2.58} and has the Lax matrix
\begin{equation}\label{9.27}
(L_N^\diamond)_{j,j} = p_j,\qquad (L_N^\diamond)_{j,j+1} = (L_N^\diamond)_{j+1,j} =  \mathrm{e}^{-((q_{j+1} - q_j)/2)}
\end{equation}
with $j = 1,\ldots,N$ in the former case, $j = 1,\ldots,N-1$ in the latter case, and $(L_N^\diamond)_{i,j} = 0$ otherwise.
 Some aspects of the scattering theory for the open Toda chain have been discussed already in Section 2.3. The explicit canonical transformation to scattering coordinates, alias action-angle variables, is stated in Eqs. \eqref{2.65} -  \eqref{2.67}. By construction, under the transformation $\Phi$, one has the identity
\begin{equation}\label{9.28}
\mathrm{tr}\big[V(L^\diamond_N)\big]\circ \Phi = \sum_{j=1}^N V(\lambda_j).
\end{equation}
Inserting from \eqref{2.65}, \eqref{2.66}, in addition one obtains
\begin{equation}\label{9.29}
\mathrm{e}^{-q_1} \circ \Phi =  \frac{\sigma_{N-1}}{\sigma_{N}}= \sum_{j=1}^{N} Y_j\mathrm{e}^{-\phi_j} , \qquad \mathrm{e}^{q_N} \circ \Phi =  \frac{\sigma_1}{\sigma_0}= \sum_{j=1}^{N} Y_j\mathrm{e}^{\phi_j}
\end{equation}
with
\begin{equation}\label{9.30}
Y_j = \prod_{i=1, i \neq j}^N|\lambda_i -\lambda_j|^{-1}.
\end{equation}

We first consider the case when the average stretch $\nu >0$. To study GGEs, the end particles are subject to suitable
boundary potentials, $V^+_{\ell,1}(q_1)$, $V^+_{\ell,N}(q_N)$, such that all particles are confined to a  volume of approximate size $\ell$ with  $\ell >0$.
Our observation above suggests to choose  the boundary potential as
\begin{equation}\label{9.31}
V^+_{\ell,1}(q_1) + V^+_{\ell,N}(q_N)= \mathrm{e}^{-\ell/2}\big(\mathrm{e}^{-q_1} + \mathrm{e}^{q_N}\big),
\end{equation}
since $\mathrm{e}^{-q_1}$ and $ \mathrm{e}^{q_N}$ are given by rather simple formulas.  Then in approximation $-\ell/2 <q_1$ and $q_N <\ell/2$, which yields the desired confinement because $\nu >0$.
In case $\nu <0$, one has to choose $\ell <0$ and the boundary potential becomes 
\begin{equation}\label{9.32}
V_{\ell,1}^-(q_1) + V_{\ell,N}^-(q_N)= \mathrm{e}^{\ell/2}\big(\mathrm{e}^{q_1} + \mathrm{e}^{-q_N}\big).
\end{equation} 
But now $Y_j$ is switched to the denominator, which makes the analysis more delicate.
In the following we restrict ourselves to the case $\nu>0$.

Next considered is a canonical type ensemble with $\ell$ and $N$ as parameters. 
Since $\Phi$ is a canonical transformation, its Jacobian determinant equals $1$ which implies for the volume elements $\mathrm{d}^N \hspace{-1pt} p\, \mathrm{d}^N \hspace{-1pt}q = \mathrm{d}^N \hspace{-1pt}\phi \,\mathrm{d}^N \hspace{-1pt}\lambda$. To the energy-like term $\mathrm{tr}\big[V(L^\diamond_N)\big]$
one adds the boundary potential  \eqref{9.31}.  
For the partition function one thereby arrives at
\begin{eqnarray}\label{9.33}
&&\hspace{-28pt} Z_{\mathrm{to},N}(\ell,V) = \int_{\Gamma_N} \mathrm{d}^N \hspace{-1pt} p\, \mathrm{d}^N \hspace{-1pt}q \exp\big(- \mathrm{tr}\big[V(L^\diamond_N)\big] -\mathrm{e}^{-\ell/2}\big(\mathrm{e}^{-q_1} + \mathrm{e}^{q_N}\big)\big)\nonumber\\
&&\hspace{20pt} = \frac{1}{N!} \int_{\mathbb{R}^{N}} \mathrm{d}^N \hspace{-1pt}\phi \,\int_{\mathbb{R}^{N}}\mathrm{d}^N \hspace{-1pt}\lambda \exp\Big(-\sum_{j=1}^NV(\lambda_j) - 
2 \mathrm{e}^{-\ell/2}Y_j \cosh \phi_j \Big).
\end{eqnarray}
Here we used that under the map $\Phi^{-1}$ the eigenvalues are ordered. Since the integrand is symmetric in the action variables,  their integration 
is extended over $\mathbb{R}^{N}$ at the expense of the factor $1/N!$.  The 
$\phi_j$-integrations can be carried out explicitly and yield 
\begin{equation}\label{9.34}
Z_{\mathrm{to},N}(\ell,V) = \frac{1}{N!} \int_{\mathbb{R}^{N}}\mathrm{d}^N \hspace{-1pt}\lambda \prod_{j=1}^N\mathrm{e}^{-V(\lambda_j)} 
\prod_{j=1}^N 2K_0\big( 2 \mathrm{e}^{-\ell/2}Y_j \big),
\end{equation}
where
\begin{equation}\label{9.35}
K_0(x) = \int_0^\infty \mathrm{d}t\exp(- x \cosh{t}) 
\end{equation}
is the  zero order modified Bessel function of the second kind. 

As a consequence, the probability distribution of eigenvalues of the Lax matrix $L_N^\diamond$ under the normalized measure
\begin{equation}\label{9.36}
Z_{\mathrm{to},N}(\ell,V)^{-1}  \exp\big(- \mathrm{tr}\big[V(L^\diamond_N)\big] - \mathrm{e}^{-\ell/2}\big(\mathrm{e}^{-q_1} + \mathrm{e}^{q_N}\big)\big) \mathrm{d}^N \hspace{-1pt} p\,\mathrm{d}^N \hspace{-1pt}q
\end{equation}
is given by 
\begin{equation}\label{9.37}
Z_{\mathrm{to},N}(\ell,V)^{-1}  
\exp\Big[- \sum_{j=1}^N V(\lambda_j) + \sum_{j=1}^N \log\Big(2K_0\big( 2 \exp\big[ - \tfrac{1}{2} N \big(\nu - \frac{2}{N} \sum_{i = 1,i\neq j}^N
\log Y_j\big)\big]\big)\Big)\Big]
\end{equation}
relative to $(1/N!)\mathrm{d}^N \hspace{-1pt}\lambda$. Note that 
\begin{equation}\label{9.38}
-\frac{2}{N} \log Y_j = \frac{1}{N}\sum_{i=1,i\neq j}^N \phi_\mathrm{to}(\lambda_j - \lambda_i)
\end{equation}
which, except for the diagonal term, equals $\phi_\mathrm{to}(\lambda_j - \cdot)$ integrated over the empirical DOS.
Only the functional in \eqref{9.37} is more complicated than before and has  an explicit $N$-dependence  in addition.

The next step is the $\ell \to \infty$ asymptotics at fixed $\nu = \ell/N$. 
The Bessel function  behaves as $K_0(x) = - \log x$ for $x \to 0$ and $ K_0(x) = (\pi/2x)^{1/2}\mathrm{e}^{-x}$ for $x \to \infty$. Hence for large $\ell$
\begin{equation}\label{9.39}
K_0(\mathrm{e}^{-a\ell}) \simeq a\ell\hspace{8pt} \mathrm{for} \hspace{3pt}a>0,\qquad \log K_0(\mathrm{e}^{-a\ell}) \simeq - \mathrm{e}^{-a\ell} \hspace{8pt}\mathrm{for} \hspace{3pt} a<0. 
\end{equation}
Let us define
\begin{equation}\label{9.40}
\Upsilon_j = \nu + \frac{1}{N} \sum_{i=1,i\neq j}^N \phi_\mathrm{to}(\lambda_j - \lambda_i).
\end{equation}
Then for large $N$
\begin{equation}\label{9.41}
\log\Big(2K_0\big( 2 \exp\big[ - \tfrac{1}{2} N \Upsilon_j\big]\big)\Big) \simeq \begin{cases}\log N + \log \Upsilon_j, \,\,\mathrm{if} \,\,\Upsilon_j >0, \\[1ex] 
-2 \mathrm{e}^{-(N\Upsilon_j/2)}, \,\,\mathrm{if}\,\, \Upsilon_j < 0.
\end{cases}
\end{equation}
Following the arguments in Section \ref{sec3.5}, the corresponding free energy functional reads 
\begin{equation}\label{9.42}
\mathcal{F}_{\mathrm{to},\mathsf{f},+}^\circ(\varrho) = \nu^{-1}\int_\mathbb{R}\mathrm{d}w\varrho(w)\Big( V(w) -1 + \log \varrho(w) - \log\Big(\nu +  \int_\mathbb{R}\mathrm{d}w'\varrho(w')\phi_\mathrm{to}(w - w')\Big)\Big) 
\end{equation}
in case $\nu +  \int_\mathbb{R}\mathrm{d}w'\varrho(w')\phi_\mathrm{to}(w - w') >0$ and $\mathcal{F}_{\mathrm{to},\mathsf{f},+}^\circ(\varrho) = \infty$ otherwise.	
The variation is over all probability densities with $\mathcal{F}_{\mathrm{to},\mathsf{f},+}^\circ(\varrho) < \infty$. Momentarily we ignore this infinite-dimensional constraint, but will return to the issue below. Denoting the minimizer by $\varrho^*$, as before, one concludes 
that 
\begin{equation}\label{9.43} 
\lim_{\ell\to\infty}- \frac{1}{\ell}\log Z_{\mathrm{to},N}(\ell,V) = F_{\mathrm{to},\mathsf{f},+}(\nu,V) = \mathcal{F}_{\mathrm{to},\mathsf{f},+}^\circ(\varrho^*).
\end{equation}

It is convenient to substitute $\nu^{-1}\varrho = \rho$. 
 Then the free energy per unit length of the Toda fluid 
 is determined by the free energy functional  
 \begin{equation}\label{9.44}
\mathcal{F}_{\mathrm{to},\mathsf{f},+}(\rho) = \int_\mathbb{R}\mathrm{d}w\rho(w)\Big( V(w) -1 + \log \rho(w) - \log\Big(1 + \int_\mathbb{R}\mathrm{d}w'\rho(w')\phi_\mathrm{to}(w - w')\Big)\Big),
\end{equation}
where the  minimization is constrained by $\rho \geq 0$ and 
 \begin{equation}\label{9.45} 
\int_\mathbb{R}\mathrm{d}w\rho(w) = \nu^{-1} .
\end{equation} 
As before the constraint is lifted by the Lagrange multiplier $\mu$. Introducing the \textit{space density} $\rho_\mathsf{s}$ by
 \begin{equation}\label{9.46} 
\rho_\mathsf{s} = 1 + T  \rho,
\end{equation}
the free energy functional can be written as 
\begin{equation}\label{9.47} 
\mathcal{F}_{\mathrm{to},\mathsf{f},+}^\bullet(\rho) =  \int_\mathbb{R} \mathrm{d}w \rho(w) \big(V(w) -1 - \mu\big) + \int_\mathbb{R} \mathrm{d}w\rho_\mathsf{s}(w)\Big(\frac{\rho}
{\rho_\mathsf{s}} \log\frac{\rho}{\rho_\mathsf{s}}\Big)(w). 
\end{equation}
The first summand can be viewed as energy-like and the second summand as the entropy of $\rho$ relative to $\rho_\mathsf{s}$. 

Provisionally we denote the minimizer of \eqref{9.47} by $\rho_\mathsf{p}$ which satisfies the Euler-Lagrange equation
\begin{equation}\label{9.48} 
V(w) -\mu +\log\frac{\rho_\mathsf{p}}{1 + T\rho_\mathsf{p}} - T\frac{\rho_\mathsf{p}}{1 + T\rho_\mathsf{p}} = 0.
\end{equation}
In view of  \eqref{3.59} and \eqref{3.47} this suggests to set
\begin{equation}\label{9.49} 
\rho_\mathsf{n} = \frac{\rho_\mathsf{p}}{1 + T\rho_\mathsf{p}}, \qquad \rho_\mathsf{n} = \mathrm{e}^{-\varepsilon},
\end{equation}
which leads to 
\begin{equation}\label{9.50} 
\varepsilon = V -\mu  - T\mathrm{e}^{-\varepsilon},
\end{equation}
in agreement with the TBA equation \eqref{3.54}. Since $\rho_\mathsf{s} = 1 + T\rho_\mathsf{p}$, 
the space density satisfies
\begin{equation}\label{9.51} 
\rho_\mathsf{s} = \varsigma_0^\mathrm{dr}, \quad \rho_\mathsf{s}\rho_\mathsf{n}= \rho_\mathsf{p}.
\end{equation}
In conclusion 
 \begin{equation}\label{9.52} 
\lim_{\ell \to \infty} -\ell^{-1}\log Z_{\mathrm{to},\mathsf{f},+,N}(\ell,\mu,V) = \mathcal{F}_{\mathrm{to},\mathsf{f},+}^\bullet(\rho_\mathsf{p}) = - \int_\mathbb{R} \mathrm{d}w \mathrm{e}^{-\varepsilon(w) },
\end{equation} 
in agreement with \eqref{9.11}.

As explained in Section \ref{sec3.5},
the convergence of the DOS to a deterministic limit follows on general grounds, but can also be confirmed by considering the first order variational derivatives of the free energy.  Returning to the infinite-dimensional constraint  below Eq. \eqref{9.42}, evaluating it at 
$\rho_\mathsf{p}$ reads 
\begin{equation}\label{9.53}
0 < 1 +  \int_\mathbb{R}\mathrm{d}w'\rho_\mathsf{p}(w')\phi_\mathrm{to}(w - w')= \rho_\mathsf{s}(w).
 \end{equation}
For $P < P_\mathrm{c}$ 
 the number density $\rho_\mathsf{n} >0$ and also 
 $\rho_\mathsf{p},\rho_\mathsf{s}$ stay positive. Hence \eqref{9.53} is equivalent to $\nu >0$, a condition which was required already 
at the very beginning of the argument. Thus, indeed, the minimizer remains unaffected by the constraint. 

The free energy for the Toda
fluid  has also been obtained by using scattering coordinates in combination with a specifically designed external potential. 
We presented two very different proofs for the generalized free energy and arrived at the same answer, as it should be.  Since the proofs are lengthy, one might worry about missing factors of $2$ or so. 
Having two independent derivations vouches for correctness. Furthermore, the availability of a second tool opens up the possibility to handle the generalized free energy for a larger class of integrable models.\bigskip\\
$\blackdiamond\hspace{-1pt}\blackdiamond$~\textit{Local macroscopic conservation laws.} 
By locally conserved we mean here the identity 
\begin{equation}\label{9.55} 
\partial_t \mathfrak{n}(x,t;w) +\partial_x\mathfrak{j}(x,t;w) =0
\end{equation}
pointwise in $w$. If GHD is considered in  a periodic interval, say $[0,\ell]$, this implies the conservation law 
\begin{equation}\label{9.56} 
\partial_t \int_0^\ell \mathrm{d}x \,\mathfrak{n}(x,t;w)  =0.
\end{equation}
According to \eqref{9.47}, a physical example is the entropy with density
\begin{equation}\label{9.54} 
s(x,t;w)  = - \rho_\mathsf{p}(x,t;w)\log\rho_\mathsf{n}(x,t;w).
\end{equation}
If the solution to GHD is free of discontinuities, the entropy  is locally conserved.  

Surprisingly, as a general property of GHD, beyond entropy a much larger family of local macroscopic conservation laws 
is obtained in replacing the $\log$ by some arbitrary function $g$, hence
\begin{equation}\label{9.57} 
 \mathfrak{n}(x,t;w) = (\rho_\mathsf{p}g(\rho_\mathsf{n}))(x,t;w).
\end{equation}
 To  verify one starts from
\begin{eqnarray}\label{9.58} 
&&\hspace{-30pt}- \partial_t(\rho_\mathsf{p}f(\rho_\mathsf{n})) = -(\partial_t\rho_\mathsf{p})f(\rho_\mathsf{n})- \rho_\mathsf{p}f'(\rho_\mathsf{n})
 \partial_t \rho_\mathsf{n}\nonumber\\[0.8ex]
 &&\hspace{39pt} = \partial_x(v^\mathrm{eff}\rho_\mathsf{p}) f(\rho_\mathsf{n}) + \rho_\mathsf{p}v^\mathrm{eff}\partial_x
 f(\rho_\mathsf{n}) =\partial_x(v^\mathrm{eff} \rho_\mathsf{p} f(\rho_\mathsf{n})),
\end{eqnarray}
 which implies
\begin{equation}\label{9.59} 
 \mathfrak{j} = v^\mathrm{eff} \rho_\mathsf{p} f(\rho_\mathsf{n}),
 \end{equation}
 as to be confirmed.
\hfill$\blackdiamond\hspace{-1pt}\blackdiamond$\bigskip
 \begin{center}
 \textbf{Notes and references}
 \end{center}
 \begin{center}
 \textbf{Section 9.1}
\end{center} 
A more detailed discussion of the Toda fluid is presented by Doyon \cite{D19,D19a}.  The high-low pressure domain wall is studied by Mendl and Spohn \cite{MS21}.\bigskip
\begin{center}
\textbf{Section 9.2}
\end{center} 
The Toda lattice with boundary potential $g_1(\mathrm{e}^{-q_1} + \mathrm{e}^{q_N}) +g_2(\mathrm{e}^{-2q_1} + \mathrm{e}^{2q_N})$ is still integrable, see van Diejen (1995) for a proof. But the respective Lax matrix is less accessible. Using scattering coordinates for the computation of the generalized free energy is anticipated in 
 Doyon  \cite{D19}, where the free energy functional \eqref{9.42} can be found already. The identity \eqref{9.33} seems to be a novel result.
 The large class of local conservation laws for the hydrodynamic equations has been noted by
 Caux et al. \cite{CDDKY19}. I learned about \eqref{9.58} through the master thesis of Deokule  \cite{D22}.
 \newpage
 \section{Hydrodynamics of soliton gases}
\label{sec10} 
\setcounter{equation}{0}

In his seminal contribution Morikazu Toda realized that a chain with exponential interactions allows for travelling waves, called \textit{soliton}
in analogy to continuum nonlinear wave equations.  The soliton is a coherent motion of many particles which is spatially localized and travels at constant 
velocity maintaining its shape. Two incoming solitons
undergo an intricate collision process, to eventually move outwards with   velocities unchanged and characteristic scattering shifts. 
But qualitatively this resembles the  dynamics  discussed for Toda quasiparticles already. Apparently the Toda dynamics possesses two distinct levels with soliton-like structure. 
To distinguish, we refer to either soliton-based scheme or particle-based scheme. The much studied prime example for a soliton-based scheme
is the Korteweg-de Vries (KdV) equation.  This topic will be taken up first because of its similarity with the soliton-based scheme of the Toda dynamics.
Allowing  for both hydrodynamic schemes is shared by Toda and Ablowitz-Ladik lattice. For other models only either one of the two schemes is known.
\subsection{Soliton gas of the Korteweg-de Vries equation}
\label{sec10.1}

The most widely studied soliton gas is based on the  KdV equation, written in its conventional form as 
\begin{equation}\label{10.1}
\partial_t u + \partial_x(3u^2 + \partial_x^2u) = 0
\end{equation}
with $u$ a real-valued wave field over $\mathbb{R}$.
Physically the Korteweg-de Vries equation describes the motion of surface waves in a shallow channel of water under gravity and in the approximation of small amplitudes. Solitary waves have been observed experimentally back in  1834 and the description in terms of a specific nonlinear wave equation was accomplished in 1895. While the study of soliton solutions strongly indicated integrability, the infinite hierarchy of local conservation laws was first obtained by 
Miura, Gardner, and Kruskal  in 1968.

As the Toda lattice, KdV admits for a Lax pair. Now $L$ is an operator,
\begin{equation}\label{10.2}
L = -\partial_x^2 - u(t),
 \end{equation}
while the partner operator reads
\begin{equation}\label{10.3}
B= -\big(4\partial_x^3 + 6u\partial_x +4\partial_xu\big).
 \end{equation}
Under the KdV time evolution
\begin{equation}\label{10.4}
\partial_t L = [B,L],
 \end{equation}
 a property signalling integrability. A different viewpoint is to consider,  
 for given $u$,
 the eigenvalue problem  
 \begin{equation}\label{10.5}
Lv = \lambda v
 \end{equation}
 and the evolution equation
 \begin{equation}\label{10.6}
\partial_t v = Bv.
 \end{equation}
 This system has solutions only if $u(t)$ satisfies the KdV equation, which thus plays the role
 of a consistency relation. 
 
 The lowest order conserved fields can still be obtained by hand with the resulting densities
\begin{equation}\label{10.7}
Q^{[0]}(x) = u(x), \quad Q^{[1]}(x) = u^2(x),  \quad Q^{[2]}(x) = u^3(x) -\tfrac{1}{2} (\partial_xu)^2(x).
\end{equation}
Here we adopted that convention that $u^n$ is the term with no derivatives. A more systematic tool comes from  the underlying Poisson structure. The formal Poisson bracket turns out to be
 \begin{equation}\label{10.8}
 \{u(x),u(x')\}_\mathrm{KV} = \delta'(x-x').
 \end{equation}
More conveniently, 
functions on phase space are replaced by functionals of $u$ and its derivatives,
\begin{equation}\label{10.9}
 F = \int  \mathrm{d}x h(u(x), \partial_x u(x),\ldots,\partial_x^n u(x))
 \end{equation}
 with some polynomial $h$. Of course, more general functionals could be considered. But for our purposes \eqref{10.9} will do. The gradient of $F$, denoted by $\nabla F(x)$, is the functional derivative of $F$ with respect to
 $u$ at $x$. The functionals $F,G$ have then the Poisson bracket
 \begin{equation}\label{10.10}
 \{F,G\}_\mathrm{KV} = \int_\mathbb{R} \mathrm{d}x \nabla F(x)\partial_x  \nabla G(x).
 \end{equation}
In particular this Poisson bracket satisfies the Jacobi identity.
Denoting the total conserved fields by  $Q^{[n]} =\int \mathrm{d}x Q^{[n]}(x)$, their respective gradients are 
\begin{equation}\label{10.11}
\nabla Q^{[0]} = 1,\quad \nabla Q^{[1]} = 2 u, \quad \nabla Q^{[2]} = 3u^2  + \partial_x^2u.
\end{equation}
Higher order conservation laws are obtained recursively through
 \begin{equation}\label{10.12}
(2n-1) \partial_x \nabla Q^{[n+1]} = (n+1)\big(u\partial_x + \partial_x u- \tfrac{1}{2} \partial^3_x \big)\nabla Q^{[n]}
\end{equation}
The prefactors result from our convention in \eqref{10.7} upon disregarding the operator $\partial_x^3$. In fact, this kind of recursion is familiar from integrable quantum many-body systems
and called \textit{boost operator}. 

For the hamiltonian we set $H_\mathrm{kv} = -Q^{[2]}$, which then
generates the time evolution as
\begin{equation}\label{10.13}
\partial_t u(x,t) = \{u(x,t), H_\mathrm{kv}\}_\mathrm{KV}.
\end{equation}
Hence $Q^{[2]}$ is called energy. 
Analogously, $Q^{[1]}$ generates a spatial shift to the right with unit speed, hence called momentum. 
The KdV equation possesses a two-parameter family of traveling wave solutions, called \textit{soliton}, more specifically one-soliton. 
Their analytic form is
 \begin{equation}\label{10.14}
 u_{\mathrm{s},1}(x,t;\eta,\delta) = 2\eta^2 \mathrm{sech}^2\big(\eta(x - 4\eta^2t -\delta)\big).
 \end{equation}
 Here $\eta$ is the spectral parameter, $\eta >0$, and  $\delta$ is a shift,  $\delta\in \mathbb{R}$. The center of the soliton moves on the straight line $x(t) =  4\eta^2t +\delta$. Hence the soliton speed is $4\eta^2$. The taller and narrower  the shape the faster the soliton moves. The conserved fields take the values $Q^{[0]}(u_{\mathrm{s},1}) = 4 \eta$, 
$Q^{[1]}(u_{\mathrm{s},1}) = (4^2/3) \eta^3$, and $Q^{[2]}(u_{\mathrm{s},1}) = (4^3/5) \eta^5$.

 We now start with localized initial data, in the sense that 
 \begin{equation}\label{10.15}
 \int_\mathbb{R}\mathrm{d}x|x||u(x)| <\infty.
 \end{equation}
 For long times, $t \to \pm\infty$,  the KdV solution splits into a soliton part, denoted by $u^\mathrm{s}$, and dispersive part which decays to zero with some power law. More precisely 
 for any $\epsilon >0$,
 \begin{equation}\label{10.16}
 \lim_{|t|\to\infty} \sup_{\{x||x|>\epsilon |t|\}} |u(x,t) - u^\mathrm{s}(x,t)| = 0.
 \end{equation}
 This behavior is reminiscent of charges coupled to the Maxwell field which asymptotically acquire straight line motion through emitting radiation. Hence the
 dispersive part is also called radiation. The soliton part consists of $N$ freely moving solitons, $N$ depending on the initial condition. In a formula
 \begin{equation}\label{10.17}
 u^\mathrm{s}(x,t) = \sum_{j=1}^N u_{\mathrm{s},1}(x,t;\eta_j,\phi_j^\pm),
 \end{equation}
 where $\pm$ refers to $t \to \pm\infty$. Solitons behave like quasiparticles. However for KdV there are only overtaking collisions. Thus for $t \to -\infty$ the incoming solitons are ordered as $x_{j}(t) < x_{j+1}(t)$ and the spectral parameters as 
$ \eta_1 >\ldots>\eta_N >0$,  while outgoing solitons travel in reverse order. The label $N$ is used in analogy to Section 2.3, where $N$ denotes the number of quasiparticles (which equals the number of physical particles).
 
 The crucial dynamical information comes from the scattering shifts, $\kappa_j$, which are written as 
 \begin{equation}\label{10.18}
 \phi_j^- \hspace{6pt}\mathrm{for}\,\, t \to -\infty, \qquad  \phi_j^+ = \phi_j^- + \kappa_j \hspace{6pt}\mathrm{for}\,\, t \to \infty,
 \end{equation}
 where
 \begin{equation}\label{10.19}
 \kappa_j = \frac{1}{\eta_j} \sum_{i=1,i\neq j}^N \mathrm{sgn}(\eta_j - \eta_i) \phi_\mathrm{kv}(\eta_j,\eta_i)
 \end{equation}
 with the KdV scattering kernel
 \begin{equation}\label{10.20}
  \phi_\mathrm{kv}(\eta,\eta')= \log \Big|\frac{\eta - \eta'}{\eta + \eta'}\Big|.
 \end{equation}
 The latter structure is already familiar from the hard rod gas and Toda lattice. As hallmark of integrability, the scattering shifts are determined by the two-soliton scattering shift. But in contrast to hard  rods, the latter shift depends on the incoming velocities. Note that $\phi_\mathrm{kv}(\eta,\eta') <0$ as for physical hard rods.
 
 There are concise formulas for $N$-soliton solutions, which are defined by requiring the radiation part to vanish identically. Considering the special case $N=2$ we can illustrate how the identity in \eqref{10.20} is obtained. The two-soliton solution can be written as
 \begin{equation}\label{10.21}
 u_{\mathrm{s},2}(x,t) = 2\partial_x^2\log\varphi_2(x,t)
 \end{equation}
 with 
 \begin{equation}\label{10.22}
 \varphi_2 =  1 + A_1\mathrm{e}^{2\theta_1} +A_2\mathrm{e}^{2\theta_2} + A_3\mathrm{e}^{2(\theta_1+ \theta_2)},
 \end{equation}
 where
 \begin{equation}\label{10.23}
 \theta_i = \eta_ix -\omega_it,\quad \omega_i = 4 \eta_i ^3, \quad i = 1,2.
 \end{equation}
 The coefficients $A_1,A_2$ are free parameters which can be converted into $\phi_i ^\pm$. But $A_3$ is determined by
 \begin{equation}\label{10.24}
\log  A_3 = \log A_1 + \log A_2 + \phi_\mathrm{kv}(\eta_1,\eta_2). 
 \end{equation}
 In \eqref{10.19} the actual scattering shift carries the prefactor $1/\eta_j$. This dependence can be traced to the solution ansatz \eqref{10.23}, where in the definition of $\theta_i $ the factor $\eta_i$ has not been bracketed out.

 Given previous experience, one may boldly jump ahead to the hydrodynamic equation for the KdV soliton gas. We introduce the counting function  
 $f(x,t;\eta)\mathrm{d}x \mathrm{d}\eta$ which is the number of  solitons with spectral parameter in the interval $[\eta, \eta + \mathrm{d}\eta]$ and location in $[x,x + \mathrm{d}x]$
 at time $t$. The counting function is governed by
 \begin{equation}\label{10.25}
 \partial_t f +\partial_x (v^\mathrm{eff}f) = 0.
 \end{equation}
 For the effective velocity, as before, we assume the collision rate ansatz which in our case becomes
 \begin{equation}\label{10.26}
 v^\mathrm{eff}(\eta) = 4 \eta^2 + \eta^{-1} \int _0^\infty \mathrm{d}\eta \phi_\mathrm{kv}(\eta,\eta')f(\eta') \big(v^\mathrm{eff}(\eta') - v^\mathrm{eff}(\eta)\big).
 \end{equation}
 
 Still missing are some explanations on how the counting function $f$ is related to local values of the conserved fields. In principle,  the counting function can be measured by the time-of-flight method. 
One considers some time $t$ and a sufficiently large spatial box, still small on the scale of macroscopic variations. The wave field $u(x)$ 
in this spacetime cell is cut out and then smoothly interpolated with the zero-field all the way to infinity. The such constructed field $u_\mathrm{b}(x)$ is evolved under the KdV dynamics. For sufficiently long times the solution consists of many spatially well separated one-solitons. In approximation, their statistics is that of an ideal gas with uniform density $\bar{\varrho} = \int_{\mathbb{R}_+} \mathrm{d}\eta f(\eta)$. Each soliton carries a spectral parameter, $\eta_j$, and these form a family of i.i.d. random variables with single point probability density $ h(\eta) = f(\eta)/\bar{\varrho}$. To determine the
average value of the conserved field $Q^{[n]}$ under such statistics, the positional average yields the integral over  $Q^{[n]}$ evaluated at $u_{\mathrm{s},1}(x,0;\eta,0))$
and the i.i.d. average results in the integral over $\eta$ weighted with $h(\eta)$, in formulas 
\begin{equation}\label{10.27}
\int_0^\infty\mathrm{d}\eta h(\eta) \bar{\varrho}\int_\mathbb{R}\mathrm{d}x Q^{[n]}(u_{\mathrm{s},1}(x,0;\eta,0)) =\frac{2^{2n+2}}{2n +1}\int_0^\infty \mathrm{d} \eta f(\eta)\eta^{2n+1}.
 \end{equation}
Thereby we connected the average of the $n$-th conserved field with the $(2n+1)$-th moment of the counting function. The numerical coefficients are known as Kruskal series.  Since $f$ vanishes on $\mathbb{R}_-$, it is uniquely determined by its moments \eqref{10.27}. 

A natural question is to have a more intrinsic characterization
of the statistics of the wave field $u_\mathrm{b}(x)$ in the prescribed cell. This is so to speak an inverse scattering problem and not much is known. For particle-based hydrodynamics the statistics of the positions and momenta of the particles in the cell 
should be given by a GGE with properly adjusted parameters. At least for the Toda fluid we know that, as regards to the conserved fields,  the data agree with those from the  time-of-flight method.  Also the Lax filter could be employed, compare with Section \ref{sec3.2}. For KdV the corresponding construction would involve bound state eigenvalues of the Schr\"{o}dinger operator \eqref{10.2} with potential given by the box field $u_\mathrm{b}(x)$.\bigskip\\
 $\blackdiamond\hspace{-1pt}\blackdiamond$~\textit{TBA revisited}.
 The Toda lattice consists of nonrelativistic particles with energy-momentum relation $E(P) = P^2/2$. 
 For relativistic models the energy-momentum relation would have to be modified. Also for solitons there is no particular reason  to have a quadratic energy-momentum relation. Therefore TBA  has to be extended  
 to cover such cases.
 
 In a hamiltonian context, if the energy-momentum relation $E(P)$ is given, then the velocity is determined by $v = \partial E/\partial P$. In case of the KdV equation the one-soliton momentum and energy are given in terms of the spectral parameter $\eta$. To stay in context we still use $\eta$, which however is now a parameter depending on the model under study.
 The functions $E(\eta)$ and $P(\eta)$ are prescribed. 
 Then, with prime referring to parameter differentiation, the velocity becomes $v = E'/P'$.
 Through the interaction between quasiparticles $E',P'$ become dressed and,
appealing to the fundamental relation \eqref{6.19}, one argues that
\begin{equation}\label{10.28} 
 v^\mathrm{eff} = \frac{(E'){^\mathrm{dr}}}{(P'){^\mathrm{dr}}}\,,
\end{equation}
the dressing transformation being defined as before. Of course, for $E(P) = P^2/2$ the original relation is recovered.

Starting from the collision rate ansatz, it is now reformulated as
 \begin{equation}\label{10.29}
 v^\mathrm{eff} = \frac{1}{P'}\big(E' + T(\rho_\mathsf{p} v^\mathrm{eff}) - (T\rho_\mathsf{p} ) v^\mathrm{eff}\big),
 \end{equation}
 where $T(\eta, \eta')$ is some symmetric kernel. As before, $\rho_\mathsf{p}$ is the particle density.
 The number density is defined by
 \begin{equation}\label{10.30}
 \rho_\mathsf{n} = \frac{ \rho_\mathsf{p}}{P' + T \rho_\mathsf{p}}.
 \end{equation}
Then the dressing operation for some function $f$  is still    
 \begin{equation}\label{10.31}
 f^\mathrm{dr} = (1 - T \rho_\mathsf{n})^{-1} f
 \end{equation}
with $\rho_\mathsf{n}$ regarded as multiplication operator.  The inverse to \eqref{10.30} reads
 \begin{equation}\label{10.32}
 \rho_\mathsf{p} = \rho_\mathsf{n}P'^{\,\mathrm{dr}},
 \end{equation}
  which identifies the space density  as
  \begin{equation}\label{10.32a} 
  \rho_\mathsf{s} = P'^{\,\mathrm{dr}}.
  \end{equation}
By a computation similar to the one in \eqref{6.20}, the effective velocity is indeed given by \eqref{10.28}. 

Applying our scheme to KdV solitons, one finds $P(\eta) = \eta^2/2$ and $E(\eta) = \eta^4$, which differ from the 
one-soliton momentum and energy. 
\hfill$\blackdiamond\hspace{-1pt}\blackdiamond$\\


\subsection{Soliton gas of the Toda lattice}
\label{sec10.2}

We have already studied the hydrodynamic scale of the Toda lattice within the particle-based scheme. Since the Toda lattice has also solitons in the more conventional sense,
one would expect to have also a hydrodynamic scale for the soliton-based scheme. This is the topic to be explored now. Of course, which scheme to use depends on the choice of initial conditions.  For sure, there will be also  initial conditions not covered by either scheme.

The dynamical evolution takes place on the infinitely extended lattice with equations of motion
\begin{equation}\label{10.33}
 \frac{\mathrm{d}^2}{\mathrm{d}t^2}r_j(t) = 2\mathrm{e}^{-r_j(t)} - \mathrm{e}^{-r_{j-1}(t)} - \mathrm{e}^{-r_{j+1}(t)}
 \end{equation}
 where $j \in \mathbb{Z}$. We consider solutions for which $|p_j(t)|, |r_j(t)| \to 0$ as $j \to \pm\infty$.  
 In Flaschka variables, the two-parameter family of one-soliton solutions is given by 
\begin{equation}\label{10.34}
 a_j^2(t;\eta, \delta) -1 = \omega^2 \mathrm{sech}^{-2}(|\eta| j - \omega t -\delta)
 \end{equation}
 and
 \begin{equation}\label{10.35}
 p_j(t;\eta, \delta) = |\sinh\eta|\big(\tanh(|\eta| j - \omega t -\delta) - \tanh(|\eta| (j-1) - \omega t -\delta)\big)     
 \end{equation}
 with  
 \begin{equation}\label{10.36}
\omega(\eta) = \sinh\eta.
 \end{equation}
 Deviating from the previous convention the soliton travels along the trajectory $j(t) = |\eta|^{-1}\omega t + |\eta|^{-1}\delta$.
The spectral parameter is $\eta$, $\eta \in \mathbb{R}$, the shift $\delta \in \mathbb{R}$, and the soliton velocity
\begin{equation}\label{10.37}
v_\mathrm{s} = |\eta|^{-1}\omega.
 \end{equation}
 For $\eta > 0$ the soliton is moving to the right and for  $\eta < 0$ to the left. Note that the velocity  is odd and monotone increasing in $\eta$, however with a jump of size $2$ at
 the origin.  Necessarily the soliton speed is supersonic.
 
 The construction of the two-soliton relies on the discrete analogue of \eqref{10.21} and the same ansatz as in \eqref{10.22}, using $\theta_i = |\eta_i| - \omega_i t$. This results in the Toda scattering kernel denoted by $\phi_\mathrm{tos}(\eta_1,\eta_2)$. One has to distinguish overtaking collisions, $\eta_1\eta_2 >0$ and  
 head-on collisions, $\eta_1\eta_2 <0$. Explicitly,   
 \begin{equation}\label{10.38}
  \phi_\mathrm{tos}^+(\eta_1,\eta_2) = 2 \log\Big |\frac{\sinh \tfrac{1}{2}\big(\eta_1 - \eta_2\big)}{\sinh \tfrac{1}{2}\big(\eta_1 + \eta_2\big)}\Big|
 \end{equation}
 for  $\eta_1\eta_2 >0$ and 
  \begin{equation}\label{10.39}
\phi_\mathrm{tos}^-(\eta,\eta') = 2 \log\Big |\frac{\cosh \tfrac{1}{2}\big(\eta_1 - \eta_2\big)}{\cosh \tfrac{1}{2}\big(\eta_1 + \eta_2\big)}\Big|
 \end{equation}
 for $\eta_1\eta_2 <0$.
Clearly, on their domain of definition,  $\phi_\mathrm{tos}^+ <0$ as hard core particles with $\mathsfit{a}
 >0$ and $\phi_\mathrm{tos}^-(\eta,\eta') >0$ as hard core particles with $\mathsfit{a} <0$. 

According to the collision rate ansatz, the effective velocity for Toda solitons is 
\begin{equation}\label{10.40}
 v^\mathrm{eff}(\eta) = |\eta|^{-1}\Big(\sinh \eta + \int _\mathbb{R} \mathrm{d}\eta \phi_\mathrm{tos}(\eta,\eta')f(\eta') \big(v^\mathrm{eff}(\eta') - v^\mathrm{eff}(\eta)\big)\Big).
 \end{equation}
 In terms of  Eq. \eqref{10.28} one infers that
 \begin{equation}\label{10.40a}
 P(\eta) = \mathrm{sgn}(\eta)\eta^2/2, \quad E(\eta) = \cosh \eta.  
  \end{equation}
  Without any changes we can now follow the blueprint of the KdV equation, in particularly the soliton-based  hydrodynamic equations are given by \eqref{10.25} with $\phi_\mathrm{kv}$ substituted by $\phi_\mathrm{tos}$. 
 
For the interpretation of the soliton counting function, we have to work out the analogue of \eqref{10.27}. As can be seen from \eqref{10.34}, in the Flaschka variables the soliton does not decay to $0$. Therefore one has to subtract the constant at infinity as
 \begin{equation}\label{10.41}
 Q^{[n]\infty} = \mathrm{tr}[L^n - (L_\infty)^n],
 \end{equation}
 $n \geq 1$, with $(L_\infty)_{i,j} = 1$ for $i = j \pm1$ and $0$ otherwise.
 We denote by $K_n$ the conserved field $Q^{[n]\infty}$ evaluated at the one-soliton solution written in \eqref{10.34},
 \eqref{10.35}. These coefficients can be obtained from the inverse scattering transform (IST). The result is not as explicit as the one in \eqref{10.27} and reads
 \begin{equation}\label{10.42}
\sum_{n=1}^\infty \frac{1}{n}\big(\frac{z}{1+z^2}\big)^nK_n(\eta) = \sum_{n=1}^\infty \frac{2}{n}z^n\sinh(n\eta)
 \end{equation}
 for $\eta \geq 0$. To extend to negative $\eta$ one notes the symmetry $K_n(\eta) = (-1)^nK_n(-\eta)$, which follows from the random walk expansion of the right hand side of \eqref{10.41}. Thus to lowest order,
 \begin{equation}\label{10.43}
 Q^{[1]\infty} = 2\sinh\eta, \qquad Q^{[2]\infty} = 2|\sinh(2\eta)|.
 \end{equation}
 Higher orders are determined recursively, at least in principle. But the basis functions are no longer polynomials as in the case of KdV.
 \subsection{Comparing soliton- and particle-based hydrodynamics}
\label{sec10.3}
This seems to be a good moment for a broader perspective,
for which purpose various integrable many-body systems are listed, some them only to be discussed 
in  later chapters. 
\begin{center}
\begin{tabular}[h]{l|c|c|l}model
&soliton-based & particle-based & method \\[1ex]
\hline
&&\vspace{-4pt}\\
hard rods& no & yes& model specific \\[1ex]
box-ball& yes& no  & model specific\\[1ex]
KdV& yes & no & IST\\[1ex]
Toda (classical)& yes &  yes& IST, scattering coordinates, RMT\\[1ex]
Calogero (classical) & (yes)& yes& scattering coordinates \\[1ex]   
continuum NLS &yes &no &IST\\[1ex]
Ablowitz-Ladik& (yes) & yes& IST, RMT\\[1ex]
$\delta$-Bose gas &no&yes& Bethe ansatz
\end{tabular}\medskip
\end{center}
\textbf{Table}: IST stands for inverse scattering transform and RMT for random matrix theory.\\

In the context of classical models, IST is the key tool for soliton-based hydrodynamics.
Clearly hard rods do not possess solitons. Box-ball is a spacetime discrete version of KdV. Configurations are infinite sequences of $0$'s and $1$'s. For a finite number of particles, the update from time $t$ to $t+1$ is given by the following rule: an empty carrier with unbounded capacity starts from the very left, moves to the right, and either picks up or drops a particle whenever such step is possible. After the carrier has moved to the very right the time $t+1$ configuration is completed. An $N$-soliton is simply 
a stretch of $N$ consecutive particles,
isolated and with no holes. The soliton speed equals $N$. Soliton-based dynamics is confirmed both through mathematical proofs and numerical simulations. Most remarkably the collision rate ansatz remains valid. Based on our discussions in Section 10.1, KdV is soliton based and so is the nonlinear Schr\"{o}dinger equation (NLS).

Less clear-cut is the Calogero model  with $1/\sinh^2$ interaction potential. Since 
the Toda lattice is a low density approximation to the Calogero model, one might have expected at least the analytic form of one-solitons. This is not the case, but promising numerical results  are recorded in the literature. 
For the Ablowitz-Ladik system, to be discussed in Chapter \ref{sec12},  $N$-soliton solutions are available through  IST and also the two-soliton scattering shift. But apparently soliton-based hydrodynamics has been left out. Presumably efforts have  been focussed more strongly on the study of the hydrodynamic scale of the KdV and continuum nonlinear Schr\"{o}dinger equation because of their relevance for experimental realizations.

For particle-based hydrodynamics, a prerequisite is the availability of a Lax matrix. But then two distinct methods are used. 
One is the transformation to scattering coordinates, as explained already for the Toda fluid. An alternative method emerges  from the 
observation that under GGE the Lax matrix becomes a random matrix. Thereby tools from random matrix theory (RMT) 
become available. This method works well for the Toda lattice and also for the Ablowitz-Ladik model.
In the latter case the eigenvalues of the Lax matrix are on the unit circle. In fact, the Ablowitz-Ladik model plays a peculiar role. 
In view of RMT, it is clearly particle-based. From the perspective of RMT it is obviously particle-based. But in contrast to the Toda lattice and Calogero fluid, the true particle trajectories are hidden.

In deriving GHD for quantum many-body systems the very first step towards an analysis is always the Bethe ansatz, which determines the eigenvalues of the many-body system,
compare with Chapter \ref{sec13}. Thereby generalized thermodynamics can be constructed.  Since in the respective  Bethe equations the two-particle phase shift appears explicitly, it is not so surprising that this scattering shift reappears in the generalized free energy. Here two-particle phase shift refers to the dynamics in the two-particle subspace. In our models this would be the Hilbert space $L^2(\mathbb{R}^2, \mathrm{d}x_1\mathrm{d}x_2)$ and quantum wave function is of the form $\psi(x_1,x_2,t)$. Thus, clearly, for integrable quantum many-body systems GHD is particle-based.

In analogy to IST, the Faddeev school developed the quantum inverse scattering method (QISM). This is a very powerful tool from which a unifying view of integrability emerged, encompassing also integrable models from two-dimensional statistical mechanics. Apparently, the method was never used to construct a soliton gas, say like the one for the KdV equation.
There is interesting work around the notion of quantum solitons. Whether these objects are really solitons in the sense of scattering theory remains to be explored. \bigskip  
\begin{center}
 \textbf{Notes and references}
 \end{center}
 \begin{center}
  \textbf{Section 10.1}
\end{center} 
The integrability of the KdV equation has been established in the late 60ies in a series of contributions by Gardner et al.  \cite{GGKM67},
Miura et al.\cite{MGK68}, and Kruskal et al. \cite{KMGZ70}. 
The hamiltonian structure is discussed  in Gardner   \cite{G71}. An important step was the development
of inverse scattering transform, which was then applied to a variety of other models. We refer to the 
review by Ablowitz et al. \cite{AKNS74} and the books  by Ablowitz  and Segur \cite{AS81} and Ablowitz et al. \cite{ABT04}. Scattering theory of the KdV equation is studied by Tanaka   \cite{T75,T76}. Whitham \cite{W74} developed his modulation theory in the context of KdV, see also Flaschka et al.  \cite{FFM80}. Soliton gases were first introduced by Zakharov \cite{Z71}, see also Zakharov \cite{Z09}. He used a low density approximation, for which 
in the collision rate asnsatz \eqref{10.26} the term $v^\mathrm{eff}$ on the right hand side is replaced by its bare value $4\eta^2$. The hydrodynamic scale was pioneered by Gennady El in his seminal contributions El \cite{E03} and El and  Kamchatnov \cite{EK05}. El uses the
Whitham modulation theory in a limit where the spectral bands  of $L(t)$ become very flat, see El \cite{E21} for a very instructive review. In fact, the collision rate ansatz appears only indirectly through the coupled equations
\begin{equation}\label{10.44} 
-Tf(\eta) + \sigma(\eta) f(\eta) = \eta,\qquad -Tv(\eta ) + \sigma(\eta) f(\eta) = 4\eta^3,
\end{equation}
where $v= v^\mathrm{eff} f $ is the spectral flux density. Eliminating $\sigma$ yields Eq. \eqref{10.26} for $v^\mathrm{eff}$.
A further confirmation comes from numerical studies of a tracer soliton and from matching with hydrodynamic reductions,
see Carbone  et al. \cite{CDE16}. The latter mean that the initial $f(x;\eta)$ is piecewise constant and takes only two or three values. Then the Euler
equations simplify to make explicit solutions available, which can be compared with numerical solutions of KdV. The initial
conditions have spatial randomness, but very little noise in the $\eta$ dependence. A further, this time mathematical confirmation is accomplished by Girotti et al. \cite{GGJM21}. The initial state is a long periodic arrangement of one-solitons, in other words the cnoidal wave with spatial cut-off. In addition a tracer soliton is incoming from the left with a sufficiently high velocity when compared to the building blocks of the cnoidal wave. The initial data are not random. Still in simulations, the tracer motion fluctuates random-like relative to a constant  drift. The outgoing tracer velocity is given by  \eqref{10.26} with $f$ referring to the cnoidal background. 

A more detailed comparison between GHD and KdV soliton-based hydrodynamics can be found in the article Bonnemain et al.  \cite{BDE22}.
Bidirectional solitons are studied by Congy et al.  \cite{CER21} and soliton condensates by  Congy  et al.\cite{CERT22}.  Experimental realizations are reported by Redor et al. \cite{RBMOM19}.\bigskip
\begin{center}
\textbf{Section 10.2}
\end{center} 
In part II of his famous papers on the Toda lattice, Flaschka  \cite{F74a} developed the inverse scattering transform for
the  Toda lattice. Hirota \cite{H73} constructed the $N$-soliton solution. A very readable account is  the article by Toda  \cite{T83},
where the derivation of scattering shift is explained and the similarity with KdV outlined. Scattering theory 
is studied by Kr\"{u}ger and  Teschl  \cite{KT09}. For the infinite lattice one considers  initial conditions such that
\begin{equation}\label{10.45} 
\sum_{j \in\mathbb{Z}} (1+|j|)\big(|a_j -1| + |p_j|\big) < \infty.
\end{equation}
In the long time limit, both future and past, the solution converges to $N$ freely propagating one-solitons with scattering shifts determined by \eqref{10.38}, \eqref{10.39}. There seems to be no concrete algorithm to determine $N$ and the respective soliton velocities for a given initial configuration.\bigskip
 \begin{center}
 \textbf{Section 10.3}
\end{center} 
 The clear distinction between particle-based and soliton-based hydrodynamics seems to be a novel insight.
 In principle, the  $N$-soliton solutions of the Calogero fluid can be constructed through a B\"{a}cklund transformation,
 also known as dual particles, see Philip \cite{P19}. Numerical solutions show convincingly the motion of a single soliton in a trap
 and collision between two solitons as reported by Gon and  Kulkarni  \cite{GK19}.
 Named as ``soliton cellular automaton" an ultra-discrete wave equation was introduced by    Takahashi and  Satsuma \cite{TS90}. Now box-ball is
the universally accepted label referring to the picture that the carrier picks up balls from filled boxes, resp. drops balls to empty boxes.
 The hydrodynamic scale of the box-ball system has been studied in considerable detail by Ferrari et al.  \cite{F18}, Croydon and Sasada \cite{CS20}, and Kuniba et al. \cite{KMP20,KMP20a}.

Korepin et al. \cite{KBI97} is the standard monograph on QISM. Faddeev \cite{F95} provides an illuminating account on the history.   His  1996 Les Houches lectures, Faddeev  \cite{F96}, still stand out as a masterful introduction.
\newpage
 \section{Calogero models}
 \label{sec11}
  \setcounter{equation}{0} 
Based on the experience with the Toda lattice, a physically natural goal is to advance towards one-dimensional  classical fluids.
The standard hamiltonian reads 
\begin{equation}\label{11.1} 
H_{\mathrm{mec},N} = \sum_{j = 1}^N\tfrac{1}{2}p_j^2 + \tfrac{1}{2}\sum_{i\neq j = 1}^N V_\mathrm{mec}(q_i - q_j),
\end{equation}
using units such that the particle mass equals $1$. The particles move on the real line, with possibly some confining mechanism to be introduced later. The interaction potential, $V_\mathrm{mec}$, is even and hence the 
hamiltonian is invariant under relabelling of particles. A proper hydrodynamic scale assumes a short range potential,
say with exponential decay. The borderline for a liquid-gas phase transition would be $V_\mathrm{mec}(x) \approx -|x|^{-2}$
for large $|x|$. But to our knowledge,
the minimal decay for hydrodynamic behavior has not been investigated systematically. At least, the free energy
has to be well-defined which essentially  is ensured by the stability property
\begin{equation}\label{11.2} 
\sum_{i\neq j = 1}^N V_\mathrm{mec}(q_i - q_j) \geq -\frac{c_0}{N}
\end{equation}
with some constant $c_0 \geq 0$ independent of $N$. For a purely attractive potential  such a condition is violated and  typically particles would clump together.

For a generic choice of $V_\mathrm{mec}$, one expects number, momentum, and energy to be the only locally conserved fields. As a consequence  the conserved fields are expected to be governed by the conventional hydrodynamic equations, which usually are written 
as
\begin{eqnarray}\label{11.3}
&&\hspace{10pt} \partial_t \rho_\mathsf{f} +\partial_x (\rho_\mathsf{f} v) =0\,,\nonumber\\
&&\hspace{10pt} \partial_t (\rho_\mathsf{f} v)+\partial_x (\rho_\mathsf{f} v^2 +P_\mathsf{f}) =0\,,\nonumber\\
&&\hspace{10pt} \partial_t (\rho_\mathsf{f} \mathfrak{e}_\mathrm{tot})+ \partial_x (\rho_\mathsf{f} \mathfrak{e}_\mathrm{tot} v +P_\mathsf{f}v)=0\,.
\end{eqnarray}
Here $\rho_\mathsf{f}$ is the particle density per unit length, $\rho_\mathsf{f} v$ the momentum density, and  $\rho _\mathsf{f}\mathfrak{e}_\mathrm{tot}$ the total energy density. $P_\mathsf{f}$ is the thermodynamic pressure that depends on $\rho_\mathsf{f}$ and the internal energy, the latter being obtained from
$\mathfrak{e}_\mathrm{tot} = \mathfrak{e}_\mathrm{int} + \tfrac{1}{2} v^2$. These equations are similar to Euler type
equations of a non-integrable chain. Only, since the potential is a sum over all pairs of particles, it is considerably more difficult to compute the pressure $P_\mathsf{f}$.

A priori,  there might no choice of $V_\mathrm{mec}$ which makes the model integrable. 
We discussed already a fluid with hard core potential, which is integrable but relies on having only contact interactions. In 1975 Francesco Calogero investigated the issue and discovered exactly two interaction potentials for which the $N$-particle dynamics is integrable: the long ranged  $1/x^2$ potential, called rational Calogero-Moser model,
and the short range repulsive $1/\sinh^2(x)$  potential, which we will refer to as Calogero fluid, resp. hyperbolic
Calogero model. \bigskip\\
$\blackdiamond\hspace{-1pt}\blackdiamond$~\textit{Two distinct Lax matrices.} 
For a hamiltonian as in \eqref{11.1}, to check integrability Calogero assumes the existence of a Lax pair, $L,M$, of the general form
\begin{eqnarray}\label{11.4} 
&&\hspace{0pt} L_{i,j} = \delta_{ij}p_j + (1- \delta_{ij})\alpha(q_i- q_j),\nonumber\\ 
&&\hspace{0pt}M_{i,j} =  \delta_{ij} \sum_{k = 1,k\neq j}^N\beta(q_j- q_k) +(1- \delta_{ij})
\gamma(q_i- q_j)
\end{eqnarray}
with yet arbitrary functions $\alpha, \beta, \gamma$. The commutator condition for the Lax pair then yields functional relations for $\alpha, \beta, \gamma$  and the potential $V_\mathrm{mec}$. These equations can be analysed with the result
$\gamma = \alpha'$, $\beta = \alpha''/(2\alpha)$, and $V_\mathrm{mec}(x) = \alpha(x)\alpha(-x)$. Now for $\alpha$
there are essentially only three choices: (i) $\alpha(x) = \mathrm{i}g/x$, (ii) $\alpha(x) = \mathrm{i}ga/\sinh(ax)$, and (iii) $\alpha(x) = \mathrm{i}ga \coth(ax)$. Here $a,g$ are real parameters. Case (i) yields the rational Calogero model, which will be discussed in Section \ref{sec11.5}, while (ii) and (iii) yield the hyperbolic Calogero model of the section below. 
In addition there are solutions corresponding to Lax pairs which result from periodizing over a ring of given size. This then generates 
the interaction potential $1/\sin x$ in case (i) and particular elliptic functions in cases (ii) and (iii).

It is surprising to have two Lax matrices, say $L^{[2]}, L^{[3]}$ for the same hamiltonian. Depending on the problem, either one of them might be of advantage. There seems to be no simple relation between $L^{[2]}$ and $L^{[3]}$. But since their eigenvalues are conserved when evolved by  the same dynamics, there must be a functional relation between either set of eigenvalues.

For chains one can carry out a similar analysis. Since a smooth potential is assumed, only the exponential potential
qualifies as being integrable. In fact there are also two distinct Lax matrices, but they are related by a simple 
similarity transform.   
\hfill$\blackdiamond\hspace{-1pt}\blackdiamond$
\begin{figure}[!t]
\centering
\includegraphics[width=0.9\textwidth]{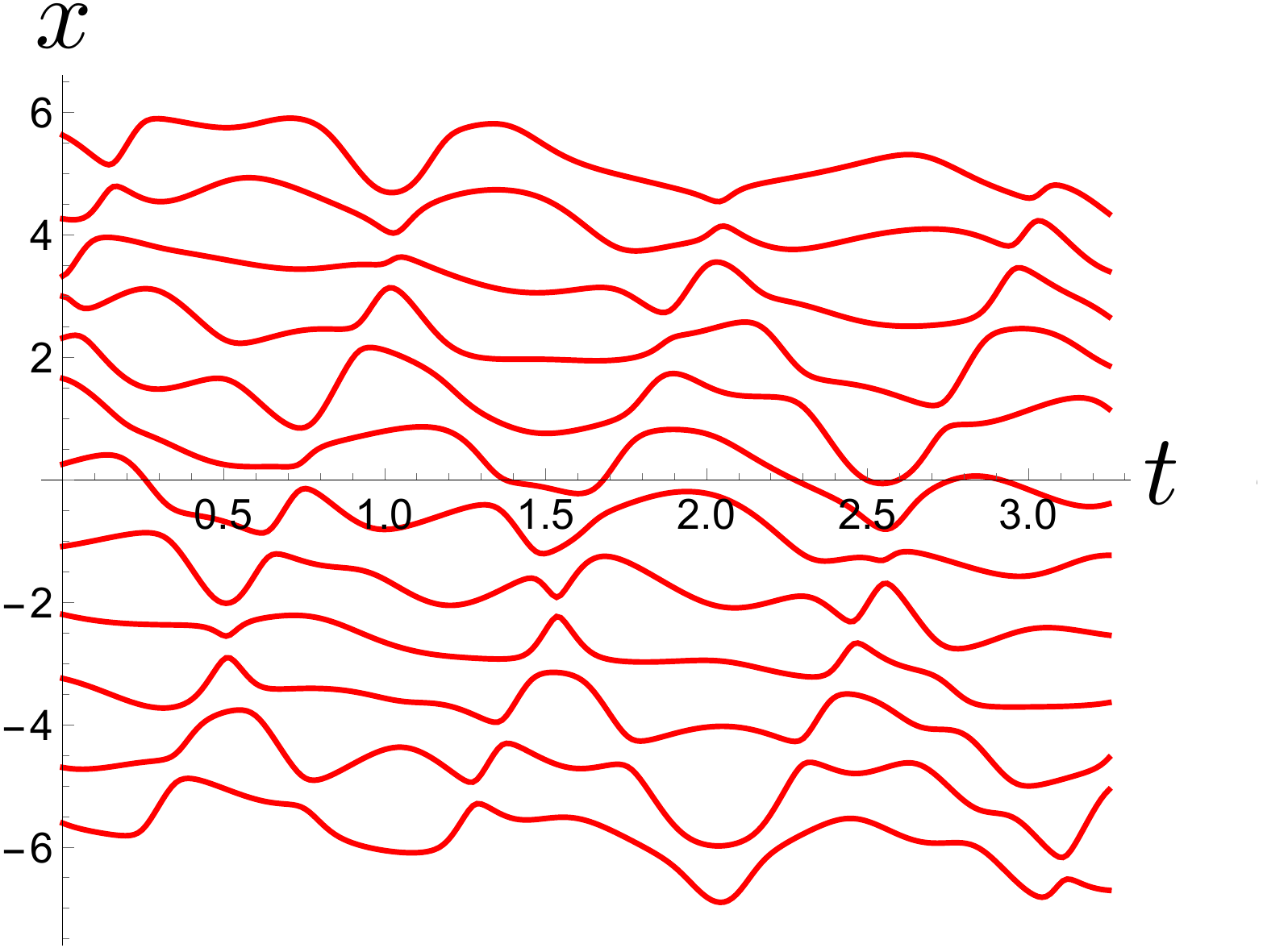}
\label{fig11}
\caption{Displayed are trajectories of 12 Calogero particles in a periodic box of size $\ell = 12$ and
at temperature $\beta = 1$. Particles repel each other  and cannot cross.  There is one fast quasiparticle 
which can be distinguished from the background. Courtesy from M. Kulkarni.}
\end{figure}
\subsection{Hyperbolic Calogero model, charges and currents}
\label{sec11.1}
The interaction potential of the Calogero fluid is given by 
   \begin{equation}\label{11.5} 
V_\mathrm{ca}(x) = \frac{1}{(2\sinh(\tfrac{1}{2}x))^2} .
\end{equation}
More commonly, one introduces length and time scales through 
 \begin{equation}\label{11.6} 
  V_\mathrm{ca}(x) = \frac{g^2a^2}{\sinh^2(ax)},
  \end{equation}
  which would make the formulas 
more lengthy. If needed, these parameters can be introduced by a linear change of spacetime coordinates. The  choice $g = 1, a = \tfrac{1}{2}$ is dictated by recovering the Toda lattice at low density. For $N$ particles on the entire real line, Newton equations of motion are
 \begin{equation}\label{11.7} 
 \frac{d^2}{dt^2} q_j = \sum_{i = 1,i\neq j}^N \tfrac{1}{4}\cosh(\tfrac{1}{2}(q_j- q_i))\big(\sinh(\tfrac{1}{2}(q_j- q_i))\big)^{-3}
 \end{equation}
for $j = 1,\ldots,N$. 
At short distances the potential has the repulsive $1/|x|^2$ singularity, which implies that the spatial ordering of particles is maintained throughout time. 
Momentarily, we will introduce a Lax matrix which is the avenue towards the particle-based scheme. The scattering shift will refer to the motion of two particles. 

The dynamics admits a Lax pair, $L,M$, through
\begin{eqnarray}\label{11.8} 
&&\hspace{-10pt} L_{i,j} = \delta_{ij}p_j + \mathrm{i}(1- \delta_{ij})\big(2\sinh(\tfrac{1}{2}(q_i- q_j))\big)^{-1},\\ 
&&\hspace{-10pt}M_{i,j} =  \mathrm{i}\delta_{ij} \sum_{k = 1,k\neq j}^N\big(2\sinh(\tfrac{1}{2}(q_j- q_k))\big)^{-2} - \mathrm{i}(1- \delta_{ij})
 \cosh(\tfrac{1}{2}(q_i- q_j))\big(2\sinh(\tfrac{1}{2}(q_i- q_j))\big)^{-2}\nonumber
\end{eqnarray}
for $i,j = 1,\ldots,N$. Referring to the Insert above, our choice is $L^{[2]}$, since $L^{[3]}$ would lead yo nonlocal conservation laws. For a lighter notation we omit the index $N$ that was used for the Toda lattice.
The Lax matrix is hermitian, hence has real eigenvalues, while the partner matrix is anti-hermitian. Then, with $L,M$ evaluated along trajectories of \eqref{11.7}, 
\begin{equation}\label{11.9} 
\frac{d}{dt} L(t) = [L(t),M(t)].
\end{equation}
Merely having  a commutator  implies that
\begin{equation}\label{11.10} 
\frac{d}{dt} \mathrm{tr}[L(t)^n] = 0.
\end{equation}
The eigenvalues of $L(t)$ do not change in time.
The conserved charges are
\begin{equation}\label{11.11} 
Q^{[n]} =  \mathrm{tr}[L^n]
\end{equation}
and their density is given by
\begin{equation}\label{11.12} 
Q^{[n]}(x) =   \sum_{j= 1}^N\delta(x - q_j)(L^n)_{j,j}
\end{equation}
with $x\in\mathbb{R}$. The total momentum corresponds to $Q^{[1]}$ and total energy to $\tfrac{1}{2} Q^{[2]}$.
As a general fact for Calogero type models, under the standard Poisson bracket,
\begin{equation}\label{11.13} 
\{Q^{[m]}, Q^{[n]}\} =0.
\end{equation}
Thus we have obtained a set of $N$ conserved fields in involution. Setting $n=0$ our labelling includes the particle number,  \begin{equation}\label{11.14} 
\mathrm{tr}[L^0] = N = Q^{[0]},\quad Q^{[0]}(x) =  \sum_{j= 1}^N\delta(x - q_j ).
\end{equation}
In the sense of hamiltonian systems the constant function is a trivial conservation law. But 
 it must be included in a  hydrodynamic context, since the particle density is a central physical observable.
 
 To obtain the currents one considers 
 \begin{equation}
 \label{11.15} 
 \frac{\mathrm{d}}{\mathrm{d}t}Q^{[n]}(f) =  \frac{\mathrm{d}}{\mathrm{d}t}\sum_{j= 1}^N f(x - q_j)(L^n)_{j,j}
= \sum_{j= 1}^N f(x - q_j)p_j (L^n)_{j,j} + 
 \sum_{i\neq j = 1}^N\big(f(q_j) - f(q_i)\big)M_{i,j}(L^n)_{j,i},
\end{equation}
which implies
 \begin{equation}\label{11.16} 
J^{[n]}(x) = \sum_{j=1}^N\big(\delta(q_j - x)p_j(L^n)_{j,j} +  \sum_{i= 1,i\neq j}^N\big(\theta(q_{j} -x) - \theta(q_{i} -x)\big)M_{i,j}(L^n)_{j,i}\big).
\end{equation}
In particular,
\begin{equation}\label{11.17} 
\partial_t Q^{[n]}(x,t) +\partial_x J^{[n]}(x,t)  = 0.
\end{equation}

Most naturally, the system is confined by 
having particles move in a periodic box $[0,\ell]$, $\ell >0$. Since the interaction potential has a tail, one has to include 
the potential  from all image particles. To introduce a notation, for some rapidly decreasing function, $f$, we define its $\ell$-periodized version through
 \begin{equation}\label{11.18} 
f_{\mathrm{per},\ell}(x) = \sum_ {m\in \mathbb{Z}} f(x +m\ell).
\end{equation}
By construction $f_{\mathrm{per},\ell}$ is $\ell$-periodic and its domain can be restricted to the box $[0,\ell]$.
The periodized interaction potential can be written in terms of the elliptic (double periodic) Weierstrass function, $\wp (z)$. Historically its two periodicities are denoted by $2\omega_1,2\omega_2$ and 
\begin{equation}\label{11.19} 
\wp (z|\omega_1,\omega_2) = \frac{1}{z^2} +  \sum_{m,n \in \mathbb{Z}}\hspace{-6pt}{'}\hspace{2pt}\Big( \big(z + 2\omega_1 m+ 2 \omega_2n\big)^{-2} -
\big(2\omega_1 m+ 2 \omega_2n\big)^{-2}\Big),
\end{equation}
where $\sum{'}$ means that the origin, $(m,n)=0$, has to be omitted  in the sum. One notes the derivative 
\begin{equation}\label{11.20} 
\wp (z|\omega_1,\omega_2)' = -\sum_{m,n \in \mathbb{Z}} 2\big(z + 2\omega_1 m+ 2 \omega_2n\big)^{-3}
\end{equation}
and the identities
\begin{equation}\label{11.21} 
\sum_{n \in \mathbb{Z}} \big(z + \mathrm{i}\pi n\big)^{-2} = \sinh^{-2}(z), \qquad - \sum_{n \in \mathbb{Z}} 2\big(z + \mathrm{i}\pi n\big)^{-3} = (\sinh^{-2}(z))'.
\end{equation}
We now set $2\omega_1 = \ell$ and $2\omega_2 = \mathrm{i}\pi$ and conclude from  \eqref{11.20} and  \eqref{11.21} that
\begin{equation}\label{11.22} 
(\sinh^{-2})_{\mathrm{per},\ell}(x) = \wp (x|\tfrac{1}{2}\ell,\mathrm{i}\tfrac{1}{2}\pi ) -\tfrac{1}{12}. 
\end{equation}
The constant $ -\tfrac{1}{12}$ has to be fixed by a separate argument and drops out when studying the dynamics. 

The equations of motion for the Calogero particles in a periodic box are still given by Eq. \eqref{11.7} with the force replaced by the periodized force. However the Lax matrix for the ring is no longer simply the periodized Lax matrix.  By the method indicated  in the Insert above, one can still construct     
a Lax pair for the periodized Calogero model in terms of particular elliptic 
functions. As for the Toda chain, the Calogero model remains integrable on the ring.
\bigskip\\
$\blackdiamond\hspace{-1pt}\blackdiamond$~\textit{Dilute limit}. If particles are far apart the interaction potential can be approximated as 
$\mathrm{e}^{-|x|}$.   Labelling particles in increasing order, the force on the $j$-th particle  
is exponentially dominated by the force from its two neighbors. Thus, in approximation,
 \begin{equation}\label{11.23} 
 \frac{d^2}{dt^2} q_j =  \mathrm{e}^{-(q_{j} - q_{j-1})} -  \mathrm{e}^{-(q_{j+1} - q_j)},
\end{equation}
which is recognized as the Toda dynamics.

One would expect that  in the dilute limit also the Lax pair converges to the Lax matrix of the Toda lattice. 
In this approximation
\begin{eqnarray}\label{11.24} 
&&\hspace{-10pt} L_{j,j} = p_j,\quad L_{j,j+1} = -\mathrm{i} a_j, \quad L_{j+1,j} = \mathrm{i} a_j,\nonumber\\[1ex]  
&&\hspace{-10pt}M_{j,j} = 0, \quad M_{j,j+1} = M_{j+1,j}=-\tfrac{1}{2} \mathrm{i} a_j ,  
\end{eqnarray}
where, as to recalled, the Flaschka variables are $a_j = \exp[-(q_{j+1} - q_j)/2]$. For the open system  all other matrix elements vanish, while for the closed system $L_{1,N} = \mathrm{i} a_N$ and $L_{N,1} = -\mathrm{i} a_N$. One checks that for the closed system there is a diagonal unitary matrix, $\tilde{U}$, such that $\tilde{U}L\tilde{U}^*  = L_N$ and 
and $\tilde{U}M\tilde{U}^*  = -B_N^\diamond$, while
for the open system there is a diagonal unitary matrix, $\tilde{U}'$, such that $\tilde{U}'M\tilde{U}'^*  = -B_N$  and  $\tilde{U}'L\tilde{U}'^*  = L_N^\diamond$ and $\tilde{U}'M\tilde{U}'^*  = -B_N^\diamond$, compare with \eqref{2.12} and \eqref{2.13}.
\hfill$\blackdiamond\hspace{-1pt}\blackdiamond$
\subsection{Scattering coordinates}
\label{sec11.2}
To compute the generalized free energy we will use action-angle variables, as in Section \ref{sec9.2} for the Toda fluid. 
This time, the map is less explicit and some stretch of linear algebra is in demand. In fact, we do
start with the Lax matrix $L$ in \eqref{11.8} for particles moving on the entire real line.  Since particles cannot cross we order them in increasing index. Then  the particle phase space is 
$(q,p) \in \Gamma_{\!N}^\triangleright  = \mathbb{W}_N \times \mathbb{R}^N$ with the Weyl chamber $\mathbb{W}_N = \{q_1 < \ldots<q_N\}$, while the phase space of scattering coordinates is $(\lambda,\phi) \in \Gamma_{\!N}^\triangleright$ with ordered eigenvalues. The eigenvalue problem for $L$ reads
\begin{equation}\label{11.25} 
L\psi_\alpha = \lambda_\alpha\psi_\alpha,
 \end{equation}
$\alpha = 1,\ldots, N$. Since $L = L^*$, there is a unitary matrix, $U$, such that
\begin{equation}\label{11.26} 
ULU^* = \mathrm{diag}(\lambda_1,\ldots,\lambda_N).
 \end{equation}
 We also introduce
\begin{equation}\label{11.27} 
A =  \mathrm{diag}(\mathrm{e}^{q_1},\ldots,\mathrm{e}^{q_N}),\qquad d_j = \mathrm{e}^{q_j/2},
 \end{equation}
 and correspondingly 
\begin{equation}\label{11.28} 
\tilde{A} =  \mathrm{diag}(\mathrm{e}^{-q_1},\ldots,\mathrm{e}^{-q_N}),\qquad \tilde{d}_j = \mathrm{e}^{-q_j/2}.
\end{equation}
Clearly $A\tilde{A} = 1$. Now
\begin{equation}\label{11.29} 
[A,L] = \mathrm{i}\big(|d\rangle\langle d| - A\big), \qquad  [\tilde{A},L] = -\mathrm{i}\big(|\tilde{d}\rangle\langle \tilde{d}|- \tilde{A}\big).
 \end{equation}
 Setting 
 \begin{equation}\label{11.30}
 Q = UAU^*, \quad |e\rangle= U|d\rangle, \qquad \tilde{Q} = U\tilde{A}U^*, \quad |\tilde{e}\rangle = U|\tilde{d}\rangle,
 \end{equation}
conjugating the identities \eqref{11.29} with $U$, and taking the $i,j$ matrix elements results in 
 \begin{equation}\label{11.31}
 Q_{i,j} = e_i \frac{\mathrm{i}}{\lambda_j - \lambda_i +\mathrm{i}} e_j^*, \qquad
 \tilde{Q}_{i,j} = \tilde{e}_i \frac{\mathrm{i}}{\lambda_i - \lambda_j +\mathrm{i}} \tilde{e}_j^*. 
 \end{equation}
 Since $Q\tilde{Q} = 1$, from the diagonal matrix elements $(Q\tilde Q)_{j,j} =1$ one concludes  
 \begin{equation}\label{11.32} 
  - e_j \tilde{e}_j^* \sum_{m=1}^N \frac{ e_m^* \tilde{e}_m}{\big(\lambda_m - \lambda_j +\mathrm{i}\big)^2} = 1.
 \end{equation}
In particular, the expansion coefficients $e_j,\tilde{e}_j$ cannot vanish. From the off-diagonal matrix elements, $(Q\tilde Q)_{i,j} =0,i \neq j$,
one infers
 \begin{equation}\label{11.33} 
 0 = \sum_{m=1}^N \frac{e_m^*\tilde{e}_m(\lambda_i - \lambda_j)}{\big(\lambda_m -\lambda_i +\mathrm{i}\big)\big(\lambda_m -\lambda_j +\mathrm{i}\big)}
 = \sum_{m=1}^N \frac{e_m^*\tilde{e}_m}{\big(\lambda_m -\lambda_i +\mathrm{i}\big)} - \sum_{m=1}^N \frac{e_m^*\tilde{e}_m}{\big(\lambda_m -\lambda_j+\mathrm{i}\big)}.
\end{equation}
 Hence the latter sum equals some constant,  $c$, independent of $j$, 
 \begin{equation}\label{11.34} 
\sum_{m=1}^N \frac{e_m^*\tilde{e}_m}{\big(\lambda_m -\lambda_j+\mathrm{i}\big)} = c.
\end{equation}
From the Cauchy-type identity in \eqref{11.43} below, it follows that 
 \begin{equation}\label{11.35}
 e_j^*\tilde{e}_j = \mathrm{i} c V_j 
\end{equation}
with
\begin{equation}\label{11.36}
V_j = \prod_{m=1,m \neq j}^N \Big(1 + \frac{\mathrm{i}}{\lambda_j - \lambda_m}\Big),
\end{equation}
while the identity \eqref{11.44} below implies
 \begin{equation}\label{11.37}
\langle e,\tilde{e} \rangle = \mathrm{i} cN.
\end{equation}
On the other hand $\langle e,\tilde{e} \rangle = N$ from the definition \eqref{11.30} and $c = -\mathrm{i}$ follows.
Hence 
 \begin{equation}\label{11.38}
  e_j^*\tilde{e}_j = V_j
\end{equation}
 and using \eqref{11.45} below one checks that Eq. \eqref{11.32} is satisfied. 
 Together with \eqref{11.31} the final result reads
 \begin{equation}\label{11.38a}
Q_{j,j} = |e_j|^2, \qquad \tilde{Q}_{j,j} =  |\tilde{e}_j|^2,\qquad  |e_j|^2 |\tilde{e}_j|^2 =  |V_j|^2.
\end{equation}

One defines a real-valued $\phi_j$ through 
 \begin{equation}\label{11.39}
Q_{j,j} = |V_j|\mathrm{e}^{\phi_j},
\end{equation}
 which indeed turns out to be equal to the scattering shift $\phi_j^+$ of the Calogero model, compare with \eqref{2.42}. Somewhat implicitly, we have obtained the scattering map $\Phi: 
 (\lambda,\phi) \mapsto (q,p)$.
 
 Since $Q_{j,j} = |e_j|^2$, the overall phase of the eigenvectors of $L$ can be chosen such that
\begin{equation}\label{11.40}
e_j = |V_j|^{1/2}\mathrm{e}^{\phi_j/2}.
\end{equation}
From the invariance of the trace under unitary transformations one concludes
\begin{equation}\label{11.41}
\mathrm{tr}[A] = \sum_{j=1}^N \mathrm{e}^{q_j} = \mathrm{tr}[Q] =  \sum_{j=1}^N|V_j|\mathrm{e}^{\phi_j}, \qquad
\mathrm{tr}[\tilde{A}] = \sum_{j=1}^N \mathrm{e}^{-q_j} = \mathrm{tr}[\tilde{Q}] =  \sum_{j=1}^N|V_j|\mathrm{e}^{-\phi_j}.
\end{equation}
$\blackdiamond\hspace{-1pt}\blackdiamond$~\textit{Cauchy-type identities}.
We consider coefficients $b_j, j = 1,\ldots,N$, satisfying the linear equations 
 \begin{equation}\label{11.42} 
 \sum_{j=1}^N \frac{b_j}{x_j - y_i} = 1,
 \end{equation}
$i = 1,\ldots,N$, for some given set of coefficients $\{x_j,y_j, j = 1,\ldots,N\}$. Then the following identities hold:
\begin{equation}\label{11.43} 
b_j = \frac{\prod_{m=1}^N(x_j -y_m)}{\prod_{m =1, m \neq j}^N(x_j - x_m)}, 
 \end{equation}
\begin{equation}\label{11.44} 
\sum_{j = 1}^N  b_j= \sum_{j=1}^N (x_j - y_j),
 \end{equation}
 
 \begin{equation}\label{11.45} 
 \sum_{j = 1}^N \frac{b_j}{(x_j - y_i)^2} = - \frac{\prod_{m =1,m \neq i}^N (y_i - y_m)}{\prod_{m = 1}^N (y_i - x_m)}.
 \end{equation}
\hfill$\blackdiamond\hspace{-1pt}\blackdiamond$
 \subsection{Generalized free energy}
\label{sec11.3}
While the periodized system is still integrable, there seems to be no analogue of the Dumitriu-Edelman change of volume elements. On the  other hand, scattering coordinates and a carefully chosen box potential can still be implemented.

To the GGE potential we add the external potential given by
\begin{equation}\label{11.46} 
 V_\mathrm{ex}(q) = \sum_{j=1}^N \mathrm{e}^{-\ell/2} \cosh(q_j),
 \end{equation}
 which has a very flat bottom over the interval $[-\ell/2,\ell/2]$ and then increases exponentially with rate $1$. In fact, for this particular choice the Calogero fluid remains integrable. While this property is not used directly, it is reflected by having a simple formula for the transformed $V_\mathrm{ex}$, perhaps 
 \begin{equation}\label{11.47} 
 V_\mathrm{ex}(q)\circ \Phi = \sum_{j=1}^N \mathrm{e}^{-\ell/2} Y_j \cosh(\phi_j)
 \end{equation}
 with
 \begin{equation}\label{11.48}
Y_j = |V_j| = \prod_{m=1, m\neq j}^N\Big(1 +\frac{1}{(\lambda_m -\lambda_j)^2}\Big)^{1/2}.
\end{equation}
Using the symmetry in $q$,  the partition function reads
\begin{eqnarray}\label{11.49}
&&\hspace{-28pt} Z_{\mathrm{ca},N}(\ell,V) =  \int_{ \Gamma_{\!N}^\triangleright } \mathrm{d}^N \hspace{-1pt} q\, \mathrm{d}^N \hspace{-1pt}p \exp\Big(- \mathrm{tr}\big[V(L)\big] - \sum_{j=1}^N \mathrm{e}^{-\ell/2} \cosh(q_j)\Big) \nonumber\\
&&\hspace{20pt} = \int_{ \Gamma_{\!N}^\triangleright } \mathrm{d}^N \hspace{-1pt}\lambda\, \mathrm{d}^N \hspace{-1pt}\phi  \exp\Big(-\sum_{j=1}^N\big(V(\lambda_j) + 
 \mathrm{e}^{-\ell/2}Y_j \cosh \phi_j \big)\Big).
\end{eqnarray}
Since the integrand is symmetric in $\lambda$, at the expense of a factor $1/N!$, the $\lambda$-integration is extended to $\mathbb{R}^N$.
The $\phi_j$-integration yields the order zero modified Bessel function of the second kind $K_0$, see \eqref{9.35}, resulting in
\begin{equation}\label{11.50}
Z_{\mathrm{ca},N}(\ell,V) = \frac{1}{N!} \int_{\mathbb{R}^{N}}\mathrm{d}^N \hspace{-1pt}\lambda \prod_{j=1}\mathrm{e}^{-V(\lambda_j)} 
\prod_{j=1}^N 2K_0\big( 2 \mathrm{e}^{-\ell/2}Y_j \big).
\end{equation}

As a consequence, the joint distribution of eigenvalues of the Lax matrix \eqref{11.8} under the normalized measure 
\begin{equation}\label{11.51}
Z_{\mathrm{ca},N}(\ell,V)^{-1}  \exp\Big(- \mathrm{tr}[V(L)]-\sum_{j=1}^N \mathrm{e}^{-\ell/2} \cosh(q_j)\Big)  \mathrm{d}^N \hspace{-1pt} p\,\mathrm{d}^N \hspace{-1pt}q
\end{equation}
is given by
\begin{equation}\label{11.52}
Z_{\mathrm{ca},N}(\ell,V)^{-1}  
\exp\Big[- \sum_{j=1}^N V(\lambda_j) + \sum_{j=1}^N \log\Big(2K_0\big( 2 \exp\big[ - \tfrac{1}{2} N \big(\nu - \frac{2}{N} \sum_{i = 1,i\neq j}^N
\log Y_j\big)\big]\big)\Big)\Big]
\end{equation}
relative to $(1/N!)\mathrm{d}^N \hspace{-1pt}\lambda$. Note that 
\begin{equation}\label{11.53}
-\frac{2}{N} \log Y_j = \frac{1}{N}\sum_{i=1,i\neq j}^N \phi_\mathrm{ca}(\lambda_j - \lambda_i)
\end{equation}
with
 \begin{equation}\label{11.54}
\phi_\mathrm{ca}(w) =  -\log \Big(1 + \frac{1}{w^2}\Big).
\end{equation} 
As discussed in the inset, $\phi_\mathrm{ca} $ is indeed the scattering shift of two Calogero particles.

We arrived at an expression very similar to the Toda lattice, except for a distinct $Y_j$. 
 Following the previous steps with
\begin{equation}\label{11.55}
\Upsilon_j = \nu +\frac{1}{N} \sum_{i=1,i\neq j}^N \phi_\mathrm{ca}(\lambda_j - \lambda_i),
\end{equation}
the free energy functional becomes
 \begin{equation}\label{11.56}
\mathcal{F}_{\mathrm{ca}}^\circ(\varrho) = \nu^{-1}\int_\mathbb{R}\mathrm{d}w\varrho(w)\Big( V(w) -1  + \log \varrho(w) - \log\big(\nu +  \int_\mathbb{R}\mathrm{d}w'\varrho(w')\phi_\mathrm{ca}(w - w')\big) \Big) 
\end{equation}
in case 
\begin{equation}\label{11.56a}
\nu +  \int_\mathbb{R}\mathrm{d}w'\varrho(w')\phi_\mathrm{ca}(w - w') >0
\end{equation}
and $\mathcal{F}_{\mathrm{ca}}^\circ(\varrho) = \infty $ otherwise. This infinite-dimensional constraint can be ignored, just  as for the Toda fluid.
The variation is over all $\varrho$ such that $\varrho>0$ and $\langle \varrho \rangle = 1$.
Substituting $\nu^{-1}\varrho = \rho$,
the free energy per unit length of the Calogero fluid 
 is determined by the free energy functional   
\begin{equation}\label{11.57}
\mathcal{F}_\mathrm{ca}(\rho) = \int_\mathbb{R}\mathrm{d}w\rho(w)\Big( V(w) -1 + \log \rho(w) - \log\big(1 + \int_\mathbb{R}\mathrm{d}w'\rho(w')\phi_\mathrm{ca}(w - w')\big)\Big). 
\end{equation}
Supporting the generic structure of GHD,  $\mathcal{F}_\mathrm{ca}$ agrees with \eqref{9.42} upon substituting the scattering shift $\phi_\mathrm{to}$ by $\phi_\mathrm{ca}$. Therefore the steps from \eqref{9.43} to \eqref{9.50} can be repeated, in particularly leading to the TBA equation for the Calogero fluid. 
\\\\
$\blackdiamond\hspace{-1pt}\blackdiamond$~\textit{Two-particle Calogero scattering shift}.
For two particles the equations of motion are
 \begin{equation}\label{11.58}
\ddot{q}_1(t) = \tfrac{1}{4}\cosh(\tfrac{1}{2}(q_1-q_2))\sinh^{-3}(\tfrac{1}{2}(q_1-q_2)), \quad \ddot{q}_2(t) = -\tfrac{1}{4}\cosh(\tfrac{1}{2}(q_1-q_2))\sinh^{-3}(\tfrac{1}{2}(q_1-q_2)),
\end{equation} 
which have to be solved with the asymptotic condition $p_1(-\infty) = p_1$,  $p_2(-\infty) = p_2$, and $p_2 < p_1$. Hence $q_1(t) < q_2(t)$. The solution can be worked out explicitly with the result
\begin{eqnarray}\label{11.59}
&&\hspace{-30pt} q_1(t) = \tfrac{1}{2}(p_1+p_2)t - \tfrac{1}{2}q(t), \qquad  q_2(t) = \tfrac{1}{2} (p_1+p_2)t + \tfrac{1}{2}q(t),\nonumber\\[1ex]
&&\hspace{-30pt}  q(t) = 2\log\Big[b\cosh(\gamma t) +\big(b^2 \cosh^2(\gamma t) -1\big)^{1/2}\Big],  \quad b^2 = 1+ (2\gamma)^{-2}
\end{eqnarray} 
with $\gamma = \tfrac{1}{2}(p_1 - p_2) >0$. The large time asymptotics is given by 
\begin{equation}\label{11.60}
q_1(t) = 
\begin{cases}p_1t  -  \log b,\\[1ex]
p_{2}t - \log b ,
\end{cases} \quad
q_2(t) = 
\begin{cases}p_2t +  \log b, &\qquad t \to -\infty,\\[1ex]
p_{1}t +  \log b, &\qquad t \to \infty.
\end{cases} 
\end{equation}
Since quasiparticle 1 is shifted by $- 2\log b$ and quasiparticle 2 by $ 2\log b$, we conclude that the two-particle scattering shift is given by
\begin{equation}\label{11.61}
\phi_\mathrm{ca}(p_1-p_2) =  -\log \Big(1 + \frac{1}{(p_1 -p_2)^2}\Big) ,
\end{equation} 
 compare with the discussion in Section \ref{sec2.3}.  For small momentum transfer, one recovers the Toda two-particle scattering shift $2 \log |p_1 -p_2|$.
Since particles do not cross each other,  $\phi_\mathrm{ca} <0$, consistent with the sign for hard rods.
 \hfill$\blackdiamond\hspace{-1pt}\blackdiamond$
 \subsection{Hydrodynamic equations} 
 \label{sec11.4}
 This discussion can be very brief. The density current is momentum, which itself is conserved.
 Therefore Eq. \eqref{9.20} is still valid. In the limit $\mu \to -\infty$ the density, and hence also the GGE averaged currents, vanish. Eq. \eqref{9.21} still holds  and as a consequence
 \begin{equation}\label{11.62} 
 v^\mathrm{eff} = \frac{\varsigma_1^\mathrm{dr}}{\varsigma_0^\mathrm{dr}}\,.
\end{equation}
Of course, the dressing transformation is now defined through the scattering shift \eqref{11.61}. As for the Toda fluid, in terms of the properly normalized density of states the hydrodynamic equations read 
\begin{equation}\label{11.63} 
\partial_t\rho_\mathsf{p}(x,t;v) + \partial_x\big(v^\mathrm{eff}(x,t;v)\rho_\mathsf{p}(x,t;v)\big) = 0
\end{equation}
in correspondence with Eq.  \eqref{9.25}.
\subsection{Classical Bethe equations}
\label{sec11.5}
Bethe equations determine the eigenvalues of a quantum hamiltonian whose eigenfunctions are obtained from employing the Bethe ansatz.  In Chapter \ref{sec13} we will discuss such method using the Lieb-Liniger $\delta$-Bose gas as prime example. 
A similar structure appears for other integrable quantum  many-body systems. An even wider class of integrable quantum models can be handled by the asymptotic Bethe ansatz, meaning that the Bethe ansatz holds only for large separation of particles.  An example is the quantum Toda lattice covered in Chapter \ref{sec14}. In this section we argue that Bethe equations can be written also for the Calogero fluid. Of course, this is not a microscopic foundation.  Of interest is merely the structural similarity. From the Bethe equations one deduces again the validity of TBA and thereby the DOS previously recorded.  

Formally one introduces the classical  ``phase shift'' of the Calogero fluid as  
\begin{equation}\label{11.64} 
\theta_\mathrm{ca}(w) = - w\log(1+w^{-2}) - 2  \arctan w, \qquad \theta_\mathrm{ca}'= \phi_\mathrm{ca}, \quad \theta_\mathrm{ca}(0) =0.
\end{equation}
 The Bethe equations are defined through
\begin{equation}\label{11.65} 
u_j = \nu N \lambda_j + \sum_{i=1}^N\theta_\mathrm{ca}(\lambda_j - \lambda_i) 
\end{equation}
for $j = 1,\ldots,N$. The input is $u \in \mathbb{W}_N$ and $(\lambda_1,\ldots,\lambda_N)$ is the output. The input is weighted uniformly by the Lebesgue measure 
$\mathrm{d}^N \hspace{-1pt}u$ and the induced output is assumed to have  the probability
\begin{equation}\label{11.66} 
Z^{-1}
\exp\Big(-\sum_{j=1}^N V(\lambda_j)\Big)\mathrm{d}^N\hspace{-1pt}u. 
\end{equation}
To define a probability measure, this expression has to be restricted to the domain for which the Jacobian matrix 
$\{\partial u_j/\partial \lambda_i\}$, $i,j = 1,\ldots,N$, is strictly positive.  To find out we differentiate \eqref{11.65} with the result
\begin{equation}\label{11.67} 
\frac{\partial u_j}{\partial \lambda_i} = \delta_{ij}\Big(\nu N + \sum_{i=1}^N\phi_\mathrm{ca}(\lambda_j - \lambda_i)\Big) 
- (1 - \delta_{ij}) \phi_\mathrm{ca}(\lambda_j - \lambda_i) = N\big(\Lambda^\circ_{i,j} + N^{-1}\Lambda^\bot_{i,j}\big),
\end{equation}
where $\Lambda^\circ$ is the diagonal part and $\Lambda^\bot$ the off-diagonal one. Then \eqref{11.66} transforms to
\begin{equation}\label{11.68} 
\frac{1}{Z}\exp\Big(-\sum_{j=1}^N V(\lambda_j)\Big) \exp\big(N\log N +  \mathrm{tr}\big[\log(\Lambda^\circ + N^{-1}\Lambda^\bot)\big]\big) \mathrm{d}^N\hspace{-1pt}\lambda.
\end{equation}
For the exponent we write
\begin{equation}\label{11.69} 
\mathrm{tr}\big[\log\big(\Lambda^\circ\big(1 + N^{-1}(\Lambda^\circ)^{-1}  \Lambda^\bot\big)\big)\big] 
= \mathrm{tr}\big[\log\Lambda^\circ\big]+ \mathrm{tr}\big[\log\big(1 + N^{-1} (\Lambda^\circ)^{-1}\Lambda^\bot\big)\big]. 
\end{equation}
Expanding the latter logarithm yields
\begin{equation}\label{11.70} 
\frac{1}{N} \mathrm{tr}\big[(\Lambda^\circ)^{-1}\Lambda^\bot\big] - \frac{1}{2N^2} \mathrm{tr}\big[\big((\Lambda^\circ)^{-1}\Lambda^\bot\big)^2\big].
\end{equation}
Since $(\Lambda^\circ)^{-1}$ is diagonal, the first summand vanishes, while the second one is of order 1, since there are $N^2$ summands each one of order 1.
The leading term is of order $N$ relative to which the remainder is small. Within this approximation one obtains 
\begin{equation}\label{11.71} 
\frac{1}{Z}\exp\Big(-\sum_{j=1}^N V(\lambda_j) +  \sum_{j=1}^N\log\Big( \nu+ N^{-1}\sum_{i=1}^N\phi_\mathrm{ca}(\lambda_j - \lambda_i)\Big)\Big)   \mathrm{d}^N\hspace{-1pt}\lambda,
\end{equation}
an expression which we have encountered already. For large $N$ the positivity of the Jacobian is ensured by  
\begin{equation}\label{11.72} 
 \nu+ N^{-1}\sum_{i=1}^N\phi_\mathrm{ca}(\lambda_j - \lambda_i) >0
 \end{equation}
and agreement with \eqref{11.56} is accomplished.
To fill in the remaining steps requires more efforts. One hurdle is the positivity of the Jacobian at fixed $N$,
 which  is connected with having a unique solution to the classical Bethe equations \eqref{11.66}. 

For quantum systems solvable by Bethe ansatz, upon substituting the appropriate phase shift, Eq. \eqref{11.65} stays intact. Only the input Lebesgue measure is replaced by the counting measure over integers  $I_1 <\ldots<I_N$, which result from quantizing the momenta  of the model with zero interaction.

\subsection{Trigonometric Calogero-Moser model}
\label{sec11.6}
For a high density of particles the potential $V_\mathrm{ca}(x) = (2\sinh(x/2))^{-2}$ can be approximated by $V_\mathrm{cm}(x) = 1/x^2$, which leads to a fluid called rational Calogero-Moser model. Its hamiltonian reads 
\begin{equation}\label{11.73}
H_{\mathrm{cm},N} = \sum_{j=1}^N \tfrac{1}{2} p_j^2+\tfrac{1}{2} \sum_{i,j=1,i\neq j}^N \frac{1}{(q_i - q_j)^2 }.
\end{equation}
This model is still integrable with Lax pair
\begin{equation}\label{11.74} 
 L_{i,j} = \delta_{ij}p_j + \mathrm{i}(1- \delta_{ij}) \frac{1}{q_i- q_j},\quad
M_{i,j} =  \mathrm{i}\delta_{ij} \sum_{m = 1,m\neq j}^N\frac{1}{(q_j- q_m)^2} - \mathrm{i}(1- \delta_{ij})
\frac{1}{(q_i- q_j)^2}.
\end{equation}
For systems in one dimension, equilibrium correlations generically have the same decay as the potential. Thus the much emphasized locality of conservation laws can no longer be
maintained. Nevertheless, as to be discussed, the Calogero-Moser model has a structure rather similar to the Toda and Calogero fluid. So to speak, integrability dominates long range interactions. 

To start with a low key, we study the two-particle scattering. The relative motion $q(t) = q_1(t) - q_2(t)$  is governed by
\begin{equation}\label{11.75}
\frac{d^2}{dt^2} q(t) = 4 q(t)^{-3}.
\end{equation}
For $v>0$ we impose $\dot{q}(0) = 0$ and $q(0) = 2/v$, which matches with $\dot{q}(\pm\infty) = \pm v$. Then the solution consists of two branches, one for $t>0$ and one for $t<0$, with
\begin{equation}\label{11.76}
q(t) = \pm\sqrt{(vt)^2 + 4v^{-2}} \simeq \pm vt \big(1 +2 v^{-4}t^{-2}\big)
\end{equation}
 for large $|t|$. Since the first order deviation from the linear motion decays as $|t|^{-1}$, the scattering shift of the rational model vanishes,
\begin{equation}\label{11.77}
\phi_\mathrm{cm}(p_1 - p_2) = 0.
\end{equation}
If GHD is still applicable, then the hyperbolic conservation laws governing the motion of the conserved fields would decouple and the Euler type dynamics cannot be distinguished from the one of an ideal gas. Of course, in a local spacetime patch the conserved fields are highly correlated and very far from a Poisson distribution.

To check whether such predictions are valid, one has to compute the generalized free energy. Because of long range interactions, the notion of finite volume has some level of arbitrariness. One physically natural procedure is to start from an infinitely extended periodic configuration of spatial period $\ell$,  $\ell >0$, and study the limit when $\ell \to \infty$. Since particles cannot cross, the order $q_j < q_{j+1}$ is maintained. The periodicity condition thus reads 
\begin{equation}\label{11.78}
q_{j+N} = q_j +\ell, \quad p_{j+N} = p_j 
\end{equation}
for all $j \in \mathbb{Z}$. The interaction potential becomes then the periodized rational potential
\begin{equation}\label{11.79}
\sum_{m\in \mathbb{Z}} \frac{1}{(x -\ell m)^2} = \frac{\pi^2}{\ell^2\sin^2(\pi x/\ell)}  = V_\mathrm{tcm}(x).
\end{equation}
Particles with interaction potential $V_\mathrm{tcm}$ are called trigonometric Calogero-Moser model. Each cell $[m\ell, (m+1)\ell]$ contains $N$ particles. A single cell can be viewed as a ring of length $\ell$, $\{z\in \mathbb{C}\big| |z| = \ell/2\pi\}$. If $q_i,q_j$ are the locations of particles on the ring, then their interaction potential is $1/d(q_i,q_j)^2$, where $d(q_i,q_j)$ denotes their cord distance.  

 The phase space of this model can be taken as $\Gamma_{{\!N,\ell}}^\triangleright  = \mathbb{W}_{N,\ell} \times \mathbb{R}^N$ with the Weyl chamber
 $\mathbb{W}_{N,\ell} = \{0\leq q_1 <\ldots<q_N < \ell \}$.  
 The trigonometric Calogero-Moser model is still integrable with Lax matrix
\begin{equation}\label{11.80}
L_{i,j} = p_j \delta_{ij} +\mathrm{i} (1- \delta_{ij})  \frac{\pi}{\ell\sin(\pi(q_i - q_j)/\ell)},
\end{equation}
which depends on $\ell$, $ L = L(\ell)$. 
Through  the canonical transformation $\check{p}_j = \ell p_j$,  $\check{q}_j = q_j/\ell $ one obtains the scaling with $\ell$ as
\begin{equation}\label{11.81}
\ell L(\ell) = L(1).
\end{equation}
As a side remark, for the Calogero fluid the interaction is $V_\mathrm{ca}(x) =g^2(\mu/2)^2 (\sinh(x/2\mu))^{-2}$. Thus the periodized potential \eqref{11.79} corresponds to
$ g=1$ and $\mu = \mathrm{i} 2\pi/\ell$, 
which is merely the analytic continuation of the potential $V_\mathrm{ca}$ to purely imaginary arguments.

Since particles move on the circle, rather than scattering coordinates, the conventional action-angle variables come into use. Phase space is foliated into invariant tori and the trajectories are quasi-periodic in time.  
The action-angle variables are constructed by an algebra very similar to the one in Section \ref{sec11.2}. 
Under the action-angle map $\Phi$, the angle variables vary as $\phi_j \in [0,\ell]$. In comparison to the hyperbolic model, the transformation to the eigenvalues is much more explicit, in the sense that the eigenvalues are merely 
 constrained as
\begin{equation}\label{11.82}
\lambda_{j+1} - \lambda_j \geq 2\pi/\ell, 
\end{equation}
which defines the set $\Omega_{N,\ell}$. The Boltzmann weight then transforms as
\begin{equation}\label{11.83}
\mathrm{e}^{-\mathrm{tr}[V(L(\ell))]} d^N\hspace{-1pt}qd^N\hspace{-1pt}p = \exp\Big(- \sum_{j=1}^N V(\lambda_j)\Big)
\chi(\Omega_{N,\ell}) d^N\hspace{-1pt}\lambda\mathrm{d}^N\hspace{-1pt}\phi,
\end{equation}
where $\chi(\Omega_{N,\ell})$ is the characteristic function of the set $\Omega_{N,\ell}$. 

We first consider the partition function, in fact its version with the additional constraint $\lambda_N \leq w$. The set of admissible eigenvalues  is then denoted by
\begin{equation}\label{11.84}
\Omega_{N,\ell}(w) = \{\lambda \in \mathbb{R}^N|\lambda_{j+1} -\lambda_j \geq  (2\pi/\ell), j = 1,\ldots,N-1, \lambda_N \leq w\}
\end{equation}
with the convention $\Omega_{N,\ell} = \Omega_{N,\ell}(\infty)$.
The constrained partition function is given by 
\begin{equation}\label{11.85}
Z_{N,\ell}(V,w)  = \int_{\mathbb{W}_{N,\ell}}\mathrm{d}^N \hspace{-1pt} q\int_{\mathbb{R}^N} \mathrm{d}^N \hspace{-1pt}p\exp\big(-\mathrm{tr}(V(L))\big) = \ell^N \int_{\Omega_{N,\ell}(w)}
\mathrm{d}^N \hspace{-1pt}\lambda \prod_{j=1}^N \mathrm{e}^{-V(\lambda_j)}, 
\end{equation}
including the cases $Z_{0,\ell}(V,w) = 1$ and $Z_{1,\ell}(V,w) = \ell \int_{-\infty}^w \mathrm{d} \hspace{-1pt}\lambda_1\exp(-V(\lambda_1))$. 
The factor $\ell^N$ results from the $\phi$-integrations. By introducing the chemical potential $\mu$, dual to the particle number, we switch to the grand canonical ensemble as 
\begin{equation}\label{11.86}
Z_\ell(V,w) = \sum_{N=0}^\infty \mathrm{e}^{\mu N}Z_{N,\ell}(V,w) 
\end{equation}
and note that
\begin{equation}\label{11.87}
Z_\ell'(V,w) = \ell \mathrm{e}^{-V(w) + \mu}Z_\ell(V,w- (2\pi/\ell))
\end{equation}
with $Z_\ell' = (\mathrm{d}/\mathrm{d}w) Z_\ell$. The constrained free energy per unit length is
\begin{equation}\label{11.88}
F_\ell(V,w) = \frac{1}{\ell}\log  Z_\ell(V,w), \qquad \lim_{\ell\to \infty}\frac{1}{\ell}\log  Z_\ell(V,w) = F(V,w).
\end{equation}
Commonly the free energy would be  $-F(V,w)$. With our definition, $F'(V,w) >0$ that will be more convenient. Inserting in \eqref{11.87}, one concludes that in the infinite volume limit
\begin{equation}\label{11.89}
F'(V,w) = \mathrm{e}^{-V(w) +\mu} \mathrm{e}^{-2\pi F'(V,w)} .
\end{equation}

The connection with the Toda fluid can now established by introducing 
\begin{equation}\label{11.90}
F'(V) = \mathrm{e}^{-\varepsilon} = \rho_\mathsf{n}.
\end{equation}
Then \eqref{11.88} turns into
\begin{equation}\label{11.91}
\varepsilon = V -\mu  + 2\pi  \mathrm{e}^{-\varepsilon}, 
\end{equation}
which one recognizes as TBA equation.  The $T$ operator has the kernel 
\begin{equation}\label{11.92}
T(w,w') = - 2\pi\delta(w - w'), 
\end{equation}
in accordance with zero scattering shift.
Observe that for hard rods the corresponding operator $T$ is the projection onto the constant function, while here $-T/2\pi$ is the identity matrix.
The $\ell \to \infty$ limit of the free energy per unit volume is given by
\begin{equation}\label{11.93}
F(V,\infty) =\int_\mathbb{R} \mathrm{d} w F'(V,w).
\end{equation}

To complete the analogy, we have to still determine the density of states of the Lax matrix per unit volume, which amounts to the average
\begin{equation}\label{11.94}
\ell^{-1} \langle \sum_{j=1}^N f(\lambda_j) \rangle_{V,\ell},
\end{equation}
average with respect to the grand canonical density as in the partition function \eqref{11.85} at $w = \infty$.
This average can be accomplished  
by perturbing  the confining potential as  $V +\delta f$. Then to first order in $\delta$  
\begin{equation}\label{11.95}
-\int_\mathbb{R} \mathrm{d} w f(w) \partial_\delta F'(V,w)|_{\delta = 0} = \int_\mathbb{R} \mathrm{d} w f(w) \rho_\mathsf{p}(w). 
\end{equation}
Using \eqref{11.89} to work out the derivative with respect to $\delta$ at $\delta = 0$ yields
\begin{equation}\label{11.96}
\rho_\mathsf{p}(w) =   \frac{ F'(V,w)}{1 + 2\pi F'(V,w)}, \qquad \rho_\mathsf{s}(w) =   \frac{1}{1 + 2\pi F'(V,w)}.
\end{equation}
Then indeed $\rho_\mathsf{p} = \rho_\mathsf{n} \rho_\mathsf{s}$ and $\rho_\mathsf{s} = 1 + T\rho_\mathsf{p}$.
As a final step, the effective velocity can be determined through \eqref{6.19} with the result
\begin{equation}\label{11.97}
v_\mathrm{eff} (w) = w,
\end{equation}
implying the completely decoupled set of hydrodynamic equation 
\begin{equation}\label{11.98}
\partial_t \rho_\mathsf{p}(x,t;v) + v\partial_x \rho_\mathsf{p}(x,t;v) = 0.
\end{equation}
\bigskip
\begin{center}
 \textbf{Notes and references}
 \end{center}
 \begin{center}
 \textbf{Section 11.0}
\end{center} 
The famous hierarchical model was introduced by Dyson \cite{D69} to study ferromagnetic Ising models in one dimension with long range interactions.  Johansson \cite{J91} established the liquid-gas transition for a one-dimensional fluid with a potential decaying 
 as $- r^{-\alpha}$, $1 \leq \alpha < 2$.
 The functional relations for $\alpha,\beta, \gamma, V_\mathrm{mec}$ are studied in Calogero \cite{C01},
 Olshanetsky and  Perelomov \cite{OP81}.  Interestingly enough, the construction of the Calogero soliton uses the Lax matrix $L^{[3]}$ Gon and  Kulkarni  \cite{GK19}. 
  \bigskip
 \begin{center}
 \textbf{Section 11.1}
\end{center} 
Calogero  \cite{C71} and Sutherland \cite{S71} independently discovered the  integrability of the \textit{quantum} many-body system
 with $1/x^2$ interaction potential.  A few years later the classical version was investigated by Calogero \cite{C75a} as part of a wider class of classical systems having a Lax matrix pair. This class also includes the two models discussed in this chapter. Independently Moser \cite{M75a} discovered the Lax matrix for the rational and trigonometric model Moser. In the rational case he proves that the $N$-particle system has zero scattering shift. For the trigonometric model
 he found concise expressions for the time-dependent solution. In honor of these early contributions, the names Calogero-Moser-Sutherland are attached to the models already listed and their vast generalizations discovered in the mean time,
as documented in the book edited by van Diejen and Vinet \cite{DV12}. We cover only two specific cases, namely the Calogero fluid, interaction potential $1/\sinh^2$, and the 
 Calogero-Moser model, interaction potential $1/x^2$ and its periodized version $1/\sin^2$.
 The 2001 monograph by Calogero \cite{C01} is a rich source on integrable classical particle models.
   The hyperbolic model is discussed  in Calogero \cite{C01}, Section 2.1.5. The connection to Lie algebras is reviewed by Olshanetsky and  Perelomov \cite{OP81}.
   When adding a fine-tuned external potential integrability is preserved, see  the contributions by Inozemtsev \cite{I83}, Polychronakos \cite{P92}, and
   Kulkarni and  Polychronakos \cite{KP17}. The strength can be arbitrary, but the decay has to match the one of the $\sinh^{-2}$ potential. The term $-\tfrac{1}{12}$ appearing in \eqref{11.22} is discussed by Ruijsenaars \cite{R99}. 
\newpage
\begin{center}
\textbf{Section 11.2}
\end{center}
The scattering coordinates for the hyperbolic Calogero model have been constructed by Ruijsenaars \cite{R88}.  Our exposition follows the very readable contribution of Bogomolny  et al. \cite{BGS11}, who provide the required algebra. The definition  of the scattering shifts through \eqref{11.40} is based on analogy with earlier contributions by Airault  et al. \cite{AMM77} and Adler  \cite{A77}. The difficult part is to establish agreement with the dynamical approach in \eqref{2.42}.  Only then the required properties of $\Phi$  can be proved, as accomplished by  Ruijsenaars \cite{R88}. Ruijsenaars\cite{R99} provides an instructive summary of his own contributions.\bigskip
\begin{center}
\textbf{Section 11.3}
\end{center}
While there is some discussion in the literature, for example  Polychronakos \cite{P11}, the variational principle for the free energy is a recent result.
   The two-particle scattering shift has been computed by Calogero \cite{C01},  Section 2.1.5.
   The Calogero fluid with external potential $g_1\cosh x +g_2\cosh 2x$, compare with (11.46),   is still integrable,
see van Diejen \cite{VD95}  for a proof. But the respective Lax matrix is less accessible.\bigskip
\begin{center}
\textbf{Section 11.5}
\end{center}
Classical Bethe equations seem to be novel. To study such a problem, a convenient starting point would be to assume a toy model defined by $\phi_\mathrm{ca} >0$. Then the Jacobian matrix is positive and Bethe equations have a unique solution.\bigskip
 \begin{center}
 \textbf{Section 11.6}
\end{center}
 The construction of action-angle variables can be found in the article Ruijsenaars \cite{R95} under model $III_{nr}$.  
 Since also other models are discussed in parallel, the results of interest in the context of our discussions are scattered throughout the article. A concise account of the algebraic steps are provided by Bogomolny  et al. \cite{BGS11}.  Our discussion is based on the clarifying contribution by Choquard   \cite{C00}, who covers also the quantized model.
In the recent contribution by Bulchandani et al. \cite{BKMC21} both the quantum and classical rational Calogero-Moser model is studied, emphasizing the aspect of quasiparticles. Presented are plots of the DOS and also time-dependent solutions.
 \newpage
 \section{Discretized nonlinear Schr\"{o}dinger equation}
 \label{sec12}
\subsection{Continuum wave equations}
\label{sec12.1}
 \setcounter{equation}{0}
 A famous integrable classical field theory in $1+1$ dimensions is the nonlinear Schr\"{o}dinger equation (NLS). 
In the defocusing case the wave field, $\psi(x,t) \in \mathbb{C}$, is governed by 
\begin{equation} 
\label{12.1}
\mathrm{i}\partial_t\psi = - \partial_x^2\psi + 2|\psi|^2 \psi.
\end{equation}
Our interest will be random initial conditions linked to the hydrodynamic scale. 
For NLS the  densities of the locally conserved fields are known. The starting entries of the list read
\begin{equation}
\label{12.2}
Q^{[0]}(x) = |\psi(x)|^2,
\qquad Q^{[1]}(x) = -\mathrm{i} \bar{\psi}(x)\partial_x \psi(x),
\qquad Q^{[2]}(x) = |\partial_x \psi(x)|^2 +  |\psi(x)|^4.
\end{equation}
 For a given bounded interval $\Lambda \subset \mathbb{R}$, with periodic boundary conditions, the total conserved quantities then become
\begin{equation}
\label{12.3}
Q_\Lambda^{[n]} =
\int_\Lambda \hspace{-3pt}\mathrm{d}x Q^{[n]}(x), 
\end{equation}
$n = 0,1,\ldots.$\,. Formally, the time-stationary generalized Gibbs ensembles are of the form
\begin{equation}
\label{12.4}
\exp\Big[ - \sum_{n=0}^\infty \mu_n Q_\Lambda^{[n]}\Big] \prod_{x \in \Lambda}\mathrm{d}^2 \psi(x),
\end{equation}
where the $\mu_n$'s are suitable chemical potentials. 
For the thermal case, $n=0,1,2,$ much work has been invested to construct  a proper probability measure. The basic idea of the construction is easily explained.
Obviously, one can lattice discretize the volume $\Lambda$. Then the product Lebesgue measure makes sense. However, the limit of zero lattice spacing is ill-defined
within standard measure theory. On the other hand the weighted measure
\begin{equation}
\label{12.5}
(Z_{\mathrm{nls},N})^{-1} \prod_{j=1}^N \mathrm{d}x_j\exp\Big(- \mu_0 \sum_{j=1}^N |\psi(j)|^2 - \mu_2 \sum_{j=1}^{N} |\psi(j+1)-\psi(j)|^2\Big), 
\end{equation}
with boundary condition $\psi(N+1) = \psi(1)$, has a well defined continuum limit, which is Gaussian and well-known as stationary $\mathbb{R}^2$-valued  Ornstein-Uhlenbeck process.
The remaining GGE piece reads
\begin{equation}
\label{12.6}
\exp\Big(- \mu_2 \int_\Lambda \hspace{-3pt}\mathrm{d}x|\psi(x)|^4\Big).
\end{equation}
If $\mu_2 >0$, it is integrable with respect to the Ornstein-Uhlenbeck process.
As a separate issue, it has to be shown that the so constructed measure is invariant under the NLS dynamics. Recently such  method has been extended to a much larger class of generalized Gibbs measures in case the sum in \eqref{12.4} is restricted to some highest 
even $n$. To provide a very rough idea,
the Gaussian a priori measure becomes now a generalized Ornstein-Uhlenbeck process with action
\begin{equation}
\label{12.7}
\int_\Lambda \hspace{-3pt}\mathrm{d}x \big(|\partial_x^{n/2} \psi(x)|^2  +|\psi{(x)}|^2 \big). 
\end{equation} 
As for $n=2$, the technical part is to establish that, for an appropriate choice of chemical potentials, the exponential of all remaining terms can be integrated with respect to this Gaussian measure. In essence, the thereby constructed measure is concentrated on $(n-2)$ times differentiable paths. Time-stationarity
is established separately. To go beyond such an existence result seems to be a challenging problem.

Given the difficulties in even obtaining the basic building blocks, a different strategy might be pursued. In numerical simulations of NLS one  discretizes the equation. While generically this would break integrability, surprisingly enough, in many cases there is one very specific discretization for which  integrability is maintained. For NLS such a discretization was discovered in 1975 
by Ablowitz and Ladik and is usually referred to as IDNLS, i.e. integrable discrete NLS. For convenience we will use here AL as acronym. 

Another example in the same spirit is the classical sinh-Gordon equation, 
\begin{equation}
\label{12.8}
\partial_t^2\phi - \partial_x^2\phi + \sinh \phi = 0,
\end{equation}
with $\phi(x,t)$ a real-valued wave-field. This dynamics is integrable, while its naive discretization is not. But there is a more subtle lattice version which is still integrable. Its construction is
based on an analytic continuation of the discretized sine-Gordon equation. 

A further much studied classical system is the Landau-Lifshitz model of a one-dimensional magnet. Here the spin field is a three-vector, $\vec{S}(x,t)$, with  
$|\vec{S}| = 1$. In the simplest case of an isotropic interaction, no external magnetic field, the continuum equations of motion are
\begin{equation}
\label{12.9}
\partial_t\vec{S} = \vec{S} \wedge \partial_x^2 \vec{S} ,
\end{equation}
which is integrable. In the naive discretization the hamiltonian would be
\begin{equation}
\label{12.10}
H_\mathrm{ll} = \sum_{j\in\mathbb{Z}} \vec{S}_j\cdot\vec{S}_{j+1},
\end{equation}
which is not integrable, while the integrable discrete model turns out to be governed by the hamiltonian
\begin{equation}
\label{12.11}
\tilde{H}_\mathrm{ll} = \sum_{j\in\mathbb{Z}} \log(1+ \vec{S}_j \cdot \vec{S}_{j+1} ).
\end{equation}

The nonlinear Schr\"{o}dinger equation and its Ablowitz-Ladik discretization are solvable through IST, compare with Section \ref{sec10.1}. 
IST is a powerful and systematic method to obtain $n$-soliton solutions. Generically the one-soliton solution is still explicit, while the asymptotics of 
the two-soliton solution yields the scattering shift. Thus the basic input for a soliton-based hydrodynamic scale is available.  For NLS the spectral parameter turns out to be 
complex, which foreshadows the possibility of spectral parameters different from $\mathbb{R}$. Here we will pursue the in spirit particle-based scheme for the Ablowitz-Ladik lattice.
\subsection{Ablowitz-Ladik discretization}
\label{sec12.2}
Upon discretization, the wave field is over the one-dimensional lattice $\mathbb{Z}$, $\psi_j(t) \in \mathbb{C}$,
 and is governed by   
\begin{equation}\label{12.12}
\mathrm{i}\frac{d}{dt}\psi_j  = - \psi_{j-1}  + 2 \psi_j - \psi_{j+1} +  |\psi_j|^2 (\psi_{j-1} + \psi_{j+1}),
\end{equation}
 hence
\begin{equation} \label{12.13} 
\mathrm{i}\frac{d}{dt}\psi_j  = - (1 - |\psi_j|^2)(\psi_{j-1} + \psi_{j+1}) + 2 \psi_j.
\end{equation}
Setting $\alpha_j(t) = \mathrm{e}^{2\mathrm{i}t}\psi_j(t)$, one arrives at the standard version
\begin{equation}\label{12.14} 
\frac{d}{dt}\alpha_j  = \mathrm{i} \rho_j^2 (\alpha_{j-1} + \alpha_{j+1}),\quad \rho_j^2 = 1 - |\alpha_j|^2.
\end{equation}
Clearly the natural phase space is $\alpha_j \in \mathbb{D}$ with the unit disk $\mathbb{D} = \{z||z| \leq 1\}$. In principle, whenever $\alpha_j (t)$
hits the boundary of $\mathbb{D}$, it freezes and thereby decouples the system. As we will discuss, a conservation law ensures that, if initially away from the boundary, the solution will stay so forever.

 For determining the generalized free energy,  the method of scattering coordinates does not seem to be available.  
 An alternative set-up would be the closed Ablowitz-Ladik model, which consists of a finite ring of $N$ sites, labelled as $j = 0,\ldots,N-1$, with periodic boundary conditions,
$\alpha_{j+N} = \alpha_j$. While the system with periodic boundary conditions is integrable, there seems to be no method for obtaining its generalized free energy 
 in the limit $N \to \infty$. However, there is an analogue of $T = L^\diamond$ from the Toda chain, in the sense that the interactions for bonds $(-1,0)$ and $(N-2,N-1)$ are modified. In analogy this is called the open Ablowitz-Ladik lattice. As for the Toda chain, to compute the free energy one has to impose a linear pressure ramp of slope $1/N$.
Closed (periodic) and open chain will be discussed separately.\bigskip\\
\textbf{Conserved fields}. We consider a ring of $N$ sites. The evolution equations are of hamiltonian form by 
regarding $\alpha_j$ and its complex conjugate $\bar{\alpha}_j$ as canonically conjugate variables and by introducing
the weighted Poisson bracket
\begin{equation}\label{12.15} 
\{f,g\}_\mathrm{AL} =\mathrm{i} \sum_{j=0}^{N-1} \rho_j^2\big(\partial_{\bar{\alpha}_j} f 
\partial_{\alpha_j} g  - \partial_{\alpha_j} f 
\partial_{\bar{\alpha}_j} g \big).
\end{equation}
The hamiltonian of the Ablowitz-Ladik system then reads
\begin{equation}\label{12.16} 
H_{\mathrm{al},N} = - \sum_{j=0}^{N-1} \big( \alpha_{j-1} \bar{\alpha}_{j} + \bar{\alpha}_{j-1} \alpha_{j}\big).
\end{equation}
One readily checks that indeed
\begin{equation} \label{12.17}
\frac{d}{dt}\alpha_j = \{\alpha_j,H_{\mathrm{al},N}\}_\mathrm{AL}   = \mathrm{i} \rho_j^2 (\alpha_{j-1} + \alpha_{j+1}) .
\end{equation}

The next step is to find out the locally conserved fields. As discovered by Irena Nenciu, the Ablowitz-Ladik model has also  a  
 Lax matrix, this time  in the form of  a Cantero-Moral-Vel\'{a}zquez (CMV) matrix. The basic building blocks are $2\times 2$ matrices,
which requires $N$ to be \textit{even} because of periodic boundary conditions. One defines
\begin{equation}\label{12.18} 
\Xi_j =
\begin{pmatrix}
\bar{\alpha}_j & \rho_j \\
\rho_j & -\alpha_j  \\
\end{pmatrix}
\end{equation}
and forms the $N\times N$ matrices
\begin{equation}\label{12.19} 
 L_N = \mathrm{diag}(\Xi_0,\Xi_2,\dots, \Xi_{N-2}), \quad M_N = \mathrm{diag}(\Xi_1,\Xi_3,\dots, \Xi_{N-3}),
\end{equation}
where $\Xi_{N-1}$ respects the periodic boundary conditions.
 More pictorially $L_N$ corresponds to the $2\times 2$ blocks $(0,1),\ldots,(N-2,N-1)$, while $M_N$ uses  $2\times 2$ blocks shifted  by one, $(1,2),\ldots,(N-1,0)$. The CMV matrix associated to the coefficients $\alpha_0,\dots,\alpha_{N-1}$ is then given by 
\begin{equation} \label{12.20}
C_N = L_N M_N.
\end{equation}
Obviously, $L_N,M_N$ are unitary and so is $C_N$. The  eigenvalues of $C_N$ are denoted by $\mathrm{e}^{\mathrm{i}\vartheta_j}$, $\vartheta_j \in [0,2 \pi]$,
$j = 1,\dots,N$.  Of course, the eigenvalues depend on $N$, which is suppressed in our notation.

Next we define for a general matrix, $A$, the $+$ operation as
\begin{equation} \label{12.21}
 (A_+)_{i,j} = \begin{cases} A_{i,j}  &\mathrm{if} \,\,i < j,\\
   \tfrac{1}{2}A_{j,j} &\mathrm{if} \,\,i = j,\\
  0 &\mathrm{if} \,\,i > j.
 \end{cases}
\end{equation}
Then one version of the Lax pair reads
\begin{equation}\label{12.22}
\{C_N,\mathrm{tr}(C_N)\}_\mathrm{AL} = \mathrm{i} [C_N,(C_N)_+],\qquad \{C_N,\mathrm{tr}(C_N^*)\}_\mathrm{AL} = \mathrm{i} [C_N,(C_{N+})^*].
\end{equation}
Since the Poisson bracket acts as a derivative, one deduces
\begin{equation} \label{12.23}
\{(C_N)^n,\mathrm{tr}(C_N)\}_\mathrm{AL} = \sum_{m=0}^{n-1} (C_N)^m\mathrm{i}[C_N,C_{N+}](C_N)^{n-1 -m} = \mathrm{i} [(C_N)^n,C_{N+}], 
\end{equation}
and similarly 
\begin{equation} \label{12.24}
\{(C_N)^n,\mathrm{tr}(C_N^*)\}_\mathrm{AL} =  \mathrm{i} [(C_N)^n,(C_{N+})^*]. 
\end{equation}
Therefore locally conserved fields are given by
\begin{equation} \label{12.25}
Q^{[n],N} = \mathrm{tr}\big[(C_N)^n\big].
\end{equation}
Their mutual Poisson brackets vanish,
\begin{equation} \label{12.26}
\{Q^{[n],N}, Q^{[m],N}\}_\mathrm{AL} = 0 \bigskip.
\end{equation}

The fields can be turned  real-valued by considering real and imaginary parts,
\begin{eqnarray}\label{12.27} 
&& Q^{[n,+],N} = \tfrac{1}{2}\mathrm{tr}\big[(C_N)^n + (C_N^*)^n\big] = \mathrm{tr}\big[\cos((C_N)^n)\big],\nonumber\\[1ex]
&& Q^{[n,-],N} = -\tfrac{1}{2}\mathrm{i}\,\mathrm{tr}\big[(C_N)^n - (C_N^*)^n\big] = \mathrm{tr}\big[\sin((C_N)^n)\big],
\end{eqnarray}
with $n = 1,\dots, N/2$. In particular $H_{\mathrm{al},N} = 2Q^{[1,+]} = \mathrm{tr}[C_N + C_N^*]$. These fields have a density, respectively given by
\begin{equation} \label{12.28}
Q_j^{[n],N}= Q_j^{[n,+],N}+\mathrm{i}Q_j^{[n,-],N} = ((C_N)^n)_{j,j}.
\end{equation}
 
 Although the matrices $L_N,M_N$ have a basic $2\times 2$ structure,  the densities of the conserved fields are shift covariant by $1$
 in the sense
\begin{equation}\label{12.29} 
Q^{[n,\sigma],N}_{j+1}(\alpha) = Q^{[n,\sigma],N}_j(\tau \alpha),\qquad \sigma = \pm1,
\end{equation}
with the shift operator $(\tau \alpha)_j = \alpha_{j+1}$ $\mathrm{mod}(N)$. To confirm, one introduces the
 unitary shift matrix $S$ through $(S^*AS)_{i,j} = A_{i+1,j+1}$ $\mathrm{mod}(N)$. Then
\begin{equation} \label{12.30}
S^*C_N({\alpha,\bar{\alpha}})S =S^*L_N(\alpha,\bar{\alpha})SS^*M_N(\alpha,\bar{\alpha}) S =  M_N(\tau \alpha,\tau \bar{\alpha}) L_N(\tau \alpha,\tau \bar{\alpha}) = C_N(\tau \alpha,\tau \bar{\alpha})^\mathrm{T},
\end{equation}
which implies  \eqref{12.29}. 

 \begin{figure}[!b]
\centering
\includegraphics[width=8cm]{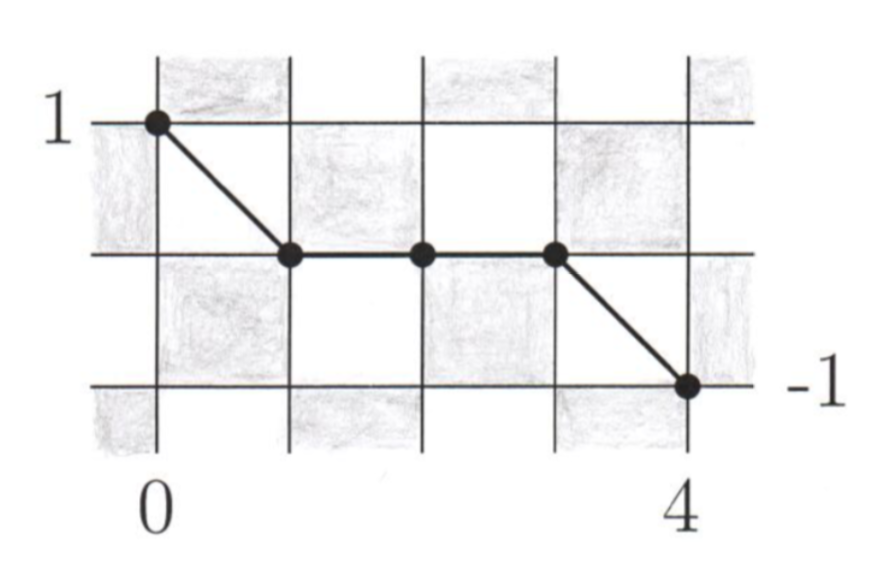}
\caption{An admissable walk from $(0,1)$ to $(4,-1)$. According to the rules its weight is given by $\rho_0(-\alpha_{-1})\bar{\alpha}_0 \rho_{-1}$.}
\label{fig12}
\end{figure}

Later on, we will consider the infinite volume limit, $N \to \infty$. This will always be understood as a two-sided limit. For example the infinite volume limit of $L_N$,
denoted by $L$, is $L  = \mathrm{diag}(\dots, \Xi_{-2},\Xi_0,\Xi_2,\dots)$ and correspondingly $M  = \mathrm{diag}(\dots, \Xi_{-1},\Xi_1,\Xi_3,\dots)$.
$L,M$ are unitary operators on the Hilbert space $\ell_2(\mathbb{Z})$, and so is $C = LM$. The traces in \eqref{12.25} have no limit, but  densities do.
The matrix elements of $C^n$  can be expanded as the sum
\begin{equation}\label{12.31}  
(C^n)_{i,j} =  \sum_{j_1\in \mathbb{Z}}\dots \sum_{j_{2n}\in \mathbb{Z}} L_{i,j_1}M_{j_1,j_2} \dots L_{j_{2n-1},j_{2n}} M_{j_{2n},j},
\end{equation}
 which consists of a finite number of terms, only. For the infinite system the index $n$ runs over all positive integers.
 The infinite volume densities, $Q^{[n,\sigma]}_j$, are strictly local functions of $\alpha$ with support of at most $2n$ sites. The sum in \eqref{12.31} can be viewed
 as resulting from a nearest neighbor $2n$ step random walk from left to right. For this purpose one considers a checkerboard on $[0,2n] \times \mathbb{R}$. The unit square with corners $(0,0),(1,0),(1,1), (0,1)$ is white. Single steps of the walk are either horizontal, $j \leadsto j$, or up-down, $j \leadsto j\pm 1$. Such diagonal steps are permitted only on white squares, compare with Figure 12.
 The matrix element $(C^n)_{i,j}$ is then the sum over all walks with  $2n$ steps starting at $i$ and ending
 at $j$. Each walk represents a particular polynomial obtained by taking the product of local weights along the walk. The weights are\bigskip\\
 \begin{tabular}{ll}
 \hspace{30pt}$\rho_j$ &  for the diagonal steps $j \leadsto j+1$ and $j+1 \leadsto j$,\\
 \hspace{30pt}$\bar{\alpha}_{j}$ & for the horizontal step $j \leadsto j$  in case its lower square is black, \\
 \hspace{30pt}$-\alpha_{j-1}$ & for the horizontal step $j \leadsto j$  in case its upper square is black. \\
 \end{tabular} 
 \bigskip\\
   As examples, $C_{j,j} = - \alpha_{j-1}\bar{\alpha}_j $ and $(C^2)_{j,j} = 
 \alpha_{j-1}^2\bar{\alpha}_j^2 - \rho_{j-1}^2\alpha_{j-2}\bar{\alpha}_j  - \rho_{j}^2\alpha_{j-1}\bar{\alpha}_{j+1}$.
 Note that densities are not unique in general, while the total conserved fields, $Q^{[n],N}$, are unique. To illustrate, in the previous formula, an equivalent density would be $\alpha_{j-1}^2\bar{\alpha}_j^2 - 2\rho_{j}^2\alpha_{j-1}\bar{\alpha}_{j+1}$.
 
 The CMV matrix misses one physically very important field namely  
\begin{equation}\label{12.32}
Q^{[0],N} =  - \sum_{j=0}^{N-1}  \log(\rho_j^2).
\end{equation}
To simplify notation we set $[0] = [0,\sigma]$ and $0 = 0\sigma$. The time-derivative of $Q^{[0],N}$ yields a telescoping sum, which vanishes on a ring. In lack of a common name we call $Q^{[0],N}$ the log intensity.
The log intensity vanishes for small amplitudes $|\alpha_j|^2$ and diverges at the maximal value, $|\alpha_j|^2 = 1$. Note also that
 \begin{equation}\label{12.33} 
\exp\big(-Q^{[0],N}\big) =  \prod_{j=0}^{N-1}\rho_j^2
\end{equation}
is conserved. Thus if initially $\exp\big(-Q^{[0],N}\big) > 0$, it stays so for all times, guaranteeing that the 
phase space boundary is never reached.\\\\
\textbf{Generalized Gibbs ensemble}. 
The natural a priori measure of the Ablowitz-Ladik model is the product measure
\begin{equation} \label{12.34}
\prod_{j=0}^{N-1}\mathrm{d}^2\alpha_j (\rho_j^2)^{P-1}= \prod_{j=0}^{N-1}\mathrm{d}^2\alpha_j (\rho_j^2)^{-1}\exp\big(-PQ^{[0],N}\big) 
\end{equation}
on $\mathbb{D}^N$. To normalize the measure, $P>0$ is  required.  The log intensity is controlled by the parameter $P$ which, in analogy to the 
Toda lattice, is called pressure. Small $P$ corresponds to maximal log intensity, i.e. $|\alpha_j|^2 \to 1$,  and large $P$ to low log intensity, i.e. $|\alpha_j|^2 \to 0$. 
In the grand-canonical ensemble, the Boltzmann weight is constructed from a linear combination of the conserved fields, which is written as
\begin{equation} \label{12.35}
 \sum_{n\in \mathbb{Z}}\mu_n \mathrm{tr}\big[(C_N)^n\big] = \mathrm{tr}\big[\mathsfit{V}(C_N)\big], \qquad \mathsfit{V}(z)=  \sum_{n\in \mathbb{Z}}\mu_n z^n.
\end{equation}
The chemical potentials, $\mu_n$, are complex and assumed to be independent of $N$.  To have the trace real-valued one imposes $\mu_n = \bar{\mu}_{-n}$.
In fact is suffices to assume that $\mathsfit{V}$ is real-valued and continuous on the circle $|z|= 1$. Combining \eqref{12.34} and \eqref{12.35} yields the GGE as
\begin{equation} \label{12.36}
Z_{\mathrm{al},N}(P,V)^{-1}\prod_{j=0}^{N-1}\mathrm{d}^2\alpha_j (\rho_j^2)^{P-1}\exp\big(-\mathrm{tr}[\hat{V}(C_N)]\big).
\end{equation}
$Z_{\mathrm{al},N}(P,V)$ is the normalizing partition function. As label, the more natural object turns out to be the Fourier transform of the sequence $\{\mu_n, n\in \mathbb{Z}\}$,
\begin{equation} \label{12.37} 
V(w) = \mathsfit{V}(\mathrm{e}^{\mathrm{i}w}).
\end{equation}
$V$ is a real-valued continuous function on $[0,2\pi]$. The corresponding object for the Toda lattice lives on $\mathbb{R}$  and was called confining potential, 
since it confines the eigenvalues of the Lax matrix. While for the CMV matrix the eigenvalues are on the unit circle, hence there is nothing to confine,  for convenience we still stick to ``confining" so to distinguish
from other potentials. 
If $\mathsfit{V}$ is given by a finite sum, then the interaction of the Gibbs measure in \eqref{12.36} is of finite range. In this case the infinite volume limit can be
controlled through transfer matrix methods. In particular,  the limit measure is expected to have  a finite correlation length.
As before, finite volume expectations with respect to 
the measure in \eqref{12.36} are denoted by $\langle \cdot \rangle_{P,V,N}$ and their infinite volume limit by $\langle \cdot \rangle_{P,V}$.

The generalized free energy, $F_\mathrm{al}$, is defined through 
\begin{equation} \label{12.38}
\lim_{N \to \infty} - \frac{1}{N} \log Z_{\mathrm{al},N}(P,V) = F_\mathrm{al}(P,V).
\end{equation}
In the hydrodynamic context of particular interest is the empirical density of states,
\begin{equation} \label{12.39}
\rho_{\mathrm{Q},N}(w)    = \frac{1}{N} \sum_{j=1}^N \delta (w - \vartheta_j) ,    
\end{equation}
where $\{\mathrm{e}^{\mathrm{i}\vartheta_j}\}$ are the  eigenvalues of $C_N$. $\rho_{\mathrm{Q},N}$ is a probability measure on $[0,2\pi]$ and has an almost sure limit as
\begin{equation} \label{12.40}
\lim_{N \to \infty} \rho_{\mathrm{Q},N}(w) = \rho_\mathrm{Q}(w).
\end{equation}
To grasp the significance of the DOS, we first introduce  the trigonometric functions $\varsigma_0(w) =1$, $\varsigma_{n-}(w) = \sin(nw)$, and $\varsigma_{n+}(w) = \cos(nw)$, $n = 1,2,\dots$ . They span the Hilbert space $L^2([0,2\pi], \mathrm{d}w)$. Then, for the trigonometric moments of   $\rho_{\mathrm{Q},N}(w)$,
\begin{equation} \label{12.41}
 \langle \rho_{\mathrm{Q},N} \varsigma_{n\sigma} \rangle = N^{-1}\langle Q^{[n,\sigma],N}\rangle_{P,V,N} 
 \end{equation}
 and
 \begin{equation} \label{12.41a}
\lim_{N \to \infty} N^{-1}\langle Q^{[n,\sigma],N}\rangle_{P,V,N} = 
 \langle Q_0^{[n,\sigma]}\rangle_{P,V} = \langle \rho_{\mathrm{Q}} \varsigma_{n\sigma} \rangle.
\end{equation}
Consistent with our previous notation, $\langle \cdot \rangle$ is simply a short hand  for the integration over $[0,2\pi]$.
The limit value can also be expressed as variational derivative of the generalized free energy per site,
\begin{equation} \label{12.42}
\frac{ \mathrm{d}}{\mathrm{d}\kappa} F_\mathrm{al}(P,V +\kappa \varsigma_{n\sigma})\big|_{\kappa = 0} = \langle Q_0^{[n,\sigma]}\rangle_{P,V}.
\end{equation}
In addition, one introduces the average log intensity, denoted by $\nu$, for which
\begin{equation} \label{12.43}
\nu =  \langle Q_0^{[0]}\rangle_{P,V} = \partial_P F_\mathrm{al}(P,V).
\end{equation}

Note that by definition $\nu > 0$. For $V=0$, one readily confirms  $F_\mathrm{al}(P,0)= \log(P/\pi)$ with log intensity $\nu(P) = P^{-1} > 0$. Hence there is no high pressure phase as encountered for the Toda lattice, see end of Section  \ref{sec9.1}. 
\subsection{Circular random matrices with pressure ramp}
\label{sec12.3}
We modify the CMV matrix at its two boundaries. As before, the number  $N$ of sites is even. $L_N$ remains unchanged, $M_N$ is modified to  
$M_N^\diamond$, where $(M_N^\diamond)_{0,0} =1$, $(M_N^\diamond)_{0,N-1}= 0$, $(M_N^\diamond)_{N-1,0}=0$, and $(M_N^\diamond)_{N-1,N-1}
= \mathrm{e}^{\mathrm{i}\phi}$, $\phi \in [0,2\pi]$. This leads to the particular CMV matrix
\begin{equation} \label{12.44}
C_N^\diamond = L_NM_N^\diamond.
\end{equation}
For the a priori measure  \eqref{12.34} the pressure $P$ is constant. To arrive at a tractable expression, the constant pressure is replaced by  a  linearly changing pressure with a yet arbitrary slope $-\tfrac{1}{2}\beta$, $\beta >0$, according to
  \begin{equation} \label{12.45}
\mathrm{d}\phi \prod_{j=0}^{N-2}(\rho_j^2)^{-1} (\rho_j^2)^{\beta(N-1-j)/2} \mathrm{d}^2\alpha_j.
\end{equation}
Surprisingly, Killip and Nenciu succeeded to compute the joint distribution of eigenvalues  of $C_N^\diamond $ under this measure.
We define the Vandermonde determinant
\begin{equation} \label{12.46}
\Delta(z_1,\ldots,z_N) = \prod_{1 \leq i < j \leq N} (z_j - z_i).
\end{equation}
Denoting the eigenvalues of $C_N^\diamond $ by $\mathrm{e}^{\mathrm{i}\vartheta_1},\dots,\mathrm{e}^{\mathrm{i}\vartheta_N}$, their joint (not normalized) distribution under the measure in \eqref{12.45}
is given by 
\begin{equation} \label{12.47}
\tilde{\zeta}_N (\beta) |\Delta(\mathrm{e}^{\mathrm{i}\vartheta_1},\ldots, \mathrm{e}^{\mathrm{i}\vartheta_N})|^\beta \prod_{j=1}^N\mathrm{d}\vartheta_j, 
\end{equation}
where the prefactor turns out to equal
\begin{equation} \label{12.47a}
 \tilde{\zeta}_N(\beta) = 2^{(1-N)} \frac{1}{N!}
\frac{\Gamma(\beta/2)^N}{\Gamma(N\beta/2)}.
\end{equation}

Since $\beta$ is a free parameter, one can choose specifically
\begin{equation} \label{12.48}
\beta = \frac{2P}{N}.
\end{equation}
Now the ramp has slope $-P/N$ and, in the limit $N \to \infty$,  close to the lattice point $(1-u)N$, $0 < u <1$, the measure of \eqref{12.45}  will converge to the product measure of \eqref{12.34} with pressure $uP$. Since
\begin{equation} \label{12.49}
\mathrm{tr}\big[\mathsfit{V}(C_N^\diamond)] = \sum_{j=1}^N V(\vartheta_j),
\end{equation}
the Boltzmann weight can be naturally included in \eqref{12.45}. Hence the partition function of the system with boundary conditions turns into
\begin{eqnarray} \label{12.50}
&&\hspace{-40pt} Z_{\mathrm{kn},N}(P,V) = \int_{[0,2\pi]^{N-1}}\prod_{j=0}^{N-2}\mathrm{d}^2\alpha_j\int_0^{2\pi} \mathrm{d}\phi \prod_{j=0}^{N-2}(\rho_j^2)^{-1} (\rho_j^2)^{P(N-1-j)/N} \exp\big(-\mathrm{tr}[V(C_N^\diamond)]\big) \nonumber\\
&&\hspace{-20pt} =\zeta_N(P) \int_{0}^{2\pi} \!\!\mathrm{d}\vartheta_1\dots  \int_{0}^{2\pi}\!\! \mathrm{d}\vartheta_N
\exp\Big( - \sum_{j=1}^N V(\vartheta_j) + P \frac{1}{N} \sum_{i,j=1,i\neq j}^N  \log|\mathrm{e}^{\mathrm{i}\vartheta_i} - \mathrm{e}^{\mathrm{i}\vartheta_j} |\Big)
\end{eqnarray}
with $\zeta_N(P) = \tilde{\zeta}_N(2P/N)$. In statistical mechanics the probability distribution 
\begin{equation} \label{12.51}
\frac{1}{Z_{\log, N}}\exp\Big( - \sum_{j=1}^N V(\vartheta_j) + P \frac{1}{N} \sum_{i,j=1,i\neq j}^N  \log|\mathrm{e}^{\mathrm{i}\vartheta_i} - \mathrm{e}^{\mathrm{i}\vartheta_j} |\Big)
\end{equation}
is known as \textit{circular log-gas}. Since the coupling strength is proportional to $1/N$, it is the mean-field version of the log-gas.  
 
 The  Ablowitz-Ladik model with linearly varying  pressure has a free energy per site defined through 
\begin{equation} \label{12.52}
F_{\mathrm{kn}}\hspace{-1pt}(P,V) = \lim_{N \to \infty} - \frac{1}{N} \log Z_{\mathrm{kn},N}(P,V).
\end{equation}
Because the pressure ramp has slope $1/N$, in the limit free energies merely add up as
\begin{equation} \label{12.53}
F_{\mathrm{kn}}\hspace{-1pt}(P,V) = \int_0^1 \mathrm{d}u F_\mathrm{al}(uP,V).
\end{equation}

Before studying the infinite volume free energy, we remark that the CMV matrix $C_N^\diamond$ is still linked to a suitably modified Ablowitz-Ladik dynamics governed by the 
hamiltonian
\begin{equation} \label{12.54}
H_N^\diamond=  \mathrm{tr}\big[C_N^\diamond + {C_N^\diamond}\hspace{-1pt}^*\big].
\end{equation}
Working out the Poisson brackets leads to the evolution equation
\begin{equation} \label{12.55}
\frac{d}{dt}\alpha_j  = \mathrm{i} \rho_j^2 (\alpha_{j-1} + \alpha_{j+1}),
\end{equation}
$j = 0,\dots, N-2$, with the boundary conditions $\alpha_{-1} = -1$ and $\alpha_{N-1} = \mathrm{e}^{\mathrm{i}\phi}$. As before 
$\mathrm{tr}\big[{(C_N^\diamond})^n\big]$ is preserved under the dynamics. However, the a priori measure \eqref{12.45} is no longer stationary.  
The long time dynamics on a ring differs qualitatively from the one with open boundary conditions.

The prefactor in \eqref{12.50} can be easily handled with the result
\begin{equation} \label{12.56}
 \lim_{N \to \infty} - \frac{1}{N} \log\zeta_{N}(P) = \log(2P) -1.
\end{equation}
The sum in the exponential of  \eqref{12.51} has a  term involving the confining potential $V$, which is linear in the empirical 
measure, $\varrho_N$,  of the $\{\vartheta_j\}$'s. The double sum of the  interaction is quadratic in  $\varrho_N$. Thus the limiting free energy 
is determined by a variational principle. We first define the free energy functional
\begin{equation} \label{12.57}
\mathcal{F}_\mathrm{kn}^\circ(\varrho) =  \int _0^{2\pi}\hspace{-6pt}\mathrm{d}w \varrho(w) \Big(V(w)  +\log \varrho(w)  
+\log P  - P\int_0^{2\pi}\hspace{-6pt}\mathrm{d}w'   
\log|\mathrm{e}^{\mathrm{i}w}-\mathrm{e}^{\mathrm{i}w'}| \varrho(w')  
\Big) .
\end{equation} 
This functional has to be minimized over all densities $\varrho$, with the constraints $\varrho(w) \geq 0$ and $\langle  \varrho \rangle = 1$. Denoting by $\varrho^\star$ the unique minimizer, one arrives at 
\begin{equation} \label{12.58}
 F_\mathrm{kn}\hspace{-1pt}(P,V) =  \log 2 -1 + \mathcal{F}_\mathrm{kn}^\circ(\varrho^\star) 
\end{equation}
and thus, using \eqref{12.53},
\begin{equation}\label{12.59} 
F_\mathrm{al}(P,V) = \partial_P(PF_\mathrm{kn}(P,V)).
\end{equation}

As for the Toda lattice, it turns out to be more convenient to absorb $P$ into $\varrho$ by setting $\rho = P\varrho$. Then 
$P \mathcal{F}_\mathrm{kn}^\circ(P^{-1}\rho)=  \mathcal{F}_\mathrm{kn}(\rho) -P\log P$ with 
the transformed free energy functional
\begin{equation}\label{12.60}
\mathcal{F}_\mathrm{kn}(\rho) =  \int _0^{2\pi}\hspace{-6pt}\mathrm{d}w \rho(w) \Big(V(w) + \log \rho(w) - \int_0^{2\pi}\hspace{-6pt}\mathrm{d}w'   
\log|\mathrm{e}^{\mathrm{i}w}-\mathrm{e}^{\mathrm{i}w'}| \rho(w') \Big).
\end{equation} 
$\mathcal{F}_\mathrm{kn}$ has to be minimized under the constraint
\begin{equation}\label{12.61}
\rho(w) \geq 0,\quad \int_0^{2\pi}\hspace{-6pt} \mathrm{d}w\rho(w) =P 
\end{equation}
with minimizer denoted by $\rho^\star$. Then
 \begin{equation}\label{12.62} 
 F_\mathrm{al}(P,V) =  \partial_P \mathcal{F}_\mathrm{kn}(\rho^\star) -1 +\log 2.
 \end{equation}
 
The constraint \eqref{12.61} is removed by introducing the Lagrange multiplier $\mu$ as
  \begin{equation}\label{12.63} 
 \mathcal{F}_\mathrm{kn}^\bullet(\rho) =  \mathcal{F}_\mathrm{kn}(\rho) - \mu \int_0^{2\pi}\hspace{-6pt}\mathrm{d}w \rho(w).
 \end{equation}
The unique minimizer of  $\mathcal{F}_\mathrm{kn}^\bullet(\rho)$ is denoted by  $\rho_{\mathsf{n},\mu}$ and determined as solution of the TBA equation
\begin{equation}\label{12.64} 
  V(w)   - \mu -  2 \int_0^{2\pi} \hspace{-6pt}\mathrm{d}w'  \log|\mathrm{e}^{\mathrm{i}w}-\mathrm{e}^{\mathrm{i}w'}| \rho_{\mathsf{n},\mu}(w') +\log \rho_{\mathsf{n}, \mu}(w) = 0.
 \end{equation}
The parameter $\mu$ has to be adjusted such that
\begin{equation}\label{12.65} 
 P =  \int _0^{2\pi}\hspace{-6pt}\mathrm{d}w  \rho_{\mathsf{n},\mu}(w).
 \end{equation}
To obtain the Ablowitz-Ladik free energy, we differentiate as
 \begin{equation}\label{12.66} 
 \partial_P \mathcal{F}_\mathrm{kn}(\rho^\star) =
   \int _0^{2\pi} \hspace{-6pt}\mathrm{d}w \partial_P\rho^\star(w) \Big(V(w) +  \log \rho^\star(w)     - 2  \int _0^{2\pi}\hspace{-6pt} \mathrm{d}w'  \rho^\star(w') \log|\mathrm{e}^{\mathrm{i}w}-\mathrm{e}^{\mathrm{i}w'}|\Big) +1.
 \end{equation}
 Integrating \eqref{12.64} against  $\partial_P\rho^\star$ one arrives at
 \begin{equation}\label{12.67} 
 \partial_P \mathcal{F}_\mathrm{kn}(\rho_{\mathsf{n},\mu}) = \mu + 1
\end{equation}
 and thus 
 \begin{equation}\label{12.68} 
  F_\mathrm{al}(P,V) =  \mu(P,V)+ \log 2.
 \end{equation}

The Ablowitz-Ladik lattice has the property that its free energy 
is determined through an explicit variational problem. At this stage,  a structure comparable to the Toda lattice has been reached. 
The modifications are minimal. The integration is over the interval $[0,2\pi]$ corresponding to the eigenvalues 
$\mathrm{e}^{\mathrm{i}w}$. The basis functions are $\varsigma_0(w) = 1$, $\varsigma_{n-}(w) = \sin(nw)$, $\varsigma_{n+}(w) = \cos(nw)$ and
the kernel of the operator $T$ is adjusted to 
\begin{equation}\label{12.69}
Tf(w) = 2 \int _{0}^{2\pi}\hspace{-6pt} \mathrm{d}w'  \log|2\sin((w-w')/2)|f(w'),\quad w \in [0,2\pi].
\end{equation}
With such modifications the results in Sections  \ref{sec3.3} and \ref{sec3.4} hold \textit{ad verbatim}.
In particular,  under the GGE in \eqref{12.36} one obtains the $N\to\infty$ limit of the DOS as
\begin{equation}\label{12.70}
\rho_\mathrm{Q} = \partial_P\rho_\mathsf{n}.\bigskip
\end{equation}

We arrived at a somewhat perplexing conclusion. Clearly the structure discussed very strongly resembles the one obtained for the Toda lattice. Our scheme is particle-based. 
Nevertheless, so far we did not identify an underlying particle structure. Their potential two-particle scattering shift arrives only indirectly through the generalized free energy. For the currents we will use the availability of a conserved currents and thus bypass the collision rate ansatz. 
 \subsection{Average currents}
\label{sec12.4}
As a fairly general fact, the average currents are related to fluctuations in the density of states. The same strategy  
will be pursued for the Ablowitz-Ladik lattice.

For the infinite lattice, the conserved fields satisfy a continuity equation of the form
\begin{equation}\label{12.71} 
\frac{\mathrm{d}}{\mathrm{d}t} Q_j^{[n,\sigma],N} = \{Q_j^{[n,\sigma],N},H_{\mathrm{al},N}\}_\mathrm{AL} = J_{j}^{[n,\sigma],N} - J_{j+1}^{[n,\sigma],N}.
\end{equation}
It will be convenient to work with the complex currents
\begin{equation}\label{12.72} 
J_{j}^{[n],N} =
J_{j}^{[n,+],N} + \mathrm{i} J_{j}^{[n,-],N}.
\end{equation}
Using Eq. \eqref{12.22} together with the fact that the matrix $C_+$ has nonvanishing matrix elements only for
$(j,j+\ell)$ with $\ell = 0,1,2$, the time derivative is obtained to
\begin{eqnarray} \label{12.73}
&&\hspace{-40pt}\{(C^n)_{j,j},H_\mathrm{al}\}_\mathrm{AL} \nonumber\\
&&\hspace{-30pt} = \sum_{\ell = 1,2}\mathrm{i}\Big(C_{j-\ell,j}(C^n)_{j,j-\ell}  -C_{j,j+\ell}(C^n)_{j+\ell,j} + \bar{C}_{j,j+\ell}(C^n)_{j,j+\ell}  - \bar{C}_{j-\ell,j}(C^n)_{j-\ell,j}\Big).  
\end{eqnarray}
The term with $\ell = 0$ does not contribute.
The remaining terms look like a shift difference, but it is not, since the off-diagonal matrix elements are only two-periodic.
 
 For $n=1$, one finds
\begin{equation} \label{12.74} 
\{C_{j,j},\mathrm{tr}(C)\}_\mathrm{AL} = \mathrm{i}\big(- \rho_{j-1}^2 \alpha_{j-2}\bar{\alpha}_{j} + \rho_{j}^2 \alpha_{j-1}\bar{\alpha}_{j+1} \big),
\end{equation}
while 
\begin{equation} \label{12.75} 
\{C_{j,j},\mathrm{tr}(C^*)\}_\mathrm{AL} = \mathrm{i}\big(\rho_{j}^2 \alpha_{j+1}\bar{\alpha}_{j+1} - \rho_{j-1}^2 \alpha_{j-2}\bar{\alpha}_{j-2} + \rho_{j}^2\rho_{j+1}^2 - 
 \rho_{j-2}^2 \rho_{j-1}^2\big)=  \mathrm{i}\big( \alpha_{j-1}\bar{\alpha}_{j-1} - \alpha_{j}\bar{\alpha}_{j}\big)
\end{equation}
for even $j$ and
\begin{equation} \label{12.76} 
\{C_{j,j},\mathrm{tr}(C^*)\}_\mathrm{AL} = \mathrm{i}\big(\rho_{j}^2 \alpha_{j-1}\bar{\alpha}_{j-1} - \rho_{j-1}^2 \alpha_{j}\bar{\alpha}_{j}  \big)=  \mathrm{i}\big( \alpha_{j-1}\bar{\alpha}_{j-1} - \alpha_{j}\bar{\alpha}_{j}\big)
\end{equation}
for odd $j$. Adding the two terms, the first current reads
\begin{equation} \label{12.77} 
J_j^{[1]} =   \mathrm{i}\big( - \rho_{j-1}^2 \alpha_{j-2}\bar{\alpha}_{j} +\alpha_{j-1}\bar{\alpha}_{j-1}\big).
\end{equation}

For higher $n$, it seems to be difficult to guess the cancellations leading to an explicit difference form. But for the moment we only need the property of currents being local, which can be ensured by an abstract argument.
\bigskip\\
$\blackdiamond\hspace{-1pt}\blackdiamond$ \textit{Existence of currents}. We want to establish that  there is a local current, $J^{[n]}_{j}$, such that
\begin{equation} \label{12.78}
\{(C^n)_{j,j},H_\mathrm{al}\}_\mathrm{AL} =J^{[n]}_{j} -  J^{[n]}_{j+1}.
\end{equation}

For this purpose, we consider a ring of size $N$ and fix $n$ such that $N > 4n$. For clarity the index $N$ is suppressed in our notation. $\{Q^{[n]}_0,H_\mathrm{al}\}_\mathrm{AL}$ is a polynomial of degree $2n+2$. This polynomial is decomposed into patterns. A pattern  consists of a specific monomial together with a collection of some of its translates. By construction patterns are distinct. The monomial defining a pattern is denoted by $\omega_1$ and its translates by $\omega_j$.  Then,
understood modulo $N$, $\{Q^{[n]}_0,H_\mathrm{al}\}_\mathrm{AL}$ is a sum of terms of the form
\begin{equation} \label{12.79}
\sum_{|\ell| \leq 2n +2}a^{(\ell)} \omega_{\ell}
\end{equation}
with some complex coefficients $a^{(\ell)}$, which may vanish. Since  $\{Q^{[n]},H_\mathrm{al}\}_\mathrm{AL} = 0$, shift invariance implies that
for every pattern
\begin{equation} \label{12.80}
0= \sum_{j=0}^{N-1}\sum_{|\ell| \leq 2n+2}a^{(\ell)} \omega_{\ell +j}= \sum_{|\ell| \leq 2n+2}a^{(\ell)} \Big(\sum_{j=0}^{N-1}\omega_{\ell +j}\Big).
\end{equation}
and as a consequence 
\begin{equation} \label{12.81}
\sum_{|\ell| \leq 2n+2}a^{(\ell)} = 0.
\end{equation}
Eliminating the nonzero $a^{(\ell)}$ with largest index, one obtains unique coefficients 
$b^{(\ell)}$ such that 
 \begin{equation} \label{12.82}
 \sum_{|\ell| \leq 2n+2} b^{(\ell)}(\omega_\ell - \omega_{\ell+1})= J^\omega_0 - J^\omega_{1}.
 \end{equation}
For the actual local current one still has to sum over all patterns.

The total fields $Q^{[n],N}$ are unique,
 but the one-shift covariant local densities  $Q^{[n],N}_j$ are not. Once the density is chosen,  
 the corresponding one-shift covariant current density is determined up to a constant.
 \hfill$\blackdiamond\hspace{-1pt}\blackdiamond$\\

To compute the GGE averaged currents, one  notes that 
\begin{equation}\label{12.82a}
J^{0}_j = 2 Q_j^{[1,-]}. 
\end{equation}
Hence  the path outlined in Section \ref{sec6.1} can be followed. Omitting all intermediate steps, the result is 
 \begin{equation}\label{12.83}
\partial_P\big(\langle J^{[n,\sigma]}_0 \rangle_{P,V} + 2P\langle \varsigma_{1-}, C^\sharp \varsigma_{n\sigma}\rangle\big) = 0,
\end{equation}
implying that the round bracket has to be independent of $P$, in particular equal to its value at $P=0$.  Since $\langle \varsigma_{1-}, C^\sharp \varsigma_{n\sigma}\rangle$ is bounded in $P$, the second summand vanishes at $P=0$. For the first summand, one notes that 
in the limit $P \to 0$, for each $j$ the a priori measure \eqref{12.34} becomes uniform  on the unit circle. Hence,  $\langle \rho_j^2\rangle_P \to 0$  as $P\to 0$ and the CMV matrix turns diagonal. Denoting 
$\alpha_j = \mathrm{e}^{\mathrm{i}\phi_j}$, $\phi_j \in [0,2\pi]$, in the limit $P \to 0$ the GGE (12.36) converges  to
\begin{equation} \label{12.84}
\frac{1}{Z_{\mathrm{al},N}(0,V)}\prod_{j=0}^{N-1}\mathrm{d}\phi_j \exp\Big( - \sum_{j=0}^{N-1} V(\phi_{j+1} - \phi_{j})\Big)
\end{equation}
with boundary conditions $\phi_N = \phi_0$. According to \eqref{12.77}, one observes that $\langle J_0^{[1]}  \rangle_{0,V} =\mathrm{i}$. By a somewhat lengthy computation  $\langle J_0^{[2]}  \rangle_{0,V} = 0$. To extend the average to general $n$ seems to be difficult, since a sufficiently explicit formula for $J_0^{[n]}$ is missing.
Let us then assume that $\langle J_0^{[n]}  \rangle_{0,V} = d_n$ with some constant $d_n$ independent of $V$.
  Next we substitute $P\varrho^* = \rho_\mathsf{n}$ with the result
\begin{equation}\label{12.85}
P\langle \varsigma_{1-}, C^\sharp \varsigma_{n\sigma}\rangle    = \langle \varsigma_{1-}, (1 -  \rho_\mathsf{n} T)^{-1}\rho_\mathsf{n} \varsigma_{n\sigma}\rangle
    -    \nu \langle \varsigma_{1-}, (1 -  \rho_\mathsf{n} T)^{-1} \rho_\mathsf{n}\rangle  \langle \varsigma_{n\sigma}, (1 -  \rho_\mathsf{n} T)^{-1} \rho_\mathsf{n}\rangle.
\end{equation}
Noting that 
$(1- \rho_\mathsf{n} T)^{-1} \rho_\mathsf{n}$ is a symmetric operator, one  finally arrives at
\begin{equation}\label{12.86}
 \langle J^{[n,\sigma]}_0 \rangle_{P,V} -d_n =  -2 \big(\langle \rho_\mathsf{n} \varsigma_{1-}^\mathrm{dr}   \varsigma_{n\sigma}\rangle -  q_{1-}\langle \rho_\mathsf{p}\varsigma_{n\sigma}\rangle\big) 
 \end{equation}
 with
 \begin{equation}\label{12.86a} 
 q_{1-} = \langle Q^{1,-}_0\rangle_{P,V}.
\end{equation}
The spectral function for the energy is $E(w) = 2\cos w$, hence  $E'(w) = - 2\sin w$. To emphasize the 
correspondence with the Toda lattice, see Eqs. \eqref{6.20} and \eqref{6.17}, we define the effective velocity through
\begin{equation}\label{12.87} 
v^\mathrm{eff} = \frac{(-2\varsigma_{1-})^\mathrm{dr}}{\varsigma_0^\mathrm{dr}},\qquad \tilde{q}_{1-}  = -2q_{1-}. 
 \end{equation}
 Then 
\begin{equation}\label{12.88} 
\langle J^{[0]}_0 \rangle_{P,V} = -\tilde{q}_{1-},\qquad \langle J^{[n,\sigma]}_0 \rangle_{P,V} - d_n  = \langle(v^\mathrm{eff} - \tilde{q}_{1-}) \rho_\mathsf{p}\varsigma_{n\sigma}\rangle.
\end{equation}
\subsection{Hydrodynamic equations} 
\label{sec12.4a}
On the hydrodynamic scale the local GGE is characterized by the log intensity $\nu$ and the CMV density of states $\nu\rho_\mathsf{p}$, both of which now become spacetime dependent. Merely inserting the average currents, 
and since $d_n$ is assumed to be a constant, one arrives at the Euler type hydrodynamic evolution equations,
\begin{eqnarray}\label{12.89} 
&&\partial_t \nu(x,t) + 2\partial_x q_{1-}(x,t) = 0,\nonumber\\[0.5ex]
&& \partial_t\big(\nu(x,t) \rho_\mathsf{p}(x,t;v)\big) + \partial_x\big(2(v^\mathrm{eff}(x,t;v) + q_{1-}(x,t))\rho_\mathsf{p}(x,t;v)\big) = 0.
\end{eqnarray}
Here we followed a standard convention, which amounts to the free dispersion relation $E(w) = 2\cos w$. Upon adopting 
$E(w) = \cos w$ the extra factors of 2 in \eqref{12.89} would be removed. 
Equation \eqref{12.89} is based on the assumption of local GGE. To actually establish such an equation from the underlying Ablowitz-Ladik lattice
seems to be a difficult task.
 
 As a general feature of generalized hydrodynamics,  the equations can be  transformed explicitly to a quasilinear 
form through
 \begin{equation}\label{12.90} 
\rho_\mathsf{n} =  \rho_\mathsf{p}(1+ (T\rho_\mathsf{p}))^{-1},
\end{equation}
compare with \eqref{6.23}. Then
Eq. \eqref{12.89} turns into the normal form
\begin{equation}\label{12.91} 
 \partial_t \rho_\mathsf{n} + 2\nu^{-1}(v^\mathrm{eff} + q_{1-})\partial_x \rho_\mathsf{n} = 0.
\end{equation}
Since $\nu >0$, no apparent singularities are encountered.

For the defocusing discrete NLS in one dimension, we established the form of the hydrodynamic equations. As a novel feature, their structure is determined 
by the mean-field version of the log gas corresponding to CUE random matrices. Thus formally expanding the circle to a line, one  recovers the 
hydrodynamic equations of the Toda lattice.  Our analysis is pretty much on the same level as the one for the Toda lattice. Only the handling of average currents in the limit $P\to 0$ is incomplete. We hope to return to this point in the future.
 \subsection{Modified Korteweg-de Vries equation}
\label{sec12.5}
Instead of the real part of $Q^{[1],N}$ one can also consider its imaginary part as hamiltonian, 
\begin{equation}
\label{12.92}
H_{\mathrm{kv},N} = - \mathrm{i}\sum_{j=0}^{N-1} \big( \alpha_{j-1} \bar{\alpha}_{j} - \bar{\alpha}_{j-1} \alpha_{j}\big) = - \mathrm{i} \,\mathrm{tr}\big[C_N - C_N^*\big].
\end{equation}
Then
\begin{equation} \label{12.93}
\frac{d}{dt}\alpha_j = \{\alpha_j,H_{\mathrm{kv},N}\}_\mathrm{AL}   =  \rho_j^2 (\alpha_{j+1} - \alpha_{j-1}) ,
\end{equation} 
which is known as Schur  flow.
Through a formal Taylor expansion, one argues that the continuum limit of Eq. \eqref{12.93} yields the modified 
Korteweg-de Vries equation,
\begin{equation} \label{12.94}
\partial_t u= \partial_x^3u - 6 u^2\partial_x u,
\end{equation} 
which is a good reason to briefly touch upon \eqref{12.93}.

As before $\alpha_j  \in \mathbb{D}$ and the conservation laws remain  unchanged. However the currents have to be modified from $J^{[n]}$ to $\breve{J}^{[n]}$.
For the log intensity current one obtains
\begin{equation} \label{12.95}
\breve{J}^{[0],N}_j =  2Q^{[1,+],N}_j= H_{\mathrm{kv},N,j}.
\end{equation} 
Thus the role of $1-$ and $1+$ are interchanged.
The arguments in Section \ref{sec12.4} can be repeated without modifications. In the hydrodynamic equations \eqref {12.88}, $q_{1-}$ is replaced by $q_{1+}$
and the effective velocity becomes 
 \begin{equation}\label{12.96} 
v^\mathrm{eff} = \frac{\varsigma_{1+}^\mathrm{dr}}{\varsigma_0^\mathrm{dr}}\,. 
 \end{equation}

These considerations badly miss that the wave field of the modified Korteweg-de Vries equation cabn be taken as real-valued. If the initial $\alpha_j$ are real, then according to  the evolution equation (12.97) this property is preserved in time. 
From the perspective of GGE, such initial conditions  amount 
to a set of measure zero and one has to reconsider the analysis. Fortunately the relevant transformation formula has been proved already by Killip and Nenciu. 
To avoid duplication of symbols, in the remainder of this section  $\bar{\alpha}_j = \alpha_j$ everywhere and hence  $\alpha_j \in [-1,1]$. The equations of motion read
\begin{equation} \label{12.97}
\frac{d}{dt}\alpha_j = \rho_j^2 (\alpha_{j+1} - \alpha_{j-1}), 
\end{equation}
where $\rho_j^2 = 1 -\alpha_j^2$ as before. While an obvious hamiltonian structure is lost, one readily checks that the a priori measure
\begin{equation} \label{12.98}
\prod_{j=0}^{N-1}\mathrm{d}\alpha_j (\rho_j^2)^{P-1}= \prod_{j=0}^{N-1}\mathrm{d}\alpha_j (\rho_j^2)^{-1}\exp\big(-PQ^{[0],N}\big) 
\end{equation}
is still stationary under the dynamics.  The densities of the conserved fields are 
\begin{equation} \label{12.99}
Q_j^{[m]} = (C^m)_{j,j},
\end{equation} 
 $m = 1,2,\ldots\,$, as before. But the imaginary part is dropped and there is no longer a distinction of $\pm$. In particular, 
\begin{equation} \label{12.100}
Q^{[0]}_j =  -\log\rho_j^2,\quad Q^{[1]}_j = -\alpha_{j-1}\alpha_j,\quad Q^{[2]}_j = \alpha_{j-1}^2\alpha_{j}^2
- \rho_{j-1}^2\alpha_{j-2}\alpha_{j} - \rho_{j}^2\alpha_{j-1}\alpha_{j+1}.
\end{equation}
 For the log intensity current, $\breve{J}^{[0]}_j = 2Q^{[1]}_j$. Thus  $\breve{J}^{[0]}$ is conserved and the Toda strategy is still applicable.

Since $C_N$ is now  a real matrix, its eigenvalues come in pairs. If $\mathrm{e}^{\mathrm{i}\vartheta}$ is an eigenvalue,  so is $\mathrm{e}^{-\mathrm{i}\vartheta}$.
For a system of size $N$, there are only $\bar{N} =N/2$ independent eigenvalues, denoted by $\mathrm{e}^{\mathrm{i}\vartheta_1},\ldots, \mathrm{e}^{\mathrm{i}\vartheta_{\!\bar{N}}}$. Rather than using a DOS reflecting such symmetry, it is more convenient to restrict the phase as $0 \leq \vartheta _j\leq \pi$. 
The empirical DOS is given by  
\begin{equation} \label{12.101}
\rho_{\mathrm{Q},\bar{N}}(w)    = \frac{1}{\bar{N}} \sum_{j=1}^{\bar{N}} \delta (w - \vartheta_j), \qquad w \in [0,\pi], 
\end{equation}
where we switched to $\bar{N}$ as size parameter. In the limit $\bar{N} \to \infty$, $\rho_{\mathrm{Q},\bar{N}}(w)$ converges to the deterministic limit $\rho_{\mathrm{Q}}(w)$.
The GGE expectations are then
\begin{equation} \label{12.102}
\langle Q_0^{[n]}\rangle_{P,V} = 2\int _{0}^{\pi}\hspace{-6pt}\mathrm{d}w\rho_{\mathrm{Q}}(w) \varsigma_n(w), \quad  \cos(nw) = \varsigma_n(w),
\end{equation}
and the confining potential becomes
\begin{equation} \label{12.103}
V(w) = \sum_{n=1}^\infty \mu_n \cos(nw) 
 \end{equation}
with real chemical potentials $\mu_n$. The confining potential is assumed to be continuous.

Under the measure 
\begin{equation} \label{12.104}
\prod_{j=0}^{N-2}(1-\alpha_j^2)^{-1}(1-\alpha_j^2)^{\beta(N-j-1)/4}(1-\alpha_j)^{a+1-(\beta/4)}(1 + (-1)^j \alpha_j)^{b +1 - (\beta/4)} \mathrm{d} \alpha_j
\end{equation}
on $[-1,1]^{N-1}$ with $\beta >0$ and $a,b > -1 +(\beta/4)$, the joint (not normalized) distribution of the eigenvalues of $C_N$, imposing $\alpha_{N-1} = 1$, is given by  
\begin{equation} \label{12.105}
\tilde{\zeta}_{\bar{N}}(\beta)2^{\kappa} |\Delta(2 \cos \vartheta_1,\ldots,2 \cos \vartheta_{\bar{N}})|^\beta \prod_{j=1}^{\bar{N}}\big(1 -  \cos \vartheta_j\big)^a\big(1 + \cos \vartheta_j)^b \sin\vartheta_j\mathrm{d}\vartheta_j
\end{equation}
on $[0,\pi]^{\bar{N}}$. 
The proportionality factor $\tilde{\zeta}_{\bar{N}}$ 
is defined in \eqref{12.47} and $\kappa = (\bar{N}-1)(-\tfrac{1}{2}\beta +a +b +2)$.  
To achieve a  pressure ramp of slope $-P/2\bar{N}$, one has to set 
\begin{equation} \label{12.106}
\beta = \frac{2P}{\bar{N}},\qquad a = b = -1 +\tfrac{1}{4}\beta.
\end{equation}
Then $\kappa = 0$ and under the measure 
\begin{equation} \label{12.107}
\prod_{j=0}^{N-2}(1-\alpha_j^2)^{-1}(1-\alpha_j^2)^{P(N-j-1)/N} \mathrm{d} \alpha_j
\end{equation}
the eigenvalues of $C_N$ have the distribution
\begin{equation} \label{12.108}
\zeta_{\bar{N}}(P) |\Delta(2 \cos \vartheta_1,\ldots,2 \cos \vartheta_n)|^{2P/\bar{N}}
 \prod_{j=1}^{\bar{N}}(\sin\vartheta_j)^{((P/\bar{N}) -1)}\mathrm{d}\vartheta_j.
\end{equation}

Now the strategy of Section \ref{sec12.3} is in force. In  \eqref{12.108} we add a confining potential. The power of $\sin\vartheta_j$ converges to $-1$ and in the limit $\bar{N} \to \infty$  the free energy functional becomes
\begin{eqnarray} \label{12.109}
&&\hspace{-86pt}\mathcal{F}_{\mathrm{kv}\triangleright}^\circ(\varrho) =  \int _{0}^{\pi}\hspace{-6pt}\mathrm{d}w \varrho(w) \Big(V(w) +   
\log\sin w  + \log \varrho(w)\nonumber\\
&&\hspace{0pt}    - P\int_{0}^{\pi}\hspace{-6pt}\mathrm{d}w'  \varrho(w') 
\log(2|\cos w  - \cos w'|) \Big).
\end{eqnarray} 
$\mathcal{F}_{\mathrm{kv}\triangleright}^\circ$ has to be minimized over all $\varrho \geq 0$ with $\int_0^\pi\mathrm{d}w \varrho(w)=1$ and boundary condition 
$\rho(-1) = \rho(1)$.
Actually our mean-field limit is somewhat singular, since at $a=b=-1+ (P/N)$ the a priori measure is barely integrable.
The quadratic energy term is repulsive, but the linear $\log\sin$ term pushes the eigenvalues towards the two end points. 
It is not so obvious, whether and how the two terms balance. 
Fortunately, the particular case $V = 0$ has been studied in detail, thereby confirming \eqref{12.109} and a normalizable DOS.

Surprisingly, deviating from the conventional script, the confining potential $V$ is corrected by the $\log \sin$ potential, which is attractive and  favors accumulation of eigenvalues at $0,\pi$. The prior computation of the average currents is carried out for fixed $V$,
which has now to be replaced by $\tilde{V}(w) = V(w) + \log(\sin w)$. Of course, $\rho_\mathsf{n}$ and $\rho_\mathsf{p}$ depend on $\tilde{V}$.
Also the operator $T$ has to be  adjusted to
\begin{equation}\label{12.110}
Tf(w) = 2 \int _0^{\pi}\hspace{-6pt} \mathrm{d}w'  \log|2(\cos w - \cos w')|f(w') 
,\quad w \in [0,\pi].
\end{equation}
With these modifications, the TBA formalism  can be taken over with no further changes. In Eq. \eqref{6.18}
the bare velocity $v$ is replaced by $2\cos v$ and the kernel $2\log|v-w|$ is substituted by $2\log|2(\cos v - \cos w)|$.
Then the effective velocity is given by
\begin{equation}\label{12.111}
v^\mathrm{eff} = \frac{\varsigma_{1}^\mathrm{dr}}{\varsigma_0^\mathrm{dr}},
\end{equation}
upon adopting the dressing operator $T$ from \eqref{12.110} and more implicitly $\tilde{V}$. The hydrodynamic equations are argued along the standard route.
The subtraction term reads $q_1 
= 2\nu \int _0^{\pi}\hspace{-2pt} \mathrm{d}w \rho_\mathsf{p}(w) \cos w$.
 
The discrete modified KdV equation is an integrable wave equation with no obvious hamiltonian structure. Compared to the Ablowitz-Ladik lattice, CUE is replaced by COE. Interestingly enough, in the standard coordinates, an additional confining potential is generated. Such a feature has not yet been encountered before.\bigskip
\begin{center}
 \textbf{Notes and references}
 \end{center}
 \begin{center}
\textbf{Section 12.1}
\end{center} 
 A most useful source of information are the monographs  by Ablowitz  and Segur \cite{AS81} and by Ablowitz et al.  \cite{ABT04}. In the book by Gr\'{e}bert and Kappeler \cite{GK14} the integrable structure of continuum NLS is discussed in great detail. Equilibrium measures were first studied by Lebowitz et al. \cite{LRS88}, see also the more recent contributions by
Oh and Quastel  \cite{OQ13} and Fr\"{o}hlich et al. \cite{FKSS17}.  More general nonlinearities are covered in Bourgain  \cite{B94}. GGEs with $V$ a finite polynomial are
studied by Zhidkov  \cite{Z01}. The  discretization discussed in this section is based on 
 Ablowitz and Ladik  \cite{AL75,AL76}.
The GHD of the continuum sinh-Gordon equation has been developed by Bastianello et al.  \cite{BDWY18}, see also De Luca and  Mussardo \cite{DM16}. The integrable lattice version is reported by Orfanidis  \cite{O78}. The integrable structure of continuum spin chain has been studied first by Takhtajan \cite{T77} and Sklyanin  \cite{S79}. Its integrable discretization is covered by Faddeev and Takhtajan   \cite{FT07}, see also the more recent  contribution by Krajnik et al.  \cite{KIPP22}. Numerical simulations of the spin chain are reported by  Das et al.   \cite{DKSD19}. The article  by Spohn and  Lebowitz \cite{LS77} is an early investigation of what is now called GGE for wave equations. 
\newpage
\begin{center}
\textbf{Section 12.2}
\end{center} 
Nenciu \cite{N05,N06}   discovered the role played by the Cantero-Moral-Vel\'{a}zquez matrices, which were originally introduced with  distinct goals, see
 Cantero et al.  \cite{CMV03,CMV05}. Further accounts are Killip and Nenciu \cite{KN07} and
 Simon \cite{S07}.\bigskip
\begin{center}
\textbf{Section 12.3}
\end{center} 
 The change of volume elements for the linear pressure ramp has been accomplished by Killip and Nenciu \cite{KN04} with
  motivation from the corresponding result by Dumitriu and Edelman \cite{DE02} for the Toda lattice. The circular log-gas is summarized in Forrester \cite{F10}
  and its mean-field version is studied by Trinh and Trinh \cite{TT20} and  Hardy and  Lambert  \cite{HL21}. The  results reported in the text are based on Spohn \cite{S21}. Independently,
  similar results are worked out by Grava and Mazzuca  \cite{GM21}. In particular, for polynomial confining potentials the free energy is proved to be the solution of the variational problem in \eqref{12.57}. Mazzuca and  Memin  \cite{MM22} extend such results to all continuous confining potentials $V$,  conceptually is an important step. More generally one would like to prove the following property of GHD. One starts from an initial profile $x \mapsto \nu\rho_\mathsf{p}(x)$, such that every $\nu\rho_\mathsf{p}(x)$ determines a unique GGE. To be shown is  propagation of such a property  in time. 
  Numerical simulations of the Ablowitz-Ladik chain are reported  by Mendl and Spohn \cite{MS15} with results to some extent resembling  Fig. 8. A comparison with GHD is still missing. 
Based on Killip and Nenciu \cite{KN07} the $N$-particle scattering is elucidated by Brollo and Spohn
\cite{BS23}.
\bigskip
\begin{center}
\textbf{Section 12.5}
\end{center} 
 From a general perspective the expressions \eqref{12.87} and \eqref{12.111} are somewhat puzzling.
 The numerator is the dressing of the derivative of energy, i.e. $-\sin w$ for  \eqref{12.87} and  $\cos w$ for \eqref{12.111}. However, according to the rules,
 the denominator should be the derivative of momentum, thus respectively $\cos w$ and  $-\sin w$, while we obtained the dressing of the constant function.  \bigskip
\begin{center}
\textbf{Section 12.6}
\end{center} 
 The notion Schur flow for the dynamics generated by \eqref{12.93} goes back to Golinski\v{i} \cite{G06}.  The transformation of measure
  is proved by Killip and Nenciu  \cite{KN04}.  In Trinh and Trinh  \cite{TT20} and Forrester and Mazzuca \cite{FM21} explicit formulas are obtained for the thermal case with $a,b >-1$.  With the methods  Trinh and Trinh  \cite{TT20} also the diagonal limit, $a=b=-1+ (P/N)$, is proved to yield the stated result. Surprisingly, the two-fold degeneracy of the spectrum of the CMV matrix results in 
asymptotically two-particle bound states, see Brollo and Spohn \cite{BS23}.
\newpage
\section{Hydrodynamics for  the Lieb-Liniger $\delta$-Bose gas}
\label{sec13}
\setcounter{equation}{0} Much of the interest related to GHD is rooted in the study of integrable \textit{quantum} many-body systems. Over the past decade the number of contributions staggered and it would be difficult to present a balanced view. Reviews and journal special volumes are available. On the other hand, without mentioning quantum models the GHD story would be utterly incomplete. 
Our compromise consists of two elements. The first acquisition is the quantized version of the Toda lattice, the topic of Chapter \ref{sec14}. 
Due to missing physical realizations this model received less attention. Still the model is instructive because a close theoretical comparison between classical and quantum is feasible. 

The by far most widely studied integrable model is the quantized nonlinear Schr\"{o}dinger equation,
better known as the $\delta$-Bose gas because the interaction between bosons is point-like.
 In their pioneering work from 1963 Elliott Lieb and Werner Liniger established that eigenvalues and  eigenfunctions of the hamiltonian can be obtained through the Bethe ansatz. This  discovery triggered the field of integrable quantum many-body  systems. For GHD  
the Lieb-Liniger model also plays a very special role. It serves as paradigm for the study of more difficult models as the XXZ quantum spin chain and the spin-$\tfrac{1}{2}$ Fermi-Hubbard model.  Even more importantly, Rubidium atoms enclosed in a narrow tube are well  modelled by bosons
interacting through an extremely short range potential. Experimental realizations are available through the amazing advances in cold atom physics, see the Insert in Section \ref{13.3}.  Thus as a second acquisition we discuss aspects of GHD for the $\delta$-Bose gas. 
In particular, we will emphasize the structural similarity with the DOS of a classical Lax matrix.

For all quantum mechanical models the hydrodynamic scale is particle-based. The most convincing evidence is the two-particle scattering shift, which is computed from the scattering of two quantum particles. Mathematically this is very different from a two-soliton solution. The quantum particles are described by a wave equation in $\mathbb{R}^2$ governing the motion of the two-particle wave function $\psi(x_1,x_2,t)$, while a two-soliton wave field refers to a wave equation over $\mathbb{R}$.  The construction of soliton-based hydrodynamics for quantum models remains as a task for the future.
\subsection{Bethe ansatz}
\label{sec13.1}
To quantize the continuum nonlinear Schr\"{o}dinger equation of the introduction to Chapter \ref{sec12}, one starts  from a scalar bosonic field $\Psi(x)$, $x \in \mathbb{R}$, with canonical commutation
relations
\begin{equation}\label{13.1} 
[\Psi(x),\Psi(x')^*] = \delta(x-x'), \quad [\Psi(x),\Psi(x')] = 0, \quad [\Psi(x)^*,\Psi(x')^*] = 0.
\end{equation}
In normal ordered form the Lieb-Liniger hamiltonian reads then 
\begin{equation}\label{13.2} 
H_\mathrm{li} = \tfrac{1}{2}\int_\mathbb{R} \mathrm{d}x \big(\partial_x \Psi(x)^*\partial_x \Psi(x) + c \Psi(x)^*\Psi(x)^*\Psi(x)\Psi(x)\big).
\end{equation}
Here $c$ is the coupling constant with $c \geq 0$ imposed.  The attractive case, $c <0$, is considerably more complicated
because of bound states. In the widely accepted standard convention  the factor $\tfrac{1}{2}$ in front of the integral is omitted. We inserted here so have an  energy-momentum relation of the free theory identical to the one of the noninteracting Toda fluid. 
The Lieb-Liniger hamiltonian is formal because of point-interactions. Fortunately, there is  the first quantized version,
which in the $N$-particle sector reads
\begin{equation}\label{13.3}
H_{\mathrm{li},N} = - \sum_{j = 1}^N  \tfrac{1}{2} (\partial_{x_j})^2 + c\sum_{1 \leq i <  j\leq N}\delta(x_i -  x_j).
 \end{equation}
Bosonic means that the operator acts on wave functions symmetric under particle exchange.
The case of interest is a spatial interval $[0,\ell]$, $\ell> 0$, with periodic boundary conditions, which will be distinguished by either 
a sub\,- or superscript $\ell$. Of course, one could also consider a finite number of particles moving on the real line, which would be the natural set-up  for multi-particle scattering.  

Let us first recall, how to handle the delta potential.  The simplest example is $N =2$, for which the relative motion, $y = x_2 - x_1$, is governed by  the hamiltonian 
\begin{equation}\label{13.4} 
H_\mathrm{rel} = - \partial_y^2 + c\,\delta(y).
 \end{equation}
Away from $y=0$, $H_\mathrm{rel} \psi(y) = - \partial_y^2\psi(y)$
with $\psi \in \mathcal{C}^2(\mathbb{R}\setminus\{0\})$, the twice continuously differentiable functions on 
$\mathbb{R}\setminus\{0\}$ with bounded derivatives. The $\delta$-potential translates to the boundary condition
\begin{equation}\label{13.5}
 \partial_y\psi(0_+) -  \partial_y\psi(0_-) = 2c \psi(0). 
\end{equation}
The same mechanism works for the $N$-particle hamiltonian $H_{\mathrm{li},\ell,N}$
acting on symmetric wave functions.
For them one can restrict the construction to the Weyl chamber $\mathbb{W}_{N,\ell} = \{0 \leq x_1 \leq \ldots \leq x_N \leq \ell\}$. By permutation symmetry the hamiltonian can then be extended to all other sectors. Away from the boundary the hamiltonian is the $N$-dimensional Laplacian,
\begin{equation}\label{13.6} 
H_{\mathrm{li},\ell,N} = - \sum_{j = 1}^N  \tfrac{1}{2}(\partial_{x_j})^2,
\end{equation}
which acts on $\psi \in \mathcal{C}^2(\mathbb{W}_{N,\ell}\setminus\partial \mathbb{W}_{N,\ell})$. Periodic boundary conditions mean
\begin{equation}\label{13.7}
\psi(0,x^\bot) = \psi(x^\bot, \ell),\quad \partial_{x_1}\psi(0,x^\bot) = \partial_{x_N}\psi(x^\bot,\ell),
\end{equation}
while the interaction is encoded by the boundary condition
\begin{equation}\label{13.8} 
(\partial_{x_{j+1}} - \partial_{x_{j}})\psi(x) = c \psi(x)\big|_{x_j = x_{j+1}},\quad j = 1,\ldots,N-1,
\end{equation}
where the limit is taken from the interior of $\mathbb{W}_{N,\ell}$. It can be shown that the Laplacian with these boundary conditions defines a unique self-adjoint operator on $L^2(\mathbb{W}_{N,\ell} )$. 

On  $\mathbb{W}_{\ell,N}$ an eigenfunction of $H_{\mathrm{li},\ell,N}$ has the Bethe ansatz form
\begin{equation}\label{13.9} 
\psi(x_1,\ldots,x_N) = \sum_{\sigma \in \mathcal{S}_N} \Big[\exp\Big(\mathrm{i} \sum_{j=1}^N k_{\sigma(j)}x_j\Big) 
\prod_{\substack{i< j\\ \sigma(i) >\sigma(j)}} A(k_i, k_j)  \Big],
\end{equation}
where the sum is over all permutations $\sigma$ of $(1,\ldots,N)$. The $k_j$'s are called rapidities, so to distinguish
from the momenta of noninteracting particles.
$\psi$ has to satisfy the boundary conditions \eqref{13.8}, which leads to
\begin{equation}\label{13.10} 
A(k_1, k_2) = \frac{\mathrm{i}(k_1 - k_2) - c}{\mathrm{i}(k_1 - k_2) + 
c}  = \mathrm{e}^{-\mathrm{i} \theta_\mathrm{li}(k_1 - k_2)}.
\end{equation}
Here  the two-particle phase shift equals
\begin{equation}\label{13.11}
 \theta_\mathrm{li}(w) = 2\arctan (w/c),
 \end{equation}
implying the scattering  shift 
\begin{equation}\label{13.12}
\theta_\mathrm{li}'(w) = \frac{2c}{w^2 + c^2} = \phi_\mathrm{li}(w). 
\end{equation}
Thus when two bosons scatter, their distance is reduced in comparison to the free particle motion.
Equivalently one can think of a delay time. When two bosons scatter, they effectively stick for a short time span which depends on the incoming momenta. 

Imposing periodic boundary conditions as in  \eqref{13.7},  the rapidities have to satisfy
\begin{equation}\label{13.13}
\mathrm{e}^{\mathrm{i} k_j\ell} = (-1)^{N-1} \prod_{i = 1}^{N}\mathrm{e}^{-\mathrm{i} \theta_\mathrm{li}(k_j - k_i)}.
\end{equation}
Equivalently $k_1,\ldots,k_N$ solve the \textit{Bethe equations} 
\begin{equation}\label{13.14} 
2\pi I_j = \ell k_j + \sum_{i = 1}^N \theta_\mathrm{li}(k_j - k_i), \quad j = 1,\ldots,N,
\end{equation}
and hence are also referred to as \textit{Bethe roots}. The quantum states are labelled by vectors $I = (I_1,\ldots,I_N)$, constrained as $I_1 < \ldots< I_N$ and with entries from $\mathbb{Z}$ in case of odd and from $\mathbb{Z}+\tfrac{1}{2}$ in case of even $N$. The set of quantum numbers is denoted by $\mathbb{I}_N$. For each such $I \in   \mathbb{I}_N$ 
the Bethe equations have a unique solution denoted by $k(I)= k = (k_1,\ldots,k_N)$ ordered as $k_1 <\ldots<k_N$. 
It is known that these eigenfunctions constitute a complete orthonormal basis in $L^2(\mathbb{W}_{N,\ell})$. 
\bigskip\\
$\blackdiamond\hspace{-1pt}\blackdiamond$~\textit{Volume factors}. If  the quantum numbers above would be redefined
to take values in $2\pi\mathbb{Z}$, resp. $2\pi\big(\mathbb{Z}+\tfrac{1}{2}\big)$, then the factors of $2\pi$ would  disappear from our 
formulas. Thereby an optically even closer similarity to the Toda fluid would be achieved. For the ease of comparison the conventional notation is adopted, however.
\hfill$\blackdiamond\hspace{-1pt}\blackdiamond$\\

Symmetric eigenfunctions on $L^2([0,\ell]^N)$ can be written as 
\begin{equation}\label{13.15} 
\psi_k(x_1,\ldots,x_N)  =  \sum_{\sigma \in \mathcal{S}_N} (-1)^{\mathrm{sgn}(\sigma)} \exp\Big(\mathrm{i} \sum_{j=1}^N k_{\sigma(j)}x_j\Big)
 \prod_{1 \leq i <j \leq N} \Big(k_{\sigma(i)} -k_{\sigma(j)}+ \mathrm{i} c \,\mathrm{sgn}(x_i - x_j)\Big)
 \end{equation}
with $\mathrm{sgn}(\sigma)$ the sign of the permutation $\sigma$. 
$\psi_k$ is not normalized, the normalized state vector being denoted by  $|k\rangle = \langle\psi_k,\psi_k\rangle^{-\frac{1}{2}}  \psi_k$. The normalization constants are known and the wave functions $|k\rangle$ span the bosonic subspace 
$L^2([0,\ell]^N)_\mathrm{sym}$. 
In particular
\begin{equation}\label{13.16} 
H_{\mathrm{li},N} = \sum_{I \in \mathbb{W}_N}\Big(\tfrac{1}{2}\sum_{j=1}^N(k_j)^2\Big) |k\rangle \langle k |
\end{equation}
with $k = k(I)$. Up to the prefactor  $\tfrac{1}{2}$, if the power $2$ is substituted by $0$, one arrives at the number operator. Similarly power $1$ results in the total momentum. This suggests 
to introduce the $n$-th conserved field (= charge) by 
\begin{equation}\label{13.17} 
Q^{[n],N} = \sum_{I \in \mathbb{I}_N}\Big(\sum_{j=1}^N(k_j)^n \Big)|k\rangle \langle k |.
\end{equation}
\subsection{Bethe root densities, free energy, TBA equations}
\label{sec13.2}
The Boltzmann weight involves a linear combination of charges. Therefore, again following the Toda lattice blueprint, we introduce  the confining potential $V$ and set
\begin{equation}\label{13.18} 
Q^{[V],N} = \sum_{I \in \mathbb{I}_N}\Big(\sum_{j=1}^N V(k_j)\Big) |k\rangle \langle k |.
\end{equation}
The precise class of confining potentials 
remains yet to be studied. A natural choice seems to be $V$ continuous and $V(w) \geq c_0 + c_1|w|$ with $c_1 >0$.  
 The unnormalized  GGE density matrix is defined by
\begin{equation}\label{13.19}
\exp\big[- Q^{[V],N}\big]
\end{equation}
as operator on $L^2([0,\ell]^N)_\mathrm{sym}$. Hence the normalizing  partition function is given by
\begin{equation}\label{13.20}
Z_{\mathrm{li},N}(\ell,V) = \mathrm{tr}\big[\mathrm{e}^{- Q^{[V],N}} \big] = \sum_{I \in \mathbb{I}_N} \exp\Big[ - \sum_{j=1}^N V(k_j)\Big], 
\end{equation}
trace over $L^2([0,\ell]^N)_\mathrm{sym}$. Correspondingly the GGE average of some operator $\mathcal{O}$ is defined by
\begin{equation}\label{13.21}
\langle \mathcal{O}\rangle_{\ell,V} =  Z_{\mathrm{li},N}(\ell,V)^{-1}\mathrm{tr}\big[\mathrm{e}^{- Q^{[V],N}} \mathcal{O}\big].
\end{equation}
The Bethe roots depend on $\ell$ and, as before, this dependence is indicated in the partition function.

Our goal is the infinite volume
\begin{equation}\label{13.22}
\ell \to \infty, \qquad N = \nu \ell
\end{equation}
with $\nu >0$ and  $\rho_\mathsf{f} = 1/\nu$ the Bose gas density.
As for previously studied models, the free energy of the $\delta$-Bose gas will be determined through a variational problem. In fact, the free energy functional will be surprisingly similar to the one of the Toda and Calogero fluid, compare with Eqs. \eqref{9.47} and \eqref{11.57}. Hence it is of interest to have a closer look at the derivation. 
The second item are GGE averaged charges $\ell^{-1} \langle Q^{[n],N}\rangle_{\ell,V}$. These observables are diagonal 
in the energy basis, which reduces our task to an unconventional problem of classical statistical mechanics. As a simplification, the scattering shift
appears already explicitly through the Bethe equations. 

We first  introduce the empirical \textit{root density}
\begin{equation}\label{13.23} 
\varrho_N(w) = \frac{1}{N} \sum_{j=1}^N \delta (w- k_j), 
\end{equation}
compare with \eqref{3.24}. Then 
\begin{equation}\label{13.24} 
\sum_{j=1}^N V(k_j) = N \int_\mathbb{R} \mathrm{d}w\varrho_N(w) V(w) = N\mathcal{E}(\varrho_N),
\end{equation}
an energy-like term, and our task is to express the sum over quantum states by a sum over root densities, at least approximately for large $N$.
For given $I$, thus Bethe roots $k(I)$, we consider the equation 
\begin{equation}\label{13.25} 
y(\lambda) =  \nu\lambda + \theta_\mathrm{li} * \varrho_N(\lambda)
\end{equation}
with the convolution
\begin{equation}\label{13.26} 
 \theta_\mathrm{li}  * \varrho_N(\lambda)
=  \int_{\mathbb{R}} \mathrm{d}w \theta_\mathrm{li}(\lambda - w)\varrho_N(w).
\end{equation}
Since the function $y(\lambda)$ is strictly increasing, the relation
\begin{equation}\label{13.27} 
y(k_m^\mathrm{vac})  =  \frac{2\pi}{N} m, \quad m \in \mathbb{Z}, \,\,\,\mathrm{resp.}\,\,m \in \mathbb{Z} +\tfrac{1}{2},
\end{equation}
defines the increasingly ordered collection of points $\{k_m^\mathrm{vac}\}$. 
This set is viewed as a collection of placeholders. Their empirical density is the empirical
\textit{space density}
\begin{equation}\label{13.28} 
\varrho_{\mathsf{s},N}(w) = \frac{1}{N} \sum_{m \in \mathbb{Z}} \delta (w- k_m^\mathrm{vac}). 
\end{equation}
Since $y(k_j) = 2\pi I_j/N$, the Bethe roots are a subset of $\{k_m^\mathrm{vac}\}$. These are  particle locations in distorted momentum space.
Particles partially fill the placeholder set $\{k_m^\mathrm{vac}\}$. 
The complementary set, denoted by $\{k_m^\mathrm{h}\}$, are hole locations in distorted momentum space. The empirical \textit{hole density} is then
\begin{equation}\label{13.29} 
\varrho_{\mathsf{h},N}(w) = \frac{1}{N} \sum_{m \in \mathbb{Z}} \delta (w- k_m^\mathsf{h}). 
\end{equation}
Obviously
\begin{equation}\label{13.30} 
\varrho_N + \varrho_{\mathsf{h},N} = \varrho_{\mathsf{s},N}. 
\end{equation}

We assume that, relative to the given GGE, with probability one the empirical densities have a deterministic limit as  $N\to \infty$. 
More precisely one picks a well localized test function, $f$, and asserts that
\begin{equation}\label{13.31} 
\lim_{N \to \infty} \frac{1}{N} \sum_{m\in \mathbb{Z}} f(k_m^\mathrm{vac}) = \lim_{N \to \infty}\int_\mathbb{R}\mathrm{d}w f(w)
\varrho_{\mathsf{s},N}(w) = \int_\mathbb{R}\mathrm{d}w f(w) \varrho_{\mathsf{s}}(w),
\end{equation}
where the expression on left side is random, while its limit is non-random.
Eq. \eqref{13.31} defines the  infinite volume space density $\varrho_{\mathsf{s}}$, correspondingly for $\varrho$ and
$\varrho_{\mathsf{h}}$. According to \eqref{13.30},
\begin{equation}\label{13.32} 
\varrho + \varrho_{\mathsf{h}} = \varrho_{\mathsf{s}}. 
\end{equation}

Next we study the limit related to \eqref{13.25}. Setting $\lambda = k_j^\mathrm{vac}$ and  summing over $f(k_j^\mathrm{vac})$ yields
\begin{equation}\label{13.33} 
\frac{1}{N} \sum_{j\in \mathbb{Z}}  f(k_j^\mathrm{vac}) \frac{2\pi }{N} j = \frac{1}{N} \sum_{j\in \mathbb{Z}}  f(k_j^\mathrm{vac})
\Big( \nu k_j^\mathrm{vac} +  \theta_\mathrm{li}  * \varrho_N(k_j^\mathrm{vac})\Big).
\end{equation}
By construction the right hand side converges to 
\begin{equation}\label{13.34} 
\int_\mathbb{R}\mathrm{d}w f(w) \varrho_{\mathsf{s}}(w)\big(\nu w + \theta_\mathrm{li}  * \varrho(w)\big).
\end{equation}
For the left hand side, setting $i<j$, we start from 
\begin{equation}\label{13.35} 
 \int_{k_i^\mathrm{vac}}^{k_j^\mathrm{vac}} \mathrm{d}w
\varrho_{\mathsf{s},N}(w)  = \frac{1}{N}(j-i).
\end{equation}
For $i \to -\infty$ the nonlinearity becomes negligible and $k_i^\mathrm{vac} = 2\pi i/N$. Therefore, by restricting the support of the test function to $[-a,\infty)$
and setting $k_i^\mathrm{vac} = -a$ with $a$ sufficiently large,
\begin{equation}\label{13.36} 
\frac{1}{N} \sum_{j\in \mathbb{Z}}  f(k_j^\mathrm{vac})\Big(\int_{-a}^{k_j^\mathrm{vac}} 
 2\pi \varrho_{\mathsf{s},N}(w')\mathrm{d}w' - a\Big)
 \to \int_\mathbb{R}\mathrm{d}w f(w)\varrho_{\mathsf{s}}(w)\Big(\int_{-a}^w 2\pi\varrho_{\mathsf{s}}(w')\mathrm{d}w' -a\Big)
\end{equation}
as $N \to \infty$. Inserting \eqref{13.34} and \eqref{13.36} in Eq.  \eqref{13.33}, one arrives at the pointwise identity
\begin{equation}\label{13.37} 
 \int_{-a}^w  2 \pi\varrho_{\mathsf{s}}(w')\mathrm{d}w' -a = \nu w +\theta_\mathrm{li} * \varrho(w),
 \end{equation}
which upon differentiating yields
\begin{equation}\label{13.38} 
2 \pi \varrho_{\mathsf{s}}(w) = \nu +\phi_\mathrm{li}* \varrho(w).
 \end{equation}
Through the root density the scattering shift rules the space density.

The final step would require more details. The energy term appeared already in \eqref{13.24}. Hence one still has to figure out the entropy, namely the number of $I$'s subject to our constraints. For this purpose considered is a small
volume element, $\mathrm{d}w$, which still contains a huge number of placeholders given  by $N\varrho_{\mathsf{s}}(w) \mathrm{d}w$.
They split into Bethe roots, $N\varrho(w) \mathrm{d}w$, and holes, $N\varrho_{\mathsf{h}}(w) \mathrm{d}w$. Merely counting the corresponding quantum numbers, the corresponding
local entropy
equals   
\begin{eqnarray}\label{13.39} 
&&\hspace{-60pt}\log \frac{(N\varrho_{\mathsf{s}}(w)\mathrm{d}w)!}{(N\varrho(w) \mathrm{d}w)! (N\varrho_{\mathsf{h}}(w) \mathrm{d}w)!}\nonumber\\
&& \approx N \mathrm{d}w \big(\varrho_{\mathsf{s}}(w)\log\varrho_{\mathsf{s}}(w) - \varrho(w)\log\varrho(w) 
 -\varrho_{\mathsf{h}}(w)\log\varrho_{\mathsf{h}}(w) \big),
  \end{eqnarray}
where the linear term vanishes because of \eqref{13.32}. With $\varrho_{\mathsf{s}}$ defined in \eqref{13.38} and  $\varrho_{\mathsf{h}}$ in \eqref{13.32}, the resulting entropy per particle is thus given by 
\begin{equation}\label{13.40} 
\mathcal{S}(\varrho) = \int_\mathbb{R}\mathrm{d}w \big(\varrho_{\mathsf{s}}\log\varrho_{\mathsf{s}} - \varrho\log\varrho 
 -\varrho_{\mathsf{h}}\log\varrho_{\mathsf{h}} \big).
 \end{equation}
In approximation one arrives at 
\begin{equation}\label{13.41} 
Z_{\mathrm{li},N}(\ell,V)  \approx \int \mathcal{D}(\varrho) \mathrm{e}^{-N\big( \mathcal{E}(\varrho) - \mathcal{S}(\varrho)\big)}
\end{equation}
involving a somewhat vague sum over all root densities. 

The exponent defines the Yang-Yang free energy functional 
\begin{equation}\label{13.42}
\mathcal{F}^\circ_\mathrm{yy}(\varrho) = \nu^{-1}\int_\mathbb{R}\mathrm{d}w \big( \varrho V +\varrho \log \varrho +  \varrho_\mathsf{h} \log \varrho_\mathsf{h} - 
\varrho_\mathsf{s}\log \varrho_\mathsf{s}\big).
\end{equation}
The free energy is per unit length, compare with \eqref{9.47} for the Toda fluid. It is convenient  to switch to $\rho  = \nu^{-1} \varrho$ and correspondingly
$\rho_\mathsf{s}  = \nu^{-1} \varrho_\mathsf{s}$, $\rho_\mathsf{h}  = \nu^{-1} \varrho_\mathsf{h}$.
We also introduce the $T$ operator
 \begin{equation}\label{13.43}
Tf(w)  = \frac{1}{2\pi} \int_\mathbb{R} \mathrm{d}w'  \phi_\mathrm{li}(w-w') f(w'). 
\end{equation}
The kernel of $2\pi T$ equals the $\delta$-Bose scattering shift. Also $T\varsigma_0 = \varsigma_0$ and  
$T \to 1$ for $c \to 0$, while $T \to 0$ for $c \to \infty$. The latter case corresponds to free fermions and the former to free bosons.
Eq. \eqref{13.32} becomes
\begin{equation}\label{13.44} 
\rho + \rho_{\mathsf{h}} = \rho_{\mathsf{s}},
\end{equation}
while \eqref{13.38} transforms to 
\begin{equation}\label{13.45} 
2\pi\rho_\mathsf{s} = 1 +  2\pi T \rho.
 \end{equation}
Then, using \eqref{13.44}, the free energy per unit length is written as
\begin{equation}\label{13.46}
\mathcal{F}_\mathrm{yy}(\rho) = \int_\mathbb{R}\mathrm{d}w \big( \rho V +\rho \log \rho +  \rho_\mathsf{h} \log \rho_\mathsf{h} - 
(\rho +  \rho_\mathsf{h})\log (\rho +  \rho_\mathsf{h})\big).
\end{equation}
The variation is over all $\rho$ with $\rho \geq 0$ and $\nu \langle \rho \rangle = 1$. Note that, deviating from classical models, the factor $2\pi$ appears as a result of momentum quantization. \bigskip\\
$\blackdiamond\hspace{-1pt}\blackdiamond$~\textit{Convexity of the Yang-Yang free energy functional}.\hspace{1pt} 
Following the original  argument we establish that the functional $\mathcal{F}_\mathrm{yy}$ is convex. We start from  two  arbitrary 
spectral densities $\rho_0,\rho_1$ of well-defined Yang-Yang free energy and consider the linear interpolation
$\rho_u = \rho_0(1-u) + u \rho_1$, $0 \leq u \leq 1$. The aim is to show that 
$\mathcal{F}_\mathrm{yy}(\rho_u)'' >0$, the prime denoting derivative with respect to $u$. Note that
  $\rho_u'  = \rho_1- \rho_0$  and, in general, $\rho' + \rho_\mathsf{h}' = T\rho'$. With this input, differentiating 
  $\mathcal{F}_\mathrm{yy}(\rho(u))$ is straightforward with the result
\begin{equation}\label{13.47} 
\mathcal{F}_\mathrm{yy}(\rho_u)''  = \int_\mathbb{R}\mathrm{d}w \Big(\frac{1}{\rho_u} + \frac{1}{\rho_{\mathsf{h},u}}\Big)\Big((\rho_1 -\rho_0)
- \frac{\rho_u}{\rho_u+ \rho_{\mathsf{h},u}}T (\rho_1- \rho_0)\Big)^2 \geq 0.
\end{equation}
While likely to be correct, strict convexity would require an extra argument.\bigskip
\hfill $\blackdiamond\hspace{-1pt}\blackdiamond$

We remove the constraint $\nu \langle \rho \rangle = 1$ by the Lagrange multiplier $\mu$. Then the free energy becomes
\begin{equation}\label{13.48}
\mathcal{F}^\bullet_\mathrm{yy}(\rho) = \int_\mathbb{R}\mathrm{d}w \big( \rho (V - \mu) +\rho \log \rho +  \rho_\mathsf{h} \log \rho_\mathsf{h} - 
(\rho +  \rho_\mathsf{h})\log (\rho +  \rho_\mathsf{h})\big).
\end{equation}
The minimizing root density is the particle density $\rho_\mathsf{p}$, which is solution of   the saddle-point 
equation
\begin{equation}\label{13.49} 
V -\mu - \log \frac{\rho_\mathsf{h}}{\rho_\mathsf{p}} - T \log \Big(1 +   \frac{\rho_\mathsf{p}}{\rho_\mathsf{h}}\Big) =0.
\end{equation}
Here it is understood that space and hole density  associated with $\rho_\mathsf{p}$ are again denoted by 
 $\rho_\mathsf{s}, \rho_\mathsf{h}$. Hence  $\rho_\mathsf{s} = \rho_\mathsf{p} + \rho_\mathsf{h}$ and, in addition,
 \begin{equation}\label{13.50}
\rho_\mathsf{s} 
 = \frac{1}{2\pi}  + T  \rho_\mathsf{p}, 
 \end{equation}
 which confirms the usage of ``space density''. For $c = 0$, i.e. $T=0$, the Bethe roots are merely the eigenvalues of the quantized momentum, which 
 in the infinite volume limit converge to the Lebesgue measure $\frac{1}{2\pi}\mathrm{d}w$. Through the interaction
 this density is modified, in the same spirit as the bare velocity is modified to $v^\mathrm{eff}$. Particle and hole density are then taken relative to such modified space density.   
 
 To solve \eqref{13.49} and \eqref{13.50}, 
we introduce the number density 
 \begin{equation}\label{13.51} 
\rho_\mathsf{n} = \frac{\rho_\mathsf{p}}{\rho_\mathsf{p}+ \rho_\mathsf{h}} =  \frac{\rho_\mathsf{p}}{\rho_\mathsf{s}},
\end{equation}
thereby yielding the quasi-energy  $\varepsilon$ through
\begin{equation}\label{13.52}
\rho_\mathsf{n} = \frac{1}{1 + \mathrm{e}^\varepsilon}, \qquad
\mathrm{e}^\varepsilon
= \frac{\rho_\mathsf{h}}{\rho_\mathsf{p}}. 
\end{equation}
Then, using the dressing transformation 
\begin{equation}\label{13.53} 
f^\mathrm{dr} = f + T \rho_\mathsf{n} f^\mathrm{dr},\quad f^\mathrm{dr} = \big(1 - T\rho_\mathsf{n}\big)^{-1} f,
\end{equation} 
one obtains 
\begin{equation}\label{13.54} 
2\pi \rho_\mathsf{s} = \varsigma_0^\mathrm{dr}.
\end{equation} 
With these conventions Eq. \eqref{13.49} turns into  the quantum TBA equation
\begin{equation}\label{13.55} 
\varepsilon = V -\mu - T\log( 1 + \mathrm{e}^{-\varepsilon}),
\end{equation}
compare with \eqref{9.50} for the Toda fluid. 

To determine the free energy of the $\delta$-Bose gas one has to evaluate the free energy functional $\mathcal{F}^\bullet_\mathrm{yy}$ at its minimizer 
$\rho_\mathsf{p}$, which leads to
\begin{equation}\label{13.56} 
 \lim_{\ell\to \infty} - \frac{1}{\ell} \log Z_{\mathrm{li},N}(\ell,\mu,V) 
 = F_\mathrm{yy}(\mu,V) =  \mathcal{F}^\bullet_\mathrm{yy}(\rho_\mathsf{p}) = -\frac{1}{2\pi} \int \mathrm{d}w
  \log( 1 + \mathrm{e}^{-\varepsilon}),
  \end{equation}
a form familiar from noninteracting fermions. Adding the small perturbation $V+ \kappa \varsigma_n$ and computing the linear response of the free energy in $\kappa$ one obtains
\begin{equation}\label{13.57} 
\lim_{\ell\to \infty} \ell^{-1} \langle Q^{[n],\ell} \rangle_{V,\ell}   = \langle \rho_\mathsf{p}\varsigma_n\rangle, \quad n = 0,1,\ldots\,.
\end{equation}

This identity suggests the proper interpretation of the minimizing particle density $\rho_\mathsf{p}$. Under the GGE  the Bethe roots
$k_1,\ldots,k_N$ with variable  $N$ are random and combine into a empirical density as
\begin{equation}\label{13.58} 
\rho_{\mathsf{p},\ell}(w) = \ell^{-1}\sum_{j=1}^N\delta(w - k_j),
\end{equation}
compare with Eq. \eqref{13.23}. Obviously, $ \rho_{\mathsf{p},\ell}(w) \geq 0$ with normalization $\langle  \rho_{\mathsf{p},\ell}\rangle = \nu^{-1}$. 
Eq. \eqref{13.57} states that on average the $n$-th moment of $\rho_{\mathsf{p},\ell}$ converges to the one of $\rho_{\mathsf{p}}$
as $\ell \to \infty$. In Chapter \ref{sec4}, we explained that typical fluctuations of the empirical density of Lax matrix eigenvalues are of  order $1/\sqrt{N}$. For the Lieb-Liniger model one expects the same behavior and
\begin{equation}\label{13.59}
\lim_{\ell \to \infty} \rho_{\mathsf{p},\ell}(w) = \rho_\mathsf{p}(w)
\end{equation}
with probability one, upon integrating against a localized smooth  test function. 
\subsection{Charge currents, hydrodynamic equations}
\label{sec13.3}
To obtain the total currents, the usual strategy is to first figure out the charge densities. They satisfy a local conservation law from which one deduces the charge current density, and thereby also the total charge currents. The total charges have been defined in \eqref{13.17}.  Then, to lowest order and on the entire real line,
\begin{eqnarray}\label{13.60}
&&Q^{[0]}(x) =  \Psi(x)^*\Psi(x),\quad   Q^{[1]}(x) = -\mathrm{i}  \Psi(x)^*\partial_x\Psi(x),  \nonumber\\[1ex]
&&Q^{[2]}(x) =  \partial_x \Psi(x)^*\partial_x \Psi(x) + c \Psi(x)^*\Psi(x)^*\Psi(x)\Psi(x), 
\end{eqnarray} 
which implies the current densities
\begin{equation}\label{13.61} 
J^{[0]}(x) = Q^{[1]}(x), \qquad J^{[1]}(x) = \partial_x\Psi(x)^*\partial_x\Psi(x) + \tfrac{1}{2}c\Psi(x)^*\Psi(x)^*\Psi(x)\Psi(x).
 \end{equation}
  Thus a natural strategy would be to start from the classical nonlinear
Schr\"{o}dinger equation and its known sequence of charge densities involving higher order field derivatives. As naive guess, imposing normal order will yield the charges of the Lieb-Liniger model in second quantized form, which would then be manifestly local.
But this scheme  works only up to $n= 3$. Already for $Q^{[4]}$ more complicated subtraction terms appear.  
For the total charge currents similar difficulties will appear. 
We still freely use $Q^{[n]}(x)$ and $J^{[n]}(x)$, assuming that at some point these difficulties can be resolved. 
For infinite volume, formally,  the charges satisfy a local conservation law of the form
\begin{equation}\label{13.62}
\partial_t Q^{[n]}(x,t) + \partial_x J^{[n]}(x,t) = 0.
\end{equation} 
Just to be clear, here $t$ refers to the unitary time evolution. For the operators at time zero, we omit the argument $t$, e.g. $J^{[n]}(x) = J^{[n]}(x,t=0)$.

To compute the average current, one starts from a GGE at infinite volume, $\ell = \infty$, average being denoted as $\langle \cdot \rangle_V$ and truncation by $\langle fg \rangle_V^\mathrm{c}
= \langle f \rangle_V\langle g \rangle_V$. Then
the charge-current correlator is given by
\begin{equation}\label{13.63}
B_{m,n} = \int_\mathbb{R} \mathrm{d}x \langle J^{[m]}(x)Q^{[n]}(0)\rangle_V^\mathrm{c}.
\end{equation} 
 While the definition looks asymmetric, in fact,
\begin{equation}\label{13.64}
B_{m,n} = B_{n,m},
\end{equation}
as has been explained for other models before, compare with \eqref{6.8}. Now, setting $\partial_0 = \partial_{\mu_0}$ and $\partial_1 = \partial_{\mu_1}$, one arrives at
\begin{equation}\label{13.65}
\partial_0 \langle J^{[n]}(0)\rangle_V = - B_{n,0} = - B_{0,n} = -\int_\mathbb{R} \mathrm{d}x \langle Q^{[1]}(x)Q^{[n]}(0)\rangle_V
= \partial_1 \langle Q^{[n]}(0)\rangle_V =  \partial_1 \langle \rho_\mathsf{p}\varsigma_n\rangle.
\end{equation}
The $\partial_0$ derivative of the average current is related to a susceptibility, i.e. a particular  second derivative of the generalized free energy. This is a doable problem and we explain a short path towards its solution. 

Using the TBA equation \eqref{13.55} to differentiate $\rho_\mathsf{n}$ with respect to $\partial_0$ and  $\partial_1$, one obtains the relations 
\begin{equation}\label{13.66}
 \partial_0\rho_\mathsf{n} = - \rho_\mathsf{n} (1-\rho_\mathsf{n} ) \varsigma_0^\mathrm{dr}, \quad \partial_1\rho_\mathsf{n}  = - \rho_\mathsf{n} (1-\rho_\mathsf{n} ) \varsigma_1^\mathrm{dr}. 
\end{equation}
Hence
\begin{equation}\label{13.67}
\varsigma_1^\mathrm{dr} \partial_0\rho_\mathsf{n}=  \varsigma_0^\mathrm{dr} \partial_1\rho_\mathsf{n}
\end{equation}
and, because  $\partial_0 \partial_1\rho_\mathsf{n}  = \partial_1 \partial_0\rho_\mathsf{n}$ for mixed derivatives,
\begin{equation}\label{13.68}
\partial_1 \varsigma_0^\mathrm{dr} =  \partial_0\varsigma_1^\mathrm{dr}.
\end{equation}
Setting 
\begin{equation}\label{13.69}
  v^\mathrm{eff} = \frac{\varsigma_1^\mathrm{dr}}{\varsigma_0^\mathrm{dr}}
\end{equation}
and using \eqref{13.67} and \eqref{13.68}, one obtains
\begin{eqnarray}\label{13.70}
 && \hspace{-50pt}  \partial_0(\rho_\mathsf{p} v^\mathrm{eff})=   \tfrac{1}{2\pi} \partial_0 (\rho_\mathsf{n}\varsigma_1^\mathrm{dr}  ) = \tfrac{1}{2\pi}\left(\varsigma_1^\mathrm{dr} \partial_0\rho_\mathsf{n}  +\rho_\mathsf{n}   \partial_0 \varsigma_1^\mathrm{dr}\right) \nonumber\\[1ex]
 &&= \tfrac{1}{2\pi}\left(\varsigma_0^\mathrm{dr} \partial_1\rho_\mathsf{n} +\rho_\mathsf{n}  \partial_1 \varsigma_0^\mathrm{dr}\right)=  \tfrac{1}{2\pi}\partial_1( \varsigma_0^\mathrm{dr}\rho_\mathsf{n})  = \partial_1\rho_\mathsf{p}.
  \end{eqnarray}
Altogether we arrived at 
\begin{equation}\label{13.71} 
\partial_0\big( \langle J^{[n]}(0)\rangle_V - \langle\rho_\mathsf{p}v^\mathrm{eff}\varsigma_n\rangle\big)=0. 
\end{equation}
Thus the round bracket is independent of $\mu_0$. To determine the respective free parameter, one observes that in the limit $\mu_0 \to -\infty$ there are no particles. The GGE average current vanishes and so does  $\rho_\mathsf{p}$ since $\nu \langle \rho_\mathsf{p}\rangle  = 1$.  
 Hence one concludes
\begin{equation}\label{13.72} 
\lim_{\ell\to \infty} \ell^{-1}\langle J^{[n],\ell}\big \rangle_{V,\ell} = \langle \rho_\mathsf{p} v^\mathrm{eff} \varsigma_n \rangle.
\end{equation}
As before, one confirms that the effective velocity can be written as solution of the collision rate ansatz
\begin{equation}\label{13.73} 
 v^\mathrm{eff}(w)= w + \int_\mathbb{R}\mathrm{d}w' \phi_\mathrm{li}(w-w')\rho_\mathsf{p}(w')\big(v^\mathrm{eff}
 (w') -  v^\mathrm{eff}(w)\big).
\end{equation}

Assuming propagation of local GGE one arrives at the hydrodynamic equation of the $\delta$-Bose gas
\begin{equation}\label{13.74} 
\partial_t\rho_\mathsf{p}(x,t;v) + \partial_x\big(v^\mathrm{eff}(x,t;v)\rho_\mathsf{p}(x,t;v)\big) = 0,
\end{equation}
which in structure is identical to the ones of the  Toda and Calogero fluid. 
The transformation to an evolution equation for $\rho_\mathsf{n}$ remains valid with the expected result
\begin{equation}\label{13.75} 
\partial_t\rho_\mathsf{n}(x,t;v) + v^\mathrm{eff}(x,t;v)\partial_x\rho_\mathsf{n}(x,t;v) = 0.
\end{equation}\\
$\blackdiamond\hspace{-1pt}\blackdiamond$~\textit{Experimental realizations}.\hspace{1pt} 
 The 2006 pioneering experiment on the $\delta$-Bose gas is widely known as quantum Newton's cradle. 
In a suitably engineered 2D optical lattice $^{87}$Rb atoms are trapped in an array of tubes, consisting of 1000 to 8000 tubes with approximately  40 to 250 atoms per tube. The motion of atoms in a single tube is well described by the Lieb-Liniger model, except for the additional harmonic trapping potential which in principle breaks integrability. One starts with a centered cloud of atoms and applies a Bragg pulse which yields the superposition of two states with a momentum shift $v_0 =\pm\hbar k$. Thus the initial rapidity density is of the form 
\begin{equation}\label{13.76} 
\tfrac{1}{2}\big(\rho_{\mathsf{p},\mathrm{therm}}(x,0;v- v_0) + \rho_{\mathsf{p},\mathrm{therm}}(x,0;v+ v_0)\big),
\end{equation}
where ``therm'' indicates a thermal rapidity density at fixed temperature. The cloud splits into left and right peak,
which start to oscillate due to the trapping potential and thus periodically interpenetrate each other. On longer time scales the peaks dephase. If the dynamics of atoms would be non-integrable,  entirely different patterns would emerge. Thus  integrability of the Bose gas is confirmed, at least qualitatively. At the time no theoretical  frame was available. This has changed and good agreement with numerical simulations of GHD equations from \eqref{13.74} has been reported.
However Eq.  \eqref{13.74} has to be modified so to include a trapping potential,
\begin{equation}\label{13.77} 
\partial_t\rho_\mathsf{p}(x,t;v) + \partial_x\big(v^\mathrm{eff}(x,t;v)\rho_\mathsf{p}(x,t;w)\big) - \omega_0^2 x\partial_v \rho_\mathsf{p} (x,t;v) = 0
\end{equation} 
with $\omega_0$ the frequency of the trap potential.

A more recent experiment uses the atom on chip technique. Suitable electrical wiring on the chip generates a floating magnetic tube, considerably longer than a tube created by optical means. As a result the experimental cloud consists of $4600 \pm 100$ ${}^{87}$Rb atoms.
The initial state is a thermal cloud corresponding to an external potential $V_\mathrm{ex}$. By additional wiring a harmonic,  $V_\mathrm{ex}(x) = x^2$,
and double bump quartic, $V_\mathrm{ex}(x) = -x^2 + x^4$, can be realized. After the initial preparation the cloud expands into the tube and the average density  is recorded for several times. Density profiles are compared with numerical solutions of GHD and
of conventional hydrodynamics (CHD), which is based on conservation of mass, momentum, and energy only. This would be the appropriate scheme for a non-integrable 1D fluid. For the single bump initial condition both simulations yield similar density profiles.
Not so surprisingly, good agreement with the experimental data is recorded. For the double bump, the CHD  solution develops steep gradients which eventually will lead to shock formation, while for GHD the two bumps flatten out smoothly. Thus CHD and GHD differ qualitatively. The experimental data agree well with GHD. 

As one drawback of the atom on chip technique, the tube looses mass, roughly $25\%$, during a typical run. This effect is not included
in  \eqref{13.76}. It took substantial efforts to understand how to incorporate to  GHD such a seemingly innocent loss term.  
\hfill $\blackdiamond\hspace{-1pt}\blackdiamond$
\subsection{Comparing  Toda fluid and $\delta$-Bose gas, generic TBA structure}
\label{sec13.4}
While our results are somewhat scattered, we reached the stage to compare in more detail integrable classical and 
quantum fluids. For the Toda fluid the free energy density is given by 
 \begin{equation}\label{13.78} 
\mathsfit{F}_\mathrm{cl}(\varepsilon) = - \mathrm{e}^{-\varepsilon}, \qquad F_\mathrm{cl}(\mu,V) = \int_\mathbb{R}\mathrm{d}w
\mathsfit{F}_\mathrm{cl}(\varepsilon(w))
\end{equation} 
and the number density is defined through 
 \begin{equation}\label{13.79} 
\rho_\mathsf{n} = \mathsfit{F}_\mathrm{cl}'(\varepsilon) = \mathrm{e}^{-\varepsilon}.
\end{equation} 
The TBA equation reads
\begin{equation}\label{13.80} 
\varepsilon = V - \mu + T\mathsfit{F}_\mathrm{cl}(\varepsilon).
\end{equation} 
Its solution is the number density $\rho_\mathsf{n}$, which is linked to the first order derivatives of the free energy as 
\begin{equation}\label{13.81} 
\langle Q^{[n]}(0) \rangle_{\mu,V} =  \langle \varsigma_0^\mathrm{dr} \rho_\mathsf{n}\varsigma_n\rangle = \langle \rho_\mathsf{p}\varsigma_n\rangle. 
\end{equation} 
The average currents are determined by the current potential with density 
\begin{equation}\label{13.82} 
\mathsfit{G}_\mathrm{cl}(w) =  w \mathsfit{F}_\mathrm{cl}(\varepsilon(w)), \qquad G_\mathrm{cl}(\mu,V) = \int_\mathbb{R}\mathrm{d}w
\mathsfit{G}_\mathrm{cl}(w).
\end{equation} 
The first order derivatives of $G_\mathrm{cl}$ are the average currents,
\begin{equation}\label{13.83} 
\langle J^{[n]}(0) \rangle_{\mu,V} = \langle \rho_\mathsf{p}v^\mathrm{eff}\varsigma_n\rangle, \qquad  v^\mathrm{eff} =
\varsigma_1^\mathrm{dr}/\varsigma_0^\mathrm{dr}.
\end{equation} 
The same formulas hold for the hard rod and Calogero fluids upon adjusting the two-particle scattering shift. In case of hard rod and Toda lattices, the free energy and the GGE averaged fields and currents are per lattice site.  Therefore the free energy has to be multiplied by the factor $\nu$ and the effective velocity has to be shifted by $q_1 $.
To be noted, 
the formulas above are specific for the single particle energy-momentum relation $E(v) =v^2/2$. 

For the Ablowitz-Ladik lattice the $w$-integration is over the interval $[0,2\pi]$ and the current potential density reads 
\begin{equation}\label{13.84} 
\mathsfit{G}_\mathrm{al}(w) = -\sin(w)  \mathsfit{F}_\mathrm{cl}(\varepsilon(w)), 
\end{equation} 
which derives from $E(w) = \cos{w}$ and hence $E'(w) = -\sin{w}$. The Schur flow has 
the anomaly of an additional self-generated confining potential.

Turning to the $\delta$-Bose gas, except for the scattering shift, the \textit{only} change is the free energy density, which now reads
\begin{equation}\label{13.85}
 \mathsfit{F}_\mathrm{qu}(\varepsilon) = - \log(1 + \mathrm{e}^{-\varepsilon}).
\end{equation}
The free energy becomes
\begin{equation}\label{13.86} 
F_\mathrm{qu}(\mu,V) = \frac{1}{2\pi}\int_\mathbb{R}\mathrm{d}w
\mathsfit{F}_\mathrm{qu}(\varepsilon(w)), 
\end{equation}
where the factor of $1/2\pi$ can be viewed as modifying the classical a priori measure $\mathrm{d} w $ to $\frac{1}{2\pi}\mathrm{d} w $.
The number density results as
\begin{equation}\label{13.87} 
\rho_\mathsf{n} = \mathsfit{F}_\mathrm{qu}'(\varepsilon) = \frac{1}{1 +\mathrm{e}^{\varepsilon}}.
\end{equation}
Finally the current potential and its density become
\begin{equation}\label{13.88} 
\mathsfit{G}_\mathrm{qu}(w) = w \mathsfit{F}_\mathrm{qu}(\varepsilon(w)), \qquad
G_\mathrm{qu}(\mu,V) = \frac{1}{2\pi}\int_\mathbb{R}\mathrm{d}w \mathsfit{G}_\mathrm{qu}(\mu,V).
\end{equation}
As  to be discussed in Chapter \ref{sec14}, the quantized Toda chain follows the same scheme.

An apparently convincing pattern emerges. For a given integrable model with particle-based hydrodynamics, one first has to identify the two-particle scattering shift and the thereby resulting space of  spectral parameters. The distinction between classical and quantum is merely reflected through adopting either $\mathsfit{F}_\mathrm{cl}$ or $\mathsfit{F}_\mathrm{qu}$.
The one-particle dispersion law enters in the definition of the current potential. 

\subsection{Gaudin matrix}
\label{sec13.5}
To determine the GGE averaged currents, an interesting  entirely disjoint approach has been developed recently, for which the quantum mechanical structure of the Lieb-Liniger model enters
in a more fundamental way. While we cannot write down so easily explicit formulas, abstractly there 
must exist the total current operator $J^{[n],\ell}$ for the total charge $Q^{[n],\ell}$ as defined in Eq. \eqref{13.17} with $\ell = \nu N$ and $N$
 the dimension of the vector $k$. For this operator
one conjectures the matrix elements $\langle k|J^{[n],\ell}|k\rangle$, which suffices for the computation of 
GGE expectations. The central object is the Gaudin matrix defined by 
 \begin{equation}\label{13.89} 
(G^\ell)_{i,j} = \delta_{ij}\big(\ell + \sum_{m=1}^N2\pi T(k_j, k_m)\big) - 2\pi T(k_i,k_j), 
\end{equation} 
where $T(k_i, k_j)$ is the kernel of the integral operator defined in \eqref{13.43}. Then the claim is
\begin{equation}\label{13.90} 
\langle k |J^{[n],\ell}|k \rangle  = \ell\sum_{i,j = 1}^N\big((G^\ell)^{-1}\big)_{i,j} (k_i)^nk_j .
\end{equation}
The prefactor $\ell$ arises because by stationarity the current density has to be independent of  $x$.  

Trusting in \eqref{13.90}, the GGE  averaged currents can be obtained. We fix a GGE with some confining potential $V$. 
Under this GGE the Gaudin matrix is a random matrix. Presumably its large $N,\ell $ limit does not exist. But we have to invert the Gaudin matrix only on very special vectors. Therefore we pick test functions $f,g$ and consider the quadratic form
\begin{equation}\label{13.91} 
\ell^{-2}\sum_{i,j=1}^N f(k_i) (G^\ell)_{i,j} g(k_j)  =   \int_{\mathbb{R}^2}\!\!\mathrm{d}w\mathrm{d}w' f(w)\mathsf{G}_\ell(w,w')
g(w') .
\end{equation}
The left hand side can be expressed as quadratic functional of the empirical density integrated against some smooth functions. Therefore, relying on \eqref{13.59}, one obtains the limit  
\begin{equation}\label{13.92} 
\lim_{\ell \to \infty}  \langle f,\mathsf{G}_\ell g\rangle =  \langle \rho_\mathsf{p} f,\mathsf{G}g\rangle
\end{equation}
with probability one, where the limiting operator is defined through
\begin{equation}\label{13.93} 
\mathsf{G} g(w) = g(w) + 2\pi \int_{\mathbb{R}}\mathrm{d}w'  T(w,w')\rho_\mathsf{p}(w')\big(g(w') - g(w)\big).
\end{equation}
Since the limit is non-random, unless badly behaved near zero, also the inverse converges to the inverse of the limit
with probability one, i.e.
\begin{equation}\label{13.94} 
\lim_{\ell \to \infty}  \langle f,(\mathsf{G}_\ell)^{-1}g\rangle =  \langle \rho_\mathsf{p} f,\mathsf{G}^{-1}g\rangle.
\end{equation}
In our particular application $f(w) = w^n$ and $g(w) = w$.
Hence 
\begin{equation}\label{13.95} 
\lim_{\ell\to \infty} \ell^{-1}\langle J^{[n],\ell}\big \rangle_{V,\ell} = \langle \rho_\mathsf{p} \varsigma_n,\mathsf{G}^{-1}\varsigma_1\rangle.
\end{equation}
Defining $\mathsf{G}^{-1} \varsigma_1 = \bar{v}$, one obtains 
\begin{equation}\label{13.96} 
 \bar{v}(w)= w + 2\pi\big( T(\rho_\mathsf{p} \bar{v})(w) -  T\rho_\mathsf{p}(w) \bar{v}(w)\big).
\end{equation}
Hence $\bar{v} = v^\mathrm{eff}$ and 
\begin{equation}\label{13.97} 
\lim_{\ell\to \infty} \ell^{-1}\langle J^{[n],\ell}\big \rangle_{V,\ell} = \langle \rho_\mathsf{p} v^\mathrm{eff} \varsigma_n \rangle,
\end{equation}
in agreement with \eqref{13.72}.
\bigskip\\
$\blackdiamond\hspace{-1pt}\blackdiamond$~\textit{Number and momentum current}.\hspace{1pt} 
In second quantization the particle current equals the
momentum,
\begin{equation}\label{13.98} 
Q^{[1],\ell}(x) = -\mathrm{i}\Psi(x)^* \partial_x\Psi(x),\quad -\mathrm{i} \int_0^\ell \!\!\mathrm{d}x \Psi(x)^* \partial_x\Psi(x)
= Q^{[1],\ell} = J^{[0],\ell}. 
\end{equation}
Considering $n =0$ in \eqref{13.90} and using  $(G_\ell1)_i = 1$ yields the result 
\begin{equation}\label{13.99} 
\langle k |J^{[0],\ell}|k \rangle  = \ell\sum_{i,j = 1}^N((G^\ell)^{-1})_{i,j}(k_i)^0 k_j  = \sum_{j=1}^N k_j,
\end{equation}
which agrees with \eqref{13.98}.

To determine the momentum current,  
the left hand part of \eqref{13.98} is integrated over some test function $g$. Working out the time derivative yields
\begin{eqnarray}\label{13.100} 
&&\hspace{-40pt}\mathrm{i} \big[H_{\mathrm{li},\ell}, -\mathrm{i}\int_0^\ell\!\! \mathrm{d}x g(x)\Psi(x)^* \partial_x\Psi(x)\big] \\
&&\hspace{-30pt}= \tfrac{1}{2}\int_0^\ell\!\!\mathrm{d}x \Big(\big(g''(x)
\Psi(x)^* + 2g'(x) \partial_x\Psi(x)^* \big)\partial_x\Psi(x) + g'(x)c\Psi(x)^*\Psi(x)^*\Psi(x)\Psi(x) \Big).\nonumber
\end{eqnarray}
By the conservation law, the total current is  obtained by setting $g' = 1$,
\begin{eqnarray}\label{13.101} 
&&\hspace{-48pt}J^{[1],\ell} = \int_0^\ell \!\!\mathrm{d}x\big(\partial_x\Psi(x)^*\partial_x\Psi(x) +\tfrac{1}{2}c\Psi(x)^*\Psi(x)^*\Psi(x)\Psi(x) \big) \nonumber\\
&&\hspace{-20pt}= 2 H_{\mathrm{li},\ell} -\tfrac{1}{2}c \int_0^\ell \mathrm{d}x\Psi(x)^*\Psi(x)^*\Psi(x)\Psi(x).
\end{eqnarray}
Now taking expectations with respect to $|k\rangle$ with $N=2$ on both sides yields 
\begin{eqnarray}\label{13.102} 
 &&\hspace{-43pt} \langle k|J^{[1],\ell} |k\rangle = 2\langle k| H_{\mathrm{li},\ell} |k\rangle-\tfrac{1}{2}c \int_0^\ell \!\!\mathrm{d}x\langle k|\Psi(x)^*\Psi(x)^*\Psi(x)\Psi(x)|k\rangle\nonumber\\
 &&\hspace{16pt}= k_1^2 +k_2^2 - \frac{1}{\ell + 2 T(k_1,k_2)}\frac{2c (k_1 - k_2)^2}{(k_1 - k_2)^2 +c^2}.
\end{eqnarray}
Since the Gaudin matrix is a $2 \times 2$ matrix,  \eqref{13.102} agrees with \eqref{13.90} for $N=2,n=1$.

To proceed to $N>2$ such a brute force computation no longer works, which is a generic experience in this area. Instead we recall that  
the momentum current equals the pressure $P$ and for energy $E(\ell)$ one has $P = -dE/d\ell$.
Surprisingly  these thermodynamic relations extend to a single eigenstate of the Lieb-Liniger model. We start from 
\begin{equation}\label{13.103} 
E(\ell) = \tfrac{1}{2} \sum_{j=1}^N k_j^2,\qquad - \frac{d}{d\ell}E(\ell) = \sum_{j,k=1}^N ((G_{N,\ell})^{-1})_{j,k}k_j k_k,  
\end{equation}
where the second identity is obtained by differentiating Eq. \eqref{13.14} with respect to $\ell$. Let us consider a volume of length $\ell$ and allow in the Lieb-Liniger hamiltonian a kinetic energy of strength $\kappa$,  as $H_{\mathrm{li},\ell,\kappa,c}
 =\kappa T_{\mathrm{kin},\ell }+ c T_{\mathrm{pot},\ell}$. By continuity, one can follow one particular vector of Bethe roots, $k$, at fixed $N$ in their dependence on $\ell,\kappa, c$.
By spatial dilation the corresponding energy then scales as $E(\ell,\kappa,c) =  E(\alpha \ell,  \alpha^{-2} \kappa, \alpha^{-1}c)$, where 
$\alpha^{-2}$ results from the Laplacian and $\alpha^{-1}$ from the first order derivative of the boundary condition in \eqref{13.8}.
To first order in the deviation from $\ell$ there is an overall factor of $-\ell^{-1}$ and a factor $2$ for the kinetic energy and a factor 1 
for the potential energy, in agreement with \eqref{13.16}. Hence
 \begin{equation}\label{13.104} 
\frac{d}{d\ell} \langle k| H_{\mathrm{li},\ell,N} |k \rangle = - \ell^{-1} \langle k |J^{[1],\ell}| k \rangle.
\end{equation}
Combining \eqref{13.103} and \eqref{13.104}, we conclude that Eq. \eqref{13.90} spezialized to $n=1$ holds for arbitrary $N$.

 It is remarkable, when averaging over a single energy eigenstate, 
one finds already a structural dependence which persists in the limit of large $\ell,N$. No such features seem to be known for the 
Toda lattice or other classical integrable many-particle systems.
\hfill $\blackdiamond\hspace{-1pt}\blackdiamond$\bigskip
\begin{center}
\textbf{Notes and references}
\end{center}
 \begin{center}
 \textbf{Section 13.0}
\end{center} 
 Transport in one-dimensional quantum lattice models, including integrable chains, is reviewed in Bertini et al. \cite{BHK20}.
 Conformal field theories out of equilibrium are investigated by Bernard and Doyon \cite{BD16}. Generalized hydrodynamics for the XXZ chain is studied by 
 Bertini  et al. \cite{BCDF16}, Bulchandani et al.  {BVKM17} and the spin-$\frac{1}{2}$ Fermi-Hubbard model by Ilievski and  De Nardis   \cite{ID17}. The lecture notes of Doyon  \cite{D19a} provide a much more extensive list of references, see also the recent special volumes, one edited by Bastianello et al. \cite{BBDV22} and a second one by Abanov et al. \cite{ADDKS22}. Up to 1996,
 the exciting progress on integrable quantum many-body systems  is well covered in Korepin et al.  \cite{KBI97}.  
 \bigskip
 \begin{center}
 \textbf{Section 13.1}
\end{center} 
 Elliott H. Lieb, together with Werner Liniger,  discovered the Bethe ansatz for the $\delta$-Bose gas and analysed ground state properties in Lieb and Liniger \cite{LL63a} and Lieb  \cite{LL63b}. For the repulsive case the completeness of Bethe wave functions has been proved by Dorlas  \cite{D93}. The attractive 
  $\delta$-Bose gas is more complicated because of bound states. An exhaustive discussion  of the structure of eigenfunctions can be found in Dotsenko   \cite{D10}.  In the attractive case completeness is established in the thesis of
 Oxford \cite{O79}, see also the contribution by  Prolhac and Spohn \cite{PS11}. The local structure of higher charges is studied in  Davies \cite{D90} and Davies and Korepin \cite{DK11}. For computations 
  the algebraic Bethe ansatz is often more powerful. An example is the norm of eigenfunctions, as discussed in Piroli and  Calabrese \cite{PC15}.  The lecture notes by Franchini \cite{F17}
are a most readable introduction.  \bigskip
\begin{center}
\textbf{Section 13.2}
\end{center} 
The variational type solution for the thermodynamics has been obtained by C.N. Yang and C.P. Yang \cite{YY69}. A proof is accomplished by 
 Dorlas, Lewis, and Pul\'{e}  \cite{DLP93} using methods from large deviations. The Bethe ansatz is very well covered in the literature, for example see  Takahashi \cite{T84},  Faddeev \cite{F96}, Sutherland \cite{S04}, and Gaudin \cite{G14}. At the time the conventional thermodynamics was in focus. The GGE is discussed by  Mossel and Caux \cite{MC12}. It would be interesting to find out whether and how the methods in Dorlas et al.  \cite{DLP93} extend to a more general class of confining potentials. There have been several attempts on improving the Yang-Yang method. A recent study is Koslov et al.  \cite{KSV18} with references to earlier work.\bigskip
\begin{center}
\textbf{Section 13.3}
\end{center} 
The GHD of the $\delta$-Bose gas is discussed by Doyon  \cite{D19a}, see also Bonnemain et al.  \cite{BDE22}. 
 For the collision rate ansatz we follow the contribution of Yoshimura and Spohn \cite{YS20}. 
 
Highly recommended reading is the comprehensive review by Bouchoule and  Dubail \cite{BD21}, see also Malvania et al. \cite{MZLDRW21}. In addition to a theoretical section, the article discusses  at length numerical schemes for solving GHD of the $\delta$-Bose gas. The second half of the review is concerned with experimental realizations. The original quantum Newton cradle experiment was invented by Kinoshita, Wenger, and Weiss \cite{KWW06}, see also the more recent version Li et al.  \cite{ZMS20}.  The GHD analysis of the experiment can be found in Caux et al.  \cite{CDDKY19}. The atom on chip techniques are described by Schemmer et al. \cite{SBDD19}. The GHD handling of  mass loss is accomplished in Bouchoule et al. \cite{BDD20}.  The distinction between GHD and CHD is studied by Doyon et al. \cite{DDKY17}. The connection of the Lieb-Liniger model to cold atom physics is reviewed by Zwerger \cite{Z22}.
\bigskip
\begin{center}
\textbf{Section 13.4}
\end{center}
Our discussion is based on the lecture notes by Doyon  \cite{D19a}. He points out three additional classes referring to classical phonons, classical radiation, and quantum bosons.
\bigskip 
\begin{center}
\textbf{Section 13.5}
\end{center} 
 For the discussion of the Gaudin matrix and its relation to currents we follow the work of Borsi et al. \cite{BPP19}. Other methods rely on long range deformations, see Pozsgay \cite{P20a}, and properties of the boost operator, see Yoshimura and Spohn \cite{YS20}. Corresponding results for spin systems are obtained by Pozsgay  \cite{P20}.
\newpage
 \section{Quantum Toda lattice}
\label{sec14}
\setcounter{equation}{0} 
The classical Toda hamiltonian is quantized according to standard rules. Thereby one obtains a system of $N$ particles with
 Hilbert space $\mathcal{H}_N =  
L^2(\mathbb{R})^{\otimes N} = L^2(\mathbb{R}^{N})$. The particles have position $x_j$ and momenta $p_j$, $ j = 1,\ldots,N$, satisfying the commutation relations 
$[p_i,x_j] = -\mathrm{i}\hbar\delta_{ij}$. In position space representation, the complex-valued wave functions  are of the form  $\psi(x_1,\ldots,x_N)$. For the $j$-th particle, the position operator
 is multiplication by $x_j$ and the momentum operator $p_j = -\mathrm{i}\hbar\partial_{x_j}$. In conventional notation the interaction term 
reads 
\begin{equation}\label{14.1} 
X_j  = \mathrm{e}^{x_j  - x_{j+1}}, \qquad X_N = \mathrm{e}^{x_N - x_1},
\end{equation}
$ j = 1,\ldots,N-1$, where the latter condition imposes periodic boundary conditions. Then the hamiltonian of the quantum Toda lattice reads
\begin{equation}\label{14.2} 
H_{\mathrm{qt},N} = \sum_{j=1}^N \big(\tfrac{1}{2} p_j^2 +  X_j\big). 
\end{equation}
Planck's constant $\hbar$ regulates the relative strength of kinetic and potential energy. In the same spirit, for the classical Toda chain we could have introduced a mass parameter, which however can be absorbed through an appropriate  rescaling of  spacetime.  The classical chain has no free model parameter. However, quantum mechanically $\hbar$ cannot be scaled and must be maintained as relevant parameter. The semi-classical limit corresponds to $\hbar \to 0$. Another common choice is to explicitly introduce an interaction strength, $\eta$,
and the inverse decay length of the potential through $\eta\,\mathrm{e}^{-\gamma x}$. Then the standard form \eqref{14.2} is recovered upon replacing $\hbar^2$ by $\hbar^2\gamma^2/\eta$.

For the harmonic lattice, i.e. $X_j = \tfrac{1}{2}\omega_0^2(x_j - x_{j+1})^2$, one usually introduces creation and annihilation operators, satisfying $[a_i,a_j^*] = \delta_{ij}$. The hamiltonian  is then quadratic in $\{a_i,a_j^*, i,j = 1,\ldots,N\}$ and describes bosonic excitations of the ground state. Formally such a transformation can be
implemented also for the Toda lattice, leading to a nonlinear interaction between bosons. For our purposes this representation does not seem to be so useful.

Our strategy is to follow the trail laid out by the classical model. With some confidence, we will arrive at the appropriate hydrodynamic Euler equations. 
As might  have been anticipated there is an evident similarity with the Lieb-Liniger model.
\subsection{Integrability, monodromy matrix}
\label{sec14.1}
While there are examples for which integrability is maintained under quantization, no guarantee can be issued. Thus our first task is to establish $N$ local conservation laws.  The elegant approach relies on the monodromy matrix, which is familiar from other integrable many-body systems.  Let us start with the classical chain governed by the hamiltonian \eqref{14.2} and its associated Lax matrix $L_N$,
see Eq. \eqref{2.12}. The characteristic polynomial of $L_N$ is given by
\begin{equation}\label{14.3}
\det(\lambda - L_N) = \lambda^N\Big(\sum_{m=0}^{N-1}  (-\lambda)^{-m} I_m  + (-1)^N(\lambda)^{-N} (I_N- 2)\Big),
\end{equation}
$I_0 = 1$. The coefficients $I_m$ are the \textit{H\'{e}non invariants} which are in involution, i.e. $\{I_m,I_n\} = 0$. Since the hamiltonian 
is proportional to $I_1^2 - I_2$, the  H\'{e}non invariants are in fact conserved. To work out the details of such an expansion, a different expression 
for the determinant is convenient. We first define the $2\times 2$ matrix
\begin{equation}\label{14.4}
L_j(\lambda) = 
\begin{pmatrix}
\lambda - p_j & \mathrm{e}^{q_j}\\
-  \mathrm{e}^{-q_j}& 0
 \end{pmatrix}
\end{equation}
and, as product of $2\times 2$ matrices, the \textit{monodromy matrix}
\begin{equation}\label{14.5} 
T_N(\lambda) = L_1(\lambda) \cdots L_N(\lambda).
\end{equation}
The parameter $\lambda$ is the \textit{spectral parameter}. To emphasize the distinction sometimes $L_N$ is called the big Lax matrix and $L_j(\lambda)$ the little Lax matrix. While $L_N$ is 
tridiagonal, the little Lax matrix operates in $\mathbb{C}^2$. However the trace yields again the characteristic polynomial
\begin{equation}\label{14.6} 
\mathrm{tr}[T_N(\lambda)] = \det(\lambda - L_N) + 2(-1)^N.
\end{equation}

For the expansion in $\lambda$, one merely has to successively differentiate the matrix product in \eqref{14.5}, which yields
\begin{equation}\label{14.7} 
I_m = \sum_{\{k+2\ell = m\}} p_{i_1}p_{i_2}\cdots p_{i_k}(-X_{j_1})\cdots (-X_{j_\ell}),
\end{equation}
where the sum is over all strings of indices $i_1,i_2,\ldots,i_k, j_1,j_1+1,\ldots, j_\ell ,j_\ell +1$. The  indices are distinct and satisfy  
the constraint $ m = k +2\ell$, $k\geq 0,\ell\geq0$. Two summands differing in the order of indices are counted only once.
More visually, one considers a ring of  $N$ sites and has available $k$ singletons and $\ell$ nearest neighbor pairs, the dominos. The ring is partially covered by singletons and dominos with no overlap allowed, which then defines the string  of indices in \eqref{14.7}. As an example, for $N=4$ one obtains
\begin{eqnarray}\label{14.8} 
&&\hspace{-20pt}I_1 = p_1+p_2+p_3 +p_4, \,\,
I_2 = p_1 p_2 + p_1p_3 + p_1p_4 + p_2p_3 + p_2p_4+ p_3p_4 - X_1 - X_2 -X_3 - X_4,\nonumber\\
&&\hspace{-20pt}I_3 = p_1p_2p_3 + p_1p_2p_4 + p_1p_3 p_4 + p_2 p_3 p_4\nonumber\\
&&\hspace{44pt} - p_1(X_2 +X_3) - p_2(X_3  +X_4) - p_3(X_4 + X_1)  - p_4(X_1 +X_2),\nonumber\\
&&\hspace{-20pt}I_4 = p_1p_2p_3p_4 - p_1p_2X_3 - p_2p_3X_4 - p_3p_4X_1 -  p_4p_1X_2  + X_1X_3  +X_2X_4. 
\end{eqnarray}
Except for $m=1$, the $I_m$'s are not local and have no meaningful density in the limit $N\to \infty$. 
To convert the H\'{e}non invariants into a local form we use the identity
\begin{equation}\label{14.9} 
\log \det(1 - \lambda^{-1}L_N) = - \sum_{m=1}^\infty\frac{1}{m} \lambda^{-m}\mathrm{tr}[(L_N)^m] =  - \sum_{m=1}^\infty\frac{1}{m} \lambda^{-m}Q^{[m], N}.
\end{equation}
As we know already from  Section \ref{sec2.1}, the $Q^{[m],N}$'s have indeed a local density.

In the quantum setting, $[p_j, X_m] = 0$, except for $j = m,m+1$. Such terms do not appear
in the sum \eqref{14.7} and no ambiguity in the operator ordering arises for the quantized version of the $I_m$'s. 
This property strongly suggests that they will
continue to be conserved quantities quantum mechanically.

Switching to the \textit{quantum} Toda chain, there seems to be no analogue of the Lax pair equation \eqref{2.13}.
However, as first observed by Sklyanin in 1985, the definition of monodromy matrix stays intact. As common usage, we switch from $\lambda$ to $u$ as spectral parameter. The entries of the matrix $L_j(u)$ become now operators. 
As a consequence the monodromy matrix $T_N(u)$, still defined as in \eqref{14.5}, is a $2\times2$ matrix with operator entries.  We introduce the $R$-matrix acting on $\mathbb{C}^2\otimes\mathbb{C}^2$ by 
\begin{equation}\label{14.10} 
R(u) = u1\hspace{-3pt}\mathrm{l}\otimes 1\hspace{-3pt}\mathrm{l} -\mathrm{i}\hbar \mathsfit{P},
\end{equation}
where $\mathsfit{P}$ permutes the indices 1 and 2, i.e. $\mathsfit{P} \varphi_1\otimes \varphi_2 = \varphi_2\otimes \varphi_1$.
Noting the commutation relations
\begin{equation}\label{14.11} 
[p_j,X_j] = -\mathrm{i} \hbar X_j, \quad [p_{j+1},X_j] = \mathrm{i} \hbar X_{j},
\end{equation}
one confirms that 
\begin{equation}\label{14.12}
R(u-v) (L_j(u)\otimes 1\hspace{-3pt}\mathrm{l})( 1\hspace{-3pt}\mathrm{l} \otimes L_j (v)) =  (1\hspace{-3pt}\mathrm{l} \otimes L_j (v))( L_j(u)\otimes 1\hspace{-3pt}\mathrm{l}) R(u-v)
\end{equation}
and therefore also for the product
\begin{equation}\label{14.13}
R(u-v) (T_N(u)\otimes 1\hspace{-3pt}\mathrm{l})( 1\hspace{-3pt}\mathrm{l} \otimes T_N(v)) =  (1\hspace{-3pt}\mathrm{l} \otimes T_N(v))( T_N(u)\otimes 1\hspace{-3pt}\mathrm{l}) R(u-v).
\end{equation}
Setting $\hat{t}_N(u) = \mathrm{tr}[T_N(u)]$, the trace of a  $2\times 2$-matrix, one obtains the commutation relations 
 \begin{equation}\label{14.14} 
 \hat{t}_N(u) \hat{t}_N(v) = \hat{t}_N(v)\hat{t}_N(u)
\end{equation}
valid for arbitrary $u,v$. $\hat{t}_N(u)$ is a polynomial of degree $N$, with operator-valued coefficients, the operators acting on $\mathcal{H}_N$. On abstract grounds, the relation \eqref{14.14} ensures that also the Taylor coefficients of $\hat{t}_N(u)$ commute with each other. 

In fact, the  family of operators can be worked out more concretely. Let us define 
\begin{equation}\label{14.15} 
F_N(u) =  \sum_{\{k+2\ell = N\}} (u-p_{i_1})(u-p_{i_2})\cdots (u- p_{i_k})(-X_{j_1})\cdots (-X_{j_{\ell}}),
\end{equation}
where the string of indices satisfies the conditions listed below \eqref{14.7}. We split as
\begin{equation}\label{14.16} 
F_N(u) =  F_N^{\diamond}(u) + F_N^{\diamond\diamond}(u),
\end{equation}
where $F_N^{\diamond\diamond}(u)$ collects all summands containing the factor $\mathrm{e}^{x_N}$.
Then, by an induction argument, 
\begin{equation}\label{14.17} 
T_N(u) =
\begin{pmatrix}
 F_N^{\diamond}(u) & \mathrm{e}^{x_N} F_{N-1}^{\diamond}(u)\\
\mathrm{e}^{-x_{N+1}}F_{N+1}^{\diamond\diamond}(u)& F_N^{\diamond\diamond}(u)
 \end{pmatrix}.
\end{equation}
In particular
\begin{equation}\label{14.18} 
 \hat{t}_N(u) = F_N(u) = u^N \sum_{m =0}^N (-1)^m u^{-m} \hat{I}_m,
\end{equation}
where $\hat{I}_m$ is the quantization of $I_m$. The $\hat{I}_m$-operators are nonlocal. To switch to local fields we follow the recipe
from the classical scheme.
Since $\{\hat{I}_1,\ldots,\hat{I}_N\}$ is  a  family of commuting operators, one can still expand  $\log(u^{-N}\hat{t}_N(u))$ at $u = \infty$ as 
\begin{equation}\label{14.19} 
-\log\big(u^{-N} \hat{t}(u) \big) = u^{-1} \hat{Q}^{[1],N} + \tfrac{1}{2}u^{-2} \hat{Q}^{[2],N}\ldots + \tfrac{1}{N}u^{-N} \hat{Q}^{[N],N} + \mathcal{O}\big(u^{-(N+1)}\big),
\end{equation}
which defines the quantized Flaschka invariants. By construction, $\hat{Q}^{[m],N}$ is a polynomial in $\hat{I}_1,\ldots,\hat{I}_m$ of maximal order $m$, which coincides with the corresponding 
classical identity.  Thus $\{ \hat{Q}^{[1],N},...,\hat{Q}^{[N],N}\}$ is a  family of commuting operators, also commuting with the
$\hat{I}_1,\ldots,\hat{I}_N$. For the hamiltonian
\begin{equation}\label{14.20} 
H_{\mathrm{qt},N} =  \tfrac{1}{2} \hat{Q}^{[2],N} = \tfrac{1}{2}\hat{I}_1^2 - \hat{I}_2,
 \end{equation}
 which implies that  the $\hat{I}_m$'s and $\hat{Q}^{[m],N}$'s are conserved.  
 
 For the particular case $N=4$ the result is 
\begin{eqnarray}\label{14.21} 
&&\hspace{-30pt}\hat{Q}^{[1],4} =  \hat{I}_1,\quad  \hat{Q}^{[2],4} = \hat{I}_1^2 - 2  \hat{I}_2, \quad  \hat{Q}^{[3],4} =   \hat{I}_1^3 - 3  \hat{I}_1\hat{I}_2 + 3  \hat{I}_3,\nonumber\\
&&\hspace{-30pt}\hat{Q}^{[4],4} =   \hat{I}_1^4 - 4 \hat{I}_1^2\hat{I}_2 + 2 \hat{I}_2^2 + 4  \hat{I}_1 \hat{I}_3 - 4 \hat{I}_4.
\end{eqnarray}
Expanding the powers the corresponding charge densities are 
\begin{eqnarray}\label{14.22} 
&&\hspace{-30pt}\hat{Q}_j^{[1]} = p_j,  \quad \hat{Q}_j^{[2]} = p_j^2 + 2X_j, \quad
\hat{Q}_j^{[3]}=   p_j^3 + 3(p_j+p_{j+1})X_j,\\
&&\hspace{-30pt}\hat{Q}_j^{[4]} =   p_j^4+ 2(p_j^2 +p_jp_{j+1} +p_{j+1}^2)X_j + 2X_j (p_j^2 +p_jp_{j+1} +p_{j+1}^2) + 2 X_j^2 +4X_jX_{j+1},
\nonumber
\end{eqnarray}
which are local, as anticipated. In fact obtained is the minimal version of these densities, compare with the discussion above \eqref{2.27}. The operator ordering is in force only starting from $\hat{Q}^{[4],N}$ onwards.

More generally, one can argue that $\hat{Q}^{[m],N}$ must be a particular operator ordering of $Q^{[m],N}$. If so, locality would follow from the fact that $Q^{[m],N}$ has a local density. Actually, the precise functional form of the higher order charges is fairly irrelevant for our purposes. As crucial information, the density of the $m$-th charge depends on local operators in a block of size at most $(m/2) +1$.
Since the hamiltonian is nearest neighbor, this property implies that the $m$-th current  density operator depends on a block of size at most $(m/2) +2$.
\subsection{Spectral properties}
\label{sec14.2}
Since the interaction depends only on relative positions, an eigenfunction of  $H_{\mathrm{qt},N}$ must be of the form
\begin{equation}\label{14.23} 
\psi_{E}(x_1,\ldots,x_N) =\tilde{\psi}_{E_\perp}(x_2-x_1,\ldots, x_N - x_{N-1}) \mathrm{e}^{\mathrm{i}\hbar^{-1}E_1N^{-1}(x_1 +\ldots+x_N)},
\end{equation}
where $E = (E_1,\ldots,E_N) = (E_1,E_\perp)$, implying $\hat{I}_1 \psi_{E} = E_1 \psi_{E}$. More abstractly one can think of a fiber decomposition of $H_{\mathrm{qt},N}$ with respect to the total momentum. $E_1$ would be then the corresponding fiber parameter. In terms of the hamiltonian one explicitly splits into kinetic energy of the  center of mass plus relative internal motion,
\begin{equation}\label{14.24} 
H_{\mathrm{qt},N} = H_\mathrm{cm} + H_{\mathrm{rel},N-1}.
\end{equation}
Thus $\tilde{\psi}_{E_\perp} \in L^2(\mathbb{R})^{\otimes {N-1}}$ is an eigenfunction of $H_{\mathrm{rel},N-1}$.
In fact, working out $H_{\mathrm{rel},N-1}$, one finds an interaction potential increasing exponentially in all directions. The eigenvalues of $H_{\mathrm{rel},N-1}$ are isolated,
even nondegenerate, and the thermal operator $\exp(- H_{\mathrm{rel},N-1})$ is trace class. 

It is of great advantage to consider simultaneously all eigenvalues of the H\'{e}non invariants, to say
\begin{equation}\label{14.25} 
\hat{I}_m \psi_{E} = E_m \psi_{E}, 
\end{equation}
$m = 1,\ldots,N$,
with $E_1$ as in \eqref{14.23}. There is an explicit formula for the joint eigenfunctions and it is known that they are complete in the sense that linear combinations of the set $\{\tilde{\psi}_{E_\perp} \}$ span the entire
Hilbert space $L^2(\mathbb{R}^{N-1})$.  Then $\psi_{E}$ is also an eigenfunction of the transfer matrix $\hat{t}(u)$ with eigenvalue
$\tau_N(u,E)$, 
\begin{equation}\label{14.26} 
\hat{t}_N(u)\psi_{E} = \tau_N(u,E)\psi_{E},
\end{equation}
where 
\begin{equation}\label{14.27} 
\tau_N(u,E) = u^N \Big(1 +  \sum_{m =1}^N (-1)^m u^{-m} E_m\Big)
\end{equation}
The $m$-th Flaschka invariant has eigenvalue $q_m$, 
\begin{equation}\label{14.28} 
\hat{Q}^{[m],N}\psi_{E}  = q_m\psi_{E}. 
\end{equation}
Since the operators commute, the $q_m$'s are determined by 
\begin{equation}\label{14.29} 
-\log\big(u^{-N} \tau_N(u,E) \big) = u^{-1} q_1 + \tfrac{1}{2}u^{-2} q_2 + \ldots + \tfrac{1}{N}u^{-N} q_N+ \mathcal{O}\big(u^{-(N+1)}\big).
\end{equation}
We now write
\begin{equation}\label{14.30} 
\tau_N(u) = \prod_{j=1}^N(u- \tau_j).
\end{equation}
Then 
\begin{equation}\label{14.31} 
-\log\big(u^{-N} \tau_N(u)\big) = -\sum_{j=1}^N \log\big(1- u^{-1}\tau_j\big) = \sum_{m=1}^N u^{-m} \sum_{j=1}^N\tfrac{1}{m} (\tau_j)^m + \mathcal{O}\big(u^{-(N+1)}\big)
\end{equation}
and
\begin{equation}\label{14.32} 
q_m = \sum_{j=1}^N (\tau_j)^m,
\end{equation}
$ m = 1,\ldots,N$.
As already familiar, thereby one introduces an empirical density through
\begin{equation}\label{14.33} 
\varrho_N(w)  = \frac{1}{N} \sum_{j=1}^N \delta(\tau_j - w)
\end{equation}
with the property that 
\begin{equation}\label{14.34} 
N^{-1} q_m = \int_\mathbb{R} \mathrm{d}w \varrho_N(w) w^m.
\end{equation}

Before continuing, we have to recall the quantum mechanical analog of the scattering shift studied in Section \ref{sec2.3.2}. The relative motion of two Toda particles is governed by
the Schr\"{o}dinger equation
\begin{equation}\label{14.35} 
\mathrm{i}\partial_t \psi_t(x) = \big(-\hbar^2\partial_x^2  + \mathrm{e}^{-x}\big)\psi_t(x).
\end{equation}
The particle representing the relative motion travels inwards from the far right with momentum $\hbar k_\mathrm{in}$, gets reflected at the potential barrier, and moves outwards with momentum
$\hbar k_\mathrm{out} = - \hbar k_\mathrm{in}$.  The phase shift accumulated during the scattering process is given by 
\begin{equation}\label{14.36} 
\theta_\mathrm{qt}(k) = k \log \hbar^2 -\mathrm{i} \log \frac{\Gamma(1 +\mathrm{i}k)}{\Gamma(1 -\mathrm{i}k)}.
\end{equation}
Surprisingly, the phase shift is independent of $\hbar$, except for the additive constant $k \log \hbar^2$.
The scattering shift is then
\begin{equation}\label{14.37} 
\theta'_\mathrm{qt}(k) =  \log \hbar^2 + \psi_\mathrm{di}(1 +\mathrm{i}k) +\psi_\mathrm{di}(1 -\mathrm{i}k) = \phi_\mathrm{qt}(k)
\end{equation}
with Digamma function $\psi_\mathrm{di} = \Gamma'/\Gamma$. From the convergent power series,
\begin{equation}\label{14.38} 
\psi_\mathrm{di}(1 +\mathrm{i}k) +\psi_\mathrm{di}(1 -\mathrm{i}k) = -\gamma_\mathrm{E} + \sum_{n = 1}^\infty \frac{k^2}{n(n^2 +k^2)},
\end{equation}
$\gamma_\mathrm{E} = 0.577\ldots$ the Euler-Mascheroni number, one concludes that $\phi_\mathrm{qt}$ has a positive curvature at $k=0$. The large $k$ behavior can be inferred from the expansion
\begin{equation}\label{14.39} 
\psi_\mathrm{di}(1 +\mathrm{i}k) +\psi_\mathrm{di}(1 -\mathrm{i}k) = \log(1+k^2) - \frac{1}{1+k^2} +     \mathcal{O}(k^{-4}).     
\end{equation}
For the semiclassical limit we set $\hbar k = w$ and then
\begin{equation}\label{14.40} 
\lim_{\hbar \to 0} \phi_\mathrm{qt}(\hbar^{-1}w )   = \log w^2 = \phi_\mathrm{to}(w),  
\end{equation}
as it should be.

In 2010 Kozlowski and Teschner obtained a remarkable result on the spectral properties of the chain, which comes handy now.
They fix the eigenvalue $E_1$ and establish that for each vector $(\tau_1,\ldots,\tau_N)$ there is
another vector $(\delta_1,\ldots,\delta_N)$ such that 
\begin{equation}\label{14.41} 
 \sum_{j=1}^N (\tau_j)^m =  \sum_{j=1}^N (\hbar \delta_j)^m  + \mathcal{O}(\mathrm{e}^{-N})
\end{equation}
for $m = 1,\ldots,N$. In addition, for every ordered set $\{n_1<\ldots<n_N, n_j \in \mathbb{Z}, j = 1,\ldots,N\}$, the roots  $(\delta_1,\dots,\delta_N)$ satisfy the identity
\begin{equation}\label{14.42} 
2 \pi n_j = N\delta_j\log \hbar^2 +\mathrm{i} \log \zeta - \sum_{i=1}^N   \mathrm{i} \log \frac{\Gamma(1 +\mathrm{i}(\delta_j - \delta_i))}{\Gamma(1 -\mathrm{i}(\delta_j - \delta_i))}
+ \mathcal{O}(\mathrm{e}^{-N}).
\end{equation}
This expression has already the flavor of \eqref{13.14}  for the $\delta$-Bose gas, except for the model dependent phase shift, of course.
For the $\delta$-Bose gas \eqref{13.14} and \eqref {13.17} are a strict identities, no error term, whereas for the quantum Toda lattice,
\eqref{14.41} and \eqref{14.42} hold approximately for large $N$. 

Actually, for both error terms there is an explicit identity. So far, the exponential bound $\mathcal{O}(\mathrm{e}^{-N})$  has been established only for particular cases.  
\subsection{GGE and hydrodynamics}
\label{sec14.3}
Given the input from \eqref{14.42} and assuming $n_1 <\ldots<n_N$, with small modifications one can follow the Yang-Yang scheme from the $\delta$-Bose gas.
By definition the parameter $\zeta$ has unit length, $|\zeta| = 1$, and hence drops out as $N \to \infty$.
The empirical density \eqref{14.33} satisfies the constraints
\begin{equation}\label{14.43} 
\varrho_N(w) \geq 0,\quad \int_\mathbb{R}\mathrm{d}w\varrho_N(w) =1, \quad \int_\mathbb{R}\mathrm{d}w\varrho_N(w)w = N^{-1} E_1 =e_1.
\end{equation}
Omitting both error terms, Eq. \eqref{14.42} becomes
\begin{equation}\label{14.44} 
\frac{1}{N} 2\pi n_j =  \frac{1}{N}\sum_{i=1}^N \theta_\mathrm{qt}(\delta_j - \delta_i) +e_1.
\end{equation}

The GGE is defined through the weight
\begin{equation}\label{14.45} 
\exp\Big[- \sum_{j=1}^N V(\delta_j)\Big]
\end{equation}
with confining potential $V$. Following the analysis of the $\delta$-Bose gas, for quantum Toda the free energy functional  per particle  reads
\begin{equation}\label{14.46}
\mathcal{F}_\mathrm{qt}^\circ (\varrho)= \int_\mathbb{R}\mathrm{d}w \big( \varrho V +\varrho \log \varrho +  \varrho_\mathsf{h} \log \varrho_\mathsf{h} - 
(\varrho +  \varrho_\mathsf{h})\log (\varrho +  \varrho_\mathsf{h})\big).
\end{equation}
This functional has to be minimized under two constraints. Firstly, number and momentum,
 \begin{equation}\label{14.47}
 \varrho \geq 0,\quad \int_\mathbb{R}\mathrm{d}w\varrho(w) =1,\quad \int_\mathbb{R}\mathrm{d}w\varrho(w)w = e_1.
\end{equation}
The second constraint comes from Eq. \eqref{14.44}. 
Defining the operator $T$ through the convolution with the quantum Toda scattering shift, 
\begin{equation}\label{14.48} 
Tf(w) =  \frac{1}{2\pi}  \int_\mathbb{R}\mathrm{d}w' \phi_\mathrm{qt}(w - w')f(w'),
\end{equation}
the additional constraint  can be written as
\begin{equation}\label{14.49} 
\varrho + \varrho_\mathsf{h} =  T\varrho. 
\end{equation} 
The constraint on the average momentum can be absorbed into a chemical potential $\mu_1$, thereby adjusting the confining potential $V(w)$ to
$V(w) -\mu_1 w$. 

The variational problem appears to be identical to the one of the $\delta$-Bose gas with volume $\ell=0$, compare with \eqref{13.45}. This should come as no surprise,
since we used periodic boundary conditions with zero tilt. In this case the stretch vanishes by construction. For the classical Toda lattice 
with this particular boundary conditions, the positions have the statistics of a random walk with step size of order $1$ and zero bias, hence a fluctuating volume of order $\sqrt{N}$.
For the classical chain, on top of the Flaschka invariants based on the Lax matrix, also stretch is conserved, respectively the particle density in the fluid picture. But so far we have not identified a corresponding control parameter in the quantum model. 
Still, physically the stretch must be included in a proper hydrodynamic description. Clearly,  one option is the cell hamiltonian $H_{\mathrm{cell},N}$, see Eq. \eqref{2.10}. Since this hamiltonian is written in canonical coordinates, the conventional quantization can be used which yields 
\begin{equation}\label{14.50} 
H_{\mathrm{cell},N, {\nu}} = -\sum_{j=1}^{N} \tfrac{1}{2} \hbar^2 \partial_{x_j}^2 +   \sum_{j=1}^{N-1} X_j + \mathrm{e}^{-\nu N}   X_N. 
\end{equation}

Before adopting this choice, we have to ensure that the tilt is properly imposed. For this purpose
we introduce the unitary
transformation 
\begin{equation}\label{14.51} 
U\psi(x_1,\ldots,x_N) = \psi(x_1-\nu,\ldots,x_N -N\nu)
\end{equation} 
thereby obtaining
\begin{equation}\label{14.52} 
U^*p_jU = p_j, \qquad U^*X_jU = \mathrm{e}^{-\nu}X_j,\qquad U^*X_NU = \mathrm{e}^{\nu (N-1)}X_N,
\end{equation} 
and 
\begin{equation}\label{14.53}
U^*H_{\mathrm{cell},N,\nu}U = H_{\mathrm{qt},N,\nu} = -\sum_{j=1}^{N} \tfrac{1}{2} \hbar^2 \partial_{x_j}^2 +  \sum_{j=1}^{N} \mathrm{e}^{-\nu}  X_j.
\end{equation}
Note that $H_{\mathrm{qt},N,0} = H_{\mathrm{qt},N}$. The boundary coupling can be spread homogeneously over the entire lattice. Because of invariance under spatial translations, the thermal operator $\exp(- H_{\mathrm{cell},N,\nu})$  is not of trace class. However for observables depending only on the internal coordinates, thermal average is well-defined and denoted by $\langle \cdot \rangle_{N,\nu}$.
Using unitary equivalence one concludes
\begin{equation}\label{14.54} 
\langle (x_{j+1} - x_j) \rangle_{N,\nu} = \nu, \quad j = 1,\ldots,N-1, \quad \langle (x_{1} - x_N) \rangle_{N,\nu} = -(N-1)\nu.
\end{equation}
At the expense of a single huge jump one indeed has induced a non-zero, constant stretch.

While convincing, one has to monitor how the higher order invariants are affected by such a modification. 
We set 
\begin{equation}\label{14.55}
M_{\nu} = 
\begin{pmatrix}
1 & 0\\
0 & \mathrm{e}^{-\nu N}
 \end{pmatrix}.
\end{equation}
Then the modified monodromy matrix reads 
\begin{equation}\label{14.56} 
T_{N,\nu} = M_\nu T_N(u)
\end{equation} 
and
\begin{equation}\label{14.57} 
\hat{t}_{N,\nu} (u) = F_N^{\diamond}(u) +  \mathrm{e}^{-\nu N}
F_N^{\diamond\diamond}(u).
\end{equation} 
Thus the H\'{e}non invariants of the modified hamiltonian are still given by \eqref{14.18}. Only, 
$X_N$ is substituted everywhere by  $\mathrm{e}^{-\nu N}X_N$.
The same rule transcribes to the Flaschka invariants. Denoting the Flaschka invariants of   $H_{\mathrm{cell},N, \nu}$ by $\hat{Q}^{[m],N}_{\mathrm{cell},\nu}$ and those of  
$H_{\mathrm{qt}, N,\nu}$ by $\hat{Q}^{[m],N}_{\mathrm{qt},\nu}$, one arrives at 
\begin{equation}\label{14.58} 
U^*\hat{Q}^{[m],N}_{\mathrm{cell},\nu} U = \hat{Q}^{[m],N}_{\mathrm{qt},\nu}.
\end{equation} 

This equation teaches us that also under a generalized Gibbs ensemble for $H_{\mathrm{cell},N,\nu}$, the average stretch equals $\nu$, except for the jump of $-\nu(N-1)$ at the bond $(N,1)$. Secondly, the generalized free energy can be computed from the Flaschka invariants of the lattice with coupling strength $\mathrm{e}^{-\nu}$. By scaling this means that in \eqref{14.36} the term $\log \hbar^2$ is replaced by 
$\log (\hbar^2 \mathrm{e}^{\nu})$. Thus, combining with the arguments leading to \eqref{14.44}, the stretch modifies this relation to 
\begin{equation}\label{14.59} 
 2\pi n_j =  \nu N\delta_j  + \sum_{i=1}^N \theta_\mathrm{qt}(\delta_j - \delta_i) +N e_1.
\end{equation}
Up to the additional term coming from the fixed total momentum, we have obtained perfect analogy to the $\delta$-Bose gas,
compare with Eq. \eqref{13.14}.
However, in contrast to the $\delta$-Bose gas, the stretch parameter $\nu N$ can be negative, just as for the classical chain.

In our argument for the average currents of the $\delta$-Bose gas we never used the specific form of the scattering shift. For the quantum Toda lattice
the density is conserved and its current is the momentum, which itself is conserved. No extra work is required. To conclude, in the fluid picture the hydrodynamic equations of the quantum Toda lattice read 
\begin{equation}
\label{14.60} 
\partial_t\rho_\mathsf{p}(x,t;v) + \partial_x\big(v^\mathrm{eff}(x,t;v)\rho_\mathsf{p}(x,t;v)\big) = 0.
\end{equation}
The effective velocity is again determined by the ratio $\varsigma_1^\mathrm{dr}/
\varsigma_0^\mathrm{dr}$ as in \eqref{13.69}. Of course, the appropriate dressing operator is the one of the quantum Toda lattice.
\bigskip
\begin{center}
 \textbf{Notes and references}
 \end{center}
\begin{center}
\textbf{Section 14.0}
\end{center}
Sutherland \cite{S78} first discussed the integrability of the quantum Toda lattice, using a low density
 approximation of the integrable quantum fluid with the $1/\sinh^2$ interaction potential. He developed the method of asymptotic Bethe ansatz, which uses properties of the eigenfunctions when particles are far apart, see Sutherland \cite{S95}. Finite volume has been introduced ad hoc 
 by imposing periodic boundary conditions on the wave numbers. Sutherland studied TBA for the ground state. In a series of papers Mertens and his group,  Theodorakopoulos and  Mertens  \cite{TM83},  Mertens    \cite{M84}, Opper  \cite{O85}, Hader and  Mertens \cite{HM86}, Gruner-Bauer and Mertens \cite{GM88}, see also Theodorakopoulos \cite{T84a}, Takayama and Ishikawa \cite{TI86}, investigated TBA at non-zero temperatures and also the semi-classical classical limit. At the time, only thermal equilibrium was in focus. A most useful overview is provided by 
 Siddharthan  and Shastry \cite{SS97}, see also Shastry and  Young \cite{SY10}.
The review by Olshanetsky and Perelomov \cite{OP83} is an early summary on the more mathematical activities in the area
 of quantum integrable systems. Highly recommended is the instructive  book on Beautiful Models by Sutherland \cite{S04}. 
    
 Gutzwiller \cite{G80,G81} studied the spectrum for few, $N=3,4$, particles, later being extended to an arbitrary number, see  Gaudin and Pasquier  \cite{GP92}. 
 Sklyanin \cite{S85}
 achieved a crucial advance by linking the Toda lattice to the flourishing field of quantum integrable systems, with notions as Yang-Baxter equations and monodromy matrix. He developed the method of separation of variables, introducing the Baxter equation  in this context. 
 He also discussed the spectrum of eigenvalues in the limit of large $N$ thereby supporting the more heuristic results in Sutherland  \cite{S78}.  
 Following a proposal by Nekrasov and Shatashvili  \cite{SN10}, Kozlowski  and Teschner  \cite{KT10}  substantially advanced such methods. \bigskip
 \begin{center}
  \textbf{Section 14.1}
\end{center}
Our discussion is based on the contributions of Sklyanin \cite{S85} and Siddharthan  and Shastry \cite{SS97}, see also L\"{u}scher \cite{L76}. Identities as \eqref{14.9} are proved in Reed and  Simon  \cite{RS79},
Section 17.\bigskip
\begin{center}
\textbf{Section 14.2}
\end{center}
 A fairly explicit formula  for the eigenfunctions of the Toda lattice have been obtained by Kharchev and Lebedev \cite{KL99,KL01}. Their completeness is proved
 by  An \cite{A09}.  
 The results \eqref{14.41} and \eqref{14.42} are quoted from Kozlowski  and Teschner  \cite{KT10}. The reader is invited to look up the precise statements. Further advances are reported by Kozlowski \cite{K15}. A more detailed account is his habilitation thesis Kozlowski  \cite{K15a}.
\bigskip
 \begin{center}
 \textbf{Section 14.3}
\end{center}
The kernel of the $T$ operator of the $\delta$-Bose gas decays to $0$ for $|w| \to \infty$, while for the Toda lattice the kernel diverges logarithmically. Hence one has to check whether the techniques from Dorlas et al.  \cite{DLP93} still carry through.
Sklyanin \cite{S85} introduced already the parameter which we call $\nu$ and established that $\nu$ appears in the   ground state TBA as the chemical potential dual to the density. The analysis of Kozlowski  and Teschner  \cite{KT10} starts with the hamiltonian $H_{\mathrm{qt},N,\nu}$  from the outset.
 The parameter $\nu$ then naturally appears  at the correct location. The short-cut discussed in the main text seems to be new. 
 
 Comparing \eqref{14.59} with \eqref{13.14} complete correspondence is noted. However, $\theta_\mathrm{li}$ is strictly increasing,
 whereas  $\theta_\mathrm{qt}$ has a decreasing and increasing branch. Thus, uniqueness of solutions 
 might be a further difficulty for an asymptotic analysis.
 \newpage
 \section{Beyond the Euler time scale}
\label{sec15}
\setcounter{equation}{0} 
As a common experience, sound waves in air propagate ballistically. When air is confined to a thin tube, an initial perturbation maintains its shape and travels with the speed of sound. There could be dispersive effects due to a nonlinear energy-momentum relation.  The respective dynamics would be still time-reversible, no entropy is produced.  But on longer time scales irreversible sound damping sets in, a phenomenon which can be traced back to molecular disorder.    
Very commonly, damping is modelled by a multiplicative diffusive factor as $\exp[-Dk^2|t|]$ in wave number space,
$D$ being the diffusion coefficient. On the level of hydrodynamic equations, dissipation  amounts to the Navier-Stokes correction of Euler equations.

It is not at all obvious whether such a conventional picture extends to integrable systems. Again, the hard rod fluid serves as a convenient guiding example. Restricting to ballistic spacetime scales, a first step is to consider the motion of a tracer quasiparticle when the fluid 
is initially in some GGE.  To lowest order the tracer  acquires an effective velocity, $v^\mathrm{eff}$, through collisions with fluid quasiparticles. 
 The effective velocity arises from
summing  over many approximately independent collisions, thereby leading to a law of large numbers.
As well known from the theory of independent random variables, a more precise description would be  a central limit theorem
of the form
\begin{equation}\label{15.1} 
q_\mathrm{tr}(t) \simeq  q_\mathrm{tr}(0) +v^\mathrm{eff} t + (D_\mathrm{tr})^{1/2} b(t),
 \end{equation}
where $q_\mathrm{tr}(t)$ is the position of the tracer quasiparticle, $b(t)$ a standard Brownian motion, and  $D_\mathrm{tr}$  the diffusion constant,  which depends on the particular GGE.
For a generic integrable system approximate statistical independence is less granted. But, as long as the notion of a tracer quasiparticle is defined, 
the asymptotics \eqref{15.1} is expected to be valid. Connecting such a motion to the conserved fields leads to the  Navier-Stokes correction, second order in $\partial_x$, whose analytical form still has to be worked out. To achieve such goal we follow a standard strategy. 
The initial attempt is to find out diffusive corrections for the propagation of a small perturbation away from the spatially homogeneous   GGE. 
This will be soft mathematics, merely providing  a general framework and  even not yet distinguishing between integrable and nonintegrable systems. 
The result is then written in a form, for which the step from global GGE to local GGE  will be compelling. 
\subsection{General framework}
\label{sec15.1}

Despite the header, we explain the scheme using the Toda lattice as example. The core of the method consists of suitable sum rules that  can be written as well for other integrable models. As discussed in Section \ref{sec7.1}, the basic spacetime field-field correlator is
\begin{equation}\label{15.2} 
S_{m,n} (j,t) = \langle Q^{[m]}_j(t)Q^{[n]}_0(0) \rangle_\mathrm{gg}^\mathrm{c}.
\end{equation}
Here $\langle\cdot \rangle_\mathrm{gg}$ stands for the expectation in some GGE, which remains fixed throughout our discussion. 
By Lieb-Robinson type bounds, for fixed $t$, the correlator $S_{m,n}(j,t) $ decays exponentially outside the sound cone.
So $j$-summations are under control, while the time integration will turn out to be more delicate. To exploit spacetime symmetry,
we redefine 
\begin{equation}\label{15.3} 
Q^{[0]}_j =    \tfrac{1}{2}(r_{j-1} + r_j).
\end{equation}
To avoid too many extra symbols, for this section only, the basic correlator is still denoted by $S$, taking \eqref{15.3} into account. 
\bigskip\\
$\blackdiamond\hspace{-1pt}\blackdiamond$~\textit{Spacetime inversion}.\hspace{1pt}
By spacetime stationarity one concludes
\begin{equation}\label{15.4} 
S_{m,n} (j,t) = S_{n,m} (-j,-t).
\end{equation}
In fact, the stronger property 
\begin{equation}\label{15.5} 
S_{m,n} (j,t) = S_{m,n} (-j,-t)
\end{equation}
holds by using invariance under spacetime inversion.

 We define time reversal through
\begin{equation}\label{15.6}
\mathcal{R}_\mathrm{tr}: (r, p) \mapsto (r,-p)
\end{equation}
and space inversion by
\begin{equation}\label{15.7}
\mathcal{R}_\mathrm{si}: (r, p) \mapsto (\tilde{r},\tilde{p}), \quad \tilde{r}_j = r_{-j-1}, \,\,\tilde{p}_j = - p_{-j}.
\end{equation}
Since $r_j$ should be viewed as bond variable, the inversion is relative to the origin. The equations of motion \eqref{2.2} generate a flow on phase space denoted by $T_t$,  $T_t: (r,p) \mapsto (r(t), p(t))$. Using the flow equations, one concludes that
\begin{equation}\label{15.8}
\mathcal{R}_\mathrm{tr}\circ T_t =  T_{-t} \circ\mathcal{R}_\mathrm{tr}
\end{equation}
and 
\begin{equation}\label{15.9}
 \mathcal{R}_\mathrm{si}\circ T_t = T_{t} \circ\mathcal{R}_\mathrm{si} .
\end{equation}
Spacetime inversion is then defined through
\begin{equation}\label{15.10}
\mathcal{R}_\mathrm{si}\circ\mathcal{R}_\mathrm{tr} = \mathcal{R}: (r, p) \mapsto (\tilde{r},\tilde{p}), \quad \tilde{r}_j = r_{-j-1}, \,\,\tilde{p}_j =  p_{-j}
\end{equation}
and hence
\begin{equation}\label{15.11}
\mathcal{R} \circ T_t =  T_{-t} \circ \mathcal{R}.
\end{equation}
In the random walk summation for $Q^{[n]}_j$, see  \eqref{2.17}, to each path from $j$ to $j$  in $n$ steps
there is a path reflected at level $j$. Hence
\begin{equation}\label{15.12}
Q^{[n]}_j \circ\mathcal{R} = Q^{[n]}_{-j}.
\end{equation}
In particular, the GGE is invariant under $\mathcal{R}$. Combining \eqref{15.11} and \eqref{15.12}, the claim  \eqref{15.5} follows.

For the currents, spacetime inversion is slightly more complicated. We introduce the down- and up-currents
\begin{equation}\label{15.13}
J^{[n]}_{j} = J^{[n]\scriptscriptstyle \downarrow}_{j} = \big(L^n L^{\scriptscriptstyle \downarrow}\big)_{j,j}, \qquad
J^{[n]\scriptscriptstyle \uparrow}_{j} = \big(L^n L^{\scriptscriptstyle \uparrow}\big)_{j,j}
\end{equation}
with $L^{\scriptscriptstyle \uparrow}=  (L^{\scriptscriptstyle \downarrow})^\mathrm{T}$,
see above Eq. \eqref{2.24}. Then
\begin{equation}\label{15.14}
J^{[n]\scriptscriptstyle \downarrow}_{j} \circ\mathcal{R}  = J^{[n]\scriptscriptstyle \uparrow}_{-j}.
\end{equation}
Correspondingly there are two current-current correlators
\begin{equation}\label{15.15}
\Gamma^{\scriptscriptstyle \downarrow(\scriptscriptstyle \uparrow)}_{m,n}  (j,t) =   
\langle J^{[m]\scriptscriptstyle \downarrow(\scriptscriptstyle \uparrow)}_j( t) J^{[n]\scriptscriptstyle \downarrow(\scriptscriptstyle \uparrow)}_0 (0) \rangle_\mathrm{gg}^\mathrm{c}
\end{equation} 
and hence spacetime inversion implies
\begin{equation}\label{15.16} 
\Gamma^{\scriptscriptstyle \downarrow}_{m,n} (j,t)= \Gamma^{\scriptscriptstyle \uparrow}_{m,n}  (-j,-t). 
\end{equation}
The down- and up-currents seem to be a special feature of the Toda lattice. \hfill $\blackdiamond\hspace{-1pt}\blackdiamond$\bigskip\\

Besides invariance under spacetime reversal, the correlator also satisfies
the conservation law
\begin{equation}\label{15.17} 
\partial_t S_{m,n} (j,t) = -\partial_j B_{m,n} (j,t),
\end{equation}
employing the difference operator  $\partial_jf(j) = f(j+1) - f(j)$. $B(j,t)$ is the charge-current spacetime correlator 
\begin{equation}\label{15.18} 
B_{m,n}(j,t) = \langle J^{[m]}_j(t)Q^{[n]}_0(0) \rangle_\mathrm{gg}^\mathrm{c},
\end{equation}
which has been encountered before, compare with \eqref{6.8}. Using both properties, one derives identities, known as sum rules, 
for the lowest moments of $S(t)$.

For the \textit{zeroth moment} one obtains  
\begin{equation}\label{15.19}
\sum_{j \in \mathbb{Z}} S_{m,n} (j,t) =  \sum_{j \in \mathbb{Z}} C_{m,n}(j) = C_{m,n},
\end{equation}
where we used the conservation law \eqref{15.17} and the decay of $S_{m,n}(j,t)$ ensuring boundary terms to vanish.
 $C_{m,n}(j)$ is the static correlator,
\begin{equation}\label{15.20} 
C_{m,n}(j) = S_{m,n} (j,0),
\end{equation}
and  the spatially summed $C_{m,n}(j)$ is the matrix of static susceptibilities, $C_{m,n} = C_{n,m}$.\bigskip\\
\textit{First moment}. By \eqref{15.5}, 
$C_{m,n}(j) = C_{m,n}(-j)$ and the sum rule
\begin{equation}\label{15.21} 
  \sum_{j \in \mathbb{Z}}j C_{m,n}(j) = 0
\end{equation}
follows. In addition, multiplying \eqref{15.17} with $j$ and summing over $j$, one obtains
\begin{equation}\label{15.22} 
\partial_t  \sum_{j \in \mathbb{Z}} jS_{m,n} (j,t) =  \sum_{j \in \mathbb{Z}} B_{m,n} (j,t) = 
\sum_{j \in \mathbb{Z}} B_{m,n} (j,0) = B_{m,n} = (AC)_{m,n} 
\end{equation}
with the flux Jacobian $A$ as encountered before, compare with \eqref{7.14}. Hence 
\begin{equation}\label{15.23} 
 \sum_{j \in \mathbb{Z}} jS_{m,n} (j,t) =  (AC)_{m,n} t. 
\end{equation}
Using time stationarity of $S$ one recovers the symmetry of $B$, equivalently 
\begin{equation}\label{15.23a}
AC = CA^\mathrm{T}.
\end{equation}\\
\textit{Second moment}.  We pick some rapidly decaying test function $f$. Then 
\begin{equation}\label{15.24} 
  \sum_{j \in \mathbb{Z}} f_j\big( Q^{[n]}_j(t) - Q^{[n]}_j(0)\big)  = 
 \int_0^t\mathrm{d}s   \sum_{j \in \mathbb{Z}} ( \partial_j f_j)J^{[n]}_j(s).
\end{equation}
Squaring and using translation invariance,
\begin{eqnarray}\label{15.25} 
&&\hspace{0pt}   \sum_{j \in \mathbb{Z}} f_j   \sum_{j' \in \mathbb{Z}}\tilde{f}_{j'}\langle\big( Q^{[m]}_j(t) - Q^{[m]}_j(0)\big)
\big( Q^{[n]}_{j'}(t) - Q^{[n]}_{j'}(0)\big) \rangle_\mathrm{gg}^\mathrm{c} \nonumber\\
&&\hspace{0pt} =  -\int_0^t\mathrm{d}s  \int_0^t\mathrm{d}s' 
 \sum_{j \in \mathbb{Z}}  \sum_{j' \in \mathbb{Z}} (\partial_j^\mathrm{T}\partial_jf_j)\tilde{f}_{j'}\langle J^{[m]}_j(s)
J^{[n]}_{j'}(s')  \rangle_\mathrm{gg}^\mathrm{c}.
\end{eqnarray}
The central tool for controlling diffusive behavior is the current-current correlator defined by 
\begin{equation}\label{15.26}
\Gamma_{m,n}  (j,t) =   \langle J^{[m]}_j(t)  
J^{[n]}_{0}(0)  \rangle_\mathrm{gg}^\mathrm{c} =\Gamma^{\scriptscriptstyle \downarrow}_{m,n}  (j,t)
\end{equation} 
and the corresponding total current-current correlation
\begin{equation}\label{15.27}
 \sum_{j \in \mathbb{Z}} \Gamma_{m,n}  (j,t) = \Gamma_{m,n}  (t).
\end{equation} 
Note that by \eqref{15.16}, $\Gamma_{m,n}  (t) = 
\Gamma^{\scriptscriptstyle \downarrow}_{m,n} (t)= \Gamma^{\scriptscriptstyle \uparrow}_{m,n}  (-t) = \Gamma_{m,n}  (-t)
$ and thus $\Gamma_{m,n}  (\infty) = 
\Gamma_{m,n}  (-\infty)$. 
Setting $f_j = \tfrac{1}{2} j^2$, $\tilde{f}_j = \delta_{0,j}$ and using spacetime inversion \eqref{15.5}, one concludes that 
\begin{equation}\label{15.28} 
  \sum_{j \in \mathbb{Z}} j^2 \big(S_{m,n} (j,t) - S_{m,n} (j,0)\big)  = \Gamma_{m,n}(\infty) t^2 + \int_0^t\mathrm{d}s  \int_0^t\mathrm{d}s' \big(\Gamma_{m,n}(s - s') -  \Gamma_{m,n}(\infty)\big) .
\end{equation}
$\Gamma_{m,n}(\infty)$ is the \textit{Drude weight}. If $\Gamma_{m,n}(t) -  \Gamma_{m,n}(\infty)$ is integrable, one can define the \textit{Onsager matrix} 
\begin{equation}\label{15.29}
\mathfrak{L}_{m,n} =  \int_\mathbb{R}\mathrm{d}t \big(\Gamma_{m,n}(t) -  \Gamma_{m,n}(\infty)\big).
\end{equation}
$\mathfrak{L}$ is a symmetric  matrix with $\mathfrak{L} \geq 0$ as covariance matrix. Hence for large times 
\begin{equation}\label{15.30} 
 \sum_{j \in \mathbb{Z}} j^2 S_{m,n} (j,t)  \simeq \Gamma_{m,n}(\infty) t^2 + \mathfrak{L}_{m,n} |t|.\\
 \end{equation}

The reader may worry that we have lost our way, lots of definitions, and not even the diffusion matrix of interest.
To appreciate our preparations, a useful example is the shear viscosity of a fluid. A standard scheme to experimentally define
the shear viscosity, $\nu$,  is the Couette  flow. One plate is fixed and another plate, distance $y$ and in parallel, is moved with constant  
velocity $u$. For small $u$ one finds the relation
\begin{equation}\label{15.31}
 \frac{F}{A} = \nu \frac{u}{y}.
\end{equation}
Here $F/A$ is the pushing force per area acting on the moving plate. The shear viscosity reappears on a more sophisticated level in the Navier-Stokes equations. For example, solving these equations for the Couette flow confirms \eqref{15.31}. But having the full equations, one can think of different setups to yield $\nu$. In our context one example would be the propagation of a small perturbation of an equilibrium state. Of course, experimental accessibility may vary. But, independently of the method, measured is always the same shear viscosity.

Returning to integrable systems, if such dissipative features are present at all, diffusion is expected to be governed by a high-dimensional matrix and a priori 
it is not so obvious whether and how familiar methods would apply. For the Euler time scale, we already identified   
the dynamic correlator as
\begin{equation}\label{15.32}
\hat{S}_{m,n}(k,t) \simeq \big(\mathrm{e}^{-\mathrm{i}kAt}C\big)_{m,n}
\end{equation}
in wave number space,  valid for small $k$ and large $t$, $kt$ fixed, compare with \eqref{7.21} and  \eqref{7.25}.  Commonly, diffusive corrections are added through
\begin{equation}\label{15.33}
\hat{S}_{m,n}(k,t) \simeq \big(\mathrm{e}^{-\mathrm{i}kAt - k^2 D |t|}C\big)_{m,n}.
\end{equation}
By definition $D$ is the infinite-dimensional \textit{diffusion matrix}. In general $D$ is not symmetric. However, to describe dissipation,
$D$ is  required to have nonnegative eigenvalues. In generic examples, $D$ has a few zero eigenvalues and hence there is no diffusive decay along some particular directions. But $[A,D] \neq 0$ and $A$ mixes the modes such that $\hat{S}(k,t)$ still decays to $0$ in the long time limit. 

The first moment sum rule is obviously satisfied. Computing the second moment in the position space version of \eqref{15.33} and comparing with the sum rule 
\eqref{15.30}, one obtains 
\begin{equation}\label{15.34}
A^2Ct^2 + DC |t| = \Gamma(\infty) t^2 + \mathfrak{L}|t|.
\end{equation}
Hence the Drude weight matrix is given by
\begin{equation}\label{15.35}
\Gamma(\infty) = A^2C = A CA^\mathrm{T} = B\frac{1}{C}B.
\end{equation}
The latter expression is a natural symmetric form, which emphasizes that the Drude weight is a property
determined by  static correlations only. In addition,  the diffusion matrix is obtained as
\begin{equation}\label{15.36}
DC = CD^{\mathrm{T}} = \mathfrak{L}, \quad D = \mathfrak{L}C^{-1},
\end{equation}
which is the generalization of the \textit{Onsager relation} to integrable systems. Since $\mathfrak{L} \geq 0$ and $C >0$, $D$ is a similarity transform of $\mathfrak{L}$ and has thus only nonnegative eigenvalues.  
\bigskip\\
$\blackdiamond\hspace{-1pt}\blackdiamond$~\textit{Drude weight}.\hspace{1pt} In the context of condensed matter physics, it is fairly common to equate a non-zero Drude weight with infinite conductivity, implicitly suggesting that no further properties have to be investigated.   In fact, according to our discussion, a non-zero Drude weight simply indicates a ballistic component of the dynamic correlator. For nonintegrable systems in one dimension,  the ballistic component consists of sharp $\delta$-peaks moving with constant velocity on the Euler scale. Diffusion is then easily detected through the broadening of the peaks.
 For integrable systems the ballistic component is extended and of the generic form $t^{-1}g(t^{-1}x)$ in position space with some smooth scaling function $g$. Now diffusive corrections become harder to detect,
 unless special care is taken to suppress the broad ballistic background. An example is the domain wall discussed in Chapter \ref{sec8}. At the contact line the profile jumps from $\rho_{\mathsf{n}-}$  to $\rho_{\mathsf{n}+}$. Diffusion will broaden the step to an error function. Another example is the 
 XXZ chain at zero magnetization (half filling). At this specific point the spin Drude weight vanishes and diffusive effects can be observed
 directly. 
 \hfill$\blackdiamond\hspace{-1pt}\blackdiamond$
\subsection{Nonintegrable chains}
\label{sec15.2}
Our preparations have been accomplished already in Section \ref{sec7.1} and, in essence, we are left with duplicating from the previous
 subsection, recalling that now there are only three fields, hence $n=0,1,2$.  It will turn out to be instructive to dwell on further details. 
 In Section \ref{sec7.1}  serif letters were used for $3\times 3$ matrices, so to distinguish from the integrable case. The same convention is followed here. Also in this section, to ease the comparison with the literature,  we switch to the physical pressure, denoted as before by $ \mathfrak{p}$, which means that $P$  in Eq. 
 \eqref{7.5} has to be substituted by $ \mathfrak{p}\beta $. It is convenient to set $u=0$. Then the Gibbs state is invariant under time-reversal, which provides additional symmetries.
 The average stretch is $\nu = \langle r_0\rangle_{ \mathfrak{p},\beta}$ and average energy $\mathsf{e} =  \langle e_0\rangle_{ \mathfrak{p},\beta} $, where we dropped the mean velocity as parameter.
By convexity, one can view $ \mathfrak{p}$ as a function of $\nu,\mathsf{e}$. The linearization matrix reads
\begin{equation}\label{15.37} 
\mathsfit{A} = 
\begin{pmatrix}
0 & -1&0 \\
\partial_\nu  \mathfrak{p}& 0 &\partial_\mathsf{e}  \mathfrak{p} \\
0&  \mathfrak{p}& 0\\
\end{pmatrix}. 
\end{equation}
The static correlator is given by
\begin{equation}\label{15.38} 
\mathsfit{C} = 
\begin{pmatrix}
 \langle r_0r_0\rangle_{ \mathfrak{p},\beta}^\mathrm{c} & 0&\langle r_0V_{\mathrm{ch,0}}\rangle_{ \mathfrak{p},\beta}^\mathrm{c} \\
0&\beta^{-1}&0\\
\langle r_0V_{\mathrm{ch,0}}\rangle_{ \mathfrak{p},\beta}^\mathrm{c}& 0& \langle e_0e_0\rangle_{ \mathfrak{p},\beta}^\mathrm{c}\\
\end{pmatrix} 
\end{equation}
and the current-charge cross correlation by  
\begin{equation}\label{15.39} 
\mathsfit{B} = \beta^{-1}
\begin{pmatrix}
0 & -1&0 \\
-1& 0& \mathfrak{p}  \\
0&  \mathfrak{p} & 0\\
\end{pmatrix}. 
\end{equation}
With this information one obtains the Drude weight
\begin{equation}\label{15.40} 
\Gamma({\infty}) = \beta^{-1}
\begin{pmatrix}
 1 & 0&- \mathfrak{p}  \\
0&c^2&0\\
- \mathfrak{p} & 0&  \mathfrak{p} ^2\\
\end{pmatrix}, 
\end{equation}
where $c$ is the isentropic speed of sound,
\begin{equation}\label{15.41}
c^2 = \frac{\beta \langle (e_0 +  \mathfrak{p} r_0)(e_0+ \mathfrak{p} r_0)\rangle_{ \mathfrak{p} ,\beta}^\mathrm{c} }
{\langle r_0r_0\rangle_{ \mathfrak{p} ,\beta}^\mathrm{c}\langle e_0e_0 \rangle_{ \mathfrak{p} ,\beta}^\mathrm{c} -
(\langle r_0 e_0\rangle_{ \mathfrak{p} ,\beta}^\mathrm{c})^2}\,.
\end{equation}

Finally we have to list the two dynamic characteristics
introduced before. The Onsager matrix turns out as
\begin{equation}\label{15.42} 
\mathsfit{L} = 
\begin{pmatrix}
 0 & 0&0 \\
0&\sigma_\mathsf{p}^2&0\\
0& 0& \sigma_\mathsf{e}^2\\
\end{pmatrix}. 
\end{equation}
The only nonvanishing matrix elements are momentum-momentum and energy-energy current. The notation is supposed to indicate that they are related to the noise strength of a Ginzburg-Landau fluctuation theory.
But the defining time integral \eqref{15.29} cannot be expected to yield an explicit expression. The $0$'s at the upper left borders of the matrix result from
the stretch current being conserved. The $1,2$ matrix element vanishes, since  momentum and energy have opposite signs under time reversal. The resulting diffusion matrix is given by
\begin{equation}\label{15.43} 
\mathsfit{D} = 
\begin{pmatrix}
 0 & 0&0 \\
0&\sigma_\mathsf{p}^2&0\\
 \alpha\sigma_\mathsf{e}^2& 0& \sigma_\mathsf{e}^2\\
\end{pmatrix}, 
\end{equation}
where $\alpha = -\langle r_0 V_{\mathrm{ch,0}}\rangle_{ \mathfrak{p} ,\beta}^\mathrm{c}\big/\langle r_0r_0 \rangle_{ \mathfrak{p} ,\beta}^\mathrm{c}$. Note that $\mathsfit{L}$, and hence $\mathsfit{D}$, has a zero eigenvalue.  Since $[\mathsfit{A},\mathsfit{D}] \neq 0$,  still  $\hat{\mathsfit{S}} (k,t) = \exp[-\mathrm{i}k\mathsfit{A}t - k^2\mathsfit{D}|t|]\mathsfit{C} \to 0$ as $t \to \infty$,
except for $k=0$. If one reinstalls a nonzero mean velocity $u$, the Gibbs state is no longer invariant under time reversal. By a Galilean transformation,
one can still figure out the various matrices. As a particular consequence,  while $\mathsfit{L}_{0,n} = 0$ for $n=0,1,2$ because of  stretch current conservation, the cross term  no longer vanishes, 
$\mathsfit{L}_{1,2} \neq 0$. Also the diffusion matrix picks up further non-zero entries. 

In position space, on the Euler scale $\mathsfit{S}(x,t) $  consists of three $\delta$-peaks, the heat peak at rest and the two sound peaks traveling with velocity $\pm c$. The diffusion term broadens the peaks as $\sqrt{t}$. While this prediction looks innocent,  it is completely off the track as noted
already in the mid-1970ies. We quietly assumed that $\Gamma_{1,1}(t)$ and  
$\Gamma_{2,2}(t)$ are integrable. A more refined theory arrives at the conclusion that both current correlations 
decay generically as $|t|^{-2/3}$, which is well confirmed by molecular dynamics simulations. Also the shape functions differ from a Gaussian. As discussed at the end of Section \ref{sec7.1}, see Figure 7,
the heat peak broadens as  $t^{3/5}$ with a shape function whose Fourier transform is given by $\exp(-|k|^{5/3}|t|)$.
In the theory of independent random variables this distribution is known as stable symmetric L\'{e}vy law with exponent $\tfrac{5}{3}$.  The
two sound peaks  broaden distinctly with the power law $t^{2/3}$. To compute their shape function is a more 
difficult enterprise and related to the Kardar-Parisi-Zhang nonlinear fluctuation theory. One considers the stochastic Burgers equation
\begin{equation}\label{15.44} 
\partial_t u(x,t) = \partial_x\big( u(x,t)^2 + \partial_x u(x,t) + \xi(x,t)\big),
\end{equation}
where $ \xi(x,t)$ is spacetime white noise. Then its stationary measures are Gaussian white noise of variance one and arbitrary mean.
The scaling function for the sound peaks turn out to be identical to  the stationary two-point function $\mathbb{E}\big(u(x,t)u(0,0)\big)^\mathrm{c}$. Actually the full picture is more complicated.
Anharmonic chains  are divided into three universality classes. For one class, both $\Gamma_{1,1}(t)$ and $\Gamma_{2,2}(t)$ have an integrable decay. In this case the claims from the linear theory are fully confirmed.
\subsection{Navier-Stokes equations}    
\label{sec15.3}
For the Toda lattice there is currently no method to compute the Onsager matrix directly on the basis of the microscopic model. 
However, using form factor expansions,  a concise formula for the Onsager matrix has become available for the $\delta$-Bose gas. By itself this is a surprising result, since to explicitly compute transport coefficients for many-particle
 systems  is a rare exception.  
By the much emphasized analogy between integrable systems, one thereby arrives also at a firm prediction for the Toda fluid. We state here merely the result and discuss some of its consequences. 
The matrix $\mathfrak{L}_{m,n}$ uses the basis consisting of monomials  $\varsigma_n$. Structurally more transparent are matrix elements computed for general functions over $\mathbb{R}$. We first introduce the kernel

\begin{equation}\label{15.45} 
K(w_1,w_2) = \rho_\mathsf{p}(w_1)\rho_\mathsf{p}(w_2) |v^\mathrm{eff}(w_1) - v^\mathrm{eff}(w_2) | |T^\mathrm{dr}(w_1,w_2) |^2,
\end{equation}
the integral operator $T^\mathrm{dr}$ being defined through 
\begin{equation}\label{15.46} 
T^\mathrm{dr} = (1 - T\rho_\mathsf{n} )^{-1}T,
\end{equation}
and its action on the constant function,
\begin{equation}\label{15.47} 
\kappa(w_1) = \int_\mathbb{R}\mathrm{d}w_2 K(w_1,w_2).
 \end{equation}
Clearly, $T^\mathrm{dr}$ is a symmetric operator and thus also $K$. The claim is that,  for general functions $f,g$ on $\mathbb{R}$, 
\begin{equation}\label{15.48} 
\langle f, \mathfrak{L} g\rangle = \frac{1}{2} \int_{\mathbb{R}^2} \mathrm{d} w_1  \mathrm{d} w_2 K(w_1,w_2)\Big(\frac{f^\mathrm{dr}(w_2)}{\rho_\mathsf{s}(w_2)} - \frac{f^\mathrm{dr}(w_1)}{\rho_\mathsf{s}(w_1)}\Big)
\Big(\frac{g^\mathrm{dr}(w_2)}{\rho_\mathsf{s}(w_2)} - \frac{g^\mathrm{dr}(w_1)}{\rho_\mathsf{s}(w_1)}\Big).
\end{equation} 
As it should be, $\mathfrak{L}$ is symmetric and $\mathfrak{L} \geq 0$. To determine possible zero eigenvalues, we note that $K(w_1,w_2) >0$. Hence 
$\mathfrak{L}f = 0$ implies
\begin{equation}\label{15.49} 
f^\mathrm{dr}(w)  = c \rho_\mathsf{s}(w)
\end{equation}
with arbitrary $c \in \mathbb{R}$, which has $f = c  \varsigma_0$ as only solution. The zero subspace of $\mathfrak{L}$ corresponds to the 
projector $|\varsigma_0\rangle \langle \varsigma_0|$. 

The diffusion matrix is obtained from the Onsager matrix as 
\begin{equation}\label{15.50} 
D = \mathfrak{L}C^{-1}. 
\end{equation}
The $C$ matrix for the Toda lattice is stated in \eqref{7.26}, respectively \eqref{7.29}. The corresponding matrices for the Toda fluid are obtained by setting $\nu =1$ and dropping 
 the comoving terms $q_n \varsigma_0$.
 The result reads 
\begin{equation}\label{15.51} 
C = (1 -   \rho_\mathsf{n} T)^{-1} \rho_\mathsf{p} (1 - T \rho_\mathsf{n} )^{-1}. 
\end{equation}
Hence 
\begin{equation}\label{15.52} 
C^{-1} = (1 - T \rho_\mathsf{n} ) \rho_\mathsf{p}^{-1} (1 -   \rho_\mathsf{n} T)
\end{equation}
and 
\begin{equation}\label{15.53} 
(C^{-1}g)^\mathrm{dr} =  \rho_\mathsf{p}^{-1}(1 -  \rho_\mathsf{n} T) g.
\end{equation}
Note that the zero eigenvalue of $D$ has the left eigenvector $\varsigma_0$ and the right eigenvector  $
(1 - \rho_\mathsf{n} T)^{-1}\rho_\mathsf{p} \rho_\mathsf{s}$.

For applications it will be more convenient to have  also the integral kernel of $D$ available. To separate into a diagonal multiplication operator 
and smooth off-diagonal integral kernel, we set
\begin{equation}\label{15.54} 
Uf(w) = \rho_\mathsf{s}(w)^{-1}( (1 - T\rho_\mathsf{n})^{-1}f)(w)
\end{equation}
and 
\begin{equation}\label{15.55} 
UC^{-1}g(w)  = ( \rho_\mathsf{s}(w) \rho_\mathsf{p}(w))^{-1}((1- \rho_\mathsf{n} T)g)(w).
\end{equation}
Using the symmetry of $K$, one arrives at 
\begin{equation}\label{15.56} 
 \langle f, D g\rangle =   \int_{\mathbb{R}^2} \mathrm{d}w_1 \mathrm{d} w_2 K(w_1,w_2) \big(Uf(w_1) UC^{-1}g(w_1)
 - Uf(w_1) UC^{-1}g(w_2)\big).
\end{equation}
Setting $\bar{K}  =   U^\mathrm{T}KUC^{-1} $, it follows that
\begin{equation}\label{15.57} 
D(w_1,w_2) = -\bar{K}(w_1,w_2) + \delta(w_1 - w_2)U^\mathrm{T}\kappa UC^{-1} (w_1,w_1),
\end{equation}
where $\kappa$ is regarded as multiplication operator.
Since the left eigenvector of $D$ is $\varsigma_0$,
if $\bar{K}(w_1,w_2) \geq 0$ and also  $U^\mathrm{T}\kappa UC^{-1} (w_1,w_1) \geq 0$, then the operator 
$-D$ has the structure of the generator of a time-continuous Markov jump process.

A control check of \eqref{15.48} is easily performed by working out the case of hard rods. Firstly,
there is an exact expression for the structure function, here denoted by $\hat{S}_{m,n}^\mathrm{hr}(k,t)$, and its asymptotic behavior 
\eqref{15.33} has been confirmed. Using a sum rule as in \eqref{15.28}, the total current correlator $\Gamma_{m,n}(t)$ can be obtained, with the result of being proportional to $\delta(t)$. Thereby the time integral trivializes and the Onsager matrix for hard rods is obtained as
\begin{equation}\label{15.58}
\mathfrak{L}_\mathrm{hr}(w_1,w_2) = (\mathsfit{a}\bar{\rho})^2 (1-\mathsfit{a}\bar{\rho})^{-1}\big(\delta(w_1-w_2) r(w_1) - |w_1- w_2|h(w_1)h(w_2)\big),
\end{equation}
where $r(w) = \int_\mathbb{R}\mathrm{d}w' h(w')|w - w'| $. Hence the diffusion matrix becomes
\begin{equation}\label{15.59}
D_\mathrm{hr}(w_1,w_2) = \mathsfit{a}(\mathsfit{a}\bar{\rho}) (1-\mathsfit{a}\bar{\rho})^{-1}\big(\delta(w_1-w_2) r(w_1) - h(w_1)|w_1-w_2|\big).
\end{equation}
Secondly, in  \eqref{15.48} and \eqref{15.56} one inserts the various  explicit kernels and functions  for hard rods, compare with
Chapter \ref{sec5}. The resulting expressions are then in agreement with \eqref{15.58} and \eqref{15.59}.

Based on our input, in analogy to classical fluids, the Navier-Stokes type equation of the Toda fluid is given by
\begin{equation}\label{15.60} 
\partial_t\rho_\mathsf{p}(x,t;v) + \partial_x\big(v^\mathrm{eff}(x,t;v)\rho_\mathsf{p}(x,t;v)\big) = \partial_xD\partial_x \rho_\mathsf{p}(x,t;v)
\end{equation}
with diffusion matrix $D$ of \eqref{15.57}. The transformation to quasilinear form can still be carried out and yields
\begin{equation}\label{15.61} 
\partial_t\rho_\mathsf{n}(x,t) + v^\mathrm{eff}(x,t)\partial_x\rho_\mathsf{n}(x,t) =
\rho_\mathsf{s}(x,t)^{-1}(1- \rho_\mathsf{n} T)\partial_x D\partial_x \rho_\mathsf{p}(x,t).
\end{equation}

As a generic physical requirement, Eq. \eqref{15.61} should yield a positive entropy production. More precisely,
the balance equation for the local entropy will have a flow term 
and a production term. By the second law of thermodynamics the latter should be positive. 
The local entropy at $(x,t)$ is defined by
\begin{equation}\label{15.62} 
s(x,t) = -\langle \rho_\mathsf{p}(x,t)\log \rho_\mathsf{n}(x,t) \rangle,
\end{equation}
where $\langle\cdot\rangle$ refers to the integral over the spectral parameter, compare with Eq. \eqref{9.56}.  For a while we fix the spectral parameter $w$ and differentiate as 
\begin{eqnarray}\label{15.63} 
&&\hspace{-20pt} -\partial_t (  \rho_\mathsf{s}\rho_\mathsf{n}\log \rho_\mathsf{n}) = 
-  (\log \rho_\mathsf{n}) \partial_t \rho_\mathsf{p} -  \rho_\mathsf{s} \partial_t \rho_\mathsf{n} \nonumber\\[1ex]
&&\hspace{0pt} =  ( \log \rho_\mathsf{n}) \partial_x (v^\mathrm{eff}\rho_\mathsf{p})
+  \rho_\mathsf{s} v^\mathrm{eff}\partial_x\rho_\mathsf{n}  -  \big(\log\rho_\mathsf{n} 
+(1- \rho_\mathsf{n} T)\big) \partial_x(D \partial_x \rho_\mathsf{p}),
\end{eqnarray}
where \eqref{15.60} and \eqref{15.61} have been inserted. First order derivative terms  contribute to the flow term. The first and second summand on the right of \eqref{15.63} combine to
\begin{eqnarray}\label{15.64} 
&&\hspace{-40pt} (\log \rho_\mathsf{n}) \partial_x (v^\mathrm{eff}\rho_\mathsf{p}) 
+  \rho_\mathsf{s} v^\mathrm{eff}\partial_x\rho_\mathsf{n}  \nonumber\\[1ex]
 &&\hspace{-20pt}
= \partial_x \big((\log \rho_\mathsf{n})  v^\mathrm{eff}\rho_\mathsf{p} \big)
-  (\partial_x \log \rho_\mathsf{n})  v^\mathrm{eff}\rho_\mathsf{p} + \  \rho_\mathsf{s} v^\mathrm{eff}\partial_x\rho_\mathsf{n} = \partial_x \big( (\log \rho_\mathsf{n})  v^\mathrm{eff}\rho_\mathsf{p}\big).
\end{eqnarray}
For the third summand we move the leftmost $\partial_x$ in front, as before. This yields a further contribution to the flow term as  
\begin{equation}\label{15.65} 
-\partial_x \big( (\log \rho_\mathsf{n} +(1 - T\rho_\mathsf{n})) D \partial_x \rho_\mathsf{p}\big)
\end{equation}
together with  the dissipative term
\begin{equation}\label{15.66} 
 \big(\rho_\mathsf{n}^{-1}\partial_x\rho_\mathsf{n} - T\partial_x\rho_\mathsf{n}\big) \big(D \partial_x \rho_\mathsf{p}\big).
\end{equation}
Differentiating $\rho_\mathsf{n} = \rho_\mathsf{p}(1 + T \rho_\mathsf{p})^{-1}$, one arrives at the identity
\begin{equation}\label{15.67} 
\rho_\mathsf{n}^{-1}\partial_x \rho_\mathsf{n} = \rho_\mathsf{p}^{-1}(1 - \rho_\mathsf{n} T)\partial_x\rho_\mathsf{p}
\end{equation}
 and notes that 
 \begin{equation}\label{15.68} 
 (1- T \rho_\mathsf{n})\rho_\mathsf{n}^{-1}\partial_x\rho_\mathsf{n}
 = (1 - T\rho_\mathsf{n})\rho_\mathsf{p}^{-1}(1 - \rho_\mathsf{n} T)\partial_x\rho_\mathsf{p} = C^{-1} \partial_x  \rho_\mathsf{p}. 
 \end{equation}
We now substitute $D = \mathfrak{L}C^{-1}$ and arrive at the dissipative term
\begin{equation}\label{15.69} 
C^{-1} (\partial_x \rho_\mathsf{p})\mathfrak{L}C^{-1}(\partial_x \rho_\mathsf{p}).
\end{equation}

Altogether the entropy balance for the Toda fluid reads
\begin{equation}
\label{15.70} 
\partial_t s(x,t) +\partial_x \mathfrak{j}_\mathrm{s}(x,t) = \sigma(x,t).
\end{equation}
The entropy current is determined to 
\begin{equation}\label{15.71} 
 \mathfrak{j}_\mathrm{s} = - \langle (\log \rho_\mathsf{n} )v^\mathrm{eff}\rho_\mathsf{p}\rangle +\langle \big(\log \rho_\mathsf{n} + (1 - T\rho_\mathsf{n})\big)D \partial_x \rho_\mathsf{p} \rangle.
\end{equation}
For the entropy production the very concise formula
\begin{equation}\label{15.72} 
\sigma = \langle (\partial_x \rho_\mathsf{p}),C^{-1}\mathfrak{L}C^{-1}(\partial_x \rho_\mathsf{p})\rangle.
\end{equation}
is unveiled.

\newpage
\begin{center}
 \textbf{Notes and references}
 \end{center}
\begin{center}
\textbf{Section 15.1}
\end{center} 
For general anharmonic chains the sum rules are discussed in Mendl and  Spohn \cite{MS16}. While time reversal is 
 a standard item, the use of spacetime reversal  I learned from De Nardis et al. \cite{DBD18,DBD19}. For hard rods an exact formula 
 for the structure factor $\hat{S}_{m,n}(k,t)$ 
 is derived by Lebowitz et al. \cite{LPS68}. With this input the small $k$ behavior is studied by  Spohn  \cite{S82}, in particular respective current correlations and the bulk diffusion matrix. The Navier-Stokes correction
 for hard rods is proved by Boldrighini  and  Suhov \cite{BS97}, with a result in complete agreement with \eqref{15.59}.
 
 Pioneering work on the Drude weight is Castella et al.  \cite{CZ95} and  Zotos \cite{Z99}. For fluids the necessity to subtract the Drude weight has been recognized in Green \cite{G54},
 see Spohn  \cite{S91} for a textbook discussion. In condensed matter physics mostly one refers to the Mazur \cite{M69}, in particularly to the Mazur bound
 which in our context means to sum in \eqref{15.35} only over a restricted number of conserved fields. The general formula is stated in 
 Doyon and  Spohn  \cite{DS17}.
 Applying  \eqref{15.35} naively to the XXZ chain, one would conclude  that the spin Drude weight vanishes. In actual fact, the Drude weight vanishes only for $\Delta >1$ in the standard units, while it is non-zero and nowhere continuous for $0 \leq \Delta \leq 1$. The resolution 
 can be traced back on our insistence on strictly local conservation laws. In actual fact, the XXZ chain has additional conservation laws whose densities have exponential tails and thus contribute to hydrodynamics, in particular to the Drude weight. We refer to Mierzejewski  et al. \cite{MPP15}, Ilievski and De Nardis  \cite{ID17a} and the recent review Ilievski  \cite{I22} for more details.   
 \bigskip
 \begin{center}
 \textbf{Section 15.2}
\end{center}
Nonintegrable anharmonic chains are studied at length by  Spohn \cite{S14}, where also the connection to the three-component stochastic Burgers equation, alias KPZ equation, is explained. The dynamical correlator is confirmed through molecular dynamics with hard shoulder and other interaction potentials, see Mendl and Spohn \cite{MS14}.  Also the nonintegrable decay of the total 
current-current correlation is convincingly observed in Mendl and  Spohn \cite{MS16}. The hydrodynamic limit of anharmonic chains is studied by
Bernardin and Olla \cite{BO20}. Transport fluctuations are investigated in Myers et al. \cite{MBHD20}. In Ganapa et al. \cite{GCD20} molecular dynamics simulations are carried out, in which initially a huge amount of energy is deposited at the origin. The results compare well with hydrodynamic predictions.
\bigskip
\begin{center}
 \textbf{Section 15.3}
\end{center}
The section is entirely based on De Nardis et al. \cite{DBD18,DBD19}, where the Navier-Stokes corrections are obtained through form factor expansions. The simple looking formula for the entropy production seems to be new.
\newpage
\addcontentsline{toc}{section}{List of Symbols}
\noindent
\textbf{\large{List of Symbols}}\\\\
conserved field\hspace{8pt}$ Q^{[n]}$,\hspace{8pt} field density\hspace{8pt}$Q_j^{[n]}$\\ 
current $J^{[n]}$,\hspace{8pt}current  density $J_j^{[n]}$ \\
commutator \hspace{8pt}$[\cdot,\cdot]$ \\
correlator \vspace*{-6pt}
\begin{tabbing}
\hspace*{10pt}$S,\mathsf{S}$\=\hspace*{20pt}spacetime charge-charge\\
\hspace*{10pt}$C,\mathsf{C}$ \>\hspace*{20pt}static charge-charge \\
\hspace*{10pt}$B,\mathsf{B}$ \>\hspace*{20pt}static charge-charge current\\
\hspace*{10pt}$A,\mathsf{A}$ \>\hspace*{20pt}flux Jacobian
\end{tabbing}
density $\rho$\vspace*{-6pt}
\begin{tabbing}
\hspace*{10pt}$\varrho$\=\hspace*{20pt}generic $\langle \varrho\rangle = 1$\\
\hspace*{10pt}$\rho$ \>\hspace*{20pt}generic $\langle \rho\rangle \neq 1$\\
\hspace*{10pt}$\rho_\mathsf{n}$ \>\hspace*{20pt}number density, TBA solution\\
\hspace*{10pt}$\rho_\mathsf{p}$ \>\hspace*{20pt}particle density, root density, TBA solution\\
\hspace*{10pt}$\rho_\mathsf{s}$ \>\hspace*{20pt}space density, TBA solution \\
\hspace*{10pt}$\rho_\mathfrak{s}$ \>\hspace*{20pt}stationary solution, Dyson Brownian motion\\
\hspace*{10pt}$\rho_\mathrm{Q}$ \>\hspace*{20pt}DOS Lax matrix\\
\hspace*{10pt}$\rho_\mathrm{J}$ \>\hspace*{20pt}current DOS
\end{tabbing}
effective velocity\hspace*{8pt}$v^\mathrm{eff}$\\\\
free energy functionals $\mathcal{F}$\vspace*{-6pt}
\begin{tabbing}
\hspace*{10pt}$\mathcal{F}^\circ(\varrho)$\=\hspace*{20pt}$\langle \varrho\rangle = 1$\\
\hspace*{10pt}$\mathcal{F}(\rho)$ \>\hspace*{20pt}other constaints on $\langle \rho\rangle$\\
\hspace*{10pt}$\mathcal{F}^\bullet(\rho)$ \>\hspace*{20pt}no constraint
\end{tabbing}
GGE average\hspace*{8pt}$\langle \cdot\rangle_{P,V}$,\hspace*{6pt}$\langle \cdot\rangle_{\rho_\mathsf{f},V}$ \\\\
hamiltonian \hspace*{4pt}$H$\vspace*{-6pt}
\begin{tabbing}
\hspace*{10pt}$H_\mathrm{aa}$\=\hspace*{20pt}action-angle\\
\hspace*{10pt}$H_\mathrm{al}$ \>\hspace*{20pt}Ablowitz-Ladik lattice\\
\hspace*{10pt}$H_\mathrm{ca}$ \>\hspace*{20pt}Calogero fluid\\
\hspace*{10pt}$H_\mathrm{cell}$ \>\hspace*{20pt}Toda cell\\
\hspace*{10pt}$H_\mathrm{ch}$ \>\hspace*{20pt}generic chain\\
\hspace*{10pt}$H_\mathrm{cm}$ \>\hspace*{20pt}Calogero-Moser model\\
\hspace*{10pt}$H_\mathrm{hr}$ \>\hspace*{20pt}hard rod lattice\\
\hspace*{10pt}$H_{\mathrm{hr},\mathsf{fl}}$ \>\hspace*{20pt}hard rod fluid\\
\hspace*{10pt}$H_\mathrm{li}$ \>\hspace*{20pt}Lieb-Liniger $\delta$-Bose gas\\
\hspace*{10pt}$H_\mathrm{mec}$ \>\hspace*{20pt}general pair interaction\\
\hspace*{10pt}$H_\mathrm{qt}$ \>\hspace*{20pt}quantum Toda fluid\\
\hspace*{10pt}$H_\mathrm{to}$ \>\hspace*{20pt}Toda lattice\\
\hspace*{10pt}$H_\mathrm{to}^\diamond$ \>\hspace*{20pt}open Toda lattice\\
\hspace*{10pt}$H_\mathrm{to,\mathsf{f}}$ \>\hspace*{20pt}Toda fluid\\
\hspace*{10pt}$H_\mathrm{XYZ}$ \>\hspace*{20pt}XYZ spin chain
\end{tabbing}
Hilbert spaces\hspace*{8pt}$L^2(\mathbb{R},\mathrm{d}x)$,\hspace*{8pt}$\ell_2(\mathbb{Z})$\\\\
integral over $\mathbb{R}$, resp. $[0,2\pi]$,\hspace{8pt} $\langle \cdot\rangle$\\\\
Lax matrix\hspace*{8pt}$L_N, L$\\\\
Lax pair\hspace*{8pt}$(L_N, B_N),\hspace*{4pt}(L,B)\,\hspace*{4pt}(L,M)$\\\\
partition function \hspace*{4pt} $Z$\vspace*{-6pt}
\begin{tabbing}
\hspace*{10pt}$Z_\mathrm{de}$ \=\hspace*{20pt}Dumitriu-Edelman\\
\hspace*{10pt}$Z_\mathrm{to}$ \>\hspace*{20pt}Toda\\
\hspace*{10pt}$Z_\mathrm{li}$ \>\hspace*{20pt}Lieb-Liniger\\
\hspace*{10pt}$Z_\mathrm{can}$ \>\hspace*{20pt}canonical\\
\hspace*{10pt}$Z_N$ \>\hspace*{20pt}system size $N$
\end{tabbing}
phase space  \hspace*{4pt} $\Gamma$\vspace*{-6pt}
\begin{tabbing}
\hspace*{10pt}$\Gamma_N = \mathbb{R}^N\times\mathbb{R}^N$\\
\hspace*{10pt}$\Gamma_N^\circ = \mathbb{R}_+^N\times\mathbb{R}^N$\\
\hspace*{10pt}$\Gamma_N^\diamond = \mathbb{R}_+^{N-1}\times\mathbb{R}^N$\\
\hspace*{10pt}$\Gamma_N^\triangleright = \mathbb{W}_N\times\mathbb{R}^N$
\hspace*{10pt}$\mathbb{T}^N$ = $N$-dimensionaler torus 
\end{tabbing}
Poisson bracket\hspace*{8pt}$\{\cdot,\cdot\}$\\\\
potential\vspace*{-6pt}
\begin{tabbing}
\hspace*{10pt}$V$ \=\hspace*{20pt}confining\\
\hspace*{10pt}$V_\mathrm{ca}$ \>\hspace*{20pt}Calogero\\
\hspace*{10pt}$V_\mathrm{ch}$\>\hspace*{20pt}chain\\
\hspace*{10pt}$V_\mathrm{hr}$ \>\hspace*{20pt}hard rod\\
\hspace*{10pt}$V_\mathrm{mec}$ \>\hspace*{20pt}pair interaction
\end{tabbing}
quasi-energy\hspace*{8pt} $\varepsilon$\\\\
scattering shift \hspace*{4pt}$\phi$\vspace*{-6pt}
\begin{tabbing}
\hspace*{10pt}$\phi_\mathrm{ca}$\=\hspace*{20pt}Calogero fluid\\
\hspace*{10pt}$\phi_\mathrm{hr}$ \>\hspace*{20pt}hard rods\\
\hspace*{10pt}$\phi_\mathrm{kv}$ \>\hspace*{20pt}KdV solitons\\
\hspace*{10pt}$\phi_\mathrm{qt}$ \>\hspace*{20pt}quantum Toda\\
\hspace*{10pt}$\phi_\mathrm{to}$ \>\hspace*{20pt}Toda particles\\
\hspace*{10pt}$\phi_\mathrm{tos}$ \>\hspace*{20pt}Toda solitons
\end{tabbing}

\end{document}